# Теорія та методика навчання математики, фізики, інформатики

М. В. Попель

# Хмарний сервіс CoCalc як засіб формування професійних компетентностей учителя математики

Спецвипуск містить монографію М. В. Попель, у якій розглянуто методичні засади формування професійних компетентностей учителів математики у закладах вищої освіти України; уточнено місце хмарного сервісу CoCalc у системі засобів навчання математичних дисциплін; виявлено особливості використання CoCalc у навчанні математичних дисциплін та розроблено модель використання хмарного сервісу CoCalc як засобу формування професійних компетентностей учителя математики; спроектовано методику використання CoCalc як засобу формування професійних компетентностей учителя математики.

Для науковців, аспірантів, викладачів математичних дисциплін та студентів педагогічних навчальних закладів, всіх, хто цікавиться застосуванням хмаро орієнтованих систем в освіті.

**Адреса редакції**: а/с 4809, м. Кривий Ріг, 50086, Україна

## ЗМІСТ

## ПЕРЕЛІК УМОВНИХ ПОЗНАЧЕНЬ

| | |
|---|---|
| ЗВО | заклад вищої освіти (заклади вищої освіти) |
| ДС | досліджувані системи |
| ЕОР | електронні освітні ресурси |
| ІКТ | інформаційно-комунікаційні технології |
| ІТН | Інформаційні технології у навчанні |
| ППЗ | педагогічний програмний засіб |
| ТЗН | технічні засоби навчання |
| ТХО | технології хмарних обчислень |

# ВСТУП

Нині до професійної підготовки вчителя, здатного активно самореалізовуватися в інформаційному суспільстві, мати компетентності, що відповідали б потребам сьогодення, висуваються досить високі вимоги. Загальновизнано, що лише компетентний вчитель математики здатен підготувати вступника до ЗВО на інженерно-технічні, природничо-математичні та соціально-економічні спеціальності, оскільки їх випускники створюють основу для матеріального добробуту та соціального розвитку суспільства. Водночас, результати моніторингових досліджень з математики (TIMSS, PISA, PIRLS) свідчать про необхідність підвищення рівня математичної обізнаності учнів як через підвищення професійної компетентності учителів, так і через використання інноваційних інформаційно-комунікаційних та педагогічних технологій.

Удосконалення змісту і складників курсів математичних дисциплін та методики їх навчання постають одними з ключових питань підвищення якості підготовки фахівців, особливо у педагогічному ЗВО: при вивченні багатьох дисциплін (диференціальної геометрії та топології, диференціальних рівнянь, теорії ймовірностей та інших) засвоєння абстрактних математичних понять викликає у студентів значні труднощі. Одним із шляхів їх вирішення є застосування наочних інтерпретацій математичних понять і тверджень. Значні дидактичні можливості у реалізації принципу наочності виникають завдяки використанню у процесі навчання ІКТ.

Перехід у навчанні майбутніх учителів математики від використання традиційних комп'ютерно орієнтованих засобів і сервісів до хмаро орієнтованих створює умови:

– для педагогічного ЗВО – це вивільнення обчислювальних потужностей, матеріальних та виробничих ресурсів шляхом переходу до ІКТ-аутсорсингу із підвищенням якості обслуговування;

– для викладачів – це забезпечення більш гнучкого і широкого доступу до якісних електронних освітніх ресурсів, формування хмаро орієнтованого середовища безперервного навчання;

– для ІКТ-підрозділів педагогічного ЗВО – це уніфікація ІКТ-інфраструктури;

– для слухачів і педагогів курсів підвищення кваліфікації – це створення професійної математичної соціальної спільноти з можливістю взаємодії з використанням хмарних сервісів у реальному часі.

У зв'язку з цим, визначення перспектив використання хмарних сервісів у навчанні математичних дисциплін, їх ролі і місця в організації навчального процесу, методичних засад їх застосування є актуальною проблемою теорії та методики використання ІКТ в освіті. Суттєвим для її розв'язання є науково-методичне обґрунтування використання провідних хмарних сервісів математичного призначення, зокрема CoCalc.

Суттєвий внесок стосовно освітньо-наукових можливостей використання ІКТ, зокрема хмарних належить таким вченим, як: Г. О. Алексанян [3], О. М. Алєксєєв [4], М. Армбруст (M. Armbrust) [323], В. Ю. Биков [27], Р. Гріффіт (R. Griffith) [323], А. М. Гуржій [60], М. І. Жалдак [70], М. Ю. Кадемія [82], Т. І. Коваль [95], О. Г. Колгатін [102], К. Р. Колос [106], О. Г. Кузьмінська [155], В. М. Кухаренко [5], С. Г. Литвинова [128], М. Міллер (M. Miller) [356], В. С. Мкртчян [143], Н. В. Морзе [156], В. В. Осадчий [176; 177], Л. Ф. Панченко [182], З. С. Сейдаметова [226], С. О. Семеріков [235], О. М. Спірін [266], А. М. Стрюк [273], К. Субраманьян (K. Subramanian) [375], Н. Султан (N. Sultan) [358], П. Томас (P. Thomas) [377], Ю. В. Триус [379; 281; 286], А. Фокс (A. Fox) [323], Ю. Хмелевський (Y. Khmelevsky) [350], В. Чанг (W. Chang) [330], М. П. Шишкіна [26] та ін.

Питання підготовки майбутніх учителів математики у вітчизняних вищих педагогічних навчальних закладах розглядали у своїх працях провідні науковці: І. А. Акуленко [2], В. Г. Бевз [14], М. І. Жалдак [70; 72], І. В. Лов'янова [129], Г. О. Михалін [72], Н. В. Морзе [61; 155], Т. О. Олійник [206], М. В. Працьовитий [114], С. А. Раков [18], Ю. С. Рамський [210], О. І. Скафа [246], З. І. Слєпкань [250], О. В. Співаковський [264], Ю. В. Триус [282; 283], В. О. Швець [297] та ін.

Окремою групою постають дослідження Т. Л. Архіпової [7], Н. В. Бахмат [12; 13], В. Ю. Бикова [23; 24], Д. Бланк (D. Blank) [359], Т. В. Зайцевої [6], У. П. Когут [97; 98; 99], Ю. Г. Лотюк [131], Дж. Маршалл (J. Marshall) [359], Н. В. Морзе [155; 156], В. П. Олексюка [211], К. Дж. О'Хара (K. J. O'Hara) [359], К. І. Словак [254], С. В. Шокалюк [312] та ін., присвячені застосуванню хмарних сервісів у процесі навчання майбутніх учителів математики, в яких виявляються перспективні напрями використання хмарних сервісів у навчальному процесі ЗВО, в управлінні навчанням, управлінні освітньою установою, у підтриманні наукових досліджень.

Наразі триває процес розроблення стандартів для вищої освіти за різними спеціальностями, що мають містити, зокрема, систему професійних компетентностей випускника. На даний момент не існує усталеного переліку професійних компетентностей, якими повинен володіти майбутній вчитель математики. У зв'язку з цим проблема використання загальнодоступного хмарного сервісу CoCalc, що є досить потужним і разом з тим вільно поширюваним, виявлення перспективних шляхів його застосування у підготовці майбутніх учителів математики потребують ґрунтовного дослідження. За наявності практичних розробок М. А. Кислової [77], О. М. Маркової [140], С. О. Семерікова [233; 235], К. І. Словак [254], С. В. Шокалюк [235; 312] та ін., що стосуються використання хмарних сервісів у навчанні математичних дисциплін, питання теоретичного обґрунтування процесу застосування хмарного сервісу CoCalc залишається у наш час недостатньо розкритим. У підході до вивчення цієї проблеми спостерігаються такі суперечності:

– між рівнем абстракції математичних об’єктів і можливостями забезпечення їх візуалізації шляхом комп’ютерної інтерпретації;

– між доцільністю широкого використання ІКТ сервісів математичного призначення у підготовці майбутніх учителів математики та недостатніми можливостями їх забезпечення ІКТ-підрозділами педагогічних ЗВО;

– між доцільністю використання ІКТ-аутсорсингу хмарної інфраструктури навчання майбутніх учителів математики в Україні та неадаптованістю зарубіжних хмарних математичних сервісів до вимог вітчизняних освітніх стандартів;

– між можливостями застосування хмарних математичних сервісів у процесі формування професійних компетентностей учителя математики та нерозробленістю відповідної методики їх впровадження.

Монографія складається із чотирьох розділів.

У першому розділі розглянуто професійну підготовку вчителів математики у ЗВО України, розкрито понятійний апарат дослідження, проаналізовано вітчизняний і зарубіжний досвід використання хмарних сервісів для формування професійних компетентностей учителя математики, досліджено місце CoCalc у системі засобів навчання математичних дисциплін.

У другому розділі визначена загальна методика дослідження, розглянуто процес проектування системи професійних компетентностей учителя математики, висвітлено особливості використання сервісу CoCalc у навчанні математичних дисциплін, представлено сучасний стан і характеристики CoCalc та розроблено модель використання хмарного сервісу CoCalc як засобу формування професійних компетентностей учителя математики.

У третьому розділі наведено структуру методики використання CoCalc як засобу формування професійних компетентностей учителя математики, засоби, форми використання CoCalc та методи навчання майбутніх учителів математики з використанням цього хмарного сервісу.

У четвертому розділі подані відомості щодо етапів дослідження,

наведені завдання та зміст експериментальної роботи, виконано кількісне та якісне опрацювання результатів констатувального та формувального етапів педагогічного експерименту.

Автор щиро дякує ініціаторам створення монографії – завідувачу відділу хмаро орієнтованих систем інформатизації освіти Інституту інформаційних технологій і засобів навчання д. пед. н., с. н. с. М. П. Шишкіній, професору кафедри інформатики та прикладної математики Криворізького державного педагогічного університету д. пед. н., проф. С. О. Семерікову, та всім співробітникам спільної науково-дослідної лабораторії з питань використання хмарних технологій в освіті Криворізького національного університету та Інституту інформаційних технологій і засобів навчання НАПН України, які вносили пропозиції щодо структури та змісту цієї роботи.

# РОЗДІЛ 1
# ТЕОРЕТИЧНІ ОСНОВИ ВИКОРИСТАННЯ ХМАРНИХ СЕРВІСІВ У НАВЧАННІ МАЙБУТНІХ УЧИТЕЛІВ МАТЕМАТИКИ

## 1.1 Професійна підготовка учителів математики у ЗВО України

Вивчення математичних дисциплін, зазвичай, поєднує в собі глибоке опанування теорії та практики. В рамках Болонського процесу і в умовах єдиного навчального простору доцільно було б використати кращий досвід з освітньої практики європейських країн у поєднанні зі здобутками української освіти, що мало б вивести навчання на новий рівень. В цьому контексті постає низка невирішених проблем.

Модернізація цілей, змісту, методів, засобів, організаційних форм навчання є основою інформатизації освіти. Технологічне переоснащення навчального процесу, поява нового змісту, методів, засобів і організаційних форм навчання є необхідністю, яка забезпечує досягнення окреслених цілей [264].

Г. Г. Швачич вважає, що в першу чергу вимагають розгляду та подальшого вирішення такі проблеми [296]:

– необхідність покращення середньої освіти;

– необхідність покращення вступних кампаній у ЗВО (особливу увагу потрібно приділити якості одержаних знань);

– скорочення аудиторних занять та їх наслідки;

– необхідність зміни та удосконалення системи оцінки знань.

Теперішній стан підготовки майбутнього вчителя математики характеризується стрімким розвитком математичної науки, численними реформами освіти, але в той же час спостерігається скорочення годин на аудиторні заняття та збільшення частки самостійної роботи студентів. Існує небезпека зниження рівня освіти, а відтак, відчувається нагальна потреба в розробці нових методичних систем навчання вищої математики на основі сучасних інформаційних технологій [264, с. 8]. Дана проблема особливо актуальна в процесі підготовки майбутніх вчителів математики, оскільки вивчення фундаментальних дисциплін відбувається із застосуванням інформаційних технологій у навчанні (ІТН) і реалізації нової парадигми освіти.

Проблеми, які виникають під час вивчення математичних дисциплін у педагогічному ЗВО можуть бути вирішені за рахунок використання хмарних сервісів. Тема використання хмарних сервісів неодноразово звучала в рамках круглих столів, конгресів, наукових конференцій.

У Законі України «Про вищу освіту» вказано, що професійна підготовка, зокрема вчителів математики, є здобуттям кваліфікації відповідної спеціальності [76]. Згідно Постанови «Про затвердження переліку галузей знань і спеціальностей, за якими здійснюється підготовка здобувачів вищої освіти» [203] в Україні підготовка вчителів математики здійснюється за спеціальністю 014 Середня освіта (предметна спеціалізація – Математика). Але на сьогодні стандарти вищої освіти знаходяться лише на стадії розробки і мають містити для кожної спеціальності окремо перелік компетентностей випускника. Тому, аналізуючи питання професійної підготовки учителів математики у ЗВО України, доцільно спиратися як на попередні стандарти, інші нормативні документи, так і на публікації щодо останніх результатів педагогічних досліджень.

На думку В. П. Головенкіна, нормативні документи, що укладаються на рівні держави, це: національна рамка кваліфікацій, стандарт вищої освіти та професійний стандарт, що знаходяться у певному співвідношенні [276] (рис. 1.1).

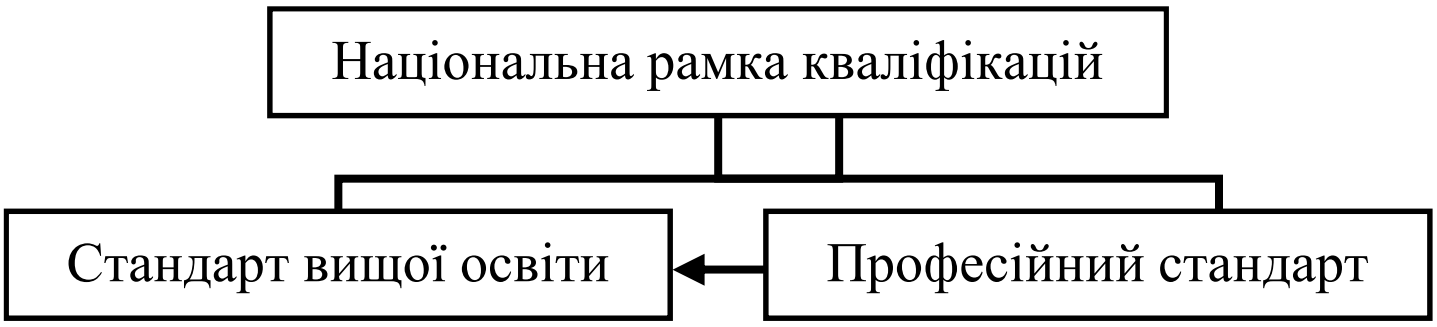

Рис. 1.1. Структура нормативних засад змісту освіти і навчання на державному рівні

В Україні підготовка вчителів математики за спеціальністю 014 Середня освіта (Математика) включає три цикли дисциплін:

1) гуманітарної та соціально-економічної підготовки;

2) природничо-наукової підготовки (диференціальна геометрія і топологія, математична логіка, теоретична фізика, теоретична механіка, екологія, валеологія, безпека життєдіяльності та ін.);

3) професійної та практичної підготовки: підцикл психолого-педагогічної підготовки, підцикл професійної науково-предметної підготовки, підцикл практичної підготовки та варіативна частина циклу.

Загальний навчальний час за спеціальністю 014 Середня освіта (Математика) показано в таблиці 1.1. Цикл природничо-наукової підготовки в нормативній частині містить три математичні дисципліни (табл. 1.2).

Цикл професійної і практичної підготовки включає в себе нормативну частину циклу та варіативну частину циклу. У нормативній частині циклу нас цікавить перш за все підцикл професійної науково-предметної підготовки (табл. 1.3), оскільки його складовими виступають математичні дисципліни.

*Таблиця 1.1*

**Загальний навчальний час підготовки бакалавра за спеціальністю 014 Середня освіта (Математика)**

| Назва циклу підготовки | Обсяг в годинах | Обсяг в кредитах |
|---|---|---|
| Гуманітарної і соціально-економічної підготовки | 870 | 29 |
| Природничо-наукової підготовки | 840 | 28 |
| Професійної і практичної підготовки | 5430 | 181 |
| Атестація | 60 | 2 |
| Загальний обсяг підготовки | 7200 | 240 |

*Таблиця 1.2*

**Нормативна частина циклу природничо-наукової підготовки (математичні дисципліни)**

| Назва дисципліни | Обсяг в годинах | Обсяг в кредитах | Форма контролю |
|---|---|---|---|
| Диференціальна геометрія і топологія | 150 | 5 | екзамен / залік |
| Комплексний аналіз | 180 | 6 | екзамен |
| Математична логіка та теорія алгоритмів | 135 | 4,5 | залік |

*Таблиця 1.3*

**Підцикл професійної науково-предметної підготовки циклу професійної і практичної підготовки**

| Назва дисципліни | Обсяг в годинах | Обсяг в кредитах | Форма контролю |
|---|---|---|---|
| Елементарна математика | 405 | 13,5 | залік |
| Математичний аналіз | 1140 | 38 | екзамен |
| Аналітична геометрія | 255 | 8,5 | екзамен/залік |
| Алгебра і теорія чисел | 240 | 8 | екзамен |
| Лінійна алгебра | 210 | 7 | екзамен/залік |
| Дискретна математика | 135 | 4,5 | екзамен |
| Диференціальні рівняння | 210 | 7 | екзамен/залік |
| Теорія ймовірностей та математична статистика | 150 | 5 | екзамен |
| Методика навчання математики | 420 | 14 | екзамен |
| Курсова робота з методики навчання математики | 60 | 2 | захист курсової роботи |

Згідно методичних рекомендацій щодо розроблення стандартів вищої освіти Міністерства освіти і науки України, «спеціальні (фахові, предметні) компетентності – компетентності, що залежать від предметної області, та є важливими для успішної професійної діяльності за певною спеціальністю» [147, с. 4].

Формування спеціальних професійних компетентностей студентів відбувається за рахунок вивчення математичних дисциплін. Спеціально професійні компетентності реалізуються через типи діяльності, типові завдання діяльності та уміння (Додаток. А: табл. А.1).

У галузевому стандарті вищої освіти України, зазначається, що: клас задачі діяльності – ознака рівня складності задач діяльності, які вирішуються фахівцем [45, с. 4].

Усі задачі діяльності розподіляються на три класи (табл. 1.4):

– стереотипні задачі діяльності – передбачають діяльність відповідно до заданого алгоритму, що характеризується однозначним набором добре відомих, раніше відібраних складних операцій і потребує використання значних масивів оперативної та раніше засвоєної інформації;

– діагностичні задачі діяльності – передбачають діяльність відповідно до заданого алгоритму, що містить процедуру часткового конструювання рішення із застосування відповідних операцій і потребує використання значних масивів оперативної та раніше засвоєної інформації;

– евристичні задачі діяльності – передбачають діяльність за складним алгоритмом, що містить процедуру конструювання рішень і потребує використання великих масивів оперативної та раніше засвоєної інформації [45, с. 4].

Дану класифікацію доречно використовувати в процесі проектування методик навчання та використання хмарних сервісів, оскільки основні типи діяльності треба будувати у відповідності до розглянутих типів задач.

*Таблиця 1.4*

**Класи діяльності, типові завдання діяльності майбутніх вчителів математики**

| Клас задач діяльності | Типове завдання діяльності |
|---|---|
| **Діагностичні** | Аналіз сучасних математичних теорій |
| | Аналіз математичної проблеми (задачі) |
| | Аналіз створеної моделі реального об’єкта |
| | Дослідження математичної моделі з використанням засобів комп’ютерної техніки |
| | Створення аксіоматичної теорії та її аналіз |

| Клас задач діяльності | Типове завдання діяльності |
|---|---|
| | Використання засобів інформаційних технологій для розв'язування математичних задач |
| | Аналіз наукового результату, оцінка його місця, ролі і значення |
| | Розв'язування задач шкільного курсу математики |
| | Використання програмного засобу навчально-виховного призначення для підтримки педагогічного процесу |
| **Евристичні** | Постановка математичної задачі |
| | Формулювання гіпотетичного твердження |
| | Доведення гіпотетичного твердження, спростування гіпотетичного твердження |
| | Вибір використання алгоритмів, методів, прийомів та способів, розв'язування математичних задач |
| | Підготовка наукової доповіді статті, реферату, звіту (наукового твору) |
| | Підготовка наукової доповіді статті, реферату, звіту (наукового твору) |

Як видно з таблиці 1.4, стереотипний клас діяльності не представлений жодним типовим завданням діяльності. Це пояснюється тим, що дуже важко виокремити в педагогічній діяльності задачі, які б підпорядковувались строгому алгоритму. Це є специфікою педагогічних задач. І. Ф. Ісаєв [80] пропонує іншу класифікацію задач діяльності: стратегічні, поточні та тактичні задачі.

При цьому під стратегічними задачами дослідник розуміє педагогічні задачі, що є похідними з цілей освіти та представляють кінцевий результат діяльності вчителя математики.

Поточні задачі, інакше кажучи оперативні – задачі, що ставить перед собою вчитель математики на даному етапі своєї педагогічної діяльності (з урахуванням найближчих цілей, в даний момент часу).

Що ж стосується тактичних задач, то І. Ф. Ісаєв [80], розглядає задачі як такі, що уточнюють стратегічні задачі, в даній практичній діяльності, вони є вирішенням того чи іншого етапу розв'язку стратегічної задачі.

## 1.2 Понятійний апарат дослідження

Для здійснення аналізу досвіду використання хмарних сервісів у навчанні майбутніх вчителів математики необхідно звернутися до уточнення основних термінів.

Якщо розглядати поняття «сервіс» у широкому розумінні, тоді будемо користуватись наступним визначенням.

*Сервіс* – система, що забезпечує одну чи декілька функцій, що мають цінність для кінцевого користувача. Прикладами є фотокопіювання, банківський сервіс, міжбібліотечний обмін, web-сервіс [265].

Спираючись на дане визначення потрібно з'ясувати, як саме розуміти поняття «система».

«Система – цілісна множина об'єктів (елементів) і відношень (зв'язки) між ними, що виділені з середовища за ознакою приналежності цих об'єктів і відношень до реалізації заданих цілей системи. Як самі об'єкти, так і відношення між ними мають певні властивості [21, с. 226]».

Тобто, система – це нова якісна характеристика утворюючих її елементів у порівнянні із безсистемною характеристикою тих самих елементів. Частина системи, яка виконує в ній визначену функцію з метою підтримання її функціонування є елементом системи.

Науковці по-різному трактують поняття «хмарний сервіс».

На думку Т. Л. Архіпової та Т. В. Зайцевої, хмарний сервіс – це особлива клієнт-серверна технологія, де потрібні для роботи ресурси користувач сприймає як віртуальний сервер, що уможливлює для нього досить просте використання ресурсів та зміну їхніх об'ємів [6, с. 102].

Група авторів в монографії «Моделювання й інтеграція сервісів хмаро орієнтованого навчального середовища» дає наступне означення: «Хмарні сервіси – це сервіси, що роблять доступними користувачеві прикладні додатки, простір для зберігання даних та обчислювальні потужності через Інтернет [151, с. 40]».

За визначенням Національного Інституту Стандартів і Технології США (NIST), хмарні обчислення – це модель надання користувачеві зручного мережного доступу до пулу обчислювальних ресурсів (таких як мережі, сервери, масиви даних, програмні приложення та послуги), які можуть бути швидко надані з мінімальними управлінськими зусиллями або взаємодією з провайдером послуг [302, с. 134].

Згідно Стандарту ISO/IEC 17788 :2014 «Інформаційні технології. Хмарні обчислення. Огляд і словник», хмарні обчисленння – це парадигма для забезпечення доступу до мережі з масштабованого і еластичного пулу розподілених фізичних або віртуальних ресурсів з самообслуговування та виділення ресурсів і управління на вимогу користувача (включають в себе сервери, операційні системи, мережі, програмне забезпечення, програми та пристрої зберігання даних) [269].

В. Ю. Биков та М. П. Шишкіна в своїх дослідженнях дають наступне визначення: *хмарні сервіси* – сервіси, що забезпечують користувачеві мережний доступ до масштабованого і гнучко організованого пулу розподілених фізичних або віртуальних ресурсів, що постачаються в режимі самообслуговування і адміністрування за його запитом (наприклад, програмне забезпечення, простір для зберігання даних, обчислювальні потужності та ін.)[26, с. 38]. Останнє визначення приймається за основне в подальшому.

Отже, можна стверджувати, що *CoCalc* – це хмарний сервіс з відкритим вихідним програмним кодом, основними складовими якого є: Web-система комп'ютерної математики Sage; система управління навчальними курсами; редактор LaTeX; інтерпретатор IPython.

Оскільки в дослідженні хмарний сервіс розглядається як засіб формування професійних компетентностей, тому потрібно визначитись з поняттям «засіб». В педагогіці поняття «засіб» розглядається дуже широко. Оскільки використання хмарного сервісу планується в процесі навчання майбутніх вчителів математики, в рамках дослідження доцільно розглянути більш вузьке поняття «засіб навчання».

Згідно проведеного аналізу у педагогічній науці існує декілька підходів до того, що розуміти під поняттям «засоби навчання». М. М. Фіцула [290] стверджує, що під засобами навчання розуміють допоміжні матеріальні засоби. П. Л. Волкова [42] розглядає в якості засобів навчання саме матеріально-технічні, яким притаманні певні дидактичні функції. На думку Н. Є. Мойсеюк [154], засоби навчання це певні матеріали навчального процесу, завдяки яким педагог здатен ефективніше досягати поставлених цілей навчання.

Під «засобом навчання» будемо розуміти: «*Засіб навчання* – сукупність предметів, ідей, явищ і способів дій, які забезпечують реалізацію навчально-виховного процесу [290]».

Одним з ключових термінів дослідження виступає поняття «компетентність».

М. І. Жалдак, Ю. С. Рамський та М. В. Рафальська «Під компетентністю розуміють комплекс знань, умінь, навичок, досвіду застосування їх для здійснення діяльності, метою якої є досягнення певних цілей, ставлення до процесу та результатів виконання цієї діяльності [71, с. 5]».

На думку О. М. Спіріна, *компетентність* – це складна інтегрована характеристика особистості, під якою розуміється сукупність знань, умінь, навичок, а також досвіду, що разом дає змогу ефективно провадити діяльність або виконувати певні функції, забезпечуючи розв'язання проблем і досягнення певних стандартів у галузі професії або виді діяльності. Компетентність розглядається як сформована якість, результат діяльності, «надбання» студента [268, с. 194-195]. Останнє означення в подальшому буде використовуватись як основне.

Проте, в даному дослідженні, в першу чергу зосередимось саме на професійних компетентностях.

І. А. Зязюн зазначає, що «професійно-педагогічна компетентність відбиває готовність і здібність людини професійно виконувати педагогічні функції згідно із прийнятими у суспільстві на цей час нормативами і стандартами. Саме тому поняття «компетентність» має конкретно-історичну визначеність і може оцінюватися лише у практичній діяльності [268, с. 195]».

Отже, узагальнюючи трактування О. М. Спіріна та І. А. Зязюна, *професійні компетентності* – це сукупність професійних здатностей особи, що дають змогу самостійно приймати рішення, виконувати професійну діяльність (в залежності від відповідної кваліфікації) та досягати певних результатів.

Окрім основних понять, потрібно визначитись з тими, які також будуть використані в процесі проведення дослідження.

Згідно Наказу «Про затвердження Методичних рекомендацій щодо розроблення Державних стандартів професійно-технічної освіти з конкретних професій на основі компетентнісного підходу» [160], можна стверджувати, що:

*Ключові компетентності* – це загальні здібності й уміння (психологічні, когнітивні, соціально-особистісні, інформаційні, комунікативні), що дають змогу особі розуміти ситуацію, досягати успіху в особистому і професійному житті, набувати соціальної самостійності та забезпечують ефективну професійну й міжособистісну взаємодію [160].

Інтегральна компетентність – узагальнений опис кваліфікаційного рівня, який виражає основні компетентністні характеристики рівня щодо навчання та/або професійної діяльності [202].

*Кваліфікація* – офіційний результат оцінювання й визнання, який отримано, коли компетентний орган встановив, що особа досягла результатів навчання за заданими стандартами. Кваліфікації поділяються на освітні (на основі освітніх стандартів) та професійні (на основі професійних стандартів) [75].

*Системи комп'ютерної математики (СКМ)* – це програмні засоби, за допомогою яких можна автоматизувати виконання як чисельних, так і аналітичних та графічних обчислень і розрахунків. В них акумульовано багатовіковий досвід розвитку математики. За допомогою СКМ користувачі математики здатні розв'язувати навіть досить складні математичні задачі [210].

*Система навчання* – дидактична система, на основі якої забезпечується цілеспрямований процес здобування знань, формування умінь, набуття навичок, засвоєння способів пізнавальної діяльності людини і сприяння її розвитку [282].

«*Хмаро орієнтовані ІКТ навчання* визначимо як сукупність методів, засобів і прийомів діяльності, що використовуються для організації і супроводу навчального процесу, збирання, систематизації, зберігання, опрацювання, передавання, подання повідомлень і даних навчального призначення та використовують динамічний масив віртуалізованих апаратних і програмних ресурсів, доступних через мережу незалежно від термінального пристрою [273, с.152]».

*Хмаро орієнтована система навчання* – це система навчання, у якій здійснення окремих дидактичних функцій підтримується завдяки доцільному, координованому та інтегрованому використанню сервісів хмарних технологій.

### 1.3 Формування професійних компетентностей учителя математики

Основна риса компетентнісного підходу полягає в тому, що під час навчання, яке базується на орієнтації освітнього процесу на досягненні навчальних результатів – у студентів формуються компетентності, необхідні для нормального життя і професійної діяльності в інформаційному суспільстві. Про відсутність будь-якої компетентності взагалі не правильно говорити, оскільки процес її формування може бути досить тривалим і попадати під вплив різних факторів: навчання в освітніх установах, професійна діяльність, міжособистісне спілкування і так далі. Отже, говорячи, що студенти здобувають певні компетентності мають на увазі формування їх певного рівня [71].

У силу специфіки кожної предметної галузі не може бути загальноприйнятого списку складників професійних компетентностей. Але, в рамках різних професійних асоціацій, міжнародних проектів, агентств із забезпечення національної якості на міжнародному рівні розроблено ряд списків для окремих спеціальностей (предметних областей), які можуть бути використані при створенні національних стандартів (стандарт результатів навчання і компетентностей), а також розробка освітніх програм конкретних університетів.

Згідно проекту Тюнінг, були проведені дослідження стосовно спеціальних компетентностей для наступних предметних галузей: Бізнес і менеджмент, Європейські студії, Історія, Математика, Науки про Землю, Освіта, Сестринська справа, Фізика, Хімія. Матеріали, що стосуються даних досліджень (Guidelines and Reference Points for the Design and Delivery of Degree Programmes in…), доступні за посиланням: http://www.unideusto.org/tuningeu/subject-areas.html. У даному проекті розміщені 42 предметні галузі: основні 9 розташовані на сайті Тюнінг, інші 33 можна знайти за розміщеними Інтернет-адресами на сторінках проекту Тюнінг [78, с. 16-17].

Робота проводилася різними групами науковців, які відображають специфічні традиції та розробку і реалізацію освітніх програм в області кожної окремої предметної області. Але при цьому кожна група враховувала методологію Тюнінга з подальшими можливостями створювати освітні програми. Так було одержано матеріал написаний однією мовою (лексика, складники), рекомендації (результати навчання і компетентності, підходи як до навчання так і оцінювання його результатів та ін.).

Ще одним джерелом щодо фахових компетентностей є матеріали Агенції забезпечення якості вищої освіти Сполученого Королівства (The Quality Assurance Agency for Higher Education, QAA, UK). Агенція затвердила набори стандартів/опорних точок (Subject benchmark statements) для 58 освітніх програм підготовки бакалавра з відзнакою (в певному наближенні – еквівалент бакалавра в Україні), 14 програм підготовки магістра та окремо для 18 програм в області охорони здоров’я. Ці матеріали містяться на Інтернет-сторінці QAA: http://www.qaa.ac.uk/assuring-standardsand-quality/the-quality-code/subject-benchmark-statements [78, с. 16-17].

Що стосується класифікації професійних компетентності, як правило, вони поділяються на три категорії: знання з предметної області, пізнавальні здібності та навички з предметної області, практичні навички в предметній області.

Розглядаючи математичну компетентність, що є складником професійних компетентностей, Я. Г. Стельмах [271] трактує її як властивість, що притаманна особистості, яка представляє готовність самостійно і з повною відповідальністю використовувати математичні інструменти, як підтвердження теоретичної та практичної готовності випускників до подальшої професійної діяльності [278, с. 20].

С. О. Скворцова [247] пропонує власну класифікацію професійних компетентностей майбутнього вчителя математики. За основу взято загально прийнятий поділ компетентностей на ключові, спеціальні та базові. При цьому кожна компетентність характеризується окремими компонентами: комунікативний, особистісний та професійно-діяльнісний.

Професійно-діяльнісний компонент складається з двох компетентностей:

– інформаційна компетентність (здібність опрацьовувати математичні факти, працювати з математичними даними, організовувати систематичний пошук та узагальнювати наявний математичний матеріал),

– предметна (наявність цілісної системи комплексних математичних знань і готовність використовувати їх в своїй професійній діяльності; вміння вирішувати типові професійні завдання з залученням математичного апарату).

Комунікативний компонент складається з комунікативної компетентності (володіння математичною термінологією; готовність передачі математичних відомостей; здатність використовувати вербальні та невербальні засоби передавання математичних даних і відомостей).

Особистісний компонент містить: рефлексивну компетентність (прагнення до удосконалення застосування математичних інструментів у професійній діяльності); творчу компетентність (використання інноваційних математичних методів у професійній діяльності) [278, с. 20].

Аналізуючи різні підходи науковців до складових професійних компетентностей можна виділити декілька напрямів. Наприклад, В. Д. Шадриков виокремлює: соціально-особистісні компетентності, спеціально та загально професійні компетентності [295]. Згідно класифікації В. І. Байденко, до складу професійних компетентностей входять: професійні, загальні та академічні [10]. Ю. В. Фролов та Д. А. Мохотін вбачають наступну структуру: методологічні, загальнокультурні та предметно-зорієнтовані [291]. За А. В. Хуторським професійні компетентності поділяються на базові, ключові та спеціальні [294]. А. К. Маркова включає до них також психологічні та педагогічні знання, педагогічні вміння, педагогічні позиції, особистісні якості [139].

Також серед професійних компетентностей виділяють загальнопрофесійні та спеціально професійні.

У цьому випадку, до спеціально професійних компетентностей відносять:

– здатність використовувати професійно профільовані знання в галузі математики (математичної статистики), для статистичної обробки експериментальних даних і математичного моделювання природних явищ і процесів;

– здатність використовувати математичний апарат для моделювання різноманітних процесів;

– здатність використовувати професійно профільовані знання й практичні навички з алгебри та теорії чисел;

– здатність використовувати професійно профільовані знання й практичні навички для оволодіння основами теорії й методами теоретичних досліджень;

– здатність використовувати професійно профільовані знання й

практичні навички з математичного аналізу;

– здатність використовувати знання, уміння й навички з аналітичної та диференціальної геометрії;

– здатність використовувати професійно профільовані знання з дискретної математики;

– здатність використовувати знання й уміння з теорії ймовірностей;

– здатність використовувати професійно профільовані знання, уміння й навички з математичної логіки та теорії алгоритмів;

– здатність використовувати знання, уміння й навички з диференціальних рівнянь;

– професійно профільовані знання й уміння з теоретичних основ інформатики й практичного використання комп'ютерних технологій;

– здатність володіти навичками роботи з комп'ютером на рівні користувача та фахівця ІКТ;

– здатність використовувати професійні знання та практичні навички з методики навчання математики в основній школі.

Наведені компетентності дозволяють вирішувати типові задачі, які постають перед випускниками педагогічного ЗВО, зокрема перед вчителями математики, під час здійснення виробничих функцій.

Аналізуючи професійні функції, типові задачі діяльності та змістову частину вмінь, доцільно виокремити певну групу професійних функцій (табл. 1.5) з урахуванням властивостей та якостей (спеціально професійних компетентностей) випускників педагогічного ЗВО, а саме вчителів математики середньої загальноосвітньої школи.

*Таблиця 1.5*

**Професійні функції, типові задачі діяльності, уміння якими повинні володіти майбутні вчителі математики**

| **Зміст виробничої функції** | **Назва типової задачі діяльності** | **Зміст уміння** |
|---|---|---|
| Проведення всіх форм занять у середніх навчально-виховних закладах | Планування занять | Скласти план проведення та конспект занять з використанням навчально-методичних документів та навчально-методичної літератури |
| | Проведення занять | Використовувати знання за фахом та спеціалізацією при проведенні занять |
| | Здійснення контролю знань, набутих учнями, | Отримувати і використовувати науково-технічну інформацію за фахом для ведення занять |

| Зміст виробничої функції | Назва типової задачі діяльності | Зміст уміння |
|---|---|---|
| | студентами-слухачами | Використовувати ЕОМ для проведення контрольних заходів |
| Навчально-методична робота | Забезпечення незалежного методичного рівня знань | Використовувати методи навчання математики та інформатики для проведення усіх форм занять |
| | Методичне забезпечення самостійної роботи учнів, студентів-слухачів | Скласти та підготувати до друку методичні вказівки, посібники з математики та інформатики з використанням навчальної літератури, навчально-методичних та інших інструктивних документів |

Як видно з таблиці 1.6, спеціально професійні компетентності не стосуються лише такої професійної функції, як навчально-громадське виховання учнів, студентів. У зазначених професійних функціях, за рахунок спеціально професійних компетентностей є можливим вирішення всіх можливих типових задач діяльності, які розкриті в комірках, що характеризують зміст уміння.

*Таблиця 1.6*

**Спеціально професійні компетентності та система умінь, що їх відображає**

| Компетентність щодо вирішення проблем і задач соціальної діяльності, інструментальних, загально-наукових і професійних задач | Зміст уміння |
|---|---|
| Користування сучасними інформаційними технологіями | Вміти використовувати наявні та вивчати нова інформаційні технології |
| Здійснювати пошук нових даних | У виробничих умовах, використовуючи ключові слова у певній галузі на базі професійно-орієнтованих (друкованих та електронних) джерел, за допомогою відповідних методів проводити пошук нових текстових даних (робота з |

| Компетентність щодо вирішення проблем і задач соціальної діяльності, інструментальних, загально-наукових і професійних задач | Зміст уміння |
|---|---|
| | джерелами навчальних, наукових та довідкових даних); пошук нових графічних, звукових та відеоданих |
| Застосовувати закони формальної логіки в процесі інтелектуальної діяльності | За допомогою формальних логічних процедур проводити аналіз наявних даних на їх відповідність умовам необхідності та достатності для забезпечення ефективної діяльності |
| | За допомогою формальних логічних процедур проводити аналіз наявних даних на їх відповідність вимогам внутрішньої несуперечності |
| | За допомогою формальних логічних процедур проводити структурування даних |
| | За результатами структурно-логічної обробки даних роботи висновок щодо їх придатності для здійснення заданих функцій |
| | На основі результатів здійсненої діяльності за допомогою певних критеріїв встановлювати якість попередньо виконаних логічних операцій |
| | За умов негативного результату діяльності знаходити помилки в структурі логічних операцій |

Якщо ж розглянути спеціально професійні компетентності з точки зору системи умінь, що їх відображає, отримаємо наступну відповідність (табл. 1.6) між спеціально професійними компетентностями та змістом умінь.

У «Методичних рекомендаціях щодо розроблення стандартів вищої освіти», спеціальні (фахові, предметні) компетентності визначаються як компетентності, що залежать від предметної області, та є важливими для успішної професійної діяльності за певною спеціальністю.

## 1.4 Вітчизняний і зарубіжний досвід використання хмарних сервісів для формування професійних компетентностей учителя математики

Освітні кваліфікаційні вимоги щодо підготовки вчителів за кордоном передбачають обов'язкове ознайомлення з основами психології та педагогіки. Психолого-педагогічний блок базової підготовки вчителів має на меті забезпечити набуття студентами (протягом принаймні двох семестрів) психологічних та педагогічних знань, які є основою для подальшої вчительської підготовки, професійної орієнтації.

Проведений аналіз надав можливість зробити висновок, що у навчальних планах, наприклад Угорщини, підготовки бакалаврів математики поєдналися знаннєво-орієнтований та компетентнісний підходи, що має забезпечити можливість вибору між відданістю традиціям та нововведенням.

Нові методики навчання базуються на використанні комп'ютерів, зокрема програмного забезпечення. М. І. Жалдак [73] вважає, що існуючі педагогічні програмні засоби (ППЗ) можна розглянути з двох позицій: ППЗ, що зменшують час спілкування викладача та студента, збільшують частку самостійної роботи та полегшують процес виконання завдань винесених на самостійне опрацювання та ППЗ, що навпаки призводять до тіснішої взаємодії між студентом та викладачем, за рахунок використання інструментів спілкування певного ППЗ. До останніх можна віднести і хмарний сервіс CoCalc, оскільки в ньому наявні інструменти спілкування в реальному часі. Тому, навчальний час, може бути використано, щоб вирішувати проблеми, встановлювати суть процесів і явищ, розробляти відповідні моделі, встановлювати зв'язки та закономірності, порівнюючи моделі, аналізуючи та синтезуючи одержані результати. Тому хмарний сервіс може бути в повній мірі використано практично в усіх видах навчальної діяльності: вивчення нового матеріалу, формування понять, знань, навичок, під час самостійної роботи, контролю, самоконтролю. При цьому будуть застосовуватись і різні методи навчання.

На думку М. І. Жалдака [73], до інтегрованого ППЗ (що можна використати одночасно і для скорочення часу спілкування між студентом та викладачем так і для покращення навчальної взаємодії) відносять GRAN1, GRAN-2D, GRAN-3D, DG, SAGE (мається на увазі попередня версія CoCalc) [385; 384; 386; 69]. При цьому, GRAN1 можна застосовувати в процесі вивчення всіх розділів елементарної математики та окремих – математичних дисциплін, різних типів та рівнів математичних завдань. Крім того, використання GRAN1 можна демонструвати в процесі опанування майбутніми вчителями математики дисципліни методика навчання математики. Використання програми будe корисним під час побудови замкнених та незамкнених ламаних ліній, обчисленні довжин ламаних, прямих та відрізків, обчислення площі та периметра, визначення об'ємів тіл, розв'язування задач на побудову, побудові графіків функцій, дослідженні функції, розв'язання рівнянь та нерівностей, їх систем, обчисленні визначених інтегралів, обчисленні ймовірностей випадкових подій, статичне опрацювання вибірки. При цьому, вказані програмні засоби не впливають на успішність та швидкість розв'язування завдань студентів з низьким та високим рівнем теоретичних знань. В даному випадку побудова моделі з використанням будь-якого з програмних засобі GRAN1, GRAN-2D, GRAN-3D, DG та SAGE не викликає жодних проблем. Студенту треба лише дослідити її та визначити певні закономірності.

Тобто студенти з низьким рівнем знань зможуть краще сприйняти матеріал, а з високим більш глибоко опанувати ту чи іншу тему математичної дисципліни. Крім того, зрозуміти як само методично грамотно використовувати подібні програмні засоби на уроках математики в подальшій професійній діяльності.

Що ж стосується питання технічних обчислень, які зазвичай досить громіздкі, то на думку М. І. Жалдака, вони будуть автоматизовані, що надасть можливість приділити більше уваги розділам математичних дисциплін, що винесені на самоопрацювання, вирішення питань які виникли в студентів в процесі їх вивчення [73].

На користь успішного використання програмних засобів GRAN1, GRAN-2D, GRAN-3D [69] свідчать низки робіт українських дослідників, білоруських [163] та польських дослідників [385; 384; 386].

Так, комплекс програм GRAN (GRAN1, GRAN-2D, GRAN-3D) в тій чи іншій мірі може використовуватись на уроках математики і частково фізики від 6-го до 11-го класу включно, при вивченні різних математичних дисциплін в педагогічному інституті (геометрія, математичний аналіз, теорія ймовірностей з елементами математичної статистики, обчислювальна математика, фізика тощо), при цьому лише в шкільних курсах математики і фізики нараховується понад 700 годин, де можуть бути використані ці програми [73, с. 9].

Крім того, в рамках системи освіти необхідною умовою виступає зростання якості навчання, яке, в свою чергу, обумовлює використання комп'ютерних засобів навчання. Використання цих засобів, можливо, сприятиме більш якісному засвоєнню матеріалу, чіткій систематизації вивченого, стимулюватиме активність мислення та надасть можливість проводити своєчасну корекцію траєкторії здобування знань, формування умінь та навичок студентів.

Організовуючи самостійну роботу студентів за підтримки ІКТ

потрібно розробити комп'ютерно-, професійно- та особистісно-орієнтовану систему дидактичних матеріалів. Розробку таких матеріалів можна здійснювати за рахунок використання хмарних технологій [317].

Значні інновації в розподілені обчислення, а також поліпшення доступу до високошвидкісного Інтернету та слабкої економії прискорили інтерес до економічно ефективних хмарних обчислень за останні роки [9, с. 82].

Поява хмарних сервісів змінює взагалі наше уявлення стосовно використання апаратного, програмного забезпечення та збереження даних.

Зростання популярності хмарних обчислень в останні роки є одним з основних трендів розвитку IT у всьому світі. Як показує практика, використання «хмар» для організації систем віддаленого доступу до корпоративних ресурсів демонструє високу ефективність. Особливу увагу привертають проблеми організації онлайн-навчання, яке здійснюється за допомогою сучасних інформаційно-комунікаційних технологій [142, с. 41].

Дійсно, при використанні хмарних технологій є можливість користуватись своїми даними, виконувати обчислення, вносити певні корективи, звертаючись до них через Інтернет. Користувачеві не має потреби перейматись стосовно встановлення і оновлення програмного забезпечення, обмеженості обсягу пам'яті, спеціальних пристроїв для збереження даних, способу збереження та оброблення внесених ним даних.

Поява хмарних обчислень змінює наше уявлення про використання апаратного й програмного забезпечення та збереження даних. Сховище даних як об'єкт, який можна відділити від окремого комп'ютера, вже стало звичайним явищем, але нині у такому сенсі почали розглядати і програмні додатки. Замість розміщення файлів і програмного забезпечення на одному комп'ютері, результати й засоби роботи поступово переносяться та розміщуються у хмарі. За таких умов програмні додатки та дані доступні з багатьох комп'ютерів, а засоби, які використовуються для вирішення певних завдань, безкоштовні або дуже дешеві [82, с. 66].

На сучасному етапі використання хмарних технологій є досить перспективним для вищих навчальних закладів України. Одним із найбільш вагомих економічних ефектів є суттєве зменшення затрат як на програмне забезпечення (офісні додатки, електронна пошта тощо) і на серверне обладнання (можна переорієнтувати, наприклад, на використання для САПР-додатків), так і зменшення затрат на обслуговуючий персонал. [58, с. 70]

Причому для навчання не потрібні будуть надпотужні пристрої чи додаткові матеріальні витрати. Для цього достатньо буде лише мати звичайний ноутбук, смартфон, чи будь-який інший пристрій, за допомогою якого користувач матиме вихід до Інтернету. Практично користувач має безкоштовний простір для збереження даних.

Технології «хмарних обчислень» вносять суттєві зміни у процес навчання будь-якої дисципліни, забезпечуючи оптимальний збір, збереження, пошук, опрацювання та представлення даних, при цьому не потребуючи внесення змін до навчальних планів закладів освіти [7, с. 72].

На практиці хмарні обчислення дають змогу розгорнути знаряддя, які за потреби можна масштабувати для обслуговування довільної кількості користувачів. Нерідко користувачі використовують хмари, навіть не підозрюючи про це [155, с. 20]. У подальшому використання хмарних сервісів у ЗВО можна розглянути в наступних ракурсах [155]:

– для користувачів;

– для ІТ-персоналу;

– для ЗВО.

Останні дослідження в США показують, що навчання за допомогою Інтернет і мультимедіа, виступаючи в якості повної заміни традиційного навчання, має в середньому щорічне збільшення чисельності студентів і охоплює трохи менше 20 % всіх студентів у період між 2002 і 2008 роками, приблизно 300000 викладачів займаються навчанням за допомогою Інтернет і мультимедіа (у тому числі у США в 2008 році від 20 до 25 % студентів реєструвалися хоча б в одному онлайн-класі). Кількість студентів, що пройшли принаймні один онлайн-курс з 1 602 970 у 2002 р. і 1 971 397 у 2003 році збільшилася до 6 700 000 у 2012 році [89, с. 40].

Певний досвід використання хмарних сервісів та хмарних технологій в навчальному процесі українських навчальних закладів вже існує. Наприклад, хмарна інфраструктура використовується у Південноукраїнському національному педагогічному університеті імені К. Д. Ушинського, хмарні сервіси Google Apps інтегровані в навчальне середовище фізико-математичного факультету Тернопільського національного педагогічного університету імені Володимира Гнатюка [293, с. 105].

Суттєвий доробок стосовно науково-дослідних можливостей використання в освіті хмарних технологій внесли такі вчені: Г. А. Алексанян [3], В. Ю. Биков [27], М. І. Жалдак [70], М. Ю. Кадемія [82], В. М. Кобися [82], О. Г. Кузьмінська [155], В. М. Кухаренко [5], С. Г. Литвинова [128], Н. В. Морзе [156], В. С. Мкртчян [143], О. С. Свириденко [222], З. С. Сейдаметова [226], С. О. Семеріков [235],

О. М. Спірін [266], Л. В. Рождественська [219], Ю. В. Триус [379; 281; 286], М. П. Шишкіна [26], Б. Б. Ярмахов [183], М. Армбруст (M. Armbrust) [323], Р. Гріффіт (R. Griffith) [323], Ю. Хмелевський (Y. Khmelevsky) [350], М. Міллер (M. Miller) [356], К. Субраманьян (K. Subramanian) [375], Н. Султан (N. Sultan) [358], В. Ю. Чанг (W. Chang) [330], П. Ю. Томас (P. Thomas) [377], А. Фокс (A. Fox) [323] та ін.

Проаналізувавши роботи українських науковців, було з’ясовано, що проводиться робота стосовно подальшого впровадження хмарних сервісів у ЗВО. Більшість досліджень зосереджені на принципах, підходах та проектуванні моделі середовища вищої освіти, до складу якої включено хмарні сервіси [293, с. 108].

Окремою групою виступають дослідження застосування хмарних сервісів у процесі навчання студентів певних спеціальностей. Дослідники Т. Л. Архіпова [7], Н. В. Бахмат [12; 13], Т. В. Зайцева [6], Ю. Г. Лотюк [131], Н. В. Сороко [298], М. А. Шиненко [298] вивчали способи застосування хмарних сервісів у навчальному процесі підготовки майбутніх учителів.

Є спроби науковців В. П. Сергієнко та І. С. Войтовича [242] об’єднання навчальних курсів середовища Moodle з одним або декількома хмарними сервісами. Особливої уваги заслуговує використання хмарних сервісів у процесі дистанційного навчання вищої математики, яке вивчалось Н. В. Рашевською [216], Ю. Г. Лотюком [131].

Зазвичай, хмарні сервіси можна використовувати для візуалізації даних та обчислень, зокрема для розв'язання задач з певної дисципліни та організації індивідуальної та колективної роботи, контролю знань студентів. На думку К. І. Словак [254], завдяки використанню таких хмарних сервісів, як CoCalc, їх роль у навчально-виховному процесі значно зростає. Завдяки використанню інструментарію хмарного сервісу можна підготувати наступні ЕОР:

– опорні конспекти лекцій,
– опорні конспекти практичних робіт,
– розробити курс лекцій,
– розробити систему самостійних та індивідуальних завдань,
– електронні книги з динамічними прикладами.

Якщо використовувати хмарні сервіси у навчанні вищої математики, то це призведе до того, що:

– представлення теоретичного матеріалу стане більш наочним;
– спростяться та автоматизуються громіздкі обчислення;
– набуде подальшого розвитку багаторівневий процес навчання;
– продуктивності навчання студентів [252, с. 22-24].

Завдяки хмарним сервісам, на думку К. І. Словак, викладач може:

– зберігати навчальний матеріал та використовувати його будь де та будь коли;

– використовувати в навчально-дослідних роботах, наприклад з математичних дисциплін;

– організовувати колективну та індивідуальну роботу з залученням відповідного інструментарію;

– застосовувати усі форми контролю та оцінювати навчальні досягнення групи студентів [254].

Найбільший потенціал щодо організації активної самостійної роботи із застосуванням мережних технологій відмічено у системі Sage.

Sage – це безкоштовне вільно поширюване мобільне математичне середовище для виконання чисельних розрахунків та символьних перетворень, а також наочної візуалізації даних [254].

Досвід використання сервісу CoCalc в процесі викладання факультативного курсу [380] викладача Загребського університету Т. Желько (Tutek Željka) в період за 2015-2016 рр., показує, що перш за все треба зорієнтувати зусилля на:

– набуття навичок самостійного використання програмного математичного забезпечення для символьних і чисельних обчислень;

– підтримування викладання математичних понять (зокрема з курсів аналітичної геометрії, лінійної алгебри та математичного аналізу).

Американські дослідники К. Дж. О'Хара (Keith J. O'Hara), Д. Бланк (Douglas Blank), Дж. Маршалл (James Marshall) [359] досліджували чотири способи використання хмарних сервісів в навчальному процесі: під час проведення лекцій (обговорень); семінарських занять; виконання домашнього (індивідуального) завдання; складання іспитів.

Використання хмарних сервісів як альтернатива традиційним презентаціям, може бути досить ефективною підтримкою під час проведення лекції. Якщо розробити лекційну демонстрацію засобами хмарного сервісу, то її можна застосовувати і під час обговорення на семінарському занятті. Аналізуючи досвід американських викладачів роботи з CoCalc, в процесі проведення лекції можна використати робочі аркуші, що містять:

– статичні демонстрації (найчастіше побудови на площині, без зміни вхідних параметрів);

– динамічні демонстрації (з змінними вхідними параметрами);

– заготовки програмного коду, без використання елементів управління (обчислення в реальному часі).

На думку авторів [359], використання CoCalc у процесі навчання математичних дисциплін надає такі переваги:

– у порівнянні з іншими математичними програмними засобами, хмарні сервіси не потрібно встановлювати на свій пристрій, а потім налагоджувати;

– матеріали, які використовувались викладачем на лекції, можна оприлюднити для групи студентів та використати на семінарському занятті;

– створені демонстрації допомагають пояснити складні теми;

– залучає студентів до використання покрокових обчислень в індивідуальних та групових дослідженнях;

– завдяки інтуїтивно зрозумілому інтерфейсу, підвищується рівень засвоєння навчального матеріалу.

В Іспанії виконувався проект «Нові безкоштовні програмні інструменти для автоматичної корекції складних вправ», в рамках якого використовувався сервіс CoCalc. Етапи, в процесі виконання яких було використано інструментарій CoCalc [336]:

1. Підготовка документів з теми та виправлення зауважень.

2. Координація викладачів та впровадження хмарного сервісу.

Використання хмарних сервісів студентами вивчалось дослідниками з Іспанії Е. Кабреро-Грандо (E. Cabrera-Granado), Е. Діас (E. Díaz), О. Г. Кальдерон (O. G. Calderón), Д. Маестре (D. Maestre), Ф. Домінгес-Адамі (F. Domínguez-Adame) [329]. Зокрема було виявлено, що CoCalc обирає більша кількість студентів в порівнянні з іншими хмарними сервісами. Це пояснюється тим, що в більшості випадків підтримка дисциплін ЗВО, зокрема математичних, відбувається за рахунок застосування інструментарію CoCalc. Але студенти не мають жодних проблем з використанням інших хмарних сервісів, під час роботи не виникає особливих труднощів. В той же час, використання хоча б одного хмарного сервісу в рамках вивчення математичної дисципліни покращує засвоєння вивченого матеріалу та поглиблює знання з більшості тем.

Що ж стосується викладачів, то більшість з них обирають CoCalc та рекомендують використання даного хмарного сервісу під час виконання індивідуальних та самостійних робіт. Викладачі досить високо оцінюють потенціал використання хмарних сервісів у процесі навчання будь-якої дисципліни взагалі та математичних зокрема.

Д. І. Кетчсон (David I. Ketcheson) [349] має досвід викладання дисциплін (починаючи з коротких – 2-3 заняття і завершуючи дисциплінами, що тривали декілька семестрів) з використанням сервісу CoCalc. На його думку, CoCalc може бути доповненням основних посібників навчальної дисципліни. Д. І. Кетчесон пропонує організовувати на початку занять засобами CoCalc короткі опитування, задля повторення основних теоретичних викладок. На його думку, це надасть додаткової мотивації для подальшого вивчення теми. Тестування, опитування можна проводити і для того, щоб переконатись, що студенти знайомі з основними методами обчислень. Подібними опитуваннями, можна завершити вивчення теми чи курсу, щоб систематизувати вивчений матеріал та перевірити якість засвоєння його студентами.

У процесі дослідження були виявлені такі переваги використання хмарних сервісів:

– економія ресурсів (зниження навантаження на аудиторний фонд, навколишнє середовище, витрат на придбання та модернізацію комп'ютерної техніки, програмне забезпечення, оплату роботи персоналу);

– мобільність доступу (заняття у міру засвоєння матеріалу в зручний час і в зручному місці);

– еластичність (надання додаткових обчислювальних ресурсів на вимогу користувача).

Недоліки використання CoCalc, на думку Д. І. Кетчесона, є такі:

1. Матеріал важко викласти в повному обсязі.

Це пов'язано з тим, що під час включення до навчального процесу хмарного сервісу значна частина часу витрачається на написання програмного коду. Проте не всі студенти вміють програмувати. Тобто кількість тем, які встигають вивчити студенти, менша. Перш ніж включати до курсу хмарний сервіс, треба бути впевненим, що студенти мають певні навички алгоритмізації та програмування. В першу чергу потрібно впровадити хмарні сервіси до курсів, у яких за програмою передбачено програмування. Краще, якщо матеріалу буде менше, але він буде зрозумілим, аніж до нього студенти втратять інтерес вже на перших заняттях.

2. Масштабованість.

Хоча хмарні сервіси орієнтовані на одночасну роботу великої групи студентів, як показує практика, найоптимальнішим складом академічної групи є приблизно 20 чоловік. Із залученням більшої кількості студентів на занятті викладачу важко індивідуально приділити увагу кожному студенту, пояснити та надати допомогу стосовно виконання завдань. Тобто треба орієнтуватись на кількість студентів, що можна розмістити в стандартній комп'ютерній аудиторії.

3. Нелінійні обчислення в робочому аркуші.

Програмний код розташовано в робочому аркуші в окремих комірках, при чому комірки між собою можуть бути пов'язані, а можуть містити розв'язки окремих завдань, які не залежать один від одного. Використання одних й тих самих змінних може призвести до плутанини, враховуючи те, що програмний код можна виконувати в будь-якому порядку, не зважаючи на попередній програмний код (на попередні комірки). Це не є серйозною проблемою, але студентів потрібно попередити про це.

4. Початок роботи з хмарним сервісом.

Для того, щоб розпочати роботу з хмарним сервісом треба витратити більше часу, аніж подвійне натискання на ярлик файлу. Замість цього потрібно відкрити на своєму пристрої браузер, авторизуватись в системі та відкрити проект, в якому зберігається робочий аркуш. Як варіант – не завершувати кожного разу сеанс роботи в системі на своєму пристрої, щоб швидше відновити сесію та роботу із завданням.

5. Тривалість виконання програмного коду.

Програмуючи онлайн немає можливості використовувати звичні налаштування текстового редактору. Якщо обчислень небагато, то на виконання коду в комірці не знадобиться багато часу. Що ж стосується громіздких обчислень, особливо коли йде паралельна робота над одним робочим аркушем групи студентів, то виконання може сповільнюватись. Треба ще враховувати швидкість та налаштування Інтернет-зв'язку.

Вирішенням даної проблеми може бути поділ громіздкого програмного коду на частини, виконання яких потрібно організувати в окремих комірках.

## 1.5 CoCalc у системі засобів навчання математичних дисциплін

У загальній педагогіці термін «засоби навчання» трактується кожним науковцем по-різному. В. А. Нікіфоров під засобами навчання розуміє педагогічну майстерність викладача та матеріальні об'єкти, які залучаються ним в процес навчання, що містять навчальний матеріал, допомагають організувати та керувати навчальною діяльністю [162, с. 89].

До питання класифікації засобів навчання різні колективи авторів мають декілька підходів. О. М. Новіков поділяє усі засоби навчання на п'ять груп: матеріальні, інформаційні, логічні, математичні та мовні засоби навчання. Перші дві групи створюються безпосередньо самим викладачем [164, с. 120-121].

Згідно досліджень В. В. Краєвського та А. В. Хуторського засоби навчання можна класифікувати по-різному в залежності від того що взяти за основу класифікації. Ними запропоновано наступні ознаки: склад об'єктів, відношення до джерела появи, складність, спосіб використання, особливості будови, характер впливу, носій даних, рівень змісту освіти, відношення до технологічного процесу [111, с. 269-270].

Так, класичною вважається класифікація, у якій усі засоби навчання поділяються на ті, що залежать від мови та дій викладача, та залежних від навчального забезпечення (рис. 1.2).

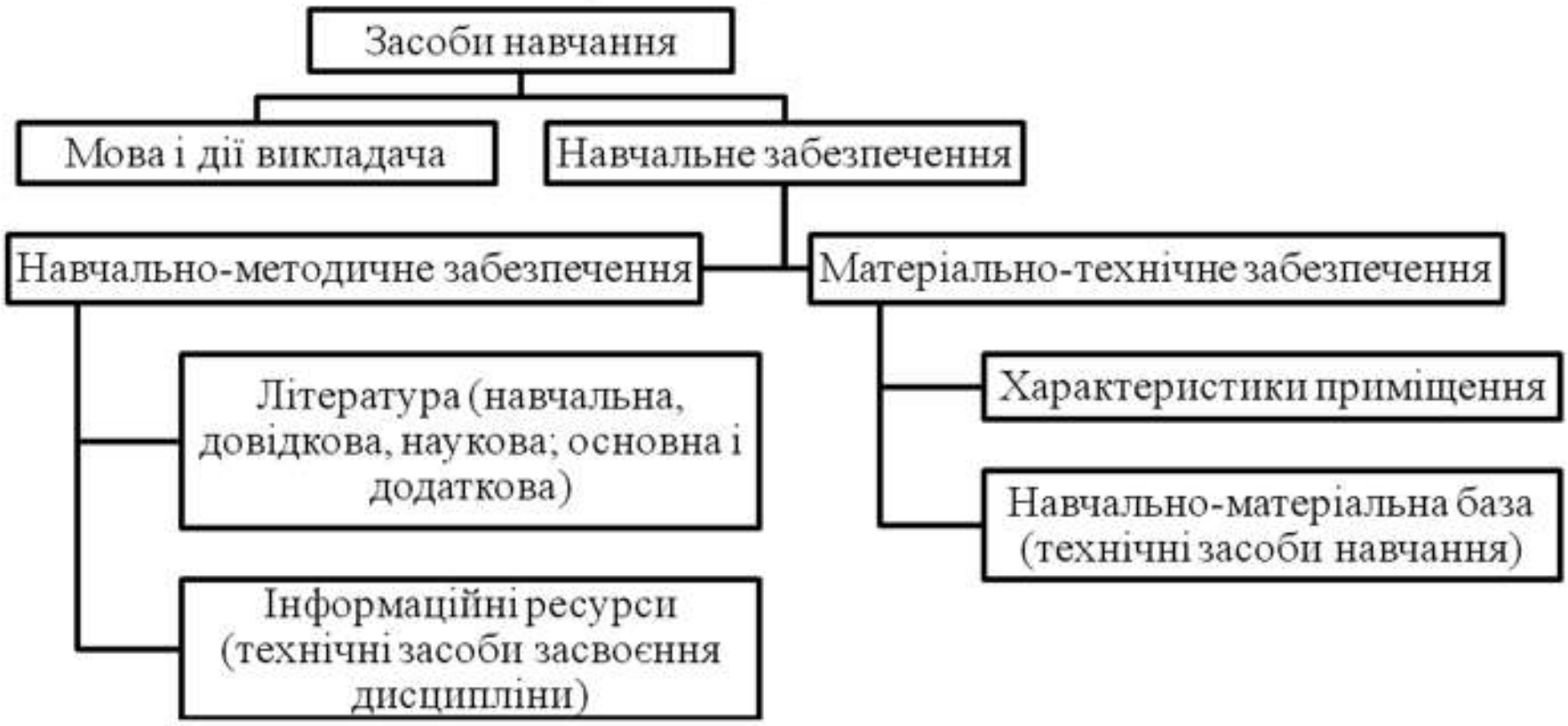

Рис. 1.2. Класифікація засобів навчання (за В. А. Нікіфоровим)

Проте класифікація В. А. Нікіфорова є досить широкою та узагальненою по відношенню до процесу вивчення математичних дисциплін. Так, в даній класифікації хмарний сервіс CoCalc можна віднести до навчально-методичного забезпечення, а саме до інформаційних ресурсів.

Оскільки, на думку Н. П. Волкової [42] під засобами навчання потрібно розглядати лише матеріальні засоби, тому запропонована нею класифікація (рис. 1.3) дещо відрізняється від попередньо розглянутих і значно звужує засоби навчання, які можна використати по відношенню до процесу навчання математичних дисциплін.

Технічні засоби навчання (ТЗН) – обладнання та технічні пристрої, які використовуються в навчальному процесі та забезпечують його ефективність.

Проте, якщо використати класифікацію Н. П. Волкової, CoCalc посяде місце серед технічних засобів навчання та охопить майже всі засоби зорової наочності: екранні засоби та технічні моделі.

На відміну від Н. П. Волкової, О. П. Буйницька [36] пропонує наступний поділ ТЗН: за призначенням (широкого і спеціального), виконуваними функціями та способом впливу. Для нашого дослідження буде цікавою остання класифікація, оскільки Н. П. Волкова в своїй класифікації за основу поділу обрала ознаку за способом впливу (візуальні, аудитивні та аудіовізуальні). Якщо розглянути яке саме місце займає хмарний сервіс CoCalc в класифікації, що представлена Н. П. Волковою, тоді буде враховано лише частину складових хмарного сервісу. Система управління навчальними курсами та редактор LaTeX залишаться позаувагою дослідження. Тому доречно буде зупинитись на класифікації навчальних засобів, що пропонує О. П. Буйницька.

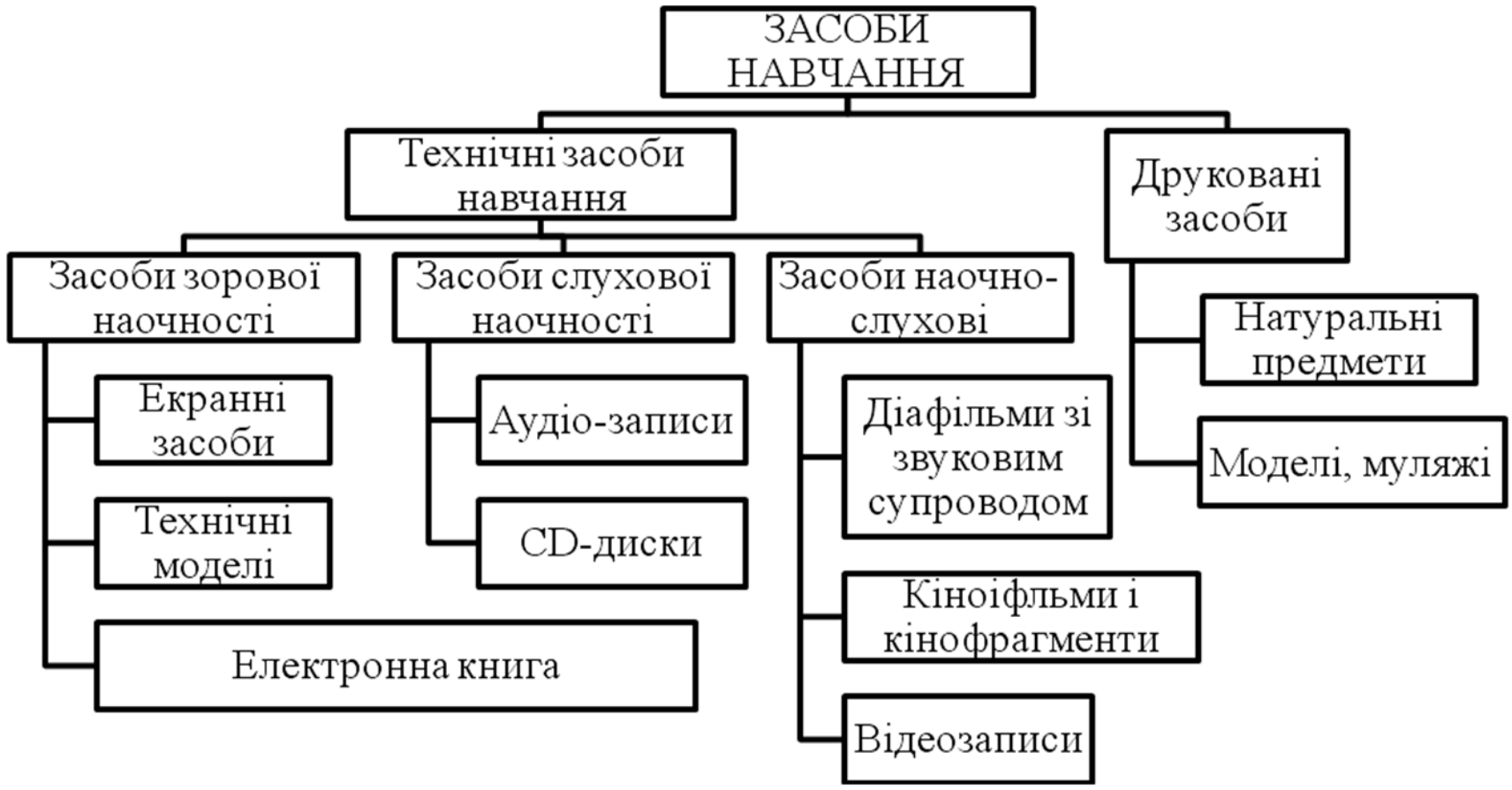

Рис. 1.3. Види засобів навчання (за Н. П. Волковою)

Проблему підготовки кваліфікованих кадрів управління освітою, а також вчителів, орієнтованих на навчання на основі ІКТ, на сьогодні навряд чи можна розглядати окремо від процесів інноваційного розвитку освітнього простору, утвореного в школі, регіоні та в освітній системі країни чи світу. У зв'язку з цим існує необхідність проведення фундаментальних досліджень з акцентом на можливі шляхи розвитку освітнього середовища освітніх установ. Варто взяти до уваги тенденції вдосконалення засобів ІКТ при пошуку нових технічних рішень і нових технологічних, педагогічних та організаційних моделей [306]. Основний акцент поставлено на перехід від масового впровадження окремих програмних продуктів, до комплексного та комбінованого середовища, яке підтримує розподілені мережні послуги і крос-платформні рішення [306].

Нові технології, інформаційно-комунікаційні мережі створюють підстави для реалізації цілісного підходу до освіти та підготовки кадрів [370]. Цілісний підхід фокусується на об'єднанні науки і практики, навчання і виробництва, фундаментальних та прикладних знань і технологічних компетентностей. Насамперед він спрямований на розвиток навичок управління в галузі освіти, які повинні бути засновані на об'єднаному підході до навчання, проектування та управління. Це – перспективний напрям для розвитку кадрового потенціалу системи освіти. Тому для організації та розвитку середовища навчання і підготовки кваліфікованих педагогічних кадрів необхідні нові підходи і моделі.

Існує проблема доступності та способів навчання і постачання ресурсів для досягнення кращого педагогічного ефекту їх використання. Ця проблема може бути частково вирішена завдяки використанню обчислювальних потужностей у хмарі. Основною перевагою даної технології є покращення доступу до якісних ресурсів (а іноді і єдино можливим способом доступу до необхідних ресурсів для всіх). Ідея полягає в тому, щоб визначити підходи до моделювання та оцінювання компонентів та обчислювальних потужностей хмари на основі вивчення потреб проектування навчального середовища і функціонування різних інструментів для його організації.

На засоби навчання суттєво впливають розвиток інформаційного суспільства і технічний прогрес. Тому останнім часом окрім класичних засобів навчання, які можна було використовувати в процесі вивчення будь-яких дисциплін з’являються нові.

Так, М. А. Кислова до складу авторського мобільного навчального середовища з вищої математики включає:

– мобільні засоби підтримки навчальної та, зокрема, математичної діяльності,

– мобільні засоби навчальної комунікації,

– мобільні засоби підтримування процесу навчання вищої математики, з використанням яких відбувається взаємодія між студентами та викладачем [85].

Окрім мобільних засобів навчання, дослідники в публікаціях останніх років розглядають хмарні сервіси в якості засобів навчання. М. А. Кислова, К. І. Словак [87] проаналізували ряд хмарних сервісів, які пропонують використовувати у навчальному процесі в поєднанні з традиційними засобами навчання: Google Apps for Education, Office 365, ThinkFree Online. Зокрема, охарактеризовано кожен хмарний сервіс, виділено його характеристики, складники та окреслено переваги його використання як засобу навчання математичних дисциплін.

Для того, щоб обрати хмарний сервіс можна скористатись такими критеріями добору [329]:

– порівняти обчислювальні ресурси (RAM, кількість доступних ядер), обсяг даних, який може обчислювати хмарний сервіс, при чому на ці характеристики не впливають обчислювальні потужності пристрою на якому працює користувач та надаються безкоштовно;

– наявність інструментів для організації навчання та його контролем (треба враховувати чи наявний розподіл вправ на рівні, можливість збирання виконаних завдань та їх оцінювання);

– можливість збільшення обчислювальних ресурсів за невелику

оплату, порівняння тарифних планів;

– відкритість програмного коду, можливість встановлення власних налаштувань та додатків (спеціальних бібліотек), окрім тих, що передбачені за замовчуванням (індивідуалізувати роботу та налаштування хмарного сервісу для окремого студента чи викладача);

– можливість спільного редагування, одночасної роботи над однією проблемою (проектом) групи студентів, ресурсів різних форматів.

За Н. В. Рашевською, до переваг хмарних сервісів можна віднести такі:

– дані доступні користувачу з будь-кого пристрою, який має вихід в Інтернет;

– користувач має змогу працювати з усім навчальним матеріалом, при цьому не встановлюючи додаткового програмного забезпечення;

– можливість працювати будь-де, а не лише в межах аудиторії;

– організація процесу навчання носить характер змішаного типу [216].

Але М. А. Кислова та К. І. Словак вважають недостатнім використання лише хмарних сервісів, в якості засобів навчання. У процесі вивчення математичних дисциплін доречно звернути увагу на спеціалізовані web-орієнтовані версії СКМ (принцип роботи яких базується на хмарних технологіях), до яких віднесено Sage. Окрім Sage, М. А. Кислова, К. І. Словак розглянули GeoGebra, MathCAD Calculation Server, MapleNet, Web-Mathematica. Кожен програмний засіб представлено у вигляді короткої характеристики, який можна використати в якості засобу навчання [87].

У процесі дослідження СКМ, web-орієнтованих версій СКМ математичного призначення, було визначено, що нині вже існують хмарні версії математичного програмного забезпечення відомих виробників, зокрема, такі як Maple Net, MATLAB web-server, WebMathematica та інші. Дані системи отримали назву «системи комп’ютерної математики», їх сутність і різновиди більш докладно висвітлені в [371]. В той же час, існує тенденція розвитку даних систем, що і раніше функціонували у мережному середовищі (так звані Web-СКМ) щодо їх поступової трасформації у хмаро орієнтовані системи. Основні відмінності Web-СКМ і хмаро орієнтованих СКМ більш докладно висвітлені в [371]. Різновидом саме такого типу систем – хмаро орієнтованих Web-СКМ – є CoCalc, хмаро орієнтована версія Web-СКМ Sage [371].

Сучасні СКМ можна розподілити на сім основних типів, але незважаючи на те, що кожна з цих СКМ має певні відмінності в своєму призначенні та архітектурі, прийнято вважати, що вони мають схожу структуру [305, с. 154]:

– центральне місце займає обчислювальне ядро системи;

– коди великої кількості скомпільованих функцій та процедур, які мають виконуватись достатньо швидко, тому зазвичай об'єм ядра прийнято максимально зменшувати;

– зручний інтерфейс, завдяки якому користувач може з легкістю звертатись до обчислювального ядра, та одержувати результат безпосередньо на екрані монітору;

– потужний графічний інструментарій розширює використання СКМ не лише для математичних обрахунків, але й завдяки йому можна ілюструвати більшість процесів нематематичного характеру;

– пакети розширень, за допомогою яких можливості використання СКМ значно зростають, за рахунок виконання більшої кількості завдань, які ставить користувач;

– бібліотеки процедур та функцій, які дають змогу використовувати менш вживані, але не менш важливі рідкісні процедури, що просто не ввійшли до складу ядра, через обмеження його розмірів;

– довідкова система, завдяки якій користувач може в будь-який момент звернутись до кожного розділу з приводу коректного використання тієї чи іншої функції, синтаксису та прикладів застосування.

У СКМ реалізовано значну кількість спеціальних математичних операцій, функцій та методів [305, с. 156]:

– розкриття дужок у символьних виразах;

– обчислення значення числового виразу;

– розклад многочленна на множники;

– обчислення значення символьного виразу, але при умові, що відомо значення змінних величини;

– зведення подібних доданків без розкриття дужок;

– розв'язання алгебраїчних рівнянь, чи системи рівнянь;

– розв'язання трансцендентних рівнянь, або наближеного значення коренів рівнянь;

– виконання операцій математичного аналізу: обчислення інтегралів, кратних інтегралів, знаходження первісних, границь функцій та числових послідовностей; розв'язання диференціальних рівнянь (аналітичним способом);

– побудова графіків функцій на площині та в просторі, побудова векторів; обчислення з розділу лінійної алгебри (множення матриць, обчислення детермінантів, піднесення квадратної матриці до будь-якого натурального степеню) та багато інших.

Перший етап розвитку СКМ (60-ті – 90-ті роки XX ст.) характеризується тим, що СКМ поширювалися у локальному варіанті –

треба було інсталювати програмне забезпечення на комп'ютер користувача і тоді можна було використовувати його у відповідності із ліцензійними умовами.

Другий етап розвитку СКМ (90-ті роки XX ст. – перше десятиріччя XXI ст.) відзначився виникненням Web-СКМ, у яких однією з основних характеристик була оснащеність Web-інтерфейсом. Web-СКМ мають такі властивості:

– не має потреби встановлювати обчислювальне ядро системи на клієнтській машині;

– виконання усіх обчислень відбувається безпосередньо на Web-сервері;

– виконання запиту та одержання результатів обчислення відбуваються за допомогою Web-браузера.

Крім цього, виокремлюють такі характеристики Web-СКМ:

– невимогливість до апаратної складової обчислювальної системи;

– індиферентність до використовуваного браузера;

– простота адміністрування;

– мобільний доступ до навчальних ресурсів, програм і даних та ін. [305, с. 156].

Сьогодні до найбільш поширених Web-СКМ відносять MathCAD Application Server (MAS), MapleNet, Matlab Web Server (MWS), webMathematica, wxMaxima та Sage. Web-СКМ оснащені зручним інтерфейсом, потужними графічним інструментарієм, в них реалізовано значну кількість виконання математичних обчислень, функцій та методів.

Серед специфічних характеристик сучасних СКМ можна зазначити:

– наявність властивих цим системам мов програмування;

– імпортування даних з інших програмних продуктів;

– засоби друку математичних текстів.

Використання Web-СКМ SAGE має ряд переваг. Серед них – найбільший арсенал засобів щодо розроблення та дослідження різноманітних математичних моделей в межах математичних дисциплін. Перша версія SAGE вийшла у лютому 2006 року. Основні характеристики Web-СКМ Sage наведені в таблиці (табл. 1.7).

Завдяки використанню Web-СКМ Sage у процесі навчання математичних дисциплін забезпечується можливість [305]:

1. Виконувати обчислення: як аналітичні, так і чисельні.

2. Подавати результати обчислень природньою, математичною мовою з використанням символіки.

3. Будувати дво- і тривимірні графіки кривих і поверхонь, гістограми і будь-які інші зображення (виключаючи анімації).

4. Поєднувати обчислення, текст і графіку в рамках одного робочого

аркуша, їх друкування, оприлюднення в мережі і спільної роботи над ними.

5. Створювати за допомогою вбудованої у Sage мови Python моделі для виконання практичних завдань, навчальних досліджень.

6. Створювати нові функції і класи мовою Python.

*Таблиця 1.7*

**Основні характеристики Web-СКМ Sage**

| Переваги | Недоліки |
|---|---|
| – відкритість системи;<br>– вільне поширення;<br>– повнофункціональний Web-сервер системи;<br>– інтеграція більше 100 математичних пакетів у єдиному середовищі, тощо. | – недостатньо вітчизняної науково-методичної літератури;<br>– не досить висока швидкодія;<br>– складність опанування, громіздкий інтерфейс;<br>– недостатньо персоніфікованого доступу. |

Третій етап розвитку СКМ (починаючи з 2009 р.) пов'язаний з виникненням хмаро орієнтованих систем. CoCalc – це безкоштовний сервіс, що існує за підтримки Університету Вашингтона, Національного наукового фонду і Google. CoCalc було розроблено спеціально для полегшення використання математичних обчислень на платформі Android. В CoCalc реалізовано усі властивості, які є у Web-СКМ SAGE, але є й певні відмінності (табл. 1.8). [305, с. 155].

*Таблиця 1.8*

**Основні характеристики CoCalc**

| Переваги | Недоліки |
|---|---|
| – покращено інтерфейс в порівнянні з Web-СКМ Sage;<br>– можливість інтеграції з іншими сервісами;<br>– обліковий запис користувача;<br>– одночасна робота в проекті більше ніж 300 користувачів;<br>– можливість розробки web-додатків з використанням мови Python;<br>– збільшено обчислювальні потужності в декілька разів;<br>– підтримка нових форматів: курси, завдання, чати;<br>– можливість колективної роботи в рамках одного проекту. | – файли не завжди завантажуються на пристрій;<br>– конвертування в файл формату PDF можливо лише за умови індивідуальної роботи з проектом;<br>– файли старого формату не підтримуються;<br>– змінено синтаксис основних функцій (попередній не працює);<br>– відсутній остаточний захист файлів користувача. |

Строго кажучи, Sage – консольна програма. І всі розрахунки цілком можна робити в консолі. Інше питання, що це незручно, і тому були створені Sage Notebook і CoCalc. Історично першим з'явився Sage Notebook. Це графічна оболонка з трохи застарілим, але зручним інтерфейсом, яка більше підходить для роботи з Sage на локальному комп'ютері. Sage Notebook йде в комплекті з Sage і є можливість виконувати розрахунки прямо в браузері. Sage Notebook підтримує практично всі можливості консольної Sage (за винятком команд налагодження коду), окрім цього можна: оформлювати аркуші, додавати зображення, відео, 2d і 3d графіку, створювати інтерактивні додатки.

Виконавши певні налаштування, можна зробити так, щоб працювати з Sage Notebook через мережу Інтернет. Існують сайти, що надають доступ до Sage Notebook, проте на них використовується стара версія Sage. Для роботи з Sage онлайн рекомендовано використання CoCalc.

Ідея створення CoCalc належить професору математики університету Вашингтона Вільяму Стейну. Більшість серверів розташовано в США на території університету Вашингтона. Є кілька серверів в Європі. На серверах встановлено операційна система Ubuntu 14.04.1 LTS (Trusty) з ядром web-СКМ Sage. Також використовується Google App Engine.

Для програм на Linux характерна така особливість: в них можна чітко виділити консольну частину, яка містить в собі всі функції програми, і графічну частину, з якої зазвичай працює користувач. При цьому легко зробити так, щоб сама програма працювала на одному комп'ютері, а її графічна частина – на іншому.

У даний час для роботи з Sage використовуються різні графічні середовища. Консольний варіант Sage зазвичай використовується для різних вузьких завдань, наприклад, для: побудови власних математичних веб-додатків, налагодження коду, конвертації файлів в різні формати.

Основним джерелом даних і ресурсом для спілкування є форум користувачів SMC. Нещодавно відкрився форум для розробників після остаточного переходу SMC в розряд відкритих проектів. Активне спілкування відбувається в чаті Gitter; створене FAQ на Github. Щомісячно кількість користувачів поступово зростає (Додаток Б).

Принцип роботи в CoCalc побудовано на створенні індивідуальних або групових проектів, наповненні їх навчальними ресурсами та роботі з окремими ресурсами чи групою ресурсів одночасно. Також в системі передбачено збереження дій користувачів, що відображається в хронологічному порядку. Можлива функція відображення історії роботи з окремим навчальним ресурсом (чи проектом) як певного користувача, так і групи користувачів. Внесення певних змін до кожного проекту призводить до резервного копіювання структури самого проекту. Усі копії зберігаються в хронологічному порядку із зазначенням автора змін.

Враховуючи вищезазначені переваги хмарних сервісів у навчанні математичних дисциплін, а також перспективи впровадження у навчальний процес хмарного сервісу CoCalc, що є вільнопоширеною, на відміну від більшості різновидів математичного програмного забезпечення інших виробників, і в той же час досить потужною, щоб забезпечувати досягнення цілей навчального процесу, застосування цієї системи було обрано предметом експериментального дослідження.

Застосування хмарних сервісів приводить до появи та розвитку форм організації навчання, орієнтованих на спільну навчальну діяльність в мережі Інтернет.

Узагальнюючи проведене дослідження, хмарні сервіси у навчанні майбутніх вчителів математики доцільно використовувати як засоби для:

– комунікації (синхронної – чати, голосовий та відеозв'язок та асинхронної – пошта, форуми);

– співпраці: доступ до даних, обмін ними та співпраця з іншими користувачами;

– зберігання та опрацювання даних.

Напрями використання CoCalc у навчанні майбутніх вчителів математики є такими:

1) організація та документування предметної навчальної комунікації;

2) підтримка індивідуальних та групових форм організації навчальної діяльності (аудиторна та позааудиторна);

3) підтримання управління навчанням;

4) забезпечення наочності шляхом різних інтерпретацій математичних моделей, візуалізації математичних абстракцій тощо;

5) забезпечення доступності та науковості через використання спільного інтерфейсу доступу до об'єктів середовища та використання надійного програмного забезпечення з відкритим кодом;

6) підвищення засобової, часової, просторової мобільності;

7) формування єдиного навчального середовища, змістова складова якого розвивається у процесі навчання [195].

## Висновки до розділу 1

1. У ході дослідження обґрунтовано ключове поняття: *професійна компетентність вчителя математики* – це здатність особи на основі знань, умінь, навичок та особистісного ставлення здійснювати професійну діяльність з навчання математики учнів та досягати певних результатів. Розглянуто загально професійні та спеціально професійні компетентності майбутніх вчителів математики. Виявлено спеціальні професійні компетентності, формування яких є доцільним з використанням хмарних сервісів: 1) здатність використовувати професійно профільовані знання у галузі математики, для статистичного опрацювання експериментальних даних і математичного моделювання природних явищ і процесів; 2) здатність використовувати математичний апарат для моделювання різноманітних процесів; 3) здатність до роботи з комп'ютером на рівні користувача та фахівця у галузі ІКТ.

2. У процесі дослідження вітчизняного та зарубіжного досвіду були виявлені такі переваги використання хмарних сервісів математичного призначення: економія ресурсів (зниження навантаження на аудиторний фонд, навколишнє середовище, витрат на придбання та модернізацію комп'ютерної техніки, програмне забезпечення, оплату роботи персоналу); мобільність доступу (заняття у міру засвоєння матеріалу в зручний час і в зручному місці); еластичність (надання додаткових обчислювальних ресурсів на вимогу користувача).

3. Застосування хмарних сервісів призводить до появи та розвитку форм організації навчання, орієнтованих на спільну навчальну діяльність в мережі Інтернет. Показано, що хмарні сервіси у навчанні майбутніх учителів математики доцільно використовувати як засоби для: комунікації (синхронної – чати, голосовий та відеозв'язок та асинхронної – пошта, форуми); співпраці (доступ до даних, обмін ними та співпраця з іншими користувачами); зберігання та опрацювання даних.

4. Визначено напрями використання CoCalc у навчанні майбутніх вчителів математики: організація навчальної комунікації; підтримування індивідуальних та групових форм організації навчальної діяльності (аудиторна та позааудиторна); підтримування управління навчанням; забезпечення наочності шляхом побудови різних інтерпретацій математичних моделей, візуалізації математичних абстракцій тощо; забезпечення доступності та науковості завдяки використанню спільного інтерфейсу доступу до об'єктів середовища та надійного програмного забезпечення з відкритим кодом; підвищення часової та просторової мобільності; формування єдиного навчального середовища, зміст якого розвивається у процесі навчання.

# РОЗДІЛ 2
# МОДЕЛЮВАННЯ ПРОЦЕСУ ВИКОРИСТАННЯ ХМАРНОГО СЕРВІСУ COCALC ЯК ЗАСОБУ ФОРМУВАННЯ ПРОФЕСІЙНИХ КОМПЕТЕНТНОСТЕЙ УЧИТЕЛЯ МАТЕМАТИКИ

## 2.1 Проектування системи професійних компетентностей учителя математики

За Г. М. Романовою, проектування – це здатність намічати, окреслювати план дій, конструювати, планувати, та здійснювати задум, намір. Це створення прототипу, прообразу передбачуваного об'єкта [220, с. 219].

Система компетентностей в освіті має ієрархічну структуру, рівні якої складають:

– ключові компетентності (міжпредметні та надпредметні компетентності) – здатність людини здійснювати складні поліфункціональні, поліпредметні, культурнодоцільні види діяльності, ефективно розв'язуючи актуальні індивідуальні та соціальні проблеми;

– загально-галузеві компетентності – компетентності, які формуються учнем впродовж засвоєння змісту тієї чи іншої освітньої галузі у всіх класах середньої школи і які відбиваються у розумінні «способу існування» відповідної галузі – тобто того місця, яке ця галузь займає у суспільстві, а також вміння застосовувати їх на практиці у рамках культурнодоцільної діяльності для розв'язку індивідуальних та соціальних проблем;

– предметні компетентності – складова загально–галузевих компетентностей, яка стосується конкретного предмету.

При цьому потрібно мати на увазі, що кожна ключова компетентність повинна «проектуватись» на загальногалузеві компетентності, які у свою чергу повинні «проектуватись» на предметні компетентності (під терміном «проектуватись» розуміється таке визначення компетентностей нижчого рівня, щоб вони у сукупності забезпечували компетентності вищих рівнів). Разом із тим, ключові компетентності не складаються просто з набору відповідних галузевих та предметних компетентностей – вони інтегрують галузеві компетентності у складну структурну компоненту, у якій елементи пов'язані між собою різноманітними зв'язками та відношеннями [208].

Ураховуючи те, що наша ціль – спроектувати систему професійних компетентностей учителя математики, спираючись на досвід С. А. Ракова можна сказати, що до неї можна включити два основних компонента: загально професійні компетентності (за С. А. Раковим – загально-галузеві) та спеціально професійні компетентності (за С. А. Раковим – предметні).

У структурі системи соціально-професійних компетентностей фахівця можна виділити чотири блоки, перші два з яких є базовими і необхідними для формування двох наступних:

1. *інтелектуальний* – сукупність сформованих у випускника прийомів розумової діяльності (аналіз, синтез, порівняння, співставлення, класифікація, систематизація, узагальнення та ін.);

2. *особистісний* – особистісні якості випускника (відповідальність, цілеспрямованість, самостійність, організованість тощо);

3. *соціально-значущі компетентності* – компетентності, володіння якими дає змогу забезпечити життєдіяльність випускника у сучасному світі та його взаємодію з іншими людьми, групою, колективом;

4. *професійні компетентності* – компетентності, набуття яких дає змогу випускнику виконувати професійну діяльність [212].

Тобто, враховуючи здобутки М. В. Рафальської, можна узагальнити, що під час визначення складників спеціально-професійних компетентностей потрібно зосередити увагу на професійній діяльності вчителя математики. На це звертає увагу і колектив науковців М. І. Жалдак, Ю. С. Рамський та М. В. Рафальська: «Таким чином, формування професійних компетентностей учителя інформатики передбачає набуття ним компетентностей у галузі інформатики та суміжних з нею дисциплін, методики навчання та дидактики, психологічних і педагогічних основ здійснення навчально-виховного процесу, дослідницької діяльності та педагогічного спілкування, що визначає якість його професійної діяльності» [71].

О. В. Співаковским [264] обґрунтовано систему критеріїв рівнів професійної готовності майбутнього вчителя математики: за здібностями та нахилами. Задля проектування структури спеціально професійних компетентностей розглянемо лише наступні критерії: здібності (організаторські, академічні та практичні) та нахили (до науково-педагогічної діяльності, до творчості та професійна спрямованість).

*Здібності*

1. Організаторські: уміти організувати вчителів для розробки, удосконалення методик і засобів навчання та виховання; науково організовувати свою працю та роботу колег і учнів.

2. Академічні: мати достатньо високий рівень володіння системою математичних знань і умінь, уявлень про ідеї і методи математики; розвинене математичне мислення; орієнтуватися в питаннях психології, педагогіки і методики навчання математики та основ інформатики; уміння поєднувати в роботі фундаментальну, наукову і практичну підготовку; достатньо високий рівень успішності.

3. Дидактичні: володіти сучасними методами та навичками навчальної та виховної роботи; викликати в учнів інтересу до навчального предмета.

*Нахили*

1. До науково-педагогічної діяльності: до удосконалення методів та засобів навчально-виховного впливу на особистість.

2. До творчості: до творчої інноваційної діяльності; неперервне поповнення своїх знань зі спеціальності; науково-дослідна робота.

3. Професійна спрямованість: прагнення займатись викладацькою діяльністю в навчальних закладах різних рівнів і типів.

Можна відмітити, що дидактичні здібності тісно переплітаються з нахилами до науково-педагогічної діяльності, організаторські – з нахилами до професійної спрямованості, академічні здібності – з творчими нахилами. Тобто можна помитити, що до складу спеціально-професійних компетентностей буде віднесено хоча б три складові, які можна охопити вищезазначеними критеріями.

Оскільки в даному дослідженні система професійних компетентностей учителя математики розглядається як сукупність двох складових: загально професійних та спеціально професійних компетентностей, то критерії: здібності (комунікативні, перцептивні та мовленнєві) та нахили (загально-культурний рівень, соціальна активність та самооцінка) будуть відповідати загально професійним компетентностям.

Але не можна розглядати вимоги до професійної підготовки вчителя лише з урахуванням вказаних критеріїв. Формування професійних компетентностей учителя математики здійснюється у процесі навчання циклів гуманітарної і соціально-економічної підготовки, природничо-наукової підготовки та професійної і практичної підготовки у педагогічному ЗВО. Причому найбільший загальний навчальний час (обсяг в годинах) відведено для циклу професійної і практичної підготовки, значна частина якого – математичні дисципліни (підцикл професійної науково-предметної підготовки циклу професійної і практичної підготовки). Формування системи професійних компетентностей учителя математики на базі опанування циклами дисциплін, що входять до програми його підготовки можна прослідкувати на прикладі саме математичних дисциплін (за рахунок яких формуються переважно спеціально професійні компетентності). Ця думка підтверджується дослідженнями М. І. Жалдака, Ю. С. Рамського та М. В. Рафальської: «формування загальнопрофесійних компетентностей учителя інформатики … значною мірою відбувається у процесі навчання природничо-математичних, психолого-педагогічних, предметно і професійно орієнтованих методичних дисциплін, а подальший їх розвиток і використання здійснюється у процесі навчання дисциплін соціально-економічної, предметної підготовки та курсів за вибором» [71].

С. А. Раков [208] до предметно-галузевих математичних компетентностей відносить наступні:

1. Процедурна компетентність – уміння розв'язувати типові математичні задачі.

2. Логічна компетентність – володіння дедуктивним методом доведення та спростування тверджень

3. Технологічна компетентність – володіння сучасними математичними пакетами.

4. Дослідницька компетентність – володіння методами дослідження соціально та індивідуально значущих задач математичними методами.

5. Методологічна компетентність – уміння оцінювати доцільність використання математичних методів для розв'язування індивідуально і суспільно значущих задач.

У разі поєднання логічної та дослідницької компетентності можна одержати складову спеціально-професійних компетентностей – наукові компетентності, які в свою чергу є складником предметних компетентностей. Технологічні компетентності можна розглянути дещо ширше, враховуючи не лише сучасні математичні пакети, але застосування ІКТ в цілому. Процедурну компетентність можна віднести до складу предметних компетентностей.

Проектувальну діяльність можна розглядати як засіб становлення фахової компетентності особистісно орієнтованого навчання учасників освітнього процесу, організації та забезпечення їх співробітництва із використанням хмарних сервісів, спільної діяльності у процесі професійної підготовки. Залучити студентів – майбутніх педагогів до індивідуальної проектної діяльності – важливе актуальне завдання. Це можна зробити можна за рахунок застосування хмарних сервісів підтримування процесів виконання науково-дослідної роботи. Здійснення проектувальної діяльності має поєднувати індивідуальні та групові форми роботи (тим більше, якщо розглядати проектувальну діяльність у контексті взаємодії та співробітництва) [220, с. 220]. Тобто проектувальну діяльність можна розглядати як складову наукових компетентностей.

Об'єктивними вимогами до сучасного вчителя є володіння ІКТ, виявлення творчої ініціативності та прагнення до самоосвіти. Як показує досвід, активність тих, хто навчається – майбутніх вчителів залежить не лише від їх професійно-методичної підготовленості, але і від їх компетентності в галузі інформаційно-комп'ютерних технологій [143, с. 16].

Розвиток ІКТ, впровадження їх у вищу систему освіти вимагає удосконалення раніше сформованих знань та вмінь від усіх суб'єктів навчального процесу: тих, хто навчає і тих, хто навчається [143, с. 16].

Одна з суттєвих перешкод на шляху впровадження технології хмарних обчислень (ТХО) полягає не тільки в недостатньо високому рівні ІКТ компетентностей викладацького складу та відсутності відповідного науково-методичного забезпечення, а в першу чергу у відсутності чіткого уявлення про можливі педагогічні моделі застосування переваг подібних технологій з метою підвищення ефективності викладання вищої математики у навчальному процесі вищих технічних навчальних закладах освіти [149, с. 117].

Ми не будемо в структурі системи окремо зазначати складову – ІКТ компетентність, оскільки на наш погляд складники ІКТ компетентності тісно поєднані з усіма спеціально професійними компетентностями вчителя математики. На противагу їй розглянемо інформаційно-технологічні компетентності, склад яких значно вужчий в порівнянні з ІКТ компетентністю.

О. М. Спірін пропонує таку структуру компенентостей майбутніх вчителів інформатики [266]:

І. Загальні компетентності:

– компетентність щодо індивідуальної ідентифікації й саморозвитку;

– міжособистісна компетентність;

– суспільно-системна компетентність.

ІІ. Професійно-спеціалізовані компетентності:

– загальнопрофесійна;

– предметно-орієнтована, або профільно-орієнтована (до її складу включено – науково-предметні компетентності та предметно педагогічні компетентності);

– технологічна (складається з компетентності в галузі педагогічних технологій та інформаційно-технологічні компетентності);

– професійно-практична (ці компетентності треба розуміти як такі, якими має володіти випускник з позицій майбутньої професійної діяльності).

При цьому потрібно зазначити, що О. М. Спірін до складу предметно-педагогічних компетентностей включає: уявлення про основні концепції навчання, розуміння основних змістових ліній шкільного курсу інформатики, готовність керувати гуртковою роботою, здатним проводити консультації тощо.

Компетентність в галузі педагогічних технологій включає в себе:

володіння методиками організації навчально-виховного процесу, готовність до вдосконалення технології навчання, досвід організації дистанційного навчання, виготовлення дидактичних матеріалів на паперових та електронних носіях, використовувати в навчанні інформаційні джерела, готовність до проведення психолого-педагогічних досліджень.

М. І. Жалдак, Ю. С. Рамський та М. В. Рафальська пропонують наступну систему професійних компетентностей учителя інформатики [71]:

1. Загальнопрофесійні компетентності:

– дидактико-методичні компетентності;

– організаційно-управлінські компетентності;

– психолого-педагогічні компетентності;

– дослідницькі компетентності;

– комунікативні компетентності;

– природничо-математичні компетентності.

2. Предметні компетентності:

– інформологічно-методологічні компетентності;

– інформаційно-технологічні компетентності;

– комп'ютерні компетентності;

– модельні компетентності;

– алгоритмічні компетентності.

У дослідженні розглянуто професійні компетентності майбутніх вчителів математики: загальнопрофесійні та спеціально професійні. При цьому було виокремлено спеціально професійні компетентності, формування яких є доцільним з використанням хмарних сервісів, а саме:

– здатність використовувати професійні знання в галузі математики, для статистичного опрацювання експериментальних даних і математичного моделювання природних явищ і процесів;

– здатність використовувати математичний апарат для моделювання різноманітних процесів;

– здатність роботи з комп’ютером на рівні користувача та фахівця в галузі ІКТ.

Узагальнюючи систему професійних компетентностей учителя математики (побудовану на основі запропонованих систем О. М. Спіріна та М. І. Жалдака, Ю. С. Рамського, М. В. Рафальської) та проведене дослідження, можна зупинитись на таких її складових (рис. 2.1):

1. Загально професійні компетентності.

2. Спеціально професійні компетентності:

– предметні компетентності: наукові компетентності, предметно-педагогічні компетентності;

– технологічні компетентності: інформологічно-методологічні компетентності, інформаційно-технологічні компетентності;

– професійно-практичні компетентності: математична компетентність, методологічна компетентність.

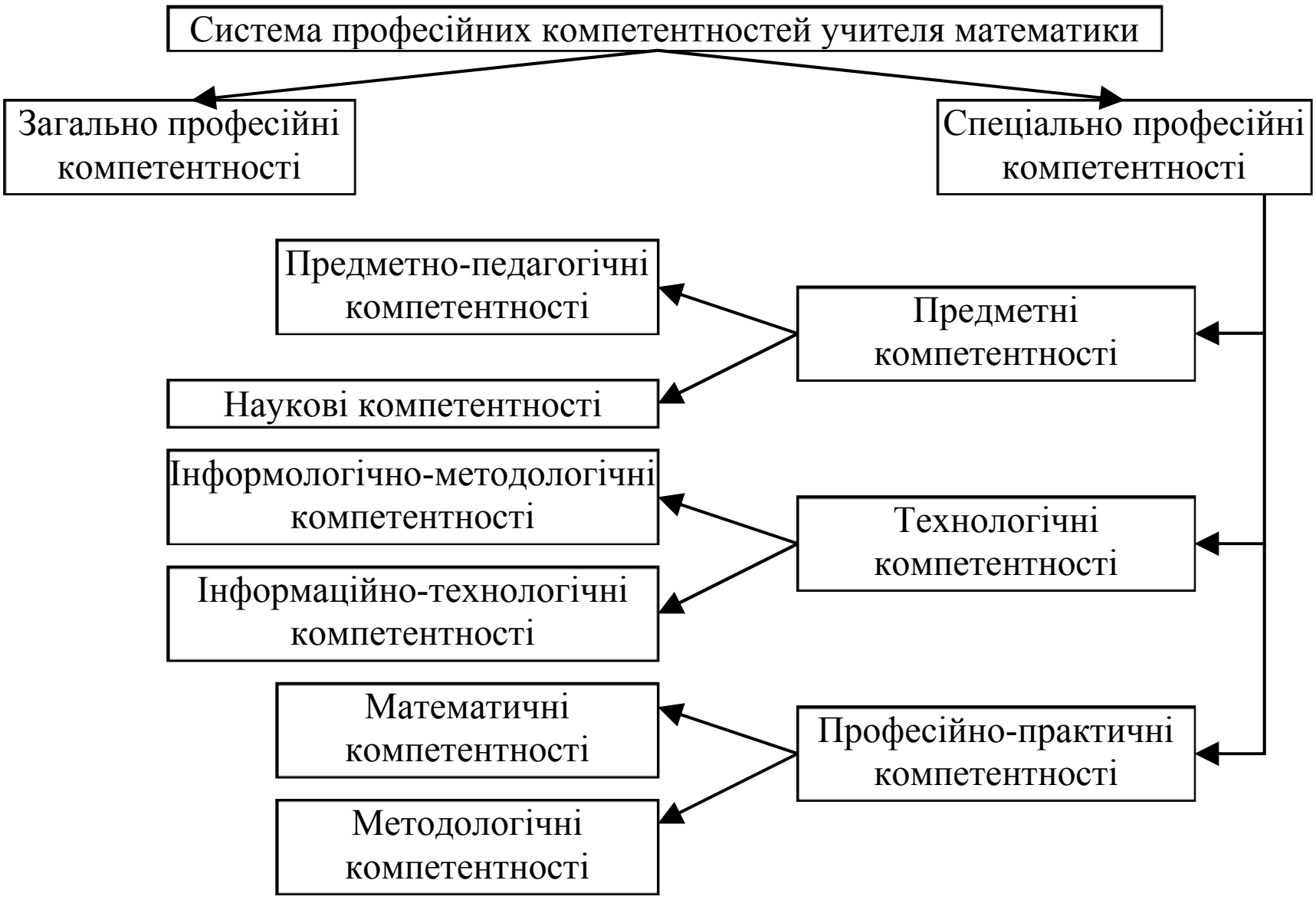

Рис. 2.1. Система професійних компетентностей учителя математики

Розглядаючи дану систему, потрібно розкрити усі складники кожного наведеного в ній компонента.

Загально професійні компетентності, як вважає І. С. Мінтій [150], враховуючи Competency Standards Modules: ICT competency standards for teacher [333], складаються з наступних компонентів:

1. спрямованість і бачення (складовими якого є: освіченість, поглиблення знань, створення знань);

2. програма й оцінка (складники: елементарні знання, застосування знань, навички XXI століття);

3. педагогіка (складники: об’єднання технологій, складне розв’язування задач, самоуправління);

4. інформаційно-комунікаційні технології (основні інструментальні засоби, складні інструментальні засоби, нові інструментальні засоби);

5. організація й адміністрування (стандартні заняття, сумісні групи, вивчення організацій);

6. розвиток професіоналізму вчителя (комп’ютерна грамотність,

управління, учитель як зразковий учень).

І. А. Колесникова вважає, що професійно-педагогічна компетентність педагога, яку можна також віднести до загально професійних компетентностей, повинна включати такі складові частини [104]:

– *гуманно-особистісну орієнтацію* як базового компоненту структури професійної компетентності;

– уміння *системно сприймати педагогічну реальність;*

– *здатність вільно орієнтуватися в предметній галузі,* на базі якої будується освітня взаємодія (технологічність);

– *уміння інтегруватися з іншим досвідом* (історичним, інноваційним, досвідом колег);

– *креативність;*

– *здібність до педагогічної рефлексії.*

Перераховані компоненти професійно-педагогічної компетентності автор розглядає як нормативно необхідні для успішної діяльності педагога в сучасних умовах. Вона відзначає, що всі вони тісно взаємозв'язані. Кожен компонент виконує свої праксеологічні функції.

Гуманно-особистісна орієнтація, що відповідає принципу відповідності діяльності природі людини, допомагає утримати цілі, зміст і результати педагогічних дій у межах взаємодії з людською рисою.

Системність бачення є умовою правильної побудови структури методів, конструювання методик, грамотне використання технологій.

Технологічність гарантує високу продуктивність дій і орієнтацію на безумовне досягнення запланованого результату.

Здатність використовувати свій і чужий педагогічний досвід стає джерелом для формування критеріїв щодо правильності і неправильності дій.

Креативність і рефлективність супроводять фахівця при вивченні і перетворенні системи професійної поведінки. Таким чином, успішність діяльності педагога може бути розглянута й оцінена тільки у межах його професійної компетенції» [182].

Буде доречним розглянути більш детально усі складники кожного компонента запропонованої системи.

І. О. Зимня [79] виділяє такі інваріантні компоненти будь-яких компетентностей:

– *готовність* до прояву компетентності (мотиваційний аспект);

– *володіння* знанням змісту компетентності (когнітивний аспект);

– *досвід* прояву компетентності в різноманітних стандартних і нестандартних ситуаціях (поведінковий аспект);

– *відношення* до змісту компетентності (ціннісно-цільовий аспект).

Отже кожен складник запропонованої системи має включати в себе компоненти, що можна розглядати як: готовність (до чогось), або володіння (чимось), або досвід (використання, навчання та ін.), або відношення (до чогось).

Крім того, М. В. Рафальска [212] вважає, що зміст професійної компетентності вчителя розкривається через 4 групи педагогічних умінь:

– постановка педагогічних задач (освітніх, розвиваючих, виховних);

– проектування дієвої педагогічної системи (планування освітньо-виховних задач, відбір змісту, форм організації, методів, засобів навчання);

– встановлення зв'язків між компонентами виховного процесу (активізація навчально-пізнавальної роботи учнів, організація спільної діяльності, забезпечення зв'язку школи з навколишнім світом);

– оцінювання результатів педагогічної діяльності (самоаналіз і аналіз навчального процесу, своєї діяльності, постановка нових педагогічних задач) [212, с. 35].

У системі мають бути відображені вказані педагогічні вміння.

*1. Предметні компетентності.*

*1.1. Наукові компетентності:*

1.1.1. Аналізувати сучасні математичні теорії: володіти уявленням про математику як науку та як про навчальний предмет, її місце в сучасному світі та системі наук; володіти поняттями даної математичної теорії; вміти підготувати огляд літератури з тематики математичного дослідження; вміти з'ясовувати склад і структуру теорії: поняття, наукові факти, закони, принципи та зв'язки між ними; вміти аналізувати теорії на предмет зв'язку з досліджуваним об'єктом та проблемою; вміти аналізувати методи теорій на предмет їх придатності на розв'язування існуючої проблеми; вміти обирати предмет та об'єкт дослідження, визначати мету дослідження та його основні завдання.

1.1.2. Вміти виконувати постановку математичної задачі: вміти раціонально і повно використовувати закони логіки; вміти аналізувати математичні факти, закономірності і теорії на предмет логічної строгості та повноти; вміти бачити логічні прогалини в обґрунтуванні математичних фактів, побудові математичних теорій; вміти використовувати методи пізнання (моделювання, аналіз, синтез, узагальнення, конкретизація, порівняння, аналогія тощо) для постановки математичної задачі; вміти будувати приклади та контрприклади, зокрема з використанням інформаційних технологій; вміти формулювати нові коректно поставлені задачі; вміти усвідомлювати застосовність існуючих методів до розв'язування поставлених проблем; вміти оцінювати перспективність розв'язування математичної задачі.

1.1.3. Аналізувати математичну проблему (задачу): вміти досліджувати коректність постановки математичної задачі; вміти аналізувати до якої галузі математичних знань належить досліджуваний об’єкт і проблема з ним пов’язана; вміти аналізувати чи має теорія, якій належить проблема, ізоморфні теорії; вміти аналізувати чи нерозв’язана дана проблема в ізоморфній теорії; вміти аналізувати взаємозв’язки досліджуваного математичного об’єкта з відомими об’єктами, а математичної проблеми з науковими фактами; вміти встановлювати ізоморфність математичних об’єктів.

1.1.4. Формулювати гіпотетичне твердження: вміти виділяти математичний об’єкт і виділяти його суттєві властивості; вміти обирати понятійний апарат, адекватний математичному об’єкту; вміти формулювати твердження в імплікативній формі; вміти формулювати твердження в еквівалентній формі; вміти формулювати твердження в формі необхідних, достатніх, необхідних і достатніх умов; вміти встановлювати протиріччя між твердженнями; вміти наводити приклади математичних об’єктів, що задовольняють умови гіпотетичного твердження.

1.1.5. Доводити гіпотетичне твердження, спростувати гіпотетичне твердження: вміти формулювати твердження, що є окремим випадком гіпотетичного твердження, і твердження більш загальне, ніж розглядуване гіпотетичне; вміти відбирати знання, необхідні для доведення або спростування гіпотетичного твердження; вміти аналізувати гіпотетичне твердження і у разі можливості розкладати його на простіші; вміти побудувати логічну схему доведення; вміти використовувати метод від супротивного при доведенні гіпотетичного твердження; вміти використовувати аналітичний метод доведення гіпотетичного твердження; вміти використовувати синтетичний метод доведення гіпотетичного твердження; вміти використовувати аналітико-синтетичний метод доведення гіпотетичного твердження; вміти обирати раціональні методи (способи, прийоми) доведення або спростування гіпотетичного твердження; вміти реалізовувати побудовану логічну схему доведення; вміти будувати контрприклади для спростування гіпотетичного твердження.

1.1.6. Аналізувати створену модель реального об’єкта: вміти добирати критерії оцінювання математичної моделі на предмет її досконалості (у відповідності до цілей моделювання); вміти встановлювати адекватність побудованої математичної моделі досліджуваному об’єкту (зокрема застосовувати критерії практики для оцінювання математичної моделі); вміти наводити приклади реальних об’єктів, моделями яких є математичні, наводити приклади задач з реальним змістом; вміти конструювати моделі проблемної (задачної) ситуації (предметні, схематичні, графічні імітаційні та ін.); вміти інтерпретувати математичні залежності в термінах конкретних математичних теорій.

1.1.7. Здатність використовувати професійно профільовані знання в галузі математики (математичної статистики), для статистичної обробки експериментальних даних і математичного моделювання природних явищ і процесів.

1.1.8. Досліджувати математичну модель з використанням засобів комп'ютерної техніки: вміти проводити комп'ютерне моделювання та чисельні експерименти для перевірки гіпотетичного твердження та його окремих випадків; вміти добирати та використовувати готові програмні засоби (математичні пакети прикладних програм) для символьно-формульного, графічного, чисельного аналізу математичних моделей реальних об'єктів; вміти при необхідності розробити алгоритм і програму для розв'язування математичної задачі; вміти виконувати чисельний експеримент з використанням комп'ютера; вміти аналізувати похибки при чисельному розв'язуванні задач.

*1.2. Предметно-педагогічні компетентності:*

1.2.1. Володіти знаннями про завдання, права і обов'язки вчителя математики у школі певного типу.

1.2.2. Володіти знаннями про навчальну роботу з учнями різних вікових категорій у цілісному педагогічному процесі.

1.2.3. Володіти знаннями про завдання і фактори соціалізації учнів різних вікових категорій.

1.2.4. Усвідомлювати і вміти аналізувати педагогічні завдання, пов'язані з виконанням обов'язків вчителя математики певного класу у школі певного типу.

1.2.5. Вміти аналізувати навчальну програму з математики для певного класу і співвідносити мету і завдання вивчення математики з цілями і завданнями вивчення кожної навчальної теми.

1.2.6. Вміти проектувати використання мікрофакторів і засобів соціалізації в цілісному педагогічному процесі.

1.2.7. Вміти враховувати при проектуванні навчально-виховного процесу правил охорони праці і норм безпеки життєдіяльності, зокрема, правила безпеки підчас проведення навчально-виховного процесу в кабінетах математики.

1.2.8. Вміти розв'язувати всі задачі і вправи, які заплановані для використання на уроці, для того, щоб оперативно перевіряти і коректувати розв'язування задач учнями.

*2. Технологічні компетентності.*

*2.1. Інформологічно-методологічні компетентності:*

2.1.1. Досвід проведення всіх форм занять у середніх навчально-виховних закладах.

2.1.2. Організація навчально-методичної роботи.

2.1.3. Застосовувати невербальні методи спілкування з учнями.

2.1.4. Проведення методичних досліджень, оформлення їх результатів.

2.1.5. Бути готовим вивчати класний колектив та окремих його членів і робити висновки про рівень загальної вихованості, дисциплінованість, успішність для цілей проектування цілісного навчально-виховного процесу.

2.1.6. Вміти складати річний план вивчення математики у певному класі школи певного типу.

2.1.7. Володіти знаннями про форми і методи навчання математики у певному класі школи певного типу.

2.1.8. Вміти аналізувати підручники, посібники, дидактичні матеріали, методичну літературу для цілей проектування цілісного навчально-виховного процесу.

2.1.9. Вміти визначити теми, педагогічні завдання, типи і орієнтовані дати окремих уроків у певному класі, виходячи з річного плану, підготовленості учнів, власних можливостей і можливостей школи.

2.1.10. Вміти проектувати зміст кожного уроку (зміст нового навчального матеріалу, повторення раніше вивченого навчального матеріалу, перевірка завдань та вмінь учнів, розв’язання задач, контрольна робота, самостійна робота, узагальнююче повторення тощо).

2.1.11. Вміти проектувати комплексне використання засобів навчання на певному уроці з математики у школі певного типу; зокрема демонстрацій, дидактичного матеріалу.

2.1.12. Вміти проектувати домашні завдання для учнів певного класу школи певного типу.

2.1.13. Вміти проектувати реалізацію зв’язків вивчення математики з вивченням фізики, хімії та інших предметів.

2.1.14. Вміти при підготовці до певного уроку конкретизувати його цілі і завдання, виділити основне для цього уроку завдання.

2.1.15. Вміти проаналізувати підручник, вибрати необхідний для засвоєння учнями фактичний матеріал.

2.1.16. Вміти виявити складні для розуміння учнями питання, проаналізувати визначення з точки зору їх точності і доступності для учнів.

2.1.17. Вміти проектувати завдання для учнів з підручником в класі і вдома.

2.1.18. Вміти відбирати з наукової і науково-популярної літератури фактичний матеріал, корисний для реалізації цілей і завдань уроку.

2.1.19. Вміти, користуючись власними знаннями та літературними методичними порадами, проектувати доцільну пізнавальну діяльність учнів для засвоєння конкретного навчального матеріалу.

2.1.20. Вміти, враховуючи рівень пізнавальних можливостей класу в цілому і окремих учнів, обрати методи, форми і засоби перевірки знань і вмінь учнів з математики.

2.1.21. Вміти виходячи з завдань уроку і програмних вимог підібрати засоби наочності.

*2.2. Інформаційно-технологічні компетентності:*

2.2.1. Володіння професійно профільованими знаннями й уміннями з теоретичних основ інформатики.

2.2.2. Підготовка і друк документу, який містить текст, таблиці, графічні об’єкти: вміти працювати з комп’ютером у якості користувача; вміти працювати з текстовим редактором; вміти працювати з глобальною мережею Інтернет; вміти на науковій основі організовувати свою працю, володіти сучасними методами і засобами збирання, зберігання, опрацювання, подання, представлення інформації, засобами підтримки інтелектуальної професійної діяльності; вміти коректно скласти документ користуючись засобами комп’ютерної технології.

2.2.3. Використання програмного засобу навчально-виховного призначення для підтримки педагогічного процесу: вміти добирати засоби та методи навчання з використанням комп’ютерної техніки; володіти методиками використання прикладних програмних продуктів для підтримки навчального процесу; вміти розробляти план вивчення навчального матеріалу з поєднанням традиційних та нових інформаційних технологій; вміти використовувати програмні засоби для обробки результатів проведених психологічних, педагогічних і методичних досліджень; вміти використовувати комп’ютерно-орієнтовані системи навчання дисциплін за фахом.

2.2.4. Базові знання в галузі інформатики й сучасних інформаційних технологій; навики використання програмних засобів і навики роботи в комп’ютерних мережах, уміння створювати бази даних і використовувати інтернет-ресурси.

2.2.5. Знання декількох мов програмування і прикладного програмного забезпечення та їх застосування для розв’язання математичних задач.

2.2.6. Використовувати ІКТ для проведення контрольних заходів.

2.2.7. Складати та підготувати до друку методичні вказівки, посібники з математики з використанням навчальної літератури,

навчально-методичних та інших інструктивних документів.

2.2.8. Організовувати позанавчальні заходи з використанням технічних, ІКТ, хмарних сервісів як засобів навчання окремо та в поєднанні з традиційними засобами навчання.

2.2.9. Вміти використовувати наявні та вивчати нові інформаційні технології (в тому числі і хмарні технології).

2.2.10. Вміти користуватись ТЗН, ІКТ; володіти методикою використання системи дидактичних засобів з використанням ІКТ, хмарних технологій; вміти виготовляти і призначати індивідуальний матеріал з використанням ІКТ для проведення навчальних занять в умовах диференціації навчання; вміти розробляти ЕОР для проведення уроків різних типів.

2.2.11. Вміти використовувати програмні засоби для вивчення учнів та учнівських колективів, вміти використовувати програмні засоби для обробки результатів проведених психологічних, педагогічних і методичних досліджень.

*3. Професійно-практичні компетентності.*

*3.1. Математичні компетентності:*

3.1.1. Вибір алгоритмів, методів, прийомів та способів, розв'язування математичних задач: вміти аналізувати відомі методи, способи, прийоми, засоби на їх придатність до розв'язування проблем; вміти використовувати індукцію, дедукцію до розв'язування математичної проблеми; вміти використовувати синтетичний, аналітико-синтетичний методи розв'язування математичної проблеми.

3.1.2. Здатність використовувати професійно профільовані знання й практичні навички з алгебри та теорії чисел: володіти теоретико-множинною і логічною символікою, основними поняттями алгебри і теорії чисел; вміти виконувати основні операції над множинами та з'ясовувати властивості множин.

3.1.3. Здатність використовувати професійно профільовані знання й практичні навички з математичного аналізу: володіти методами і прийомами обчислення границь, диференціального числення функцій, інтегрального числення функцій; вміти досліджувати властивості границь та рядів.

3.1.4. Здатність використовувати знання, уміння й навички з аналітичної та диференціальної геометрії: володіти векторним методом розв'язування задач, методами дослідження ліній другого порядку, методом геометричних перетворень до розв'язування задач, методами досліджень поверхонь другого порядку та методами перерізів, геометричних місць точок, геометричних перетворень, рухів, подібностей, інверсії та алгебраїчним методом при розв'язуванні задач на побудову; вміти застосовувати теорію прямих до розв'язування задач, використовувати метод координат для задання і дослідження геометричних об'єктів і до розв'язування задач, застосовувати теорію площин до розв'язування задач.

3.1.5. Здатність використовувати знання, уміння й навички з лінійної алгебри: володіти методами, прийомами і способами розв'язування та дослідження систем лінійних рівнянь, методами теорії многочленів; вміти виконувати основні операції над матрицями та знаходити їх основні характеристики, виконувати операції над лінійними операторами та з'ясовувати їх основні характеристики, описувати усі можливі групові операції, застосовувати конгруенції до розв'язування задач.

3.1.6. Здатність використовувати знання, уміння й навички з диференціальних рівнянь: володіти методами, прийомами і способами розв'язування диференціальних рівнянь та систем диференціальних рівнянь.

3.1.7. Здатність використовувати знання й уміння з теорії ймовірностей: володіти понятійним апаратом теорії ймовірностей і математичної статистики, методами, прийомами, способами і засобами розв'язування основних задач стохастики, статистичного опрацювання експерементальних даних; мати уявлення про метод статистичного моделювання.

3.1.8. Здатність використовувати професійно профільовані знання з дискретної математики: володіти понятійним апаратом та методами комбінаторного аналізу, теорії графів, вміти застосовувати графи до розв'язування задач.

3.1.9. Здатність використовувати професійно профільовані знання, уміння й навички з математичної логіки: володіти основами алгебри висловлень та предикатів; вміти застосовувати елементи математичної логіки в логіко-математичній практиці; мати чітке уявлення по основні числові системи і їх будову.

3.1.10. Використання засобів інформаційних технологій для розв'язування математичних задач: володіти знаряддєвим застосуванням комп'ютера, системи опрацювання текстової, числової та графічної інформації, бази даних і знань; володіти технічними засобами інформаційних технологій; володіти системним та сервісним програмним забезпеченням.

*3.2. Методологічні компетентності:*

3.2.1. Розв'язування задач шкільного курсу математики: володіти всіма методами доведення при розв'язуванні задач із курсів алгебри та геометрії основної школи; вміти виконувати операції над множинами, розв'язувати задачі на подільність ті відсотки, обчислювати цілі, дробові, ірраціональні та тригонометричні числові вирази та виконувати їх тотожні перетворення, розв’язувати найпростіші комбінаторні та стохастичні задачі, розв’язувати текстові задачі різними способами, розв’язувати рівняння та системи рівнянь, нерівності та системи нерівностей шкільного курсу математики різними способами, досліджувати функції та будувати їх графіки, розв’язувати задачі планіметрії на обчислення, побудову та доведення і володіти методиками навчання розв’язування задач шкільного курсу математики.

3.2.2. Проведення уроків з математики, алгебри та геометрії різних типів: володіти методикою організації і проведення; слідкувати за ходом уроку та здійснювати корекцію навчального процесу; вміти сприймати реакцію учнів на свої вимоги, запитання та пояснення і перебудовуватись протягом уроку.

3.2.3. Проведення навчальної консультації: володіти методикою організації і проведення навчальної консультації для слабких учнів, тих, хто навчається на підвищеному або поглибленому рівні; вміти обирати організаційні форми, методи проведення навчальної консультації, вміти добирати навчальний матеріал, враховуючи його обсяг та зміст.

3.2.4. Організація і проведення шкільної предметної олімпіади: вміти складати завдання, розв’язувати завдання підвищеної складності, володіти методикою навчання учнів виконувати олімпіадні завдання, вміти здійснювати перевірку олімпіадних робіт, класифікувати зауваження до виконаної роботи, оцінювати перевірену роботу.

3.2.5. Розробка і використання дидактичних засобів: вміти вибирати навчальні таблиці за їх дидактичними функціями для проведення уроків різних типів; виготовляти засоби наочності, роздавальний матеріал; виготовляти моделі для унаочнення матеріалу; володіти методикою використання систем дидактичних засобів.

3.2.6. Вміти використовувати засоби комп’ютерної алгебри при доведенні теорем, тверджень.

3.2.7. Вміти, враховуючи рівень пізнавальних можливостей класу в цілому і окремих учнів та завдання уроку і програмні вимоги, відібрати різнорівневі задачі і вправи, щоб мати можливість здійснити на уроці диференційований підхід.

3.2.8. Вміти скласти варіанти завдань для контрольної або самостійної роботи з певної теми, враховуючи індивідуальні особливості учнів.

3.2.9. Вміти проектувати домашнє завдання з певної теми, передбачаючи окрім загального і індивідуальні завдання (розв’язати завдання підвищеної складності, підготувати повідомлення тощо).

У реальній математичній діяльності зазвичай використовуються одночасно багато, можливо, навіть усі ці уміння. Таким чином, спроби вимірювати окремі уміння ведуть тільки до штучних задач і не можуть давати оцінки математичних компетентностей.

З метою проаналізувати аспект математичних компетентностей через конструювання завдань та тестів, доцільно структурувати уміння, що були наведені вище, у три більші класи компетентностей.

1. Клас компетентностей: репродукція, визначення, обчислення

2. Клас компетентностей: структуризація та інтеграція для розв'язування задач

3. Клас компетентностей: математичне мислення, узагальнення та інсайт»

Задля визначення критеріїв професійних компетентностей розглянемо запропоновані С. А. Раковим [208; 207] критерії математичних компетентностей:

1. концептуалізація поняття – засвоєння концептуальних ідей, що лежать в основі поняття, наприклад, типових ситуацій, в яких його доцільно використовувати (для поняття похідної функції, наприклад, це традиційні задачі обчислення миттєвої швидкості та кутового коефіцієнту дотичної до кривої);

2. властивості поняття – засвоєння основних властивостей поняття (для поняття похідної функції можна вибрати лінійність операції диференціювання та диференціювання добутку, частки, похідна композиції функцій);

3. застосування поняття – вміння «бачити" «поняття в типових ситуаціях (для поняття похідної функції, наприклад, можна вибрати вміння застосовувати поняття похідної для визначення кутового коефіцієнта дотичної до кривої або швидкості зміни деякої величини);

4. систематизація поняття – узагальнення поняття, зв'язок з іншими поняттями, межі поняття (для поняття похідної функції, наприклад, можна вибрати зв'язок диференційованості з неперервністю та іншими властивостями функцій) [208, с. 41].

М. В. Рафальска [212, с. 39-40] визначає три рівня сформованості професійно-інформатичних компетентностей учителя: базовий (елементарний), середній (функціональний) та просунутий (системний). Кожному рівню відповідають певні критерії:

– для базового (елементарного) рівня вказано критерії, які зазначаються як: володіння прийомами і методами роботи у локальній і глобальній комп'ютерних мережах, з персональним комп'ютером, з даними без використання засобів ІКТ та інших технологій;

– критерії середнього (функціонального) рівня сформульовані наступним чином: володіння компетентностями у галузі використання

дистанційних форм навчання та вміння створювати мережеві освітні ресурси, педагогічні програмні засоби, дидактичні матеріали, знаходити актуальні відомості і методичні матеріали за допомогою глобальної комп'ютерної мережі, розробляти електронні дидактичні матеріали для проведення різних форм занять;

– що ж стосується просунутого (системного) рівня, вказані такі критерії: володіння компетентностями у галузі розробки і підтримки дистанційних курсів та використання дистанційних форм навчання, дидактичними, психолого-педагогічними і методичними прийомами, різноманітними засобами ІКТ, методикою використання ІКТ в навчальному процесі.

Отже, кожна складова запропонованої системи професійних компетентностей може мати чотири рівні: високий, достатній, середній, низький.

Кожному рівню відповідатимуть декілька критеріїв, які стосуватимуться відповідних компонентів:

– предметні компетентності (предметно-педагогічні та наукові компетентності);

– технологічні компетентності (інформологічно-методологічні та інформаційно-технологічні компетентності);

– професійно-практичні компетентності (математична та методологічна компетентності).

Рівні та показники сформованості предметних компетентностей, технологічних компетентностей та професійно-практичних представлені в табл. 2.2, табл. 2.3 та табл. 2.4.

*Таблиця 2.2*

**Показники сформованості предметних компетентностей учителя математики**

| Рівень | Предметно-педагогічні компетентності | Наукові компетентності |
|---|---|---|
| Високий | Вміти проектувати використання мікрофакторів і засобів соціалізації в цілісному педагогічному процесі. Вміти враховувати при проектуванні навчально-виховного процесу правил охорони праці і норм безпеки життєдіяльності, зокрема, правила безпеки під час проведення навчально- | Вміти аналізувати методи теорій на предмет їх придатності на розв'язування існуючої проблеми. Доводити гіпотетичне твердження, або спростувати. Досліджувати математичну модель з використанням засобів комп'ютерної техніки. |

| Рівень | Предметно-педагогічні компетентності | Наукові компетентності |
|---|---|---|
| | виховного процесу в кабінетах математики. | |
| Достатній | Володіти знаннями про завдання та фактори соціалізації учнів різних вікових категорій. Вміти аналізувати навчальну програму з математики для певного класу і співвідносити мету і завдання вивчення математики з цілями і завданнями вивчення кожної навчальної теми. | Вміти з'ясовувати склад і структуру математичної теорії. Формулювати гіпотетичне твердження. Аналізувати створену модель реального об'єкта. |
| Середній | Володіти знаннями про навчальну роботу з учнями різних вікових категорій у цілісному педагогічному процесі. Усвідомлювати і вміти аналізувати педагогічні завдання, пов'язані з виконанням обов'язків вчителя математики певного класу у школі певного типу. | Вміти підготувати огляд літератури з тематики математичного дослідження. Аналізувати математичну проблему (задачу). |
| Низький | Володіти знаннями про завдання, права й обов'язки вчителя математики у школі ЗНЗ певного типу. Вміти розв'язувати всі задачі і вправи, які заплановані для використання на уроці, для того, щоб оперативно перевіряти і корегувати розв'язування задач учнями. | Володіти уявленням про математику як науку та навчальний предмет, її місце в сучасному світі та системі наук; володіти поняттями даного шкільного предмету. Вміти виконувати постановку математичної задачі. |

*Таблиця 2.3*

**Показники сформованості технологічних компетентностей учителя математики**

| Рівень | Інформологічно-методологічні компетентності | Інформаційно-технологічні компетентності |
|---|---|---|
| Високий | Вміти проектувати комплексне використання засобів навчання на певному уроці з математики у школі певного типу. Вміти проектувати реалізацію зв'язків вивчення математики та інших предметів. Вміти, користуючись власними знаннями та літературними методичними порадами, проектувати доцільну пізнавальну діяльність учнів для засвоєння конкретного навчального матеріалу. | Вміти розробляти план вивчення навчального матеріалу з поєднанням традиційних та нових інформаційних технологій; вміти використовувати програмні засоби для обробки результатів проведених психологічних, педагогічних і методичних досліджень; вміти використовувати комп'ютерно-орієнтовані системи навчання дисциплін за фахом. |
| Достатній | Володіти знаннями про форми та методи навчання математики у певному класі школи певного типу. Вміти добирати з наукової і науково-популярної літератури фактичний матеріал, корисний для реалізації цілей і завдань уроку. | Вміти добирати засоби та методи навчання з використанням комп'ютерної техніки; володіти методиками використання прикладних програмних продуктів для підтримки навчального процесу. |
| Середній | Вміти складати календарно-тематичний план вивчення математики у певному класі школи певного типу. Вміти аналізувати зміст підручників, вибрати необхідний для засвоєння учнями фактичний матеріал. Вміти при підготовці до певного уроку конкретизувати його цілі і завдання, виділити основне для цього уроку завдання. | Вміти на науковій основі організовувати свою працю, володіти сучасними методами і засобами збирання, зберігання, опрацювання, подання, передавання даних, засобами підтримки інтелектуальної професійної діяльності; вміти коректно скласти документ користуючись засобами комп'ютерної технології. |

| Рівень | Інформологічно-методологічні компетентності | Інформаційно-технологічні компетентності |
|---|---|---|
| Низький | Бути готовим до проведення всіх форм занять у середніх навчально-виховних закладах. Вміти проектувати завдання для учнів з підручником в класі і вдома. | Підготовка і друк документу, що містить текст, таблиці, графічні об'єкти: вміти працювати з комп'ютером у якості користувача; вміти працювати з текстовим редактором; вміти працювати з глобальною мережею Інтернет, із системами опрацювання текстових запитів, сервісами та ресурсами. |

*Таблиця 2.4*

**Показники сформованості технологічних компетентностей компетентностей учителя математики**

| Рівень | Математичні компетентності | Методологічні компетентності |
|---|---|---|
| Високий | Використання засобів інформаційних технологій для розв'язування компетентнісно орієнтованих математичних задач: володіти знаряддєвим застосуванням комп'ютера, системи опрацювання текстових, числових та графічних даних, бази даних і знань; володіти технічними чи апаратними засобами інформаційних технологій; володіти знаннями з налаштування та встановлення системного та сервісного програмного забезпечення. | Організація і проведення шкільної олімпіади з математики. Вміти, враховуючи рівень пізнавальних можливостей класу в цілому й окремих учнів, завдання уроку та програмні вимоги, добирати різнорівневі задачі та вправи для реалізації різнорівневої диференціації на уроці. |
| Достатній | Вибір, використання алгоритмів, методів, прийомів та способів, розв'язування математичних задач: вміти аналізувати відомі | Вміти проектувати домашнє завдання з певної теми, передбачаючи окрім загального і індивідуальні |

| Рівень | Математичні компетентності | Методологічні компетентності |
|---|---|---|
| | методи, способи, прийоми, засоби на їх придатність до розв'язування проблем; вміти використовувати індукцію, дедукцію до розв'язування математичної проблеми; вміти використовувати синтетичний, аналітико-синтетичний методи розв'язування математичної проблеми. | завдання. |
| Середній | Використовувати професійно профільовані знання й практичні навички з алгебри та теорії чисел, математичного аналізу, аналітичної та диференціальної геометрії, лінійної алгебри, диференціальних рівнянь, теорії ймовірностей та математичної логіки. | Проведення уроків з математики, алгебри та геометрії різних типів. Вміти скласти варіанти завдань для контрольної або самостійної роботи з певної теми, враховуючи індивідуальні особливості учнів. |
| Низький | Використовувати основні прийоми та методи розв'язування основних задач з алгебри та теорії чисел, математичного аналізу, аналітичної та диференціальної геометрії, лінійної алгебри, диференціальних рівнянь, теорії ймовірностей та математичної логіки. | Розв'язування задач шкільного курсу математики. Проведення навчальних консультацій. |

## 2.2 Особливості використання CoCalc у навчанні математичних дисциплін

Існує декілька напрямів розвитку хмарних технологій, які можна ефективно застосовувати в організації самостійної роботи студентів.

Software-as a Service – цей сервіс створює для студентів доступ до електронної пошти, операційних систем, програм розпізнавачів спаму й непотрібної кореспонденції, забезпечує студентів і дослідників з необхідності спеціалізованим програмним забезпеченням, а також програмним забезпеченням та обладнанням, яке потребує багато обробки й обчислень.

Використання цього сервісу є дуже важливим під час підготовки майбутніх вчителів математики, тому що «програмне забезпечення як послуга» дає можливість розв’язати проблеми ліцензійного використання дорогого програмного забезпечення, яке потребує постійного оновлення. З цим пов’язаний процес підтримки та становлення, відстежування термінів ліцензій, що потребує значних коштів і спеціальних працівників. У разі застосування хмарних обчислень схему ліцензування спрощено, а коштів необхідно менше за рахунок того, що сплачується саме послуга, а не купівля програмного забезпечення. Також, аргумент на користь технології хмарних обчислень для аспірантів і магістрів дослідники вбачають у тому, що навчальний процес найбільше за інші види діяльності потребує пошуку й експериментування [317, с. 177].

Platform as a Service – платформа як сервіс. Може надаватись інтегрована платформа для розроблення, тестування та підтримки веб-додатків, створених на основі хмарних обчислень. Під час підготовки студентів цей різновид послуг може бути застосований для керування освітніми проектами, здійснення спільних досліджень, наприклад, створення віртуальних лабораторій спільного доступу для проведення експериментів у галузі моделювання тощо. Для наших досліджень це є дуже ефективно у тому разі, коли для організації реальної лабораторії не має матеріальної бази.

Hardware as a Service – сервіс, за яким надається послуга апаратних можливостей, наприклад, певний обсяг пам’яті, процесорний час, пропускна здатність. Під час моделювання певних фізичних процесів іноді потрібні потужні комп'ютери, у купівлі яких для декількох експериментів немає необхідності, у таких випадках можна скористатись можливостями цього сервісу.

Infrastructure as a Service – до цього сервісу належать апаратні засоби; операційні системи й системне програмне забезпечення; програмне забезпечення зв’язку між системами (наприклад, інтеграції в мережі, керування обладнанням). За допомогою цієї технології створюється можливість купівлі, нарощування серверного часу, дискового простору, мережевої пропускної здатності, що відбувається динамічно тоді, коли це потрібно для функціонування певного додатка.

Communication as a Service – як сервіс надаються послуги зв’язку, наприклад, IP-телефонія, пошта, чат. Паралельно з електронними адресами надається цілий комплекс корисних додатків, наприклад, текстові редактори, електронні таблиці, презентації, які можуть бути використані у груповій роботі молодих науковців, коли з’являється можливість спільно користуватися документами через мережу. Разом з тим, надається значний обсяг віртуального дискового простору, де можна зберігати великі мультимедійні або графічні файли. Ще однією перевагою цього сервісу є те, що студенти можуть користуватися поштою дистанційно у будь-якому місті, використовуючи мобільні пристрої [317, с. 178-179].

Desktop as a Service – користувачі одержують як сервіс абсолютно готове для роботи віртуалізоване робоче місце. Користувач одержує доступ не до однієї програми, а до певного програмного середовища. Переваги цієї технології у тому, що вимоги до обладнання мінімальні й це дає можливість значно знизити затрати, що передбачають закупівлю комп'ютерної техніки і програмного забезпечення. Витрати на користування віртуальним робочим місцем значно менші завдяки тому, що платить клієнт саме за те, що йому необхідно, і тоді, коли це необхідно. Крім того, є й інша перевага, яка полягає в тому, що доступ до робочого місця користувач може мати де завгодно, через будь-який комп'ютер, де є доступ до Інтернет, а також через мобільні пристрої [317, с. 179].

Застосування хмарних обчислень у ЗВО у більш широкому аспекті:

– для користувачів (викладачів, студентів): персональний набір програмного забезпечення залежно від спеціалізації, курсу тощо, збереження персональних даних значних обсягів – незалежність від пристрою, мобільність;

– для ІТ-персоналу: централізація та гнучкість управління, мінімізація потреби в обслуговуванні, економія коштів на придбання нового обладнання, гнучкість у розгортанні нових систем;

– для ЗВО: персональне середовище студента протягом всього терміну навчання, доступ до власного середовища з будь-якого місця у будь-який час, мобільність та збереження сеансу (Hot Desking), автоматичний розподіл пакетів програмного забезпечення відповідно до навчальних планів, наукових потреб тощо [155, с. 28].

Мережна система комп'ютерної математики Sage (web-СКМ Sage) є одним з етапів розвитку хмарного сервісу CoCalc (режими доступу: cloud.sagemath.com або CoCalc.com). Наявний інструментарій web-СКМ Sage версії 4.6 (останньої версії, що передувала появі CoCalc) не був достатнім для організації усіх видів навчальної діяльності за умов дистанційного навчання або його елементів [233; 235]. При цьому доводилося або організовувати навчання із залученням двох систем – web-СКМ Sage та будь-якої системи підтримки дистанційного навчання, зокрема Moodle, або здійснювати їх інтеграцію [312]. Перший спосіб виявився незручним ані для викладачів, ані для студентів, другий спосіб – не встиг набути масового застосування, а з появою та удосконаленням

CoCalc взагалі втрачає актуальність.

Повний перелік складових CoCalc поточної версії можна отримати за однією з команд – ls /usr/share/applications або $sudo dpkg --get-selections (рис. 2.2).

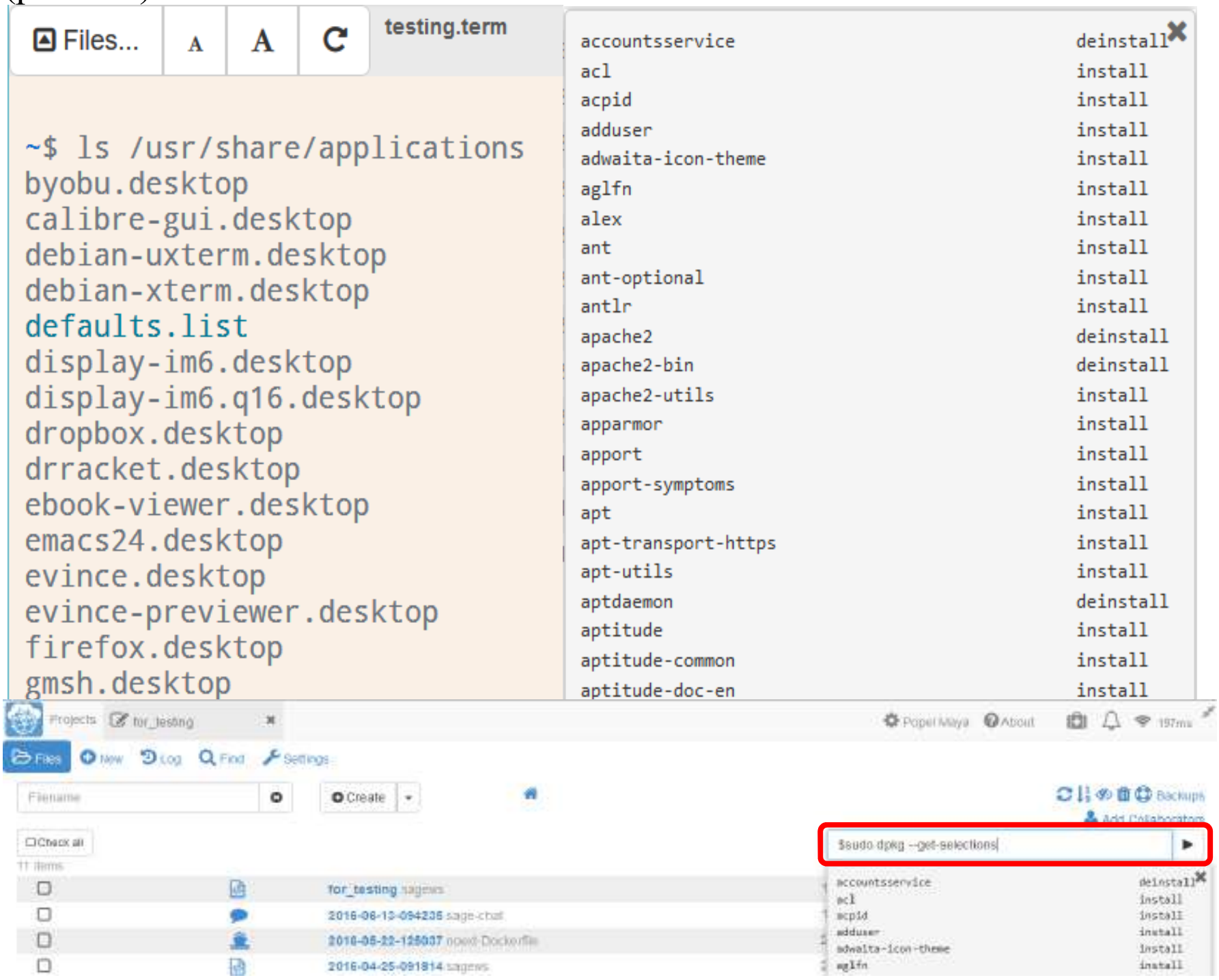

Рис. 2.2. Перегляд компонентів CoCalc

У таблиці 2.5 наведено структурований перелік основних складових CoCalc.

*Таблиця 2.5*

**Основні складові (компоненти, програмні засоби) CoCalc**

| **Вид засобу** | **Назва засобу** |
|---|---|
| *Системні програмні засоби* | |
| FTP-клієнт | CFTP |
| VNC-сервер | X11vnc |
| Архіватор | 7-ZIP, gzip, tar |
| Безкоштовна утиліта командного рядка, для стиснення даних | bzip2 |
| Збирач сміття | The Boehm-Demers-Weiser |

| Вид засобу | Назва засобу |
|---|---|
| Оболонка для GNU Screen и Tmux (додаток) | Byobu |
| Оболонка для Python бібліотеки GD | gdmodule |
| Програма для виводу списку запущених процесів | htop, ps |
| Сервер Notebook SageMath | SageMathNB |
| Операційна система | Debian GNU/Linux |
| *Прикладні програмні засоби загального призначення* | |
| Аналог screen для графічних програм | Xpra |
| База даних комбінаторних графів | graphs |
| Бібліотека для растеризації шрифтів і операцій над ними | FreeType |
| Бібліотека для роботи з растровою графікою у форматі PNG | libpng |
| Браузер підказок для GNOME | Yelp |
| Система управління файлами та спільної роботи з ними | Mercurial |
| Електронний словник (тезаурус) | WordNet |
| Засіб перегляду зображень | GPicView |
| Інтерактивний редактор та підтримка макросів | Prerex |
| Програми для порівняння вмісту текстових файлів та каталогів | Meld, diff |
| Сервіси для читання електронних книг | Calibre, Evince |
| Система опрацювання відкритого тексту документації в форматах HTML, LaTeX або документа XML | Docutils |
| Системи управління базами даних | RethinkDB, sqlite3 |
| Текстові редактори | GNU Emacs, Vim, nano, mcedit |
| Утіліта для знаходження відмінностей між файлами | GNU patch |
| Хмарне сховище файлів | Dropbox |
| *Прикладні програмні засоби спеціального призначення* | |
| Автоматичний генератор сітки для геометричних побудов | Gmsh |
| Бібліотека для виконання завдань з теорії чисел | FLINT |
| Бібліотека для динамічної роботи з зображеннями | GD Graphics Library (GD) |
| Бібліотека для опрацювання відео- та аудіофайлів | Ffmpeg |
| Бібліотека для роботи з графами та іншими мережевими структурами | NetworkX |

| **Вид засобу** | **Назва засобу** |
|---|---|
| Бібліотека для розв'язку задач лінійного програмування | GLPK |
| Бібліотека для розв'язку задач опуклого програмування | CVXOPT |
| Бібліотека, призначена для проведення прикладних та наукових математичних розрахунків | GNU Scientific Library (GSL) |
| Бібліотеки для визначення і обчислення еліптичних кривих, визначених над полем раціональних чисел | eclib |
| Векторний графічний редактор | Inkscape |
| Версії Sage | Sage.7, Sage.8, Sage.9, Sage.10 |
| Клієнт для Git-репозитарія | SparkleShare |
| Математична бібліотека | Cephes |
| Математична бібліотека для виконання дій над комплексними числами | GNU MPC |
| Набір бібліотек, що розширюють функціональність C++ | Boost |
| Розширення SageTeX пакету | SageMathTeX |
| Пакет програм для генерації тривимірних моделей | GenModel |
| Пакет програм для наукових розрахунків | Scilab |
| Пакети програм для побудови філогенетичних дерев | Phylip |
| Система для математичних обчислень | GNU Octave |
| Системи комп'ютерної алгебри | Gias/Xcas, Axiom, GAP |
| Система комп'ютерної математики | Maxima |
| *Інструментальні програмні засоби* | |
| Інтерактивна оболонка для програмування | Jupyter Notebook |
| Інтерпретатор об'єктно-орієнтованої мови програмування | Python |
| Інтерпретатори | CPython, Java, Perl, bash |
| Компілятори | Mono, Embeddable Common Lisp |
| Середовища для функціонального програмування | DrRacket, Scheme |
| Середовище для статистичних обчислень, аналізу та представлення даних в графічному вигляді | R |

Докладні відомості про зазначені у таблиці 2.5 та інші компоненти CoCalc (на момент публікації) можна отримати за прямим посиланням http://www.sagemath.org/links-components.html на офіційному сайті проекту Sage (рис. 2.3).

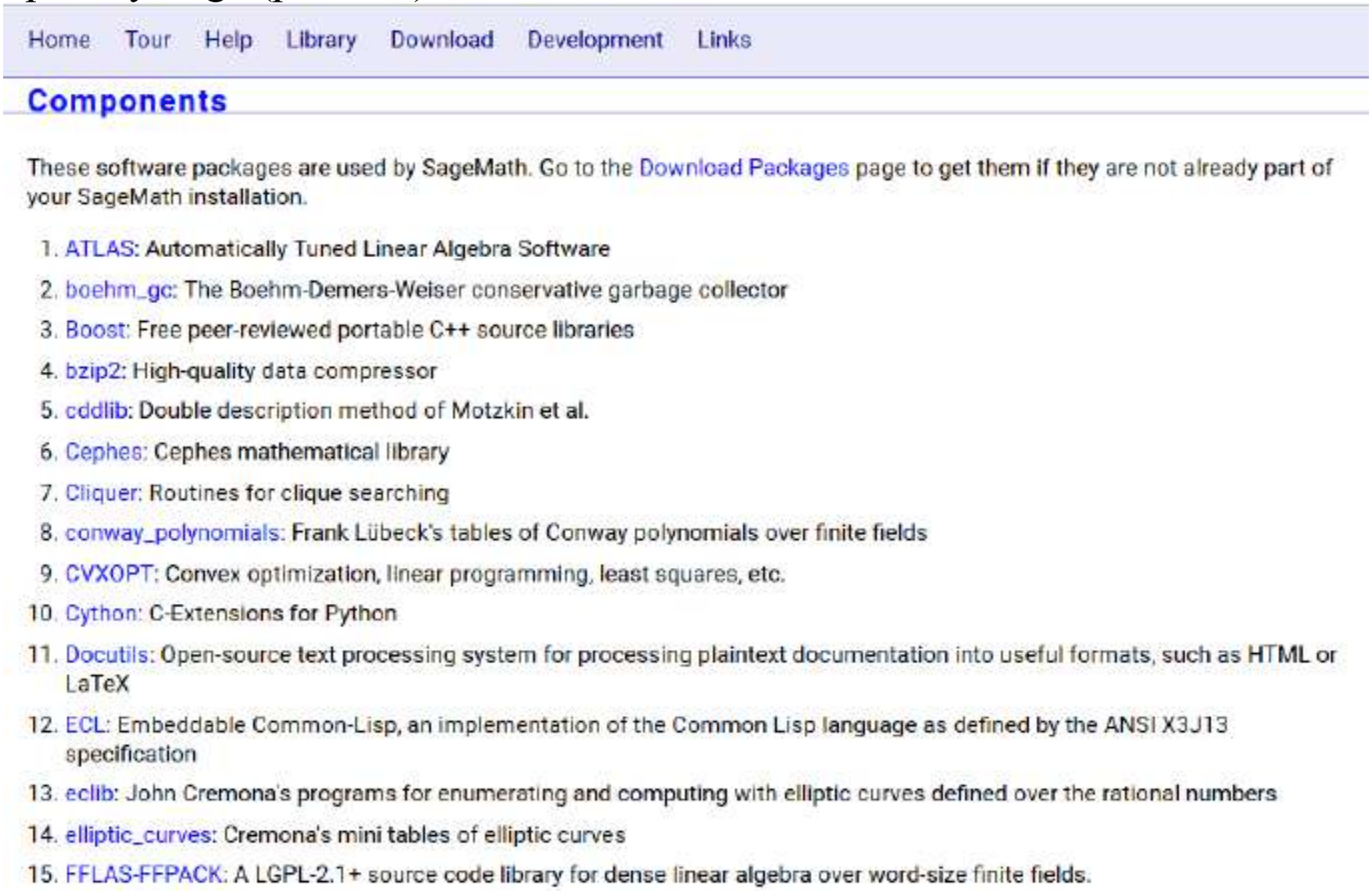

Home Tour Help Library Download Development Links

**Components**

These software packages are used by SageMath. Go to the Download Packages page to get them if they are not already part of your SageMath installation.

1. ATLAS: Automatically Tuned Linear Algebra Software
2. boehm_gc: The Boehm-Demers-Weiser conservative garbage collector
3. Boost: Free peer-reviewed portable C++ source libraries
4. bzip2: High-quality data compressor
5. cddlib: Double description method of Motzkin et al.
6. Cephes: Cephes mathematical library
7. Cliquer: Routines for clique searching
8. conway_polynomials: Frank Lübeck's tables of Conway polynomials over finite fields
9. CVXOPT: Convex optimization, linear programming, least squares, etc.
10. Cython: C-Extensions for Python
11. Docutils: Open-source text processing system for processing plaintext documentation into useful formats, such as HTML or LaTeX
12. ECL: Embeddable Common-Lisp, an implementation of the Common Lisp language as defined by the ANSI X3J13 specification
13. eclib: John Cremona's programs for enumerating and computing with elliptic curves defined over the rational numbers
14. elliptic_curves: Cremona's mini tables of elliptic curves
15. FFLAS-FFPACK: A LGPL-2.1+ source code library for dense linear algebra over word-size finite fields.

Рис. 2.3. Компоненти CoCalc
(*sagemath.org* – Меню *Links* – Команда *SageMath Components*)

Слід відмітити, що одні програмні засоби (структурні компоненти CoCalc) явно доступні користувачу як вбудовані хмарні сервіси, інші ж використовуються системою неявно.

Оснащеність CoCalc прикладними сервісами спеціального призначення стане у нагоді під час організації комп'ютерно орієнтоване навчання класичних математичних дисциплін, таких як:

– «Лінійна алгебра та числові системи»;

– «Аналітична та диференціальна геометрія»;

– «Математичний аналіз»;

– «Теорія ймовірностей та математична статистика» тощо.

Як спеціальні прикладні, так і інструментальні сервіси можуть бути успішно використані на підтримку вивчення математичних дисциплін – «Чисельні методи/Методи обчислень/обчислювальна математика», «Дискретна математика», «Дослідження операцій», «Математичне програмування» тощо.

У даний час CoCalc – найпотужніше і найзручніше середовище для роботи з Sage.

Застосовуючи CoCalc викладачі та студенти зможуть:

– працювати з IPython Notebook;

– працювати з документами LaTeX;

– використовувати можливості Інтернету через HTML, CSS, JS, CoffeeScript, markdown;

– працювати над проектом спільно з іншими користувачами і бачити їх редагування в режимі реального часу;

– запускати команди консолі як в терміналі, так на самому аркуші з обчисленнями;

– встановити потрібний пакет Python через PIP;

– написати свій математичний додаток з власним веб-сервером (наприклад, з використанням Flask);

– вставляти скрипти з інших мов програмування, таких як: Perl, Ruby, Fortran,Go;

– вставляти скрипти з інших математичних систем, таких як: Maple, Mathematica, Magma, Matlab, Macsyma, Octave, Scilab і ін.

Враховуючи зазначені чинники, можна вказати основні типи програмних засобів, що спрямовані на активізацію пізнавальної діяльності студентів при вивченні курсу вищої математики у педагогічному ВНЗ за допомогою ІКТ.

Лекційні демонстрації – програми з графічним інтерфейсом і напівавтоматичним управлінням, що ілюструють теоретичні поняття, теореми, методи тощо. Працюючи з цими програмами, користувач має можливість не просто відтворити зображення, показані на рисунках до задач, а й, вводячи свої числові або символьні дані, отримувати нові результати, що можуть слугувати підтвердженням того чи іншого математичного означення, правила теореми, тощо. За рахунок такої можливості стає реальним відповідне розширення змісту лекційного матеріалу за більшістю дисциплін математичної підготовки. Цей тип програмного забезпечення реалізує один із головних дидактичних принципів – принцип наочності, що передбачає створення у студентів чуттєвого уявлення про об’єкт вивчення, сприяє переходу від сприйняття конкретних об’єктів до сприйняття абстрактних понять про них, а також надає можливість полегшити розуміння змісту математичних методів та алгоритмів. Правильний добір засобів наочності сприяє усвідомленню сприйняття, підвищенню пізнавального інтересу, активізує мислення.

Динамічні моделі різноманітних класів (видів) математичних задач – програми з графічним інтерфейсом і напівавтоматичним управлінням, що реалізують принцип моделювання. Використання та дослідження таких моделей дозволяє значно легше зрозуміти математичну, фізичну чи економічну суть методів та алгоритмів; глибше усвідомити новий матеріал та створити змістову основу для розв'язання прикладних задач» [252, с. 5-6].

Перевага динамічних моделей полягає в тому, що студент може вибирати різні режими роботи програми, змінювати параметри досліджуваних об'єктів чи процесів, спостерігати та аналізувати результати, робити висновки на основі своїх спостережень. Вони забезпечують умови для осмислення задач, дослідження закономірностей на основі формування гіпотез з їх наступною експериментальною перевіркою. Таким чином, у студента з'являються великі можливості для здійснення дослідницької та творчої діяльності, що сприяє розвитку пізнавального інтересу тощо.

Тренажери – програми, основне призначення яких полягає у поданні всіх етапів розв'язування математичної задачі. У процесі вивчення дисциплін математичного циклу помітну роль відіграє застосування теоретичних положень до розв'язання навчальних задач прикладного характеру. При цьому в міру одержання навичок рішення типових задач здійснюється перехід до задач підвищеної складності для творчого оволодіння предметом. Однак через обмежений час, що відводиться на вивчення дисципліни, складність теоретичного матеріалу, недостатню підготовку студентів і інше доводиться витрачати значний час на розв'язання саме типових задач. Тому доцільно винести цю складову навчального процесу на самостійну роботу студентів з комп'ютерною програмою-тренажером, що виступає як засіб формування та удосконалення практичних навичок, перевірки досягнутих результатів та розраховані на повторення та закріплення навчального матеріалу [252, с. 6-7].

Методичні рекомендації:

– доцільно і методично грамотно використовувати CoCalc, що в подальшому активізує діяльність студентів і тим самим покращує результати навчання;

– CoCalc можна в першу чергу використовувати для самостійної роботи студентів, поглиблення знань, перевірки гіпотез, дослідження та виявлення нових властивостей математичних об'єктів;

– вміло поєднувати традиційні та інноваційні методи навчання із застосуванням хмарних технологій, здійснюючи новий сучасний підхід до навчання студентів.

Можна виокремити такі умови організації навчального процесу з використанням CoCalc:

1. Подання навчального матеріалу має бути лаконічним, доступним і науковим.

2. Використовувати комп'ютер лише за умови, коли вивчення нового

поняття потребує більшої наочності, або ж прискорить темп заняття.

3. Використання CoCalc має бути дозованим.

4. Забезпечити усі необхідні умови роботи студентів на занятті. (Не допустимо, щоб один комп'ютер використовували одночасно два студенти).

Для робити з CoCalc необхідно, щоб у викладача робоче місце було обладнане комп'ютером (ноутбуком, нетбуком, планшетом) чи хоча б він мав власний пристрій. Для роботи на практичному занятті достатньо буде використання смартфону, але для підготовки до лекційного заняття, для попередньої роботи з моделями, їх вдалим застосуванням під час проведення заняття потрібно забезпечити викладача комп'ютером (ноутбуком, нетбуком, планшетом) з доступом до мережі Інтернет (не має значення чи то буде мережа Wi-Fi, чи то буде кабельне підключення). Не менш важливим постає питання вільного підключення до наявної мережі Інтернет. Якщо це буде підключення за допомогою Wi-Fi, то параметри пропускної швидкості Інтернету та технічних характеристик роутера також відіграють важливу роль, бо одночасне підключення цілої групи студентів до мережі Інтернет сповільнить його роботу. Експериментальний майданчик має бути забезпечений достатньою кількістю комп'ютерних аудиторій, щоб мати можливість частину практичних занять проводити в них.

### 2.3 Модель використання хмарного сервісу CoCalc як засобу формування професійних компетентностей учителя математики

Модель – це деяке подання (аналог, образ) системи, що моделюється в якому відображається, враховується, характеризується і можуть відтворюватися такі особливості й властивості цієї системи, які забезпечують досягнення цілей побудови та використання моделі [38].

В. Ю. Биков вважає, що «при вивченні складних систем, до класу яких відносяться системи освіти і навчання, тільки завдяки коректному моделюванню їх статики і динаміки можна визначити сукупність властивостей модельованої системи, які повинна мати модель, тобто від мети моделювання залежить вибір і врахування в моделі тих чи інших особливостей і суттєвих параметрів досліджуваної системи-оригіналу, а тому і ступень деталізації її моделі. Враховуючи цілі дослідження, при моделюванні досліджуваних систем (ДС) та їх елементів дослідник має дати чіткі відповіді на такі питання: які аспекти функціонування ДС підлягають вивченню, до якої глибини має бути декомпозована ДС, які суттєві взаємозв'язки між виділеними об'єктами ДС повинні бути відображені в її моделі? Відповіді на ці питання багато в чому залежать від того, як дослідник розуміє зміст основних категорій, на яких базується модельне подання ДС, зокрема категорій методу, способу, механізму і технології функціонування систем» [22, с. 79-80].

На основі спроектованої системи професійних компетентностей майбутніх учителів математики була розроблена модель використання хмарного сервісу CoCalc як засобу формування професійних компетентностей учителя математики (рис. 2.4).

Модель складається з цільового компоненту, стимулюючо-мотиваційного, змістового, операційно-діяльнісного та оціночно-регулятивного компонентів. Модель охоплює три етапи формування професійних компетентностей: I етап: пропедевтичний, II етап: формувальний, III етап: розвивальний.

На формувальному етапі відбувається використання хмарного сервісу CoCalc в процесі вивчення математичних дисциплін. Зміст навчання математичних дисциплін розписано по семестрах та включає чотири курси підготовки майбутнього вчителя математики.

Таким чином, формування професійних компетентностей учителя математики передбачає набуття ним компетентностей у галузі математики та суміжних з нею дисциплін, методики навчання та дидактики, психологічних і педагогічних основ здійснення навчально-виховного процесу, дослідницької діяльності та педагогічного спілкування, що визначає якість його професійної діяльності.

При створенні моделі формування професійних компетентностей учителя математики будемо дотримуватися таких принципів: науковості, професійної спрямованості, самореалізації міждисциплінарної інтеграції, варіативності.

Використання комп’ютера розширює можливості індивідуального навчання, охоплюючи більшу частину студентів. Також, використовуючи комп’ютер, можна здійснити рефлексивне управління, ширше охопити пізнавальні процеси кожного студента, в результаті чого – надати індивідуальну допомогу на основі виявлених здібностей.

Реалізація моделі безпосередньо залежить від достатньої обладнаності комп’ютерних аудиторій у ЗВО, наявності у студентів та викладачів власних комп’ютерних пристроїв та доступу до мережі Інтернет.

Основу моделі складає мета, яку сформульовано наступним чином: формування професійних компетентностей учителя математики здатного до ефективного використання хмарних ІКТ у професійній діяльності. На мету в першу чергу впливають: суспільне замовлення на підготовку компетентного вчителя математики, спрямування на ІКТ-аутсорсинг засобів навчання та розвиток ІКТ.

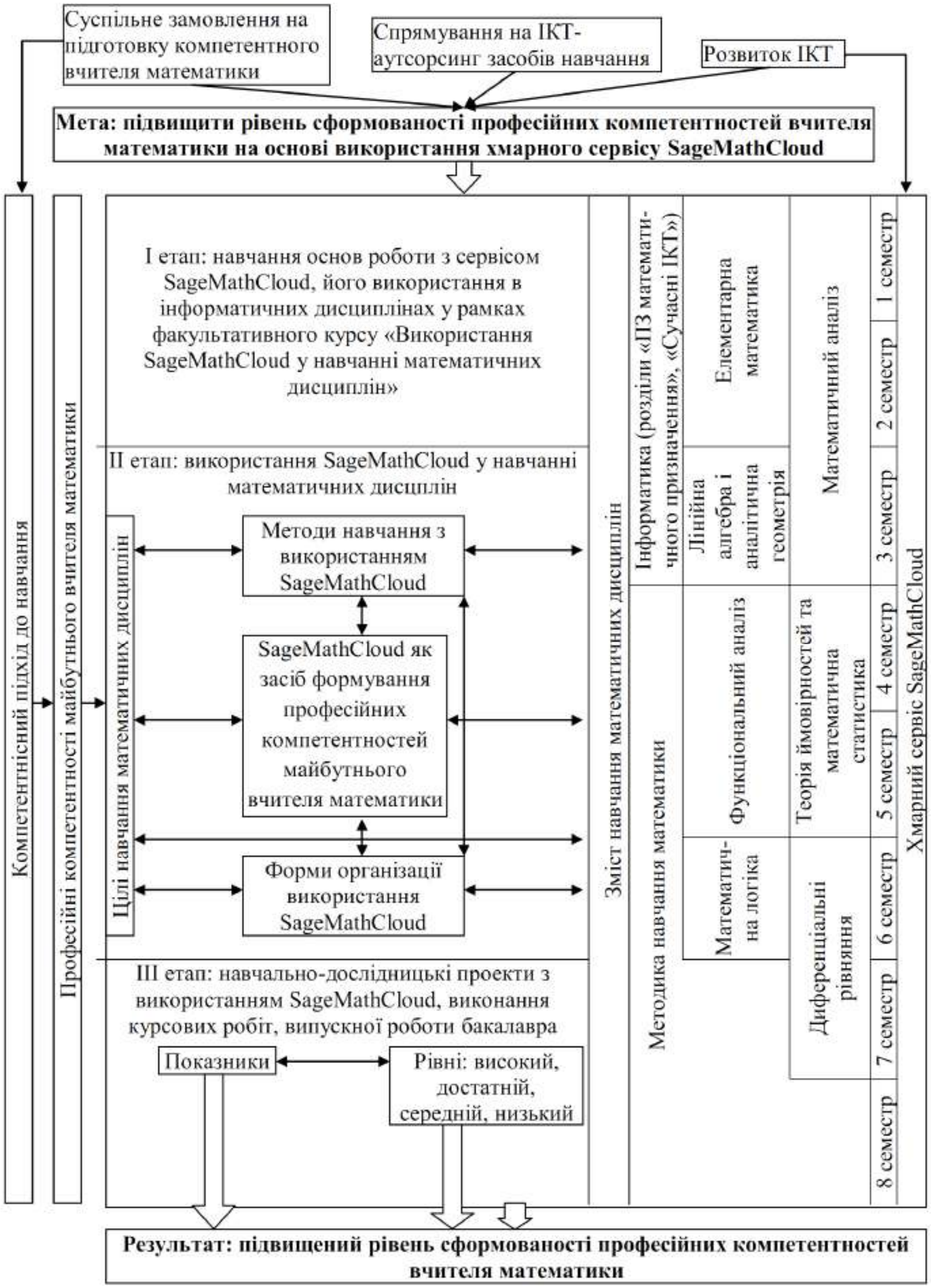

Рис. 2.4. Модель використання хмарного сервісу CoCalc як засобу формування професійних компетентностей учителя математики

Треба підкреслити, що суспільне замовлення на підготовку компетентного вчителя математики впливає не лише на мету запропонованої моделі, але й на компетентнісний підхід до навчання. Розвиток ІКТ впливає на появу хмарних сервісів навчання, які в свою чергу входять до змісту навчання математичних дисциплін.

Зміст поєднує в собі, окрім хмарних сервісів навчання, ще й перелік основних математичних дисциплін, які розміщені згідно порядку їх вивчення на кожному курсі, на протязі кожного семестру. Оскільки було раніше виявлено, що професійні компетентності вчителя математики формуються практично в процесі вивчення математичних дисциплін, тому кожен з етапів реалізації моделі тісно пов’язано з основними математичними дисциплінами, які вивчають студенти на І – IV курсів.

Модель побудована таким чином, що перший етап її реалізації охоплює перші два семестри (І курс). До складу цього етапу включено такі курси як інформатика, елементарна математика та математичний аналіз. На даному етапі передбачається вивчення студентами основ використання хмарного сервісу, безпосереднє застосування інструментарію, що входять до CoCalc та організації власної роботи засобами, які представлені в хмарному сервісі. Перше знайомство з хмарним сервісом відбувається в процесі вивчення інформатики, зокрема розділи «ПЗ математичного призначення», «Сучасні ІКТ». Альтернативою може виступати включення до навчального процесу факультативного курсу «Використання CoCalc у навчанні математичних дисциплін» в процесі якого студенти опанують основні види роботи в хмарному сервісі. Практичне застосування та відпрацювання навичок студенти одержують в рамках вивчення математичних дисциплін: елементарна математика та математичний аналіз.

Окрім того, треба зазначити, що на реалізацію кожного з етапів моделі опосередковано впливають компетентнісний підхід до навчання та безпосередньо – професійні компетентності вчителя математики, охоплюючи усі три етапи.

З іншого боку, кожен з етапів в більшій чи меншій мірі вносять зміни до змісту навчання математичних дисциплін. До змісту навчання включено хмарні сервіси навчання, які охоплюють весь період підготовки вчителя математики (з І по IV курси).

ІІ етап методики охоплює 3-6 семестри (ІІ та ІІІ курси). До другого етапу реалізації методики використання хмарного сервісу CoCalc як засобу формування професійних компетентностей учителя математики включено наступні математичні дисципліни: лінійна алгебра те аналітична геометрія, математичний аналіз, функціональний аналіз, теорія ймовірностей та математична статистика, математична логіка, диференціальні рівняння та методика навчання математики.

II етап методики містить наступні складники: цілі навчання математичних дисциплін, методи навчання з використанням CoCalc, в якості основного засобу навчання виступає CoCalc (як провідний засіб формування професійних компетентностей майбутнього вчителя математики) та форми організації використання CoCalc. Кожен складник взаємопов'язаний один з одним, впливають безпосередньо на зміст навчання математичних дисциплін. При цьому, цілі навчання математичних дисциплін залежать безпосередньо від професійних компетентностей учителя математики. В центрі розміщено хмарний сервіс CoCalc, в якості засобу навчання математичних дисциплін. Під його впливом дещо зміняться традиційні методи навчання математичних дисциплін.

Якщо ж розглянути потенціал використання комп'ютера в проблемному навчанні, то бачимо, що студент виступає в ролі дослідника, який відкриває для себе щось нове. Звичайно, факти, які розглядає студент, вже відомі науці. Комп'ютерні засоби, що задіяні в проблемному навчанні, надають можливість користувачу описувати проблему (задачу), як природною мовою, так і мовою з даної предметної галузі. За таких умов з'являється можливість не лише обговорення правильної відповіді, але й раціональність кожного окремого розв'язку задачі. Роль викладача поступово змінюється. Постає питання стосовно перегляду форм навчальної роботи: збільшення самостійної роботи, зменшення ролі пояснювально-ілюстративного методу навчання, збільшення кількості практичних і лабораторних робіт дослідницького характеру [264].

Щодо III етапу реалізації методики, то він включає в себе диференціальні рівняння та методику навчання математики. На даному етапі студенти завершують навчання та готують навчально-дослідницькі проекти з використанням CoCalc, виконують курсові роботи, випускні роботи бакалавра. Із залученням інстументарію CoCalc студенти мають змогу виконати всю роботу з використанням одного хмарного сервісу. Окрім того, на даному етапі визначені критерії та інструментарій вимірювання рівнів сформованості професійних компетентностей учителя математики, визначається відповідність результату: сформованість професійних компетентностей учителя математики. Треба зазначити, що перевірка згідно критеріїв та рівнів кожного складника системи професійних компетентностей учителя математики відбувається неперервно, та охоплює усі етапи запропонованої моделі. Однак, в якості результуючої перевірки відповідності результату може виступати державний екзамен, який виступає в якості комплексної оцінки рівня сформованості професійних компетентностей учителя математики. Як видно з моделі, результат повністю узгоджено з поставленою метою.

**Висновки до розділу 2**

1. Узагальнюючи систему професійних компетентностей учителя математики та результати дослідження, було виокремлено такі складники системи професійних компетентностей майбутніх учителів математики:

1) загально професійні компетентності;

2) спеціально професійні: предметні (наукові, предметно-педагогічні), технологічні (інформологічно-методологічні, інформаційно-технологічні) та професійно-практичні компетентності (математичні, методичні).

Кожен складник запропонованої системи професійних компетентностей охарактеризовано згідно чотирьох рівнів їх сформованості: високого, достатнього, середнього, низького.

2. На основі спроектованої системи професійних компетентностей майбутніх учителів математики була розроблена модель використання хмарного сервісу CoCalc як засобу формування професійних компетентностей учителя математики. При створенні моделі формування професійних компетентностей учителя математики було враховано такі принципи: принцип науковості, професійної спрямованості, самореалізації міждисциплінарної інтеграції, варіативності.

Модель складається з цільового, стимулювально-мотиваційного, змістового, операційно-діяльнісного та оціночно-регулятивного компонентів. Модель охоплює три етапи формування професійних компетентностей: пропедевтичний, формувальний, розвивальний.

Основу моделі складає мета, яка формується під впливом суспільного замовлення на підготовку компетентного вчителя математики, що обумовлено компетентнісним підходом до навчання та спрямуванням на ІКТ-аутсорсинг засобів навчання та розвиток ІКТ. Розвиток ІКТ сприяє появі хмарних сервісів навчального призначення, питання вивчення яких, у свою чергу, входять до змісту навчання математичних дисциплін. Зміст їх навчання розписано по семестрам та охоплює чотири курси підготовки майбутнього вчителя математики. У змісті передбачено перелік основних математичних дисциплін та предметне навчання використання хмарного сервісу CoCalc. Кожен з етапів реалізації моделі тісно пов’язаний з основними математичними дисциплінами, які вивчають студенти освітнього рівня «бакалавр».

На основі компетентнісного підходу забезпечується поєднання різних етапів формування професійних компетентностей учителя математики у єдине ціле у межах запропонованої моделі. Це призводить до зміни змісту навчання математичних дисциплін.

Підготовка вчителя математики передбачає набуття ним компетентностей у галузі математики та суміжних з нею дисциплін, методики навчання та дидактики, психологічних і педагогічних основ здійснення навчально-виховного процесу, дослідницької діяльності та педагогічного спілкування, що визначає якість його професійної діяльності.

На формувальному етапі відбувається використання хмарного сервісу CoCalc у процесі вивчення математичних дисциплін.

# РОЗДІЛ 3
# МЕТОДИКА ВИКОРИСТАННЯ COCALC ЯК ЗАСОБУ ФОРМУВАННЯ ПРОФЕСІЙНИХ КОМПЕТЕНТНОСТЕЙ УЧИТЕЛЯ МАТЕМАТИКИ

## 3.1 Структура методики використання CoCalc як засобу формування професійних компетентностей учителя математики

Методика використання CoCalc як засобу формування професійних компетентностей учителя математики – це система взаємозв'язаних форм організації, методів і засобів навчання, що викладач використовує для використання хмарного сервісу на всіх етапах формування професійних компетентностей вчителів математики, що приводить до заздалегідь визначеного очікуваного результату.

Використання сервісу CoCalc в процесі викладання математичних дисциплін, потрібно перш за все зорієнтувати на:

– набуття навичок самостійного використання програмного математичного забезпечення для символьних і чисельних обчислень;

– підтримування викладання математичних понять (зокрема з курсів аналітичної геометрії, лінійної алгебри та математичного аналізу).

Перший етап даної методики передбачає два шляхи:

1. Навчання за програмою факультативу «Використання CoCalc у навчанні математичних дисциплін», як елементу змісту навчання інформатичних дисциплін [304].

2. Через систему тренінгів, семінарів, вебінарів, індивідуальних консультацій.

Рекомендується використовувати в рамках проведення факультативного курсу незначну кількість індивідуальних робочих аркушів хмарного сервісу (приблизно 15-20). В процесі виконання завдань студенти набувають навичок роботи з редактором програмного коду та окремими його частинами – комірками. На початку потрібно ознайомити аудиторію з основами роботи в CoCalc, типовими завданнями що включають: рівняння та нерівності, вектори, точки, лінії і площини, диференціювання, інтегрування, аналіз властивостей функцій, побудова графіків функцій на площині та у просторі, матриці, лінійні системи, лінійні оператори, власні значення та власні вектори. Індивідуальні завдання мають складатись з графічного та чисельного розв'язків.

На I етапі формування професійних компетентностей згідно запропонованої методики до варіативної частини освітньо-професійної програми підготовки бакалавра математики доцільно включити факультативний курс «Використання CoCalc у процесі вивчення математичних дисциплін». Факультатив спрямований на врахування міждисциплінарних зв'язків (математичних та інформатичних дисциплін професійно науково-предметної підготовки).

Факультатив спрямований на врахування міждисциплінарних зв'язків (математичних та інформатичних дисциплін професійно науково-предметної підготовки).

На II етапі формування професійних компетентностей відбувається у межах вивчення нормативних математичних навчальних дисциплін. III етап охоплює виконання навчально-дослідницьких проектів з використанням CoCalc, курсових робіт, випускної роботи бакалавра.

Головною метою методики використання хмарного сервісу CoCalc як засобу формування професійних компетентностей учителя математики є: підвищення рівня сформованості його професійних компетентностей.

Цільова група: майбутні вчителі математики (студенти).

Очікуваний результат методики використання хмарного сервісу CoCalc:

– розширити сучасні погляди на інформаційні процеси, їх роль у вивченні математичних дисциплін;

– навчитися успішно застосовувати інструментарій CoCalc для вирішення практичних завдань з математичних дисциплін;

– набути досвід роботи в колективі (за рахунок використання інструментарію CoCalc);

– розв'язувати практичні завдання доступними способами та подавати одержані результати;

– набути уміння оцінювати та систематизувати здобуті знання з математичних дисциплін.

Змістовий компонент методики використання хмарного сервісу CoCalc включає предметне навчання цього сервісу, педагогічно обґрунтовані, логічно впорядковані та текстуально зафіксовані в навчальних програмах наукові відомості про матеріал, що доцільно вивчати із застосуванням CoCalc.

Форми організації навчання із використанням хмарного сервісу CoCalc: діалогічні форми, індивідуальні та групові консультації, самостійна робота, практична робота, індивідуальна робота, парна робота, фронтально-колективна робота, диференціально-групова робота, колективні та індивідуальні проекти.

Провідні методи навчання математичних дисциплін з використанням хмарного сервісу CoCalc:

а) методи організації й здійснення навчальної діяльності (словесні, наочні, практичні репродуктивні й проблемні, самостійної роботи);

б) методи стимулювання й мотивації навчання (методи формування обов'язковості й відповідальності в навчанні: пред'явлення педагогічних вимог);

в) методи контролю й самоконтролю (письмовий контроль, лабораторні й практичні роботи, фронтальний і диференційований контроль, поточний і підсумковий).

Засоби формування професійних компетентностей учителя математики, що передбачені із використанням хмарного сервісу:

– робочі аркуші, на яких студенти виконують дії з побудови та дослідження математичних моделей;

– чат-кімнати, що використовуються для обговорення процесу та результатів моделювання;

– засоби підтримки навчальної діяльності (ресурси типу course, tasks);

– засоби для створення математичних текстів (tex) та гіпертекстів (html);

– мобільний доступ до інших засобів підтримки математичної діяльності.

Додатковими засобами є навчальний посібник «Організація навчання математичних дисциплін у CoCalc», web-сайт з методичними рекомендаціями для майбутніх вчителів математики з використання хмарного сервісу CoCalc у процесі навчання математичних дисциплін та проекти з використанням CoCalc для підтримки навчання.

Основними положеннями [349], якими мають слідувати викладачі під час використання хмарного сервісу CoCalc, мають бути наступними.

1. На початку вивчення курсу.

Викладач має бути впевненим, що студенти на самому початку вивчення дисципліни складають програмний код самостійно, вивчаючи синтаксис основних команд. Звичайно в подальшому студенти можуть копіювати окремі частини програмного коду, але на перших практичних заняттях викладач має слідкувати за ручним набором усіх команд без виключення. Це нормально, якщо студенти роблять лише певні висновки з даних обчислень. Але погано, коли вони не зможуть самостійно виконати навіть елементарні дії без посібника чи довідника. На практиці студенти рідко читають докладно пояснення, написані викладачем, намагаючись якнайшвидше виконати те, що від них вимагають, особливо не запам'ятовуючи синтаксис головних функцій. Тобто в подальшому студенти не зможуть самостійно виконати складні обчислення не лише з використанням хмарного сервісу, але й письмово. Якщо ж привчити студентів прописувати усі нові команди по декілька разів вручну (не застосовуючи автоматичного доповнення назви функції), це значно підвищує рівень розуміння та запам'ятовування вивченого матеріалу. Тому на перших практичних заняттях рекомендується використання порожніх робочих аркушів, в той час як усі основні моменти викладені в лекційному матеріалі та опорних конспектах.

2. Потрібно допомогти студентам визначити поняття самостійно.

Це основний принцип навчання з використанням хмарних сервісів. Студенти більш вмотивовані отримати глибше розуміння теми та засвоїти вивчений матеріал, коли вони самостійно виявляють властивості математичних понять та їх закономірності, складають алгоритми виконання певних обчислень. Цей процес може займати досить значний проміжок часу, але з використанням хмарного сервісу він значно спрощується. Наприклад, вивчаючи чисельні методи, традиційний спосіб викладення матеріалу, це – проілюструвати алгоритм (формулу) задля виконання обчислень, похибку даного методу та навести приклад, який до того ж може бути не досить вдалим. Можна студентам запропонувати виконати завдання, застосовуючи нестабільні, неточні методи обчислення. При цьому студенти можуть обговорити одержані результати, коментувати хід виконання роботи. Доречним буде з'ясувати можливість удосконалення даного методу. Процес навчання за цим принципом буде кардинально відрізнятись, оскільки основні труднощі будуть у виявленні основних теоретичних положень експериментальних досліджень. Але знання, одержані дослідним шляхом будуть набагато ширшими та краще засвоєними.

3. Обговорення труднощів з іншими студентами.

Якщо студент довгий час не може знайти рішення поставленої проблеми, з часом він починає вважати її занадто важкою та втрачає до неї інтерес. Тому треба постійно обговорювати на семінарських заняттях проблеми, які виникають у процесі розв’язання завдань. Для студентів на початку обговорення можуть мати певні утруднення, враховуючи різний рівень підготовки та індивідуальні особливості. Для студентів, що в змозі опанувати більш високий рівень викладач може надати декілька індивідуальних завдань. Виконання завдань більш високого рівня покращить рівень знань студентів, поглибить теоретичне підґрунтя, теми вивчається.

4. Поступово підвищувати рівень складності.

Можливо, що студенти одразу зможуть сприймати матеріал більш високого рівня. Але вивчати теоретичний матеріал на досить високому рівні і одразу використовувати основні теоретичні положення в хмарному сервісі буде занадто складно. Тут можливо або нехтуванням строгості викладеного теоретичного матеріалу, або ж недостатнім розумінням застосування його інструментами CoCalc.

5. Анімації треба приділяти більше уваги.

Рішення нестандартних задач краще представити у вигляді анімації.

Друковані посібники, методичні рекомендації, звичайно супроводжуються ілюстраціями та графічними побудовами, але якщо викладач планує використовувати хмарний сервіс на підтримку вивчення певної дисципліни, краще проілюструвати розв'язок задачі за допомогою анімації. Мотивація студентів зростає, коли вони мають змогу візуалізувати результати своєї роботи.

Постає питання створення анімації, яка буде залежати від параметрів, що можна задавати стандартними елементами управління. В CoCalc реалізовано цей механізм. Змінюючи вхідні параметри програмного коду з використанням певного елементу управління, можна створити нову анімацію на основі декількох графічних побудов.

### 3.2 Методика навчання факультативу «Організація навчання математичних дисциплін у CoCalc»

Серед варіативної частини освітньо-професійної програми підготовки бакалавра математики може бути доцільним факультативний курс «Використання CoCalc у навчанні математичних дисциплін». Із запровадженням запропонованого курсу [301] студенти під час виконання індивідуальних завдань глибше розумітимуть матеріал, що відводиться на самопідготовку; зможуть отримати відповіді на питання, які виникли під час вивчення програмного матеріалу; удосконалити навички розв'язування практичних завдань.

Мета вивчення факультативного курсу: формування професійних компетентностей майбутніх учителів математики із використання CoCalc.

Завдання вивчення факультативного курсу [301]:

1. Поглиблення та систематизація знань з математичних дисциплін.
2. Вирішення проблем та питань, які виникають під час самостійного опрацювання програмного матеріалу математичних дисциплін.
3. Набуття навичок практичного застосування математичних понять та методів в процесі розв'язання задач.
4. Побудова і аналіз інтерпретацій розв'язків завдань за допомогою хмарних сервісів.
5. Розвиток ІКТ компетентності майбутніх учителів математики.

Очікуваний освітній результат від курсу:

– одержати загальне уявлення про хмарні математичні сервіси;

– навчитися застосовувати інструментарій CoCalc для вирішення практичних завдань з математичних дисциплін;

– набути досвід роботи в колективі (за рахунок використання інструментарію CoCalc);

– розв'язувати практичні завдання доступними способами та

подавати одержані результати;

– уміння оцінювати та систематизувати одержані знання з математичних дисциплін.

Для роботи у CoCalc необхідно володіти наступними вміннями та навичками, які набувають студенти на I етапі:

– вміти реєструватись та авторизуватись у системі, виконувати налаштування облікового запису (Додаток В);

– вміти створювати навчальні ресурси;

– вміти працювати з sagews-аркушами (включаючи найпоширеніші режими, знати основи мов: LaTeX, Python, HTML);

– вбудовувати відео, аудіо, анкети, графічні файли в ресурс «sagews»;

– спілкуватись у чатах навчальних ресурсів та в ресурсі типу «sage-chat»;

– вміти працювати з навчальним ресурсом типу «tex»;

– вміти завантажувати нові ресурси з електронних носіїв.

<u>Загальні вимоги до знань та вмінь, які повинен мати студент перед початком вивчення факультативу.</u>

Студент має знати:

– основи математичної логіки;

– основи алгоритмізації та програмування;

– основи мови HTML.

Студент повинен вміти:

– працювати з комп'ютером в якості користувача;

– працювати з глобальною мережею Інтернет;

– оцінювати місце, роль і значення отриманого результату в загальній системі математичних знань.

Орієнтований тематичний план занять факультативу (табл. 3.1, табл. 3.2) має включати:

1. Кількість годин, відведених на лекційні заняття. З лекційного матеріалу студенти мають орієнтуватись в першу чергу в формулюванні теорем, основних формул, розуміти їх практичне застосування.
2. Кількість практичних занять. Орієнтуватись потрібно більшою мірою на практичні заняття, методи та прийоми які використовуються під час розв'язання того чи іншого завдання.
3. Матеріал відведений на самостійне опрацювання.

За основу побудови кожного заняття факультативу можна взяти:

– індивідуальну роботу, яку виконують студенти протягом вивчення певної математичної дисципліни;

– домашні завдання, якщо виконання індивідуальної роботи не передбачено в рамках вивчення математичної дисципліни (чи вона складається суто з опрацювання теоретичного матеріалу).

У залежності від того, які саме практичні завдання будуть реалізовані засобами CoCalc, потрібно обрати набір функцій задля подальшого детального вивчення студентами.

*Таблиця 3.1*

**Орієнтований тематичний план факультативних занять**

| № | Тема | Кількість годин |
|---|---|---|
| 1 | Реєстрація. Початок роботи з CoCalc | 2 |
| 2 | Основи програмування: оголошення змінних, списки, умовний оператор, цикли | 2 |
| 3 | Функції виведення | 2 |
| 4 | Основи використання LaTeX | 2 |
| 5 | Робота з довідковими джерелами | 4 |
| 6 | Побудова графічних примітивів, графіків, областей, поверхонь. Створення анімацій | 4 |
| 7 | Розробка графічного інтерфейсу | 4 |
| **Разом за планом** | | **20** |

*Таблиця 3.2*

**Змістове наповнення кожної теми**

| № | Тема | Зміст |
|---|---|---|
| 1. | Реєстрація. Початок роботи з CoCalc | Ознайомлення студентів з CoCalc, опис інструментарію. Реєстрація. Надання доступу до проекту. Ознайомлення з інтерфейсом. Робота з проектами. Створення папки, файлів. Збереження файлів. Вивчення панелі інструментів у робочих аркушах. Робота в груповому чаті. |
| 2. | Основи програмування: оголошення змінних, списки, умовний оператор, цикли | Вивчення сполучень клавіш. Коментарі. Способи оголошення змінних. Оголошення функцій однієї та багатьох змінних. Виконання найпростіших обчислень. Умовний оператор: загальний вигляд та скорочений. Способи задання циклів. Особливості використання. |
| 3. | Функції виведення | Функції print(), show() та html(). Їх порівняння: особливості та недоліки. Параметри функції show(): текстові (виведення результатів обчислення) та графічні (графічні побудови). Використання функції html(),основні теги форматування тексту. Виведення результатів |

| № | Тема | Зміст |
|---|---|---|
| | | обчислень за допомогою функції html(). |
| **4.** | Основи використання LaTeX | Спеціальні символів мови LaTeX (стрілки, грецький та латинський алфавіт). Використання однорядкової математичної символіки. Використання багаторядкової математичної символіки. Спільне використання LaTeX та функції html() для візуалізації результатів обчислень. |
| **5.** | Робота з довідковими джерелами | Вивчення функцій, які стосуються саме обраної математичної дисципліни (Додаток Д). Вивчення усіх параметрів спеціальних функцій. Дослідження значень параметрів в залежності від індивідуального завдання (домашнього завдання тощо). |
| **6.** | Побудова графічних примітивів, графіків, областей, поверхонь. Створення анімацій | Використання функції show() задля побудови графічних зображень. Функція animate(). Основні параметри функції animate(). Фіксація осей, збільшення частоти анімації. |
| **7.** | Розробка графічного інтерфейсу | Основи створення інтерактивних моделей згідно з графічним інтерфейсом. Основні елементи управління та їх властивості. Використання CSS, HTML та LaTeX у процесі розробки графічного інтерфейсу. Групування елементів управління за допомогою списків. |

Перед початком проведення факультативу слід:

1. Отримати список студентів та їх електронних адрес.

2. Створити проект за назвою математичної дисципліни, в рамках якої буде проводитись факультатив.

3. У проекті необхідно:

a. створити загальну структуру каталогів. Вирішити чи усі матеріали будуть розташовані в окремому каталозі, чи вони будуть завантажені безпосередньо в проект. Організувати структуру для студентів (чи кожен студент матиме свою власну папку, де зберігатимуться його завдання, файли, чи буде існувати окрема папка для завдань, а окремо для їх виконання);

b. створити загальний чат для запитань, які виникатимуть у студентів під час роботи над кожним із завдань та обговорення виконаних завдань. Викладач може одразу надіслати повідомлення з основними посиланнями на додаткові дані в чат. Пояснити, що саме тут студенти можуть спілкуватись за темою робіт, запитувати, консультуватись як з викладачем, так і один з одним;

c. завантажити файл зі списком завдань, що охоплюватимуть весь зміст факультативу (або загальний, або індивідуально для кожного студента окремо). Якщо для кожного студента буде власний список завдань, тоді в якості назви файлу краще використати прізвище студента;

d. завантажити файли з короткими теоретичними відомостями по кожному заняттю факультативу;

e. завантажити файли-приклади, які будуть висвітлювати форму подання результатів виконання завдань. З коментарями та декількома прикладами. В подальшому треба звернути увагу студентів, що дані приклади не є універсальними. І тому кожен матиме свій результат, який одержить в процесі дослідження;

f. завантажити за необхідності, додаткову літературу;

g. створити курс (за потреби).

Також потрібно враховувати матеріально-технічне забезпечення ЗВО та групи студентів, у яких будуть проходити факультативні заняття. Заздалегідь потрібно з'ясувати:

1. В якій мірі оснащені комп'ютерні аудиторії.
2. Чи забезпечено вільний доступ до мережі Інтернет.
3. Наявність мережі Wi-Fi та загального доступу до неї.
4. Чи зможуть студенти позааудиторно виконувати завдання (наявність власного ПК, ноутбука, планшета, смартфона тощо, та можливості підключення до мережі Інтернет).

Попередити студентів, що для виконання завдань факультативу ними мають буди засвоєні в достатній мірі теоретичні відомості, тому до занять з факультативу треба готуватись заздалегідь, повторювати лекційний матеріал (основні формули, теореми, означення).

У процесі проведення курсу студенти активно використовували власні ноутбуки та Wi-Fi підключення до мережі Інтернет.

Організація спільної роботи з навчальними ресурсами проекту

Організувати спільну роботу з ресурсами CoCalc-проекту можна або на рівні окремо взятого ресурсу, зокрема робочого аркуша (sagews), або на рівні проекту в цілому.

Відкриття спільного доступу на рівні окремо взятого ресурсу є нічим іншим, як web-оприлюдненням (рис. 3.1) вмісту ресурсу у режимі «лише для читання» для всіх користувачів мережі Інтернет, які мають посилання на даний ресурс.

Недоліками такої публікації є те, що користувач-«читач» не має можливості управляти обчисленнями на sagews-аркуші, навіть якщо автор використав стандартні елементи управління у ньому. Проте, у разі необхідності, даний sagews-ресурс може бути скопійований або завантажений (рис. 3.2).

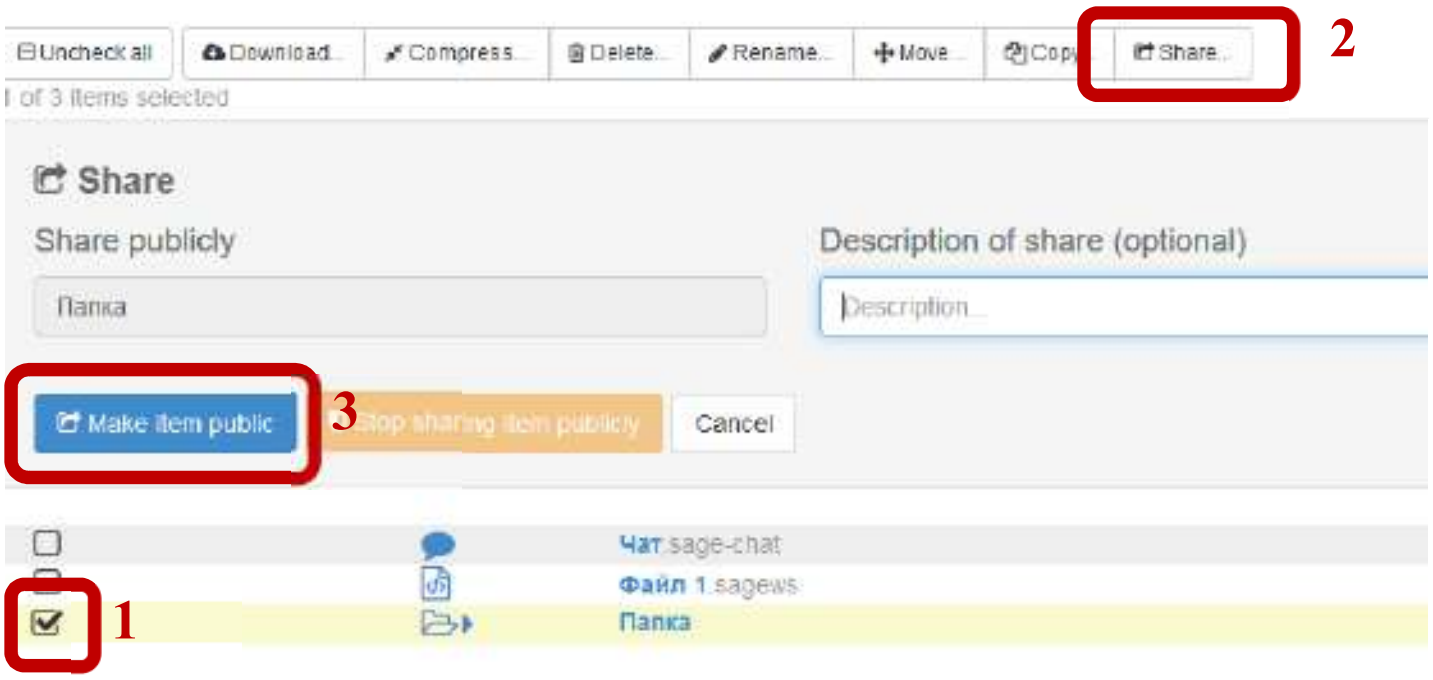

Рис. 3.1. Налаштування спільного доступу до sagews-ресурсу проекту

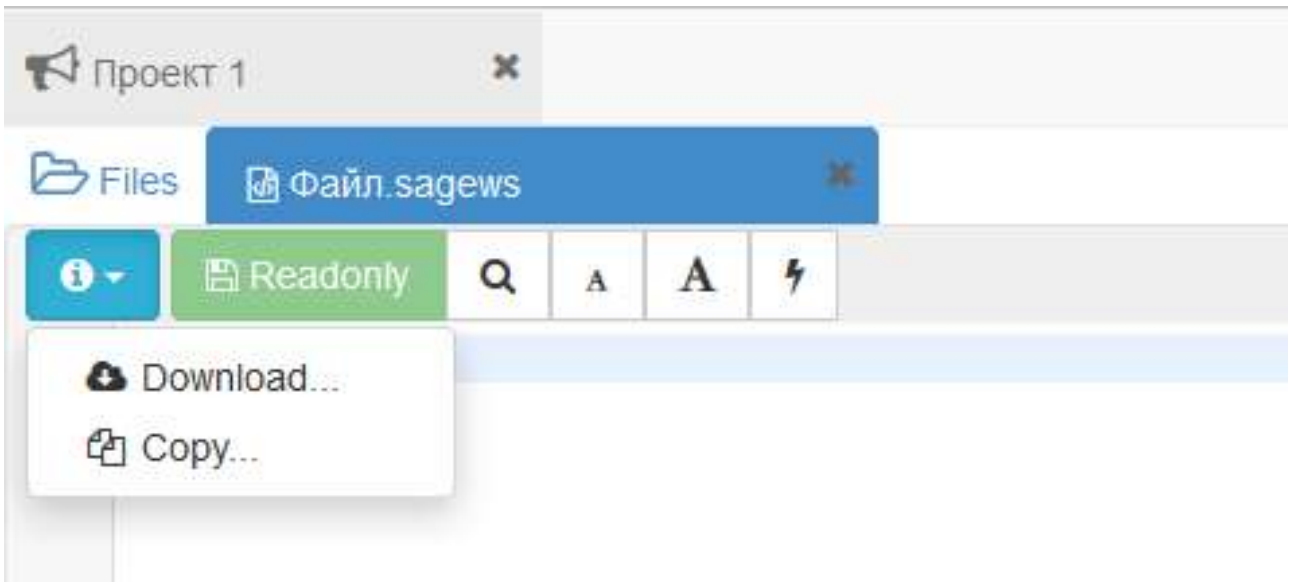

Рис. 3.2. Сторінка оприлюдненого sagews-ресурсу

Організація спільної роботи на рівні проекту в цілому можлива як без використання ресурсу типу course, так і з ним.

Перший спосіб передбачає підключення до проекту учасників, які матимуть можливість спільно працювати з вже існуючими навчальними ресурсами проекту або додавати нові, запрошувати інших учасників, спілкуватись за допомогою текстового та/або відео чатів у рамках спільного проекту.

Внесок кожного учасника спільного проекту у вирішення його завдань може бути переглянутий на сторінках історії роботи з проектом «Log» (рис. 3.3) або на сторінках резервних копій проекту «Backups» (рис. 3.4).

Резервні копії проекту зберігаються кожні 5 хвилин і не можуть бути остаточно видалені користувачем.

Одним із основних засобів зворотнього зв’язку у спільному проекті є файл-ресурс типу sage-chat (рис. 3.5, кнопка «Create a Chatroom»).

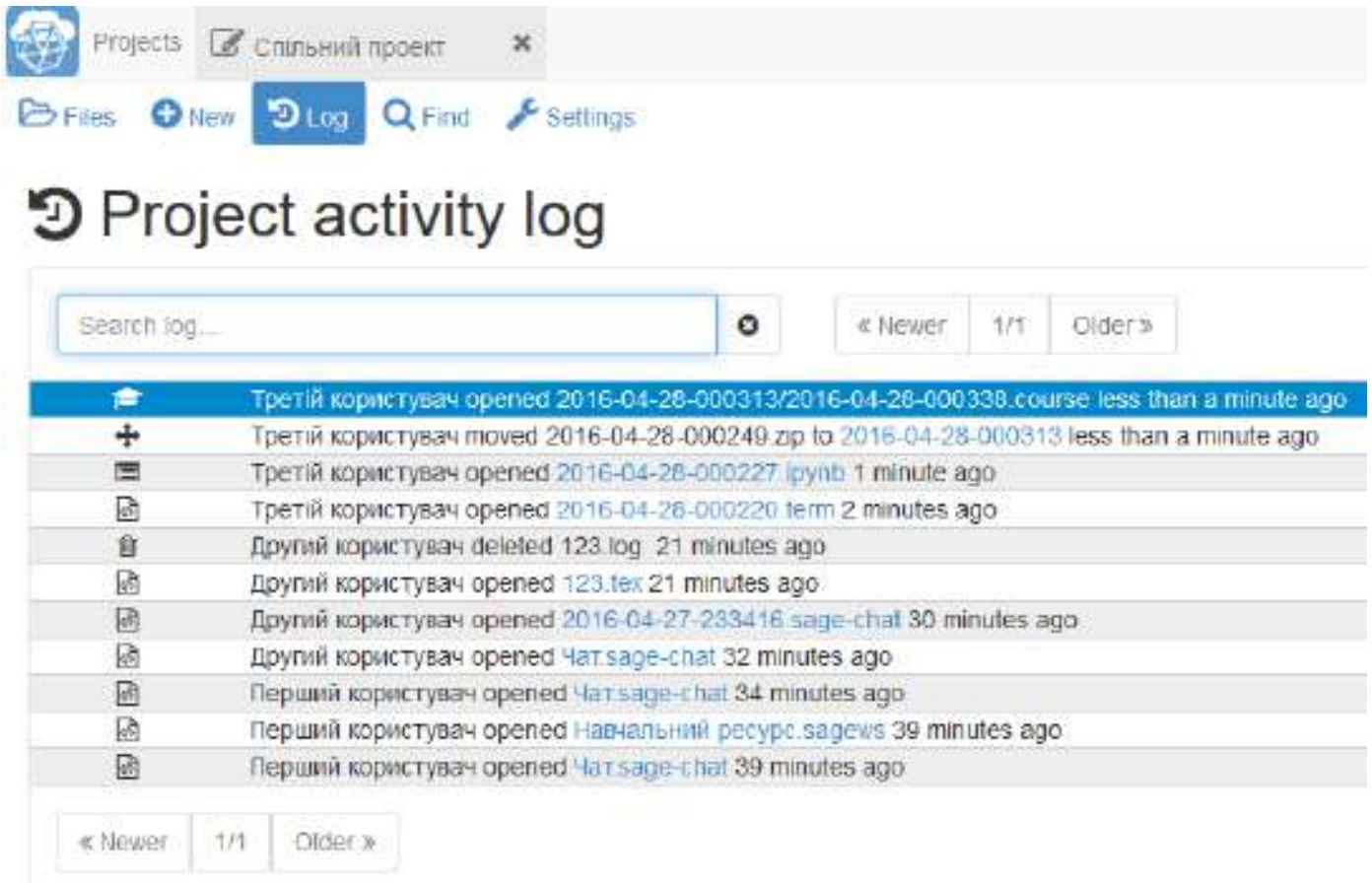

Рис. 3.3. Сторінка історії роботи з проектом

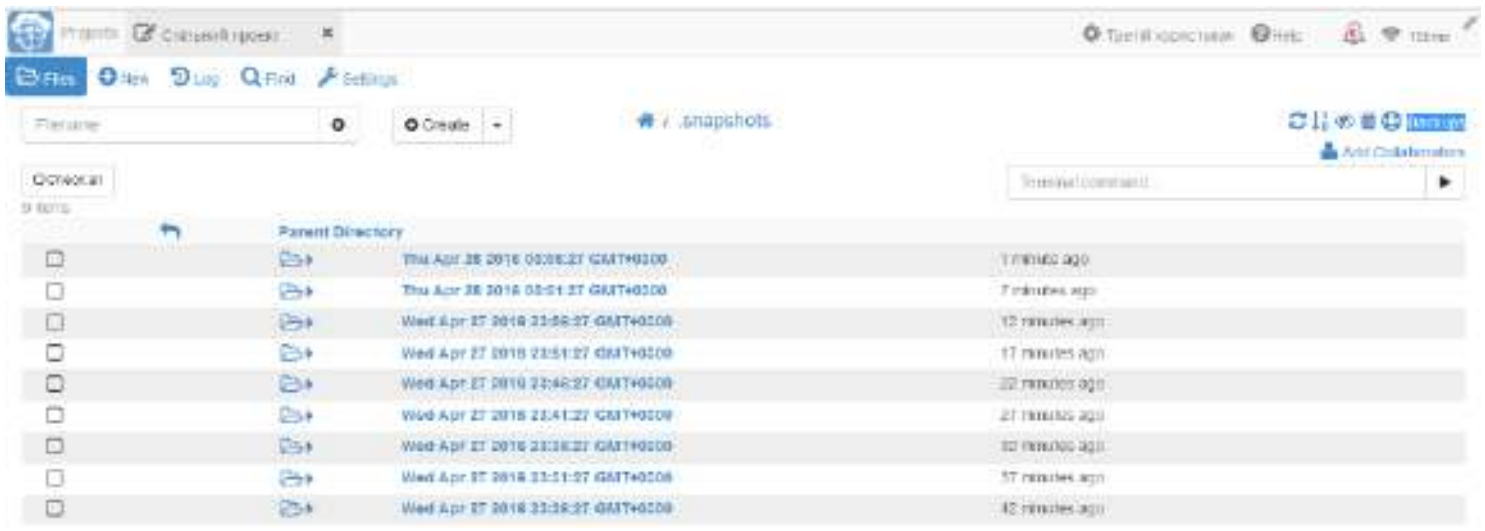

Рис. 3.4. Сторінка резервних копій проекту

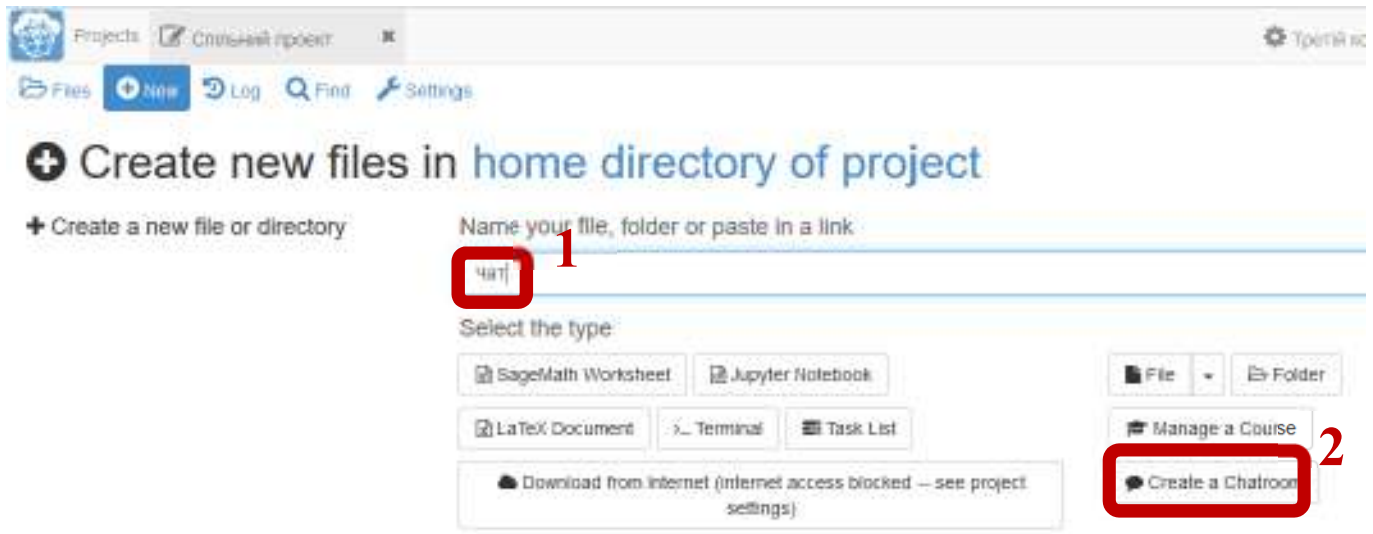

Рис. 3.5. Створення ресурсу типу sage-chat

Даний ресурс представляє собою текстовий чат, сторінку якого показано на рис. 3.6. Текст повідомлення можна форматувати за допомогою HTML-тегів та спеціальних команд розмітки (табл. 3.3).

Повідомлення математичного змісту може бути подане у звичній математичній нотації за допомогою команд LaTeX.

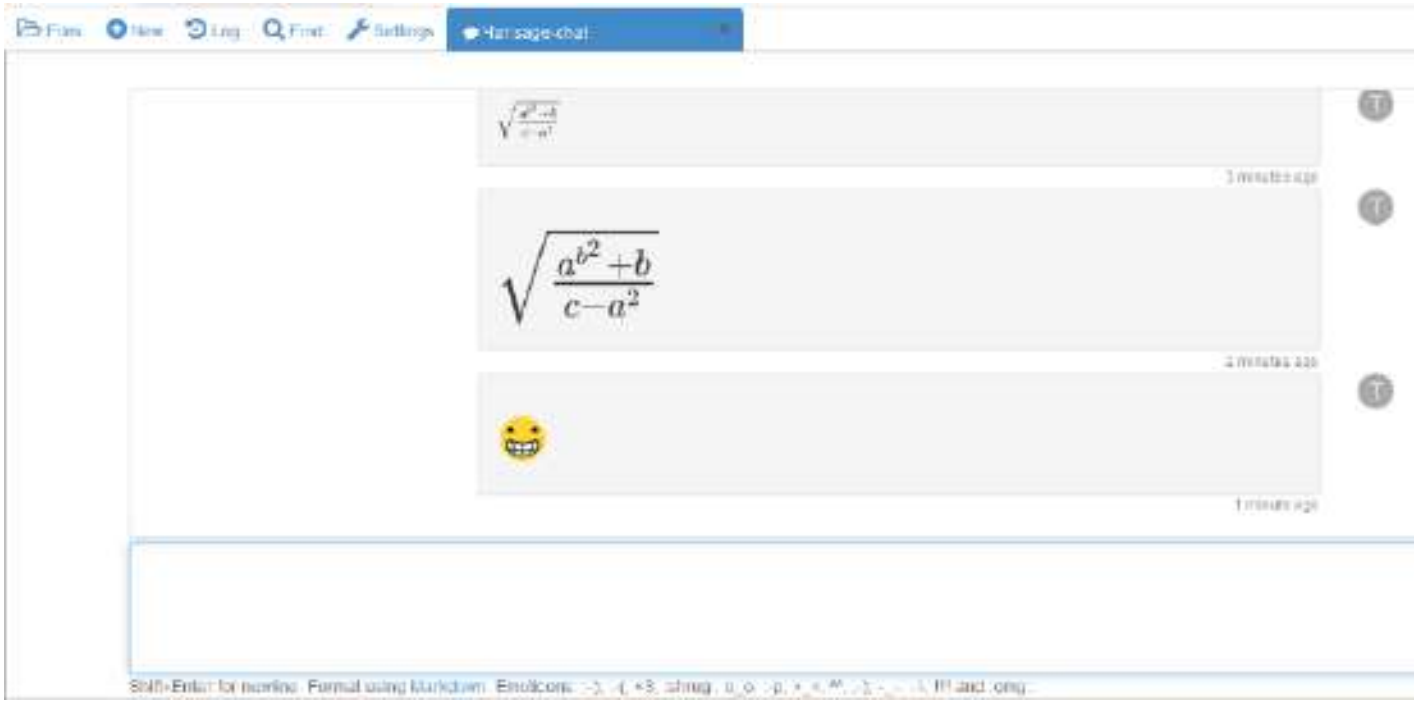

Рис. 3.6. Сторінка текстового чату

*Таблиця 3.3*

**Спеціальні команди розмітки**

| Команда | Призначення |
|---|---|
| #Текст | форматувати текст як заголовок |
| >Текст | форматувати текст як цитату |
| *Текст* | форматувати текст курсивним накресленням |
| **Текст** | форматувати текст напівжирним накресленням |
| - Текст | форматувати текст як маркований список |
| 'Текст' | форматувати текст шрифтом однакової ширини |
| Shift + Enter | перейти на наступний рядок |

Сповіщення про нове повідомлення в чаті (навіть у разі відсутності користувача у чат-«кімнаті» через роботу з ресурсам даного або навіть іншого проекту) надходить миттєво та відображається червоним кольором на піктограмі дзвоника (рис. 3.7). Переглянути сповіщення можна натиснувши на піктограму.

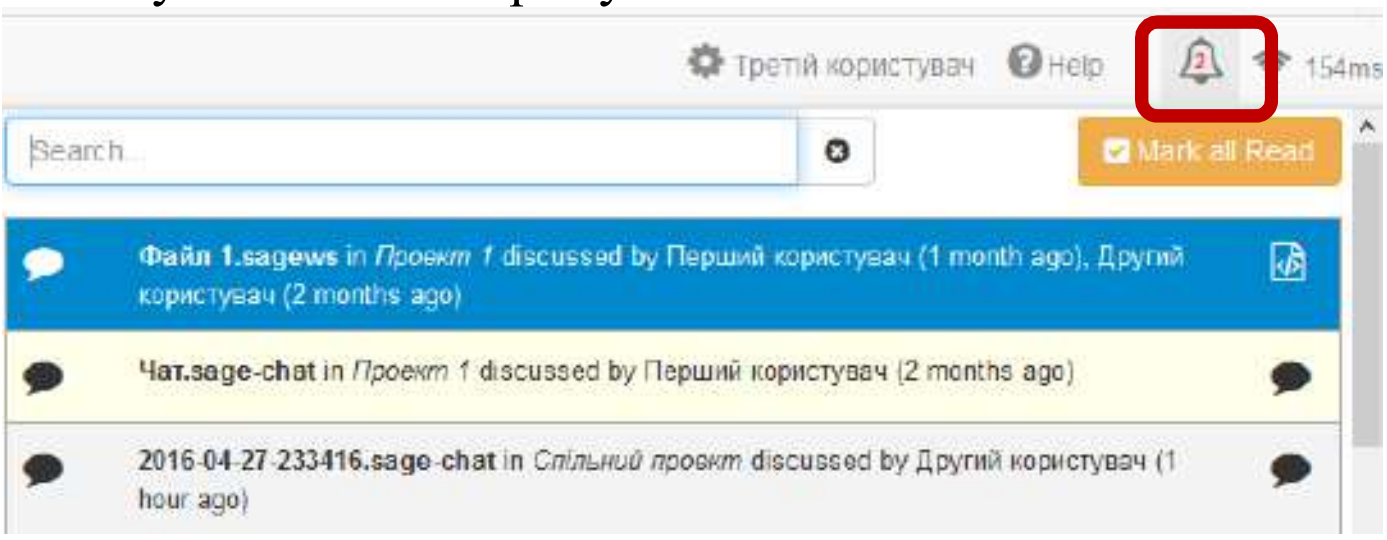

Рис. 3.7. Сповіщення про нові повідомлення в чаті

У чаті не відображається повне ім’я користувача, а лише його граватар. При наведенні курсору миші на граватар, з’явиться спливаюча підказка стосовно повного імені користувача.

Робочий аркуш SageMath є основним файлом проекту, в ньому відбувається введення команд користувача (Додаток Е) та відображаються результати обчислень. Спілкування у «кімнатах» текстового чату (рис. 3.8) або відеочату ресурсу типу sagews можливе лише для учасників проекту.

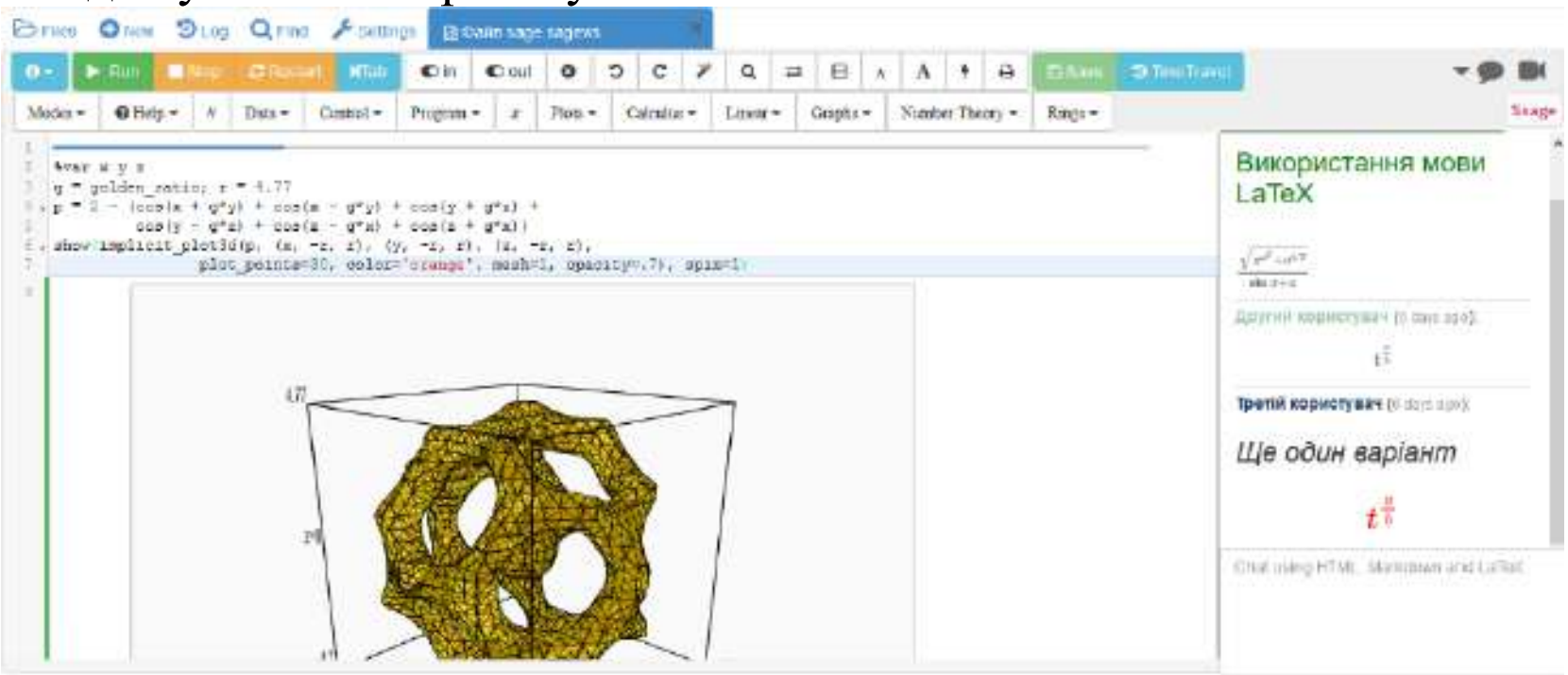

Рис. 3.8. Ілюстрація змісту текстового чату

Принцип роботи «кімнати» чату ресурсу типу sagews аналогічний до вище описаного.

Переглянути всю історію змін вмісту sagews-ресурсу можна на сторінці, що відкриється після натискання однойменної кнопки «Time Travel» (рис. 3.9). В історії зберігаються записи про всі зміни, що здійснені або самим автором проекту, або будь-ким із учасників. Пересуваючи повзунок, можна проглянути всі зміни у файлі. Початкова позиція повзунка відповідає моменту створення файлу. Якщо файл не змінювали і не редагували, в ній буде зазначено «Revision 0». Біля кожної зміни вказано дату та час. Остання позиція повзунка – останні зміни, що були здійснені під час редагування файлу.

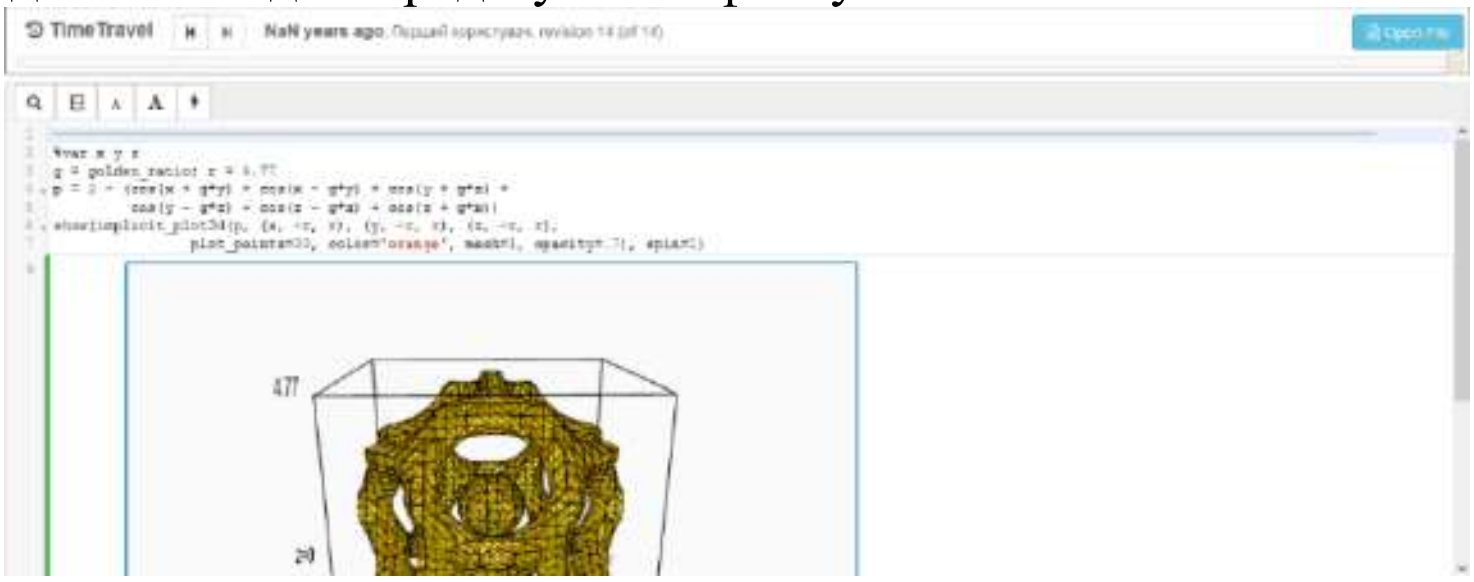

Рис. 3.9. Сторінка історії роботи з файлом

Працюючи над одним навчальним ресурсом, користувачі мають змогу бачити розташування курсорів один одного та піктограми тих учасників проекту, котрі в даний момент звернулись до файлу (рис. 3.10).

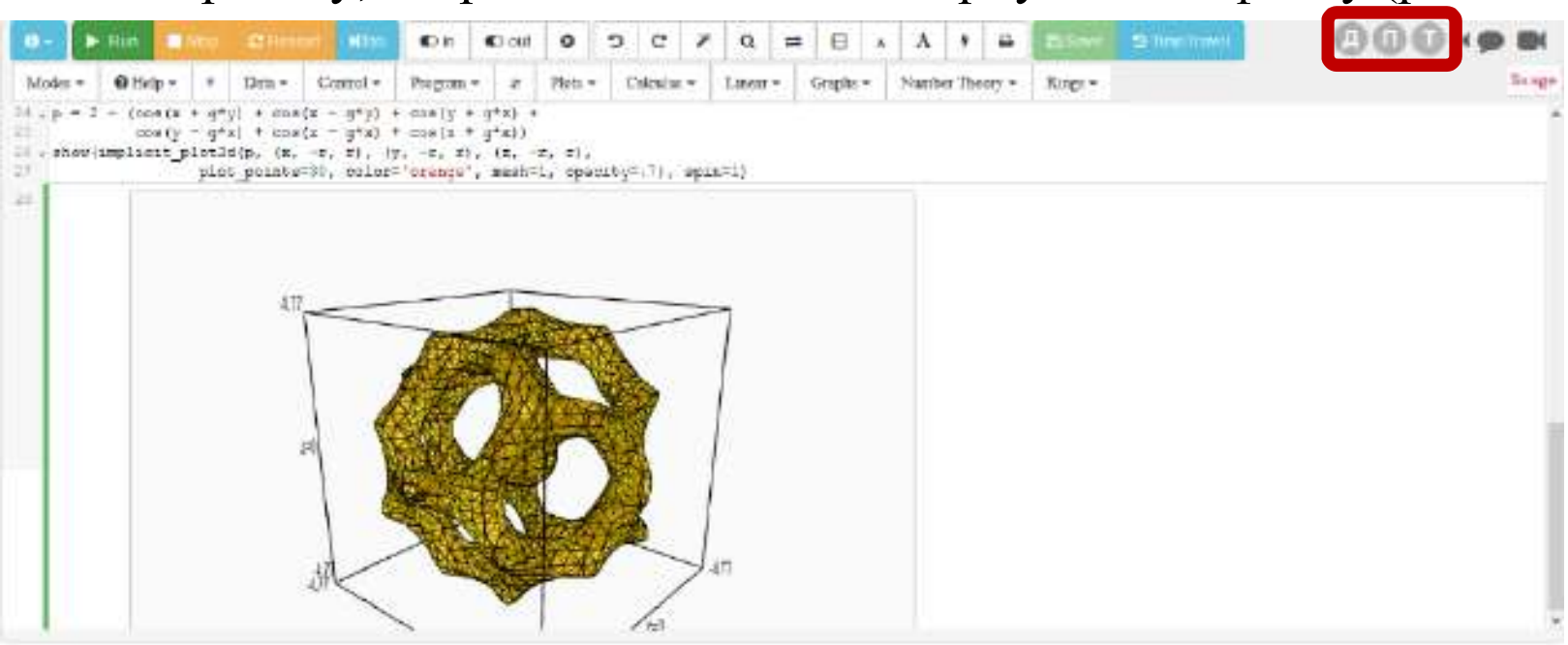

Рис. 3.10. Колективна робота над ресурсом типу sagews

Для того, щоб організувати індивідуальну роботу з кожним студентом окремо, треба створити для кожного студента групи проект, наповнити його навчальними ресурсами та відкрити доступ лише для окремого користувача (студента). В результаті викладачу треба створити таку кількість проектів, скільки студентів в кожній академічній групі, при цьому повторюючи одні й ті самі дії десятки разів. Для автоматизації даного процесу у CoCalc передбачено спеціальні ресурси (рис. 3.11) – навчальні курси (ресурси типу course).

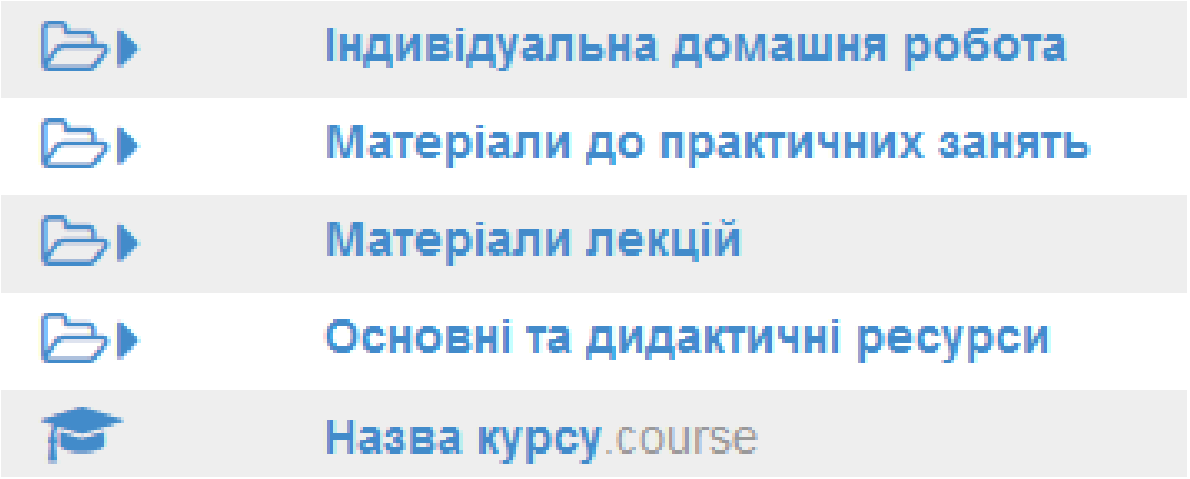

Рис. 3.11. Проект, що містить course-ресурс

Даний спосіб організації навчання у CoCalc дозволить викладачам контролювати процес виконання завдань студентами та здійснювати їх оцінювання.

Для організації навчання у CoCalc в рамках навчального курсу рекомендується створити окремий проект, наповнити його навчальними матеріалами з дисципліни та додати навчальний курс як ресурс «Manage a Course» (рис. 3.12, 3.13).

Після створення навчального курсу настає черга одного з двох етапів

– або додавання слухачів курсу, або визначення наявних у даному проекті навчальних матеріалів ресурсами курсу (порядок дій не є принциповим). Це можна виконати на відповідних сторінках після відкриття навчального курсу.

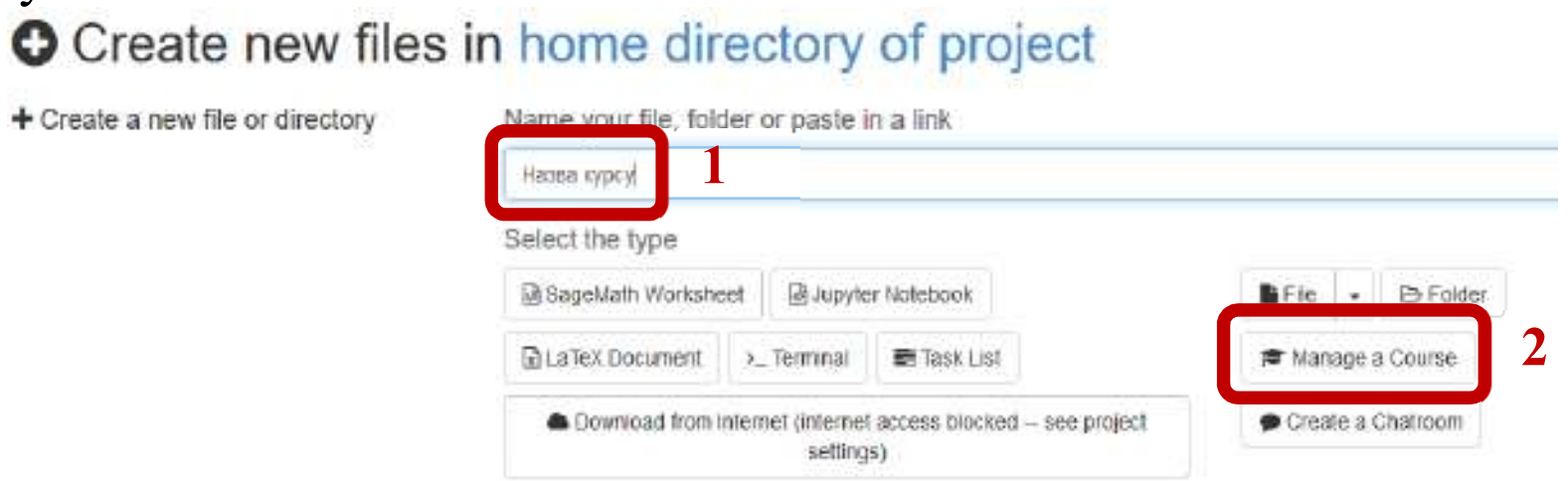

Рис. 3.12. Сторінка створення навчального курсу

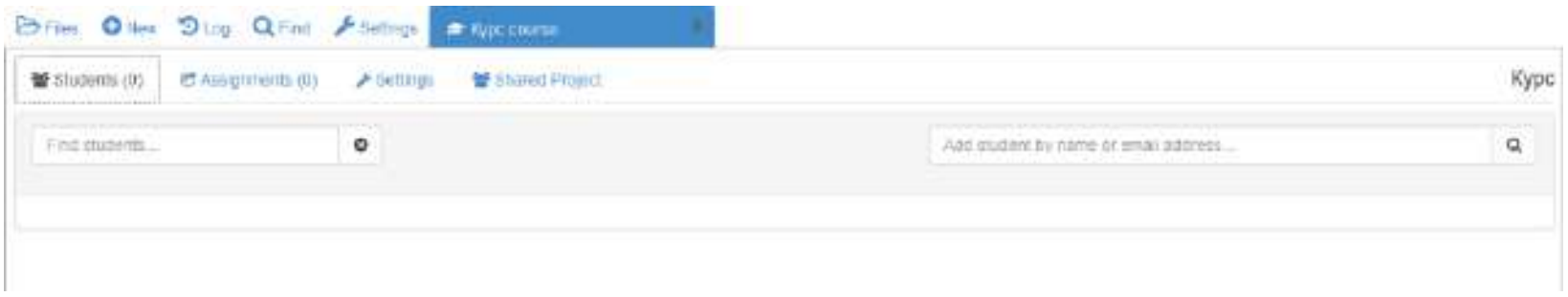

Рис. 3.13. Сторінка навчального курсу

Сторінка навчального курсу має вигляд, що представлено на рис. 3.13:

– ярлик сторінки «Students» (відкривається за замовчуванням, для управління слухачами курсу);

– ярлик сторінки «Assignments» (для управління навчальними ресурсами курсу);

– ярлик сторінки «Settings» (для налаштування параметрів навчального курсу);

– ярлик сторінки «Shared Project» (для створення проекту, спільного для всіх студентів).

Для додавання користувача до навчального курсу достатньо організувати його пошук за ім'ям (рис. 3.14) або адресою електронної поштової скриньки. У разі успішного пошуку (шукана особа має обліковий запис у CoCalc) із запропонованого списку потрібно обрати відповідний обліковий запис та додати його власника до переліку користувачів навчального курсу (рис. 3.15), натиснувши кнопку «Add selected student». Користувачів, які не мають облікового запису в CoCalc, також можна додати до навчального курсу за адресою їх електронної поштової скриньки. На вказану адресу автоматично надійде повідомлення стосовно їх включення до навчального курсу. Текст повідомлення можна змінити на сторінці «Settings».

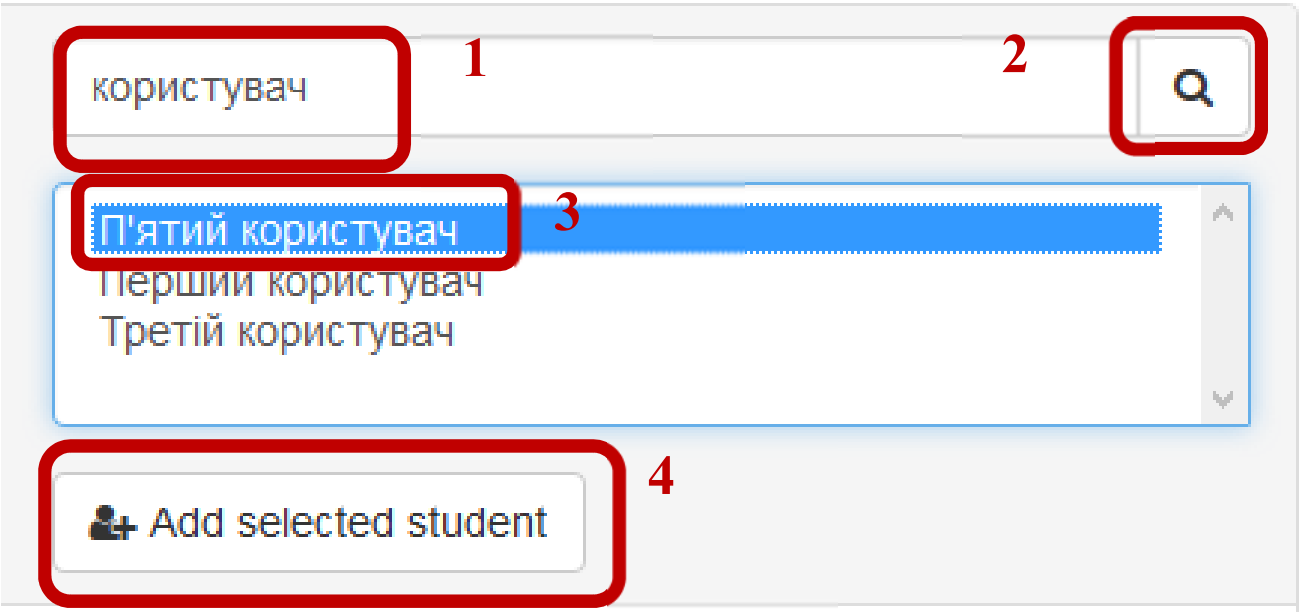

Рис. 3.14. Долучення користувача до курсу

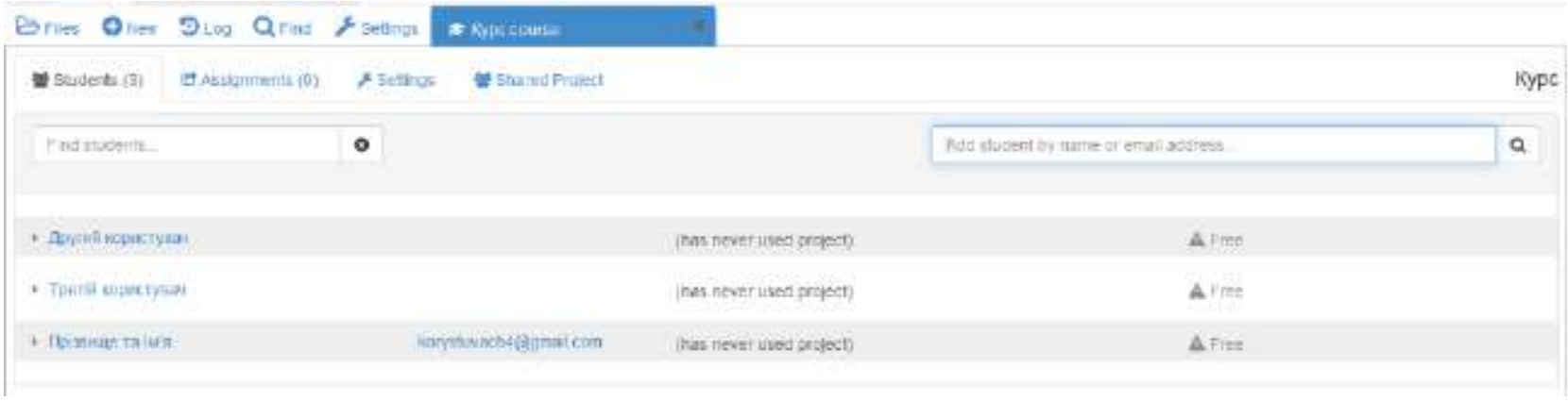

Рис. 3.15. Головна сторінка курсу: список студентів

Список користувачів курсу (сторінка «Students») містить такі відомості – прізвище та ім’я, електронна адреса, час останнього звернення та тип хостингу (рис. 3.15).

Поле «Прізвище та ім’я» є гіперпосиланням, за допомогою якого можна відкрити вікно управління індивідуальним проектом студента (коли користувача включено до курсу в якості студента, автоматично створюється особистий проект, доступ до якого відкрито для викладача та студента) та його окремими навчальними матеріалами. Назва особистого проекту студента складається з імені користувача (студента) та назви проекту в якому розміщено навчальний курс (наприклад, Другий користувач - Назва курсу). Поле «Електронна адреса» буде заповнене лише у випадку, коли пошук користувача під час додавання до курсу, відбувався за адресою електронної скриньки. В полі «Час останнього звернення» за замовчуванням зазначено – «has never used project» (ніколи не працював з проектом). У останньому полі списку за замовчуванням вказано значення «Free», що означає використання безкоштовного тарифного плану, в інших випадках – відповідні відомості про передплачений план користувача курсу.

Наповнення навчального курсу матеріалами (в подальшому – управління ними) здійснюється на сторінці «Assignments» (рис. 3.16). Процедура додавання навчальних матеріалів до курсу аналогічна до додавання слухачів курсу.

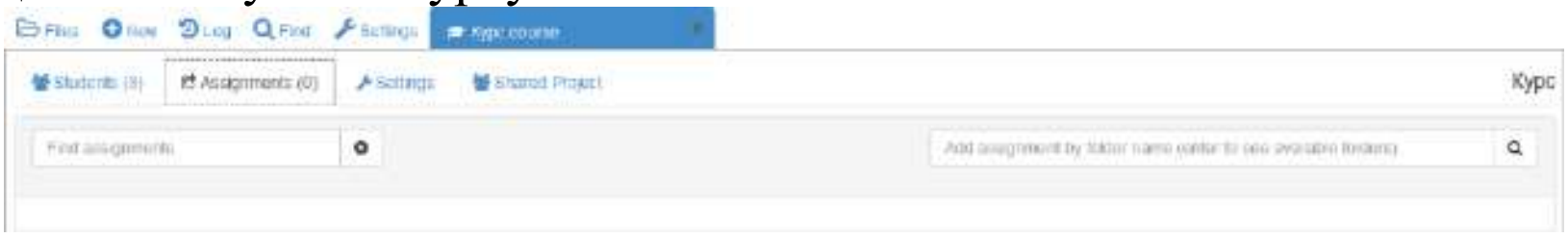

Рис. 3.16. Зовнішній вигляд сторінки «Assigenments»

У результаті буде сформовано список навчальних матеріалів. Натиснувши на назву навчального матеріалу відкриється вікно його основних відомостей та дій із ним (рис. 3.17):

– кнопка «Open» – переглянути вміст навчального матеріалу;

– поле «Due» – зазначити дату та час виконання завдань (студент може переглянути кінцевий термін виконання завдання у файлі DUE_DATE.txt, що автоматично створюється та розміщується у папці із відповідним завданням);

– прапорець «Peer Grading» – управління можливістю студентів оцінювати один одного;

– кнопка «Delete» – видалити папку навчального матеріалу з курсу;

– кнопка «Assign» – виконати копіювання папки усім студентам курсу;

– список користувачів курсу, для персоналізованого управління навчальним ресурсом (папкою);

– поле для введення «Private Assignment Notes» – за замовчуванням порожнє, але його вміст доступний лише викладачам.

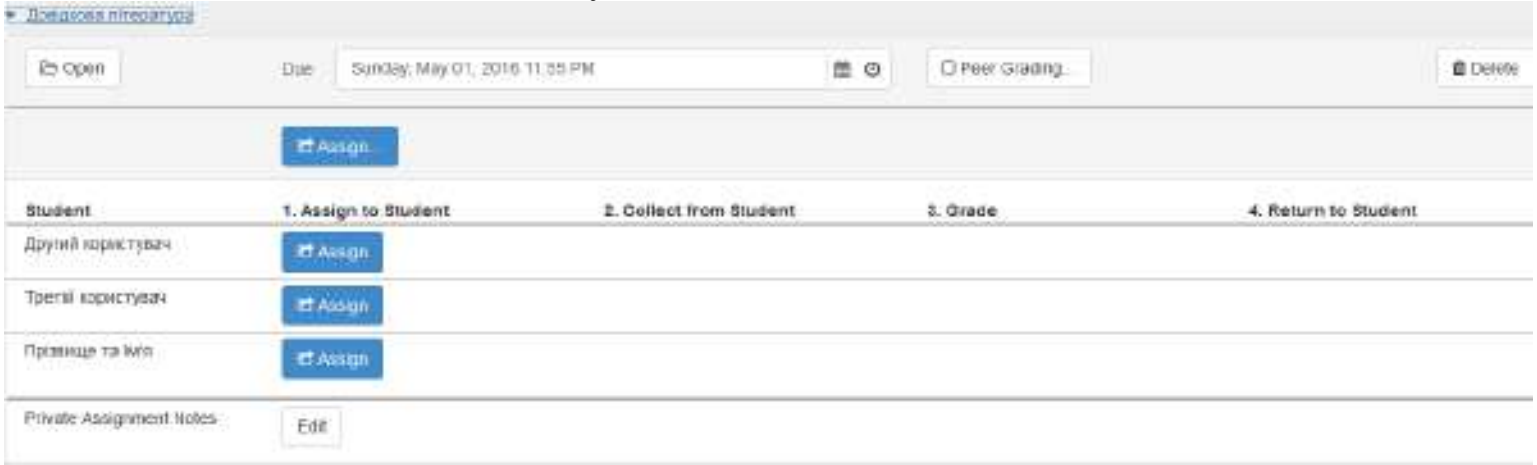

Рис. 3.17. Вікно основних відомостей папки курсу

Скопіювати навчальні матеріали усім студентам у власні проекти одночасно можна натиснувши кнопку «Assign», або окремо, навпроти певного студента. У разі призначення навчального матеріалу одночасно усім студентам курсу у вікні з'явиться діалог стосовно підтвердження дій (рис. 3.18).

Після копіювання завдання хоча б одному студенту курсу, над списком студентів у вікні загальних налаштувань папки з'являться кнопки, використання яких дозволить одночасно призначати завдання усім студентам курсу, збирати виконані завдання та повідомляти про оцінку (рис. 3.19).

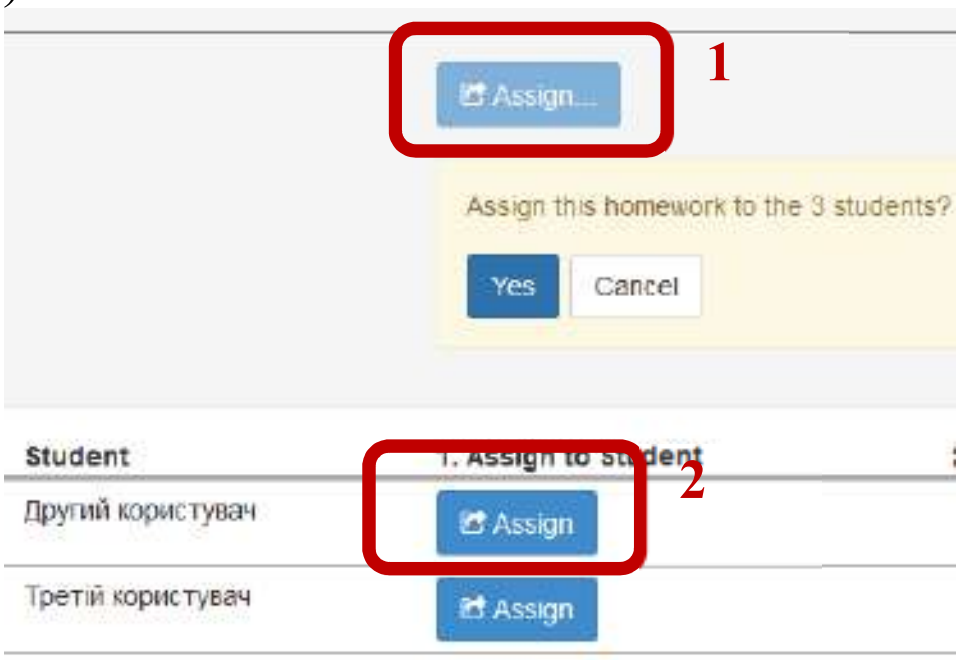

Рис. 3.18. Копіювання папки з навчальними ресурсами
(1 – всім студентам курсу, 2 – окремому студенту)

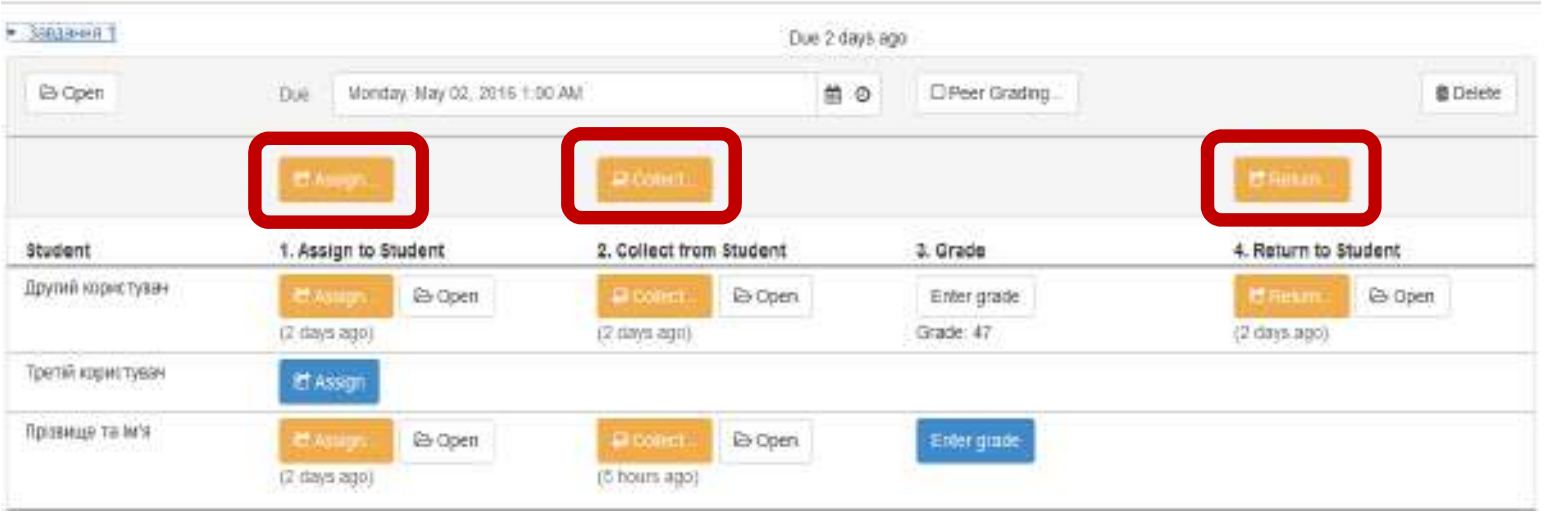

Рис. 3.19. Кнопки виконання дії одночасно для всіх студентів курсу
Assign – скопіювати/оновити папку навчальних матеріалів у індивідуальний проект кожного студента
Collect – зібрати/скопіювати папку навчальних матеріалів з виконаними завданнями кожного студента
Return – повідомити кожного студента стосовно поставленої оцінки

Значення полів даного списку повторюються у вікні управлінням індивідуальним проектом студента та його окремими навчальними матеріалами, сторінки «Students» (рис. 3.20), що містить кнопки «Open student project» та «Delete», список навчальних матеріалів та примітки.

Натиснувши кнопку «Open student project», можна одержати доступ до індивідуального проекту студента.

Список навчальних матеріалів складається з наступних полів (поля, що мають нумерацію в подальшому матимуть кнопки, за допомогою яких можна виконувати відповідні дії, що стосуються папки певного навчального матеріалу).

«Assignment» – список назв папок навчальних матеріалів (їх можна переглянути на сторінці «Assigenments»):

«1. Assign to Student» – скопіювати/оновити папку навчальних матеріалів студенту до його власного проекту;

«2. Collect from Student» – зібрати/скопіювати папку навчальних матеріалів із власного проекту студента;

«3. Grade» – виставити оцінку;

«4. Return to Student» – повідомити студента про оцінку.

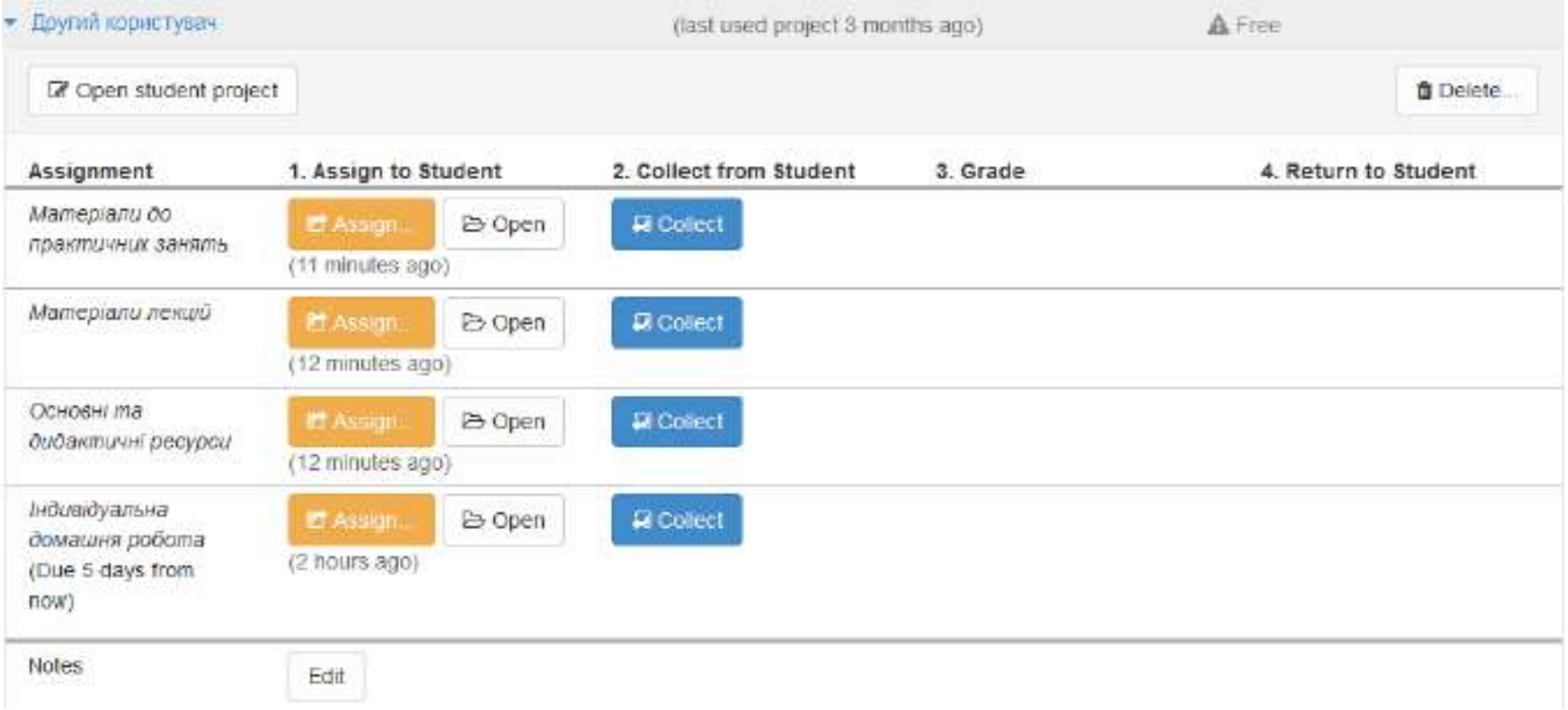

Рис. 3.20. Головна сторінка курсу: основні відомості студентів

Поле для введення «Notes» за замовчуванням порожнє, його вміст доступний лише викладачу/викладачам.

У списку студентів на сторінці «Assignments» в кожному полі з’являться аналогічні кнопки (рис. 3.21), що слугують для виконання тих же самих дії, але для певного студента/студентів (рис. 3.22), за виключенням кнопки «Enter grade».

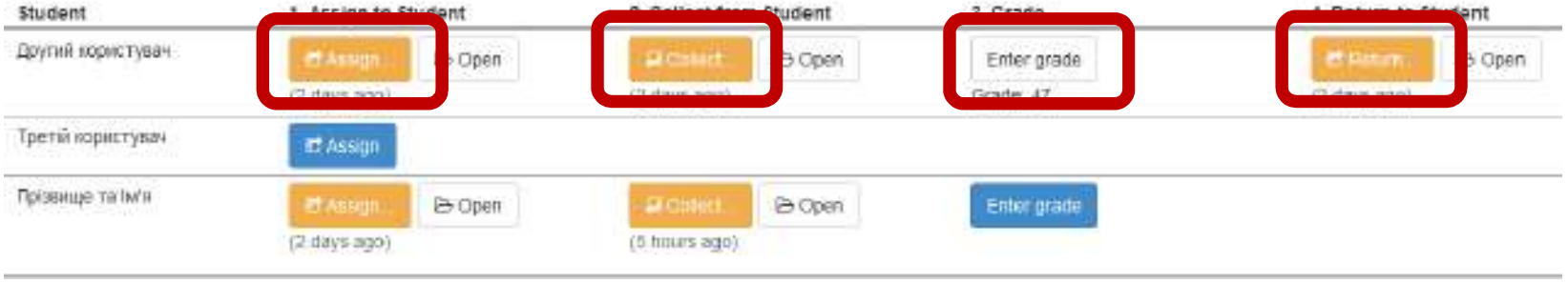

Рис. 3.21. Кнопки виконання дій для одного студенту курсу

Assign – призначити/скопіювати папку у власний проект студента

Collect – зібрати/скопіювати папку з виконаними завданнями студента

Enter grade – поставити оцінку студенту

Return – повідомити студента стосовно поставленої оцінки

Усі кнопки можна застосовувати повторно (вміст кожної папки матеріалів навчального курсу за потреби можна змінювати, студент може доопрацювати завдання тощо). Кнопки з’являються послідовно згідно покрокового виконання дій (не можна призначити завдання студенту та одразу оцінити його, пропустивши етап перевірки виконаного завдання). Крім того, в списку студентів в кожному полі (за виключенням «3. Grade») існує кнопка «Open», що відкриває зміст папки на кожному етапі дій (рис. 3.22).

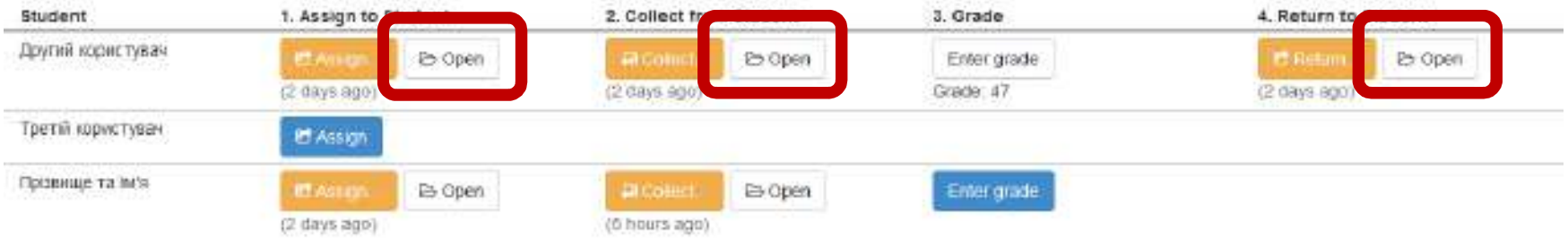

Рис. 3.22. Кнопки «Open»

Навчальні матеріали, що були вже скопійовані, можна за потребою оновити. Процедура оновлення аналогічна до процедури копіювання. Якщо вміст папки змінено (в тому числі було внесено зміни до вже існуючих ресурсів), то старі версії навчальних ресурсів будуть збережені з наступною назвою: назва.розширення~ .

Після натискання кнопки «Collect» в структурі проекту автоматично з’явиться папка з назвою, що відповідає назві ресурсу типу course та містить наприкінці «-collect». Її структура складатиметься з системи папок з виконаними завданнями та файлом, за яким можна встановити належність файлів власнику, що відповідатимуть певному студенту навчального курсу.

Коли ж викладач повідомляє студенту оцінку (поставити оцінку можна, натиснувши кнопку «Enter grade») за виконану роботу (кнопка «Return»), в проекті студента автоматично генерується папка з назвою папки завдань та в кінці додається «-graded». Зміст папки складатимуть ресурси, що переглянув викладач та автоматично створений файл з оцінкою.

На сторінці «Settings» загальних налаштувань ресурсу викладачеві можуть стати у нагоді наступні блоки:

– «Export grades» (рис. 3.23) – натиснувши на відповідну кнопку «CSV file…» або «Python file…», в структурі проекту буде створено файл із зазначенням усіх оцінок студентів обраного типу;

– «Customize email invitation» (рис. 3.24) – наведено текст повідомлення про включення користувача до курсу, що надійде на електронну поштову скриньку користувачу, якого включено до курсу, але який не зареєстрований в системі (за потреби текст повідомлення можна змінити натиснувши на кнопку «Edit»).

На сторінці «Shared Project» можна створити спільний проект для усіх студентів навчального курсу, в якому кожен студент автоматично буде учасником (рис. 3.24).

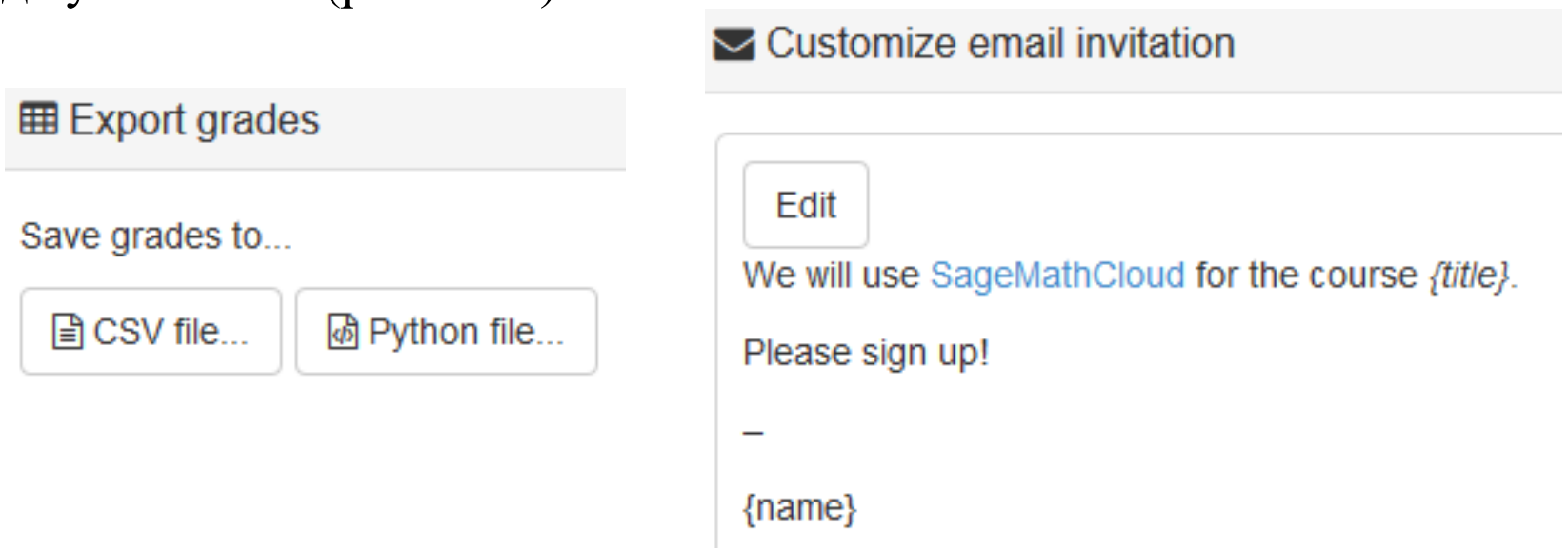

Рис. 3.23. Сторінка загальних налаштувань курсу:
блоки «Export grades» та «Customize email invitation»

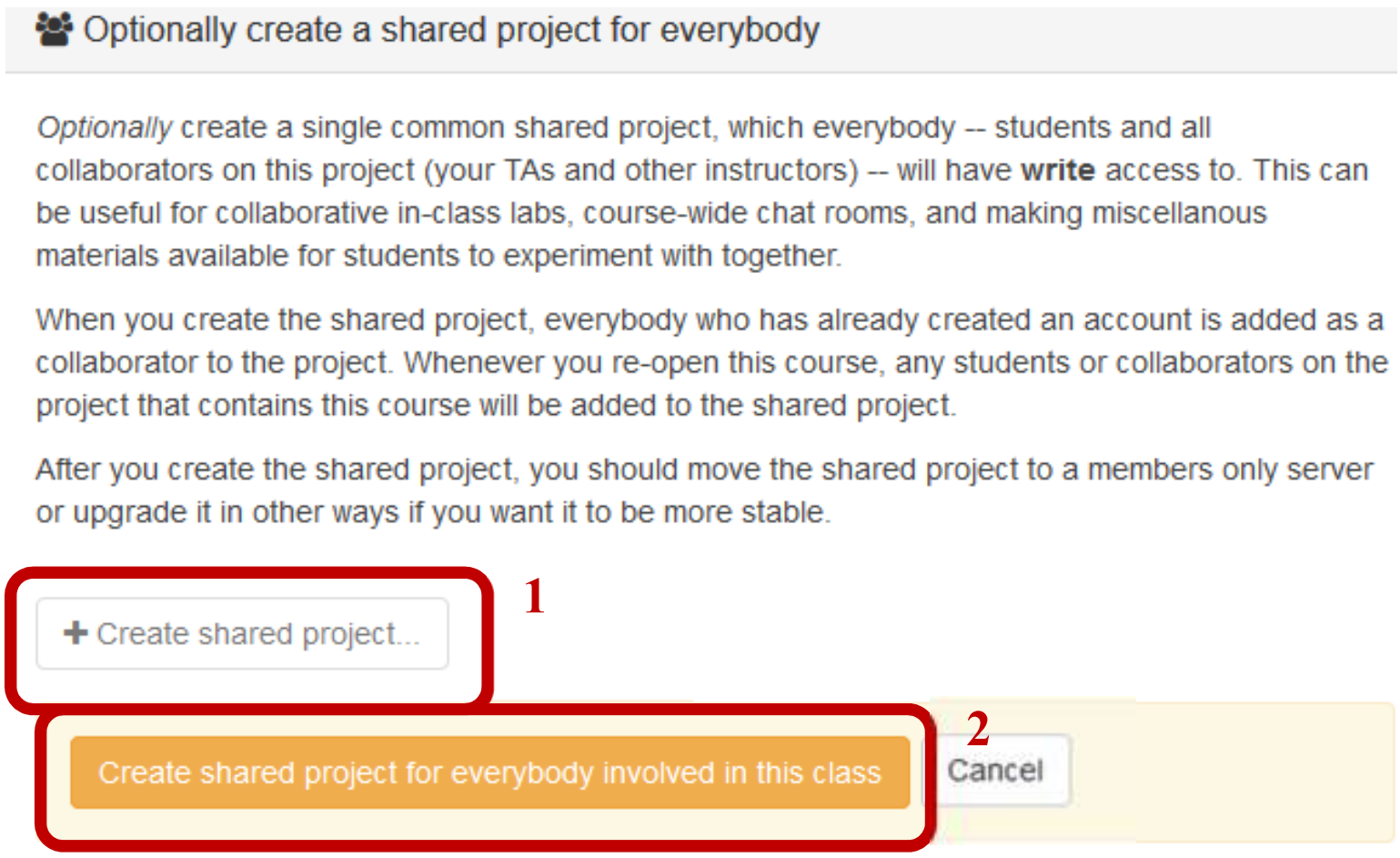

Рис. 3.24. Створення спільного проекту студентів ресурсу course

## 3.3 Форми організації навчання з використання CoCalc у формуванні професійних компетентностей майбутніх вчителів математики

До форм організації начального процесу можна віднести: лабораторні та практичні роботи, семінари [91]. Ми пропонуємо використовувати під час вивчення основ комплексного аналізу із застосуванням CoCalc наступні форми організації навчального процесу: практичні та лабораторні заняття. Практичне заняття покликано поглибити знання з даної дисципліни. Практичні заняття розвивають наукове мислення й мову, надають можливість перевірити знання студентів і постають як засіб оперативного зворотного зв'язку [91].

Можна запропонувати чотири способи використання хмарних сервісів в навчальному процесі:

– під час проведення лекцій (обговорень);

– семінарських занять;

– виконання домашнього (індивідуального) завдання;

– складання іспитів.

Використання хмарних сервісів як альтернатива традиційним презентаціям, може бути досить ефективною підтримкою під час проведення лекції. Якщо розробити лекційну демонстрацію засобами хмарного сервісу, то її можна застосовувати і під час обговорення на семінарському занятті. В процесі проведення лекції можна використати робочі аркуші, що містять:

– статичні демонстрації (найчастіше побудови на площині, без зміни вхідних параметрів);

– динамічні демонстрації (з змінними вхідними параметрами);

– заготовки програмного коду, без використання елементів управління (обчислення в реальному часі).

Презентацію можна також створити з використанням інструментарію хмарного сервісу. Єдина відмінність такої презентації буде в тому, що до слайдів є можливість включити програмний код та виконати певні обчислення [359]. Крім того, робочі аркуші можна експортувати до формату PDF та роздрукувати заздалегідь підготовлені матеріали. Особливо це буде зручно для студентів, оскільки виконані індивідуальні роботи можна буде в подальшому включити до власної презентації не витрачаючи час та зусилля.

Пропонуємо практичне заняття поділити на декілька частин: 50-60 хвилин – виконання обчислень (практичної та дослідної частини) із використанням хмарного сервісу, 20-30 хвилин (із урахуванням того, що заняття розраховано на 80 хвилин) – пояснення ходу розв'язання, обговорення методів, пропонуючи нові ідеї та рішення. Завершити заняття можна опитуванням (анкетуванням) тривалістю 5 хвилин. Протягом проведення практичного заняття викладач має приділяти час індивідуально кожному студенту: оцінити виконану роботу або ж допомогти.

Ураховуючи застосування CoCalc під час проведення практичного заняття ми пропонуємо дещо змінити його структуру (табл. 3.4).

Найбільш розповсюджені форми організації навчальної діяльності: розв'язання задач, тренувальні вправи, спостереження, експерименти.

Самостійна робота поєднує відтворювальні й творчі процеси в діяльності студента. Тому, враховуючи застосування хмарного сервісу,

три рівні самостійної діяльності студентів дещо видозміняться, набудуть іншого характеру (рис. 3.25).

*Таблиця 3.4*

**Структура проведення практичного заняття**

| Класична | З використанням CoCalc |
|---|---|
| а) вступ викладача;<br>б) відповіді на запитання студентів щодо незрозумілого матеріалу;<br>в) практична частина як планова;<br>г) заключне слово викладача. | а) повторення основних теоретичних положень;<br>б) колективна співпраця студентів (відповіді на запитання один одного засобами чату щодо незрозумілого матеріалу);<br>в) робота групи студентів з одним навчальним ресурсом;<br>г) практична/дослідна частина як планова (включає самостійну роботу);<br>д) виправлення помилок, оцінювання виконаної роботи. |

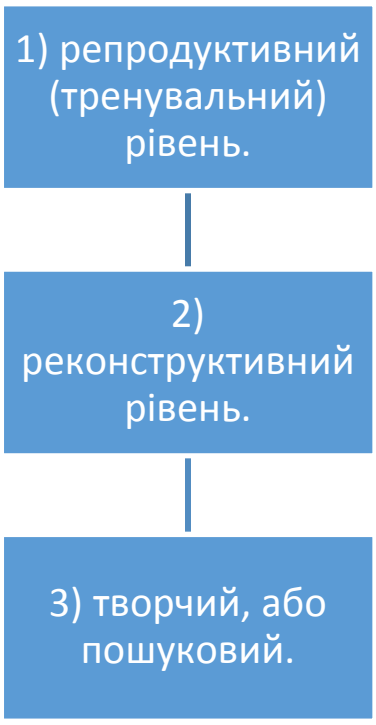

Рис. 3.25. Етапи виконання індивідуального завдання/самостійної роботи

1) репродуктивний (тренувальний) рівень (перший етап самостійної роботи). Тренувальні самостійні роботи виконуються за зразком розв'язання завдань (як правило, перші дві-три роботи). Зразки краще подавати у вигляді готових файлів з виконанням аналогічних завдань, з численними коментарями. При цьому треба чергувати теоретичні блоки завдання з практичними. Пізнавальна діяльність студента проявляється в пізнанні, осмисленні, запам'ятовуванні. Ціль таких робіт – закріплення знань, формування умінь і навичок;

2) реконструктивний рівень (другий етап самостійної роботи). Студенти використовують попередні знання та вивчають нові функції,

методи використовують довідку, додаткову літературу. Новий матеріал подається в стислій формі;

3) творчий, або пошуковий (останній етап самостійної роботи). Творча самостійна робота вимагає аналізу індивідуального завдання, отримання нових даних. Студент повинен самостійно обрати засоби та методи, функції використовуючи довідку та додаткові джерела.

Слід відмітити, що в процесі вивчення дисципліни усі студенти проходять кожен з вищеописаних етапів. При чому цей процес являється висхідним: від першого до третього (найвищого, останнього).

Що ж стосується контролю та оцінювання успішності студентів в навчальному процесі, то в системі CoCalc реалізовано значні потужності задля виконання поставлених завдань.

До основних завдань діагностування навчальних досягнень студента належать:

– визначення рівня сформованості професійних компетентностей студентів;

– виявлення, перевірка та оцінювання рівня здобутих знань, умінь та навичок студентів і якості засвоєння ними навчального матеріалу з конкретної дисципліни;

– порівняння фактичних результатів навчально-пізнавальної діяльності із запланованими;

– оцінювання відповідності змісту, форм, методів і засобів навчання меті і завданням підготовки фахівців відносно галузевій компоненті державних стандартів освіти з певної спеціальності;

– стимулювання систематичної самостійної роботи та пізнавальної активності студентів;

– виявлення і розвиток творчих здібностей, підвищення зацікавленості у вивченні навчального матеріалу;

– оцінювання ефективності самостійної роботи студентів;

– виявлення кращого досвіду та розроблення заходів для підвищення якості навчання шляхом впровадження у навчальний процес інноваційних технологій.

Досягти поставлених завдань дозволить контроль як з боку викладача за діями окремого студента, так і взаємоконтроль студентів, перегляд дій у спільному/груповому проекті, перегляд дій з навчальним ресурсом окремого студента, коментування виконаних завдань, обговорення способів їх виконання і, самі одержані результати.

Використання робочих аркушів під час проведення іспитів вимагає подальших досліджень. Студент, проходячи, як і будь-який обчислювальний іспит онлайн, може мати проблеми під час розв’язання часткових випадків одного з завдань. Проте, можливість скласти іспит не виходячи з дому, вирішує проблему обмеження студентів у часі. Але в цьому випадку стане під сумнів самостійність виконаних завдань. Тому на даній стадії розвитку хмарних сервісів, краще обмежитись виконанням індивідуальних та домашніх робіт [359].

### 3.4 Хмарний сервіс CoCalc як засіб навчання математичних дисциплін

Одним зі шляхів інтенсифікації навчального процесу, в зв'язку зі збільшенням частки навчального матеріалу, що відводиться на самоопрацювання є застосування комп'ютерних засобів. Використання ІКТ у самостійній роботі студентів збільшує кількість способів та методів подання навчальних завдань, призначених для самостійного опрацювання. Зокрема, з'явилася можливість використання студентами у самостійній роботі спеціальних завдань на планування та контроль самостійної навчальної діяльності, в яких прямим продуктом є формування вміння визначати стратегію розв'язання, планувати процес виконання діяльності, контролювати його, знаходити й виправляти помилки [317, с. 177]. Комп'ютерні засоби навчання дають змогу студентам раціонально розподіляти час на самопідготовку та самостійне опрацювання тем. Як відомо, самоосвіта передбачає не тільки неперервне засвоєння знань особистістю, а й неперервний педагогічний вплив на неї з боку інших суб'єктів, де особливе значення має зв'язок навчальної діяльності з суспільно-практичною діяльністю [184, с. 94].

Основна функція у навчальному процесі ЗВО залишається за традиційними методами навчання. Комп'ютерні засоби навчання виконують лише допоміжну роль. Навіть якщо окреслити потенційні можливості використання комп'ютерних засобів (сприяють розвитку мислення, надають можливість використати ігрові моменти під час виконання творчих завдань, стимулюють інтерес до пошуку даних, мотивують до вивчення дисципліни тощо) у практиці ЗВО, вони, на жаль, повною мірою не використовуються [184].

Найголовнішою перевагою використання CoCalc, в порівнянні з традиційними засобами навчання є те, що студенти можуть вільно спілкуватись як з викладачем, так і з іншими своїми колегами, користувачами в процесі виконання завдань, не обмежуючи це спілкування лише тими годинами, що відводяться на проведення практичних чи лекційних занять. Група студентів може обговорювати між собою в чат-кімнаті хмарного сервісу основні проблеми в ході вивчення нової теми як у навчальному закладі, так і в себе вдома. Чат-кімнати можна використати під час організації викладачем групових форм роботи (наприклад робота групи студентів відбувається в межах одного робочого аркуша, в той час як інші студенти в чаті коментують кожну дію своїх колег). Іншим варіантом організації колективної роботи студентів є обмін робочими аркушами між групами, на які поділено потік студентів (академічну групу) та обговорення засобами комунікації успішності виконання обчислень, вдалого застосування того чи іншого прийому, методу.

Поєднання комбінованого тексту, програмного коду та розв'язку завдання в межах одного робочого аркуша кардинально відрізняється від нинішніх парадигм і варто дослідити більш детально, як новий засіб навчання, в першу чергу математичних дисциплін.

Використання CoCalc сприяє розвитку нових способів управління навчальними матеріалами, оскільки студентам не треба додатково встановлювати та налагоджувати програмне забезпечення спеціального призначення, достатньо авторизуватись та розпочати роботу. Викладач зможе організувати проведення дослідження одночасно декількох груп студентів. При цьому одночасна реєстрація студентів групи відбувається легко, не вимагає додаткових витрат технічних та програмних ресурсів.

Наприклад, якщо розглядати чисельні методи, то вони є практично найпоширенішими, проте найчастіше використовують лише аналітичні методи. Це пояснюється тим, що аналітичні методи мають простіший числовий розв'язок. Використання CoCalc задля розв'язання нелінійних алгебраїчних рівнянь, чисельного інтегрування на початковому етапі вивчення математичної дисципліни надасть змогу глибше опанувати теми, розглянути більшу кількість прикладів. Завдяки використанню хмарного сервісу, студенти отримують більш реалістичне уявлення про діапазон аналітичних методів. Не зважаючи на те, що в редакторі програмного коду є можливість використання мови Python, спеціально вивчати мову програмування немає необхідності. В цілому в процесі роботи використовується близько сімдесяти команд Sage. Крім того, можна використовувати автодоповнення назви команди, якщо користувач пам'ятає лише перші літери функції. Стосовно окремої команди можна викликати довідку (з декількома прикладами використання) та переглянути синтаксис, поставивши наприкінці знак питання.

Робоча область ресурсу типу sagews складається з прямокутних комірок програмного коду, які незалежні одна від одної (сприймаються як окремі частини) та блоків відображення результатів виконання програмного коду відповідних комірок (рис. 3.26).

Поточною будемо вважати таку комірку, в які розташовано курсор миші. Над прямокутними комірками розміщені горизонтальні смуги (рис. 3.27: мітка а-в), а ліворуч від блоку виведення результатів виконання програмного коду, – зелені вертикальні (рис. 3.27: мітка г).

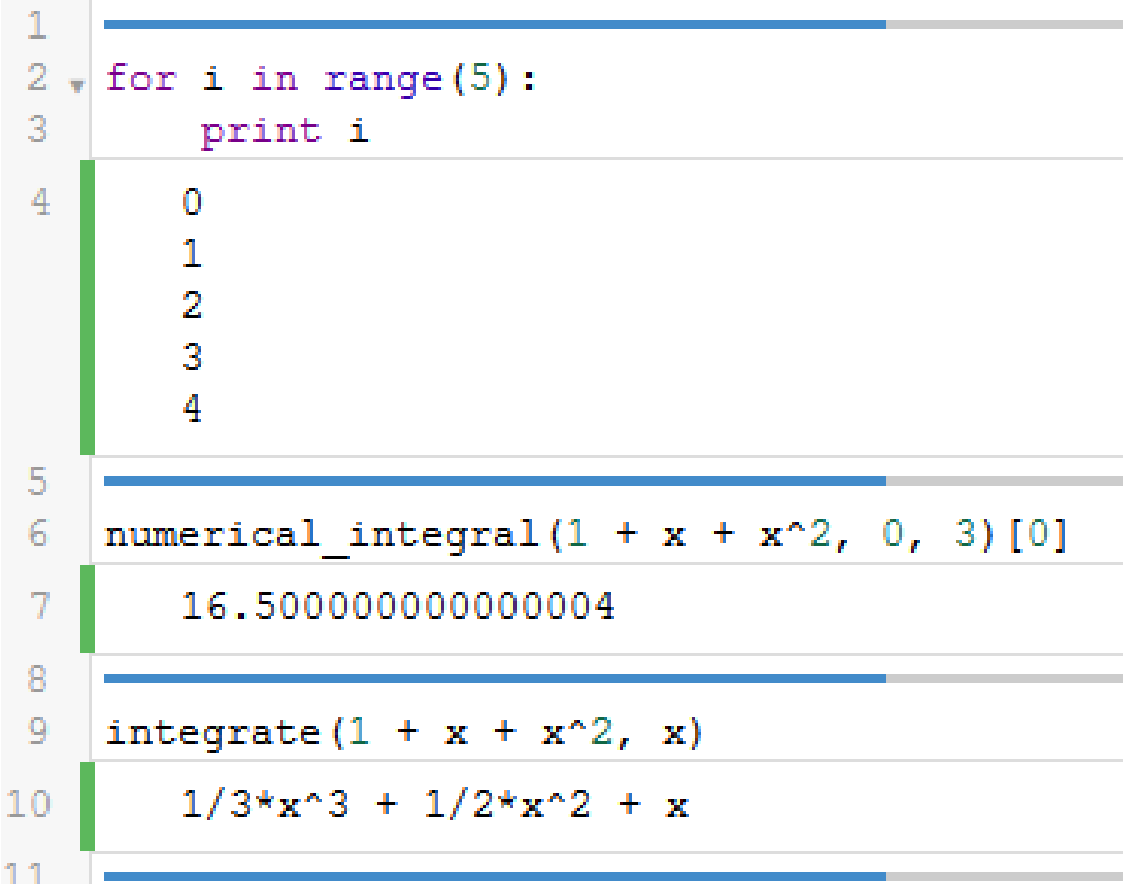

Рис. 3.26. Робоча область ресурсу типу sagews

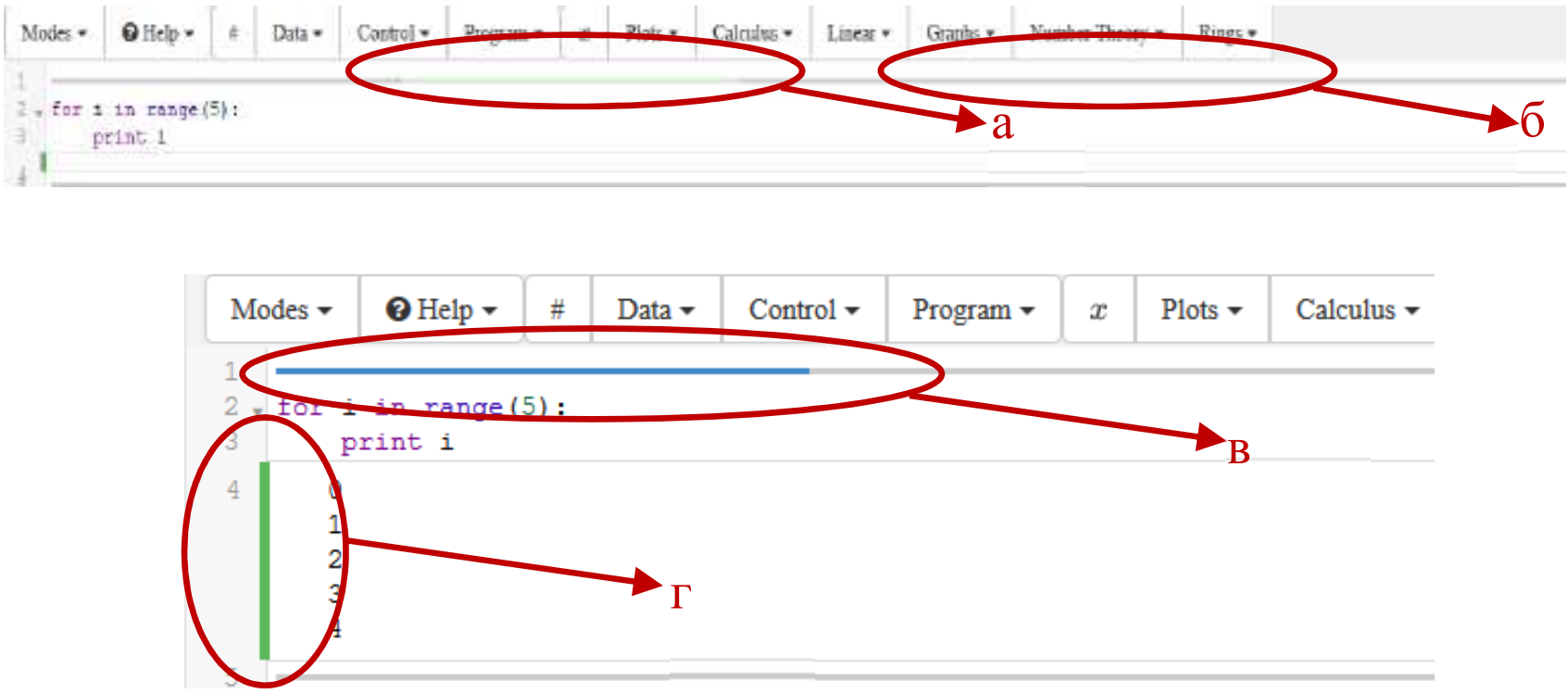

Рис. 3.27. Кольорові смуги

а – зелена горизонтальна смуга інформує, що відбувається виконання програмного коду в поточній комірці

б – сіра смуга надає можливість розділити поточну комірку на дві

в – синя смуга інформує, що виконання програмного коду в поточній комірці завершено

г – зелена вертикальна смуга обмежує блок відображення результатів виконання програмного коду в поточній комірці

Автоматичне доповнення назви функції в CoCalc (Додаток Г, табл. Г.1) значно спрощує процес введення команд: достатньо вказати перші літери назви команди та натиснути сполучення клавіш Ctrl +

пропуск, користувач отримає список команд, що починаються з послідовності вказаних символів (рис. 3.28).

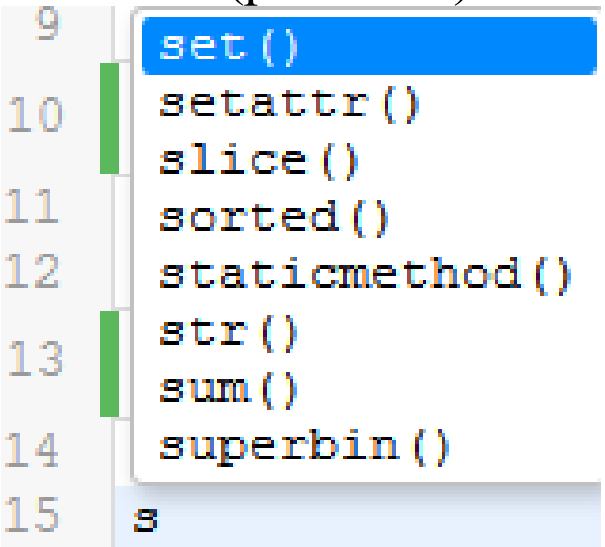

Рис. 3.28. Автоматичне доповнення назви функції: список команд

Автоматичне доповнення команд SageMath (Додаток Г, табл. Г.1) можливе за рахунок зазначення перших символів назви команди SageMath та натискання клавіші Tab (рис. 3.29). Користувачу буде запропоновано список, який складатиметься із послідовності введених символів.

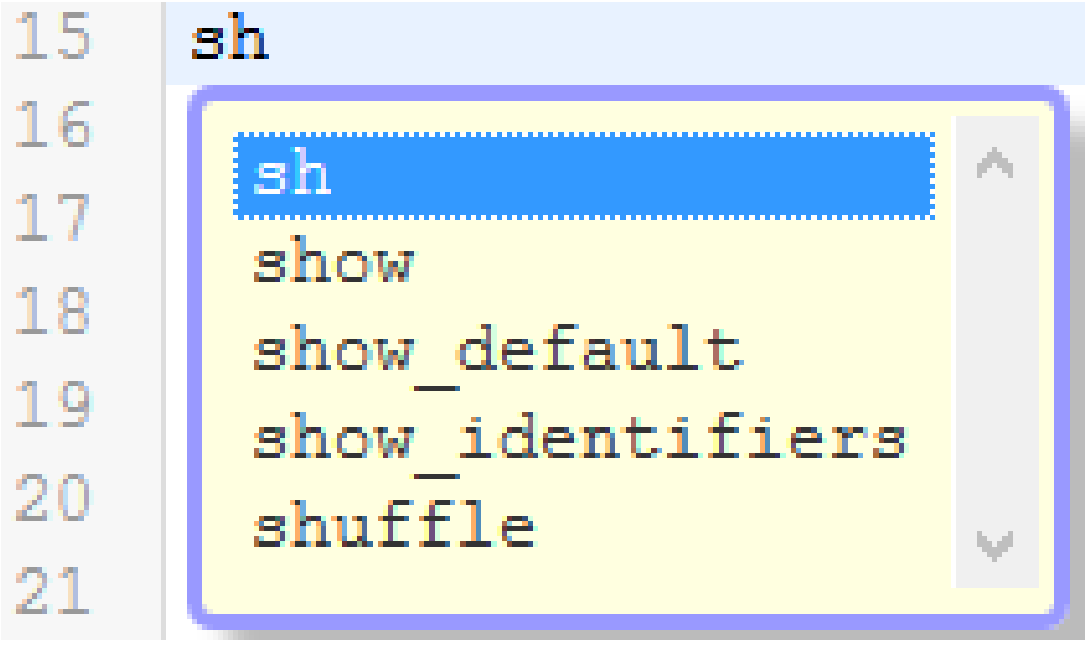

Рис. 3.29. Автоматичне доповнення команд Sage

Виконати програмний код в комірці можна двома способами:

1. Натиснувши комбінацію гарячих клавіш Shift + Enter (Додаток Г, табл. Г.1).
2. Натиснувши кнопку «Run», яка була розглянута раніше.

Керувати виконанням програмного коду можна використовуючи відповідні кнопки стандартної панелі інструментів (рис. 3.30: мітки 2-5).

Для того, щоб переглянути довідку стосовно синтаксису певної функції, треба внести її повну назву в комірку для введення та знак питання: назва_команди?. Після цього запустити виконання програмного коду одним із способів, щоб отримати довідку в блоці відображень результатів виконання програмного коду, або натиснути клавішу Tab, задля виведення довідки в окремому вікні (рис. 3.31).

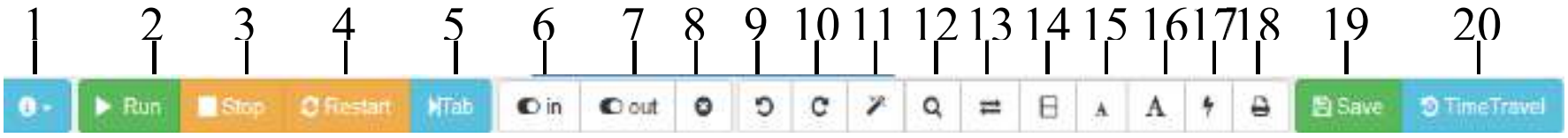

Рис. 3.30. Стандартна панель інструментів

1 – кнопка-список дій над файлом
2 – *Run* – кнопка виконати обчислення в поточній комірці
3 – *Stop* – кнопка зупинити обчислення в поточній комірці
4 – *Restart* – кнопка перезавантажити обчислення в комірці
5 – *Tab* – кнопка виставити табуляцію
6 – *in* – кнопка відобразити (приховати) програмний код поточної комірки
7 – *out* – кнопка відобразити (приховати) результати виконання програмного коду поточної комірки
8 – кнопка видалити результати виконання програмного коду поточної комірки
9 – кнопка відмінити останню дію
10 – кнопка повернути останню дію
11 – кнопка автоматичну табуляцію виділеного фрагменту
12 – кнопка виконати пошук за ключовим словом
13 – кнопка виконати пошук та замінити виділений фрагмент
14 – кнопка дублювати вікно редактора (вертикально, горизонтально) / відмінити режим дублювання вікна
15 – кнопка збільшити шрифт
16 – кнопка зменшити шрифт
17 – кнопка перейти до рядка за вказаним номером
18 – кнопка конвертувати файл в формат .pdf (лише за умови одноосібного використання проекту) з можливістю подальшого друку
19 – *Save* – кнопка зберегти
20 – *Time Travel* – кнопка показати історію роботи з файлом

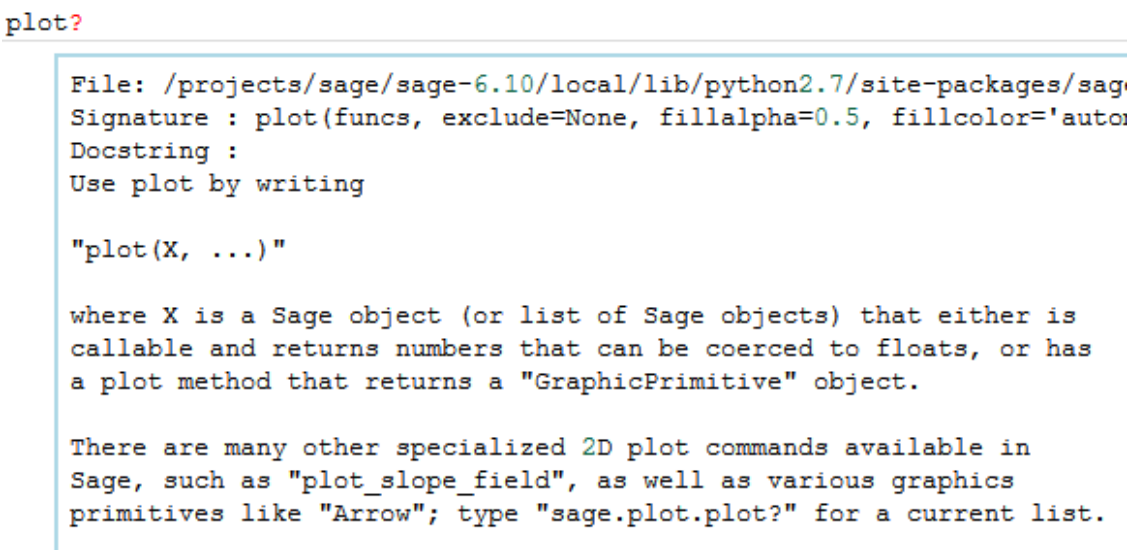

```
plot?

File: /projects/sage/sage-6.10/local/lib/python2.7/site-packages/sag
Signature : plot(funcs, exclude=None, fillalpha=0.5, fillcolor='auto
Docstring :
Use plot by writing

"plot(X, ...)"

where X is a Sage object (or list of Sage objects) that either is
callable and returns numbers that can be coerced to floats, or has
a plot method that returns a "GraphicPrimitive" object.

There are many other specialized 2D plot commands available in
Sage, such as "plot_slope_field", as well as various graphics
primitives like "Arrow"; type "sage.plot.plot?" for a current list.
```

Рис. 3.31. Довідка з функції у полі відображення результатів

Увівши після назви функції два знаки питання ?? та запустивши програмний код (або натиснувши клавішу Tab), можна переглянути її програмний код (рис. 3.33).

```
plot?

File: /projects/sage/sage-6.10/local/lib/python2.7/site-packages/sage/misc/decorators.py
Signature : plot(funcs, exclude=None, fillalpha=0.5, fillcolor='automatic', detect_poles=False, plot_points=200, thickness=1, adaptive_tolerance
Docstring :
Use plot by writing

"plot(X, ...)"

where X is a Sage object (or list of Sage objects) that either is
callable and returns numbers that can be coerced to floats, or has
a plot method that returns a "GraphicPrimitive" object.
```

Рис. 3.32. Довідка з функції у окремому вікні

```
G_kwds = Graphics._extract_kwds_for_show(kwds, ignore=['xmin', 'xmax'])

original_opts = kwds.pop('__original_opts', {})
do_show = kwds.pop('show',False)

from sage.structure.element import is_Vector
if kwds.get('parametric',False) and is_Vector(funcs):
    funcs = tuple(funcs)


if hasattr(funcs, 'plot'):
    G = funcs.plot(*args, **original_opts)
# if we are using the generic plotting method
else:
    n = len(args)
    # if there are no extra args, try to get xmin,xmax from
    # keyword arguments or pick some silly default
    if n == 0:
        xmin = kwds.pop('xmin', -1)
        xmax = kwds.pop('xmax', 1)
        G = _plot(funcs, (xmin, xmax), **kwds)
```

Рис. 3.33. Сторінка фрагменту програмного коду функції plot()

Після виведеного результату на робочому листі автоматично з'являється нова комірка для введення команди. Додати нову комірку до робочої області аркуша можна натиснувши на сіру лінію, яка постійно знаходиться над будь-якою коміркою (рис. 3.27: мітка б). Вилучення зайвої порожньої комірки виконується клавішею Backspace (видалити попередню комірку по відношенню до курсора) або Delete (видалити наступну комірку по відношенню до курсора).

За замовчування робочий лист налаштований на роботу з командами SageMath. Вибрати інший режим можна у списку додаткової панелі інструментів для команд SageMath (кнопка-список «Modes») або командою виду %назва_режиму (перелік усіх можливих режимів роботи можна переглянути, звернувшись до елементу списку списку «Help» Magic mode commands).

Для того, щоб приховати/показати програмний код комірки,

результатів виконання програмного коду, видалити програмний код на стандартній панелі інструментів передбачена група кнопок (рис. 3.30: мітки 6-8).

За замовчуванням усі дії, виконані в робочому аркуші, зберігаються автоматично. На випадок технічних проблем передбачена кнопка «Save».

Використовуючи логічні конструкції, цикл чи функцію користувача, можна за допомогою сполучення гарячих клавіш Ctrl + Q, або використовуючи відповідні кнопки (рис. 3.34) згорнути/розгорнути блоки програмного коду.

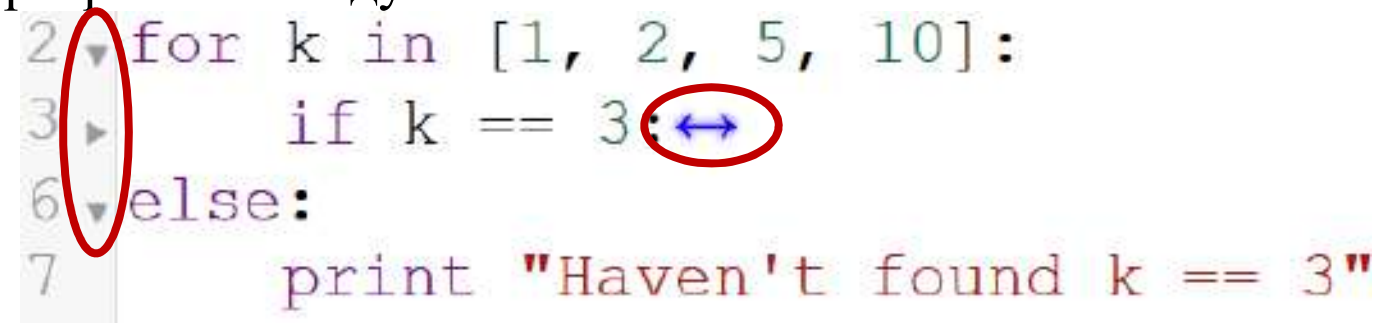

Рис. 3.34. Кнопки згортання/розгортання блоків програмного коду

Також передбачена можливість роботи з відображенням декількох курсорів окремо – гарячі клавіші Ctrl + ЛКМ (ЛКМ – ліва кнопка миші). Інші сполучення гарячих клавіш для роботи в редакторі будь-якого файлу, в тому числі і в робочому аркуші розглянуто в Додатку Г.

Оформлення робочого аркуша передбачає використання мови HTML. Існує два способи додати форматований текст мовою HTML до робочого аркуша:

1) обрати відповідний режим роботи з випадаючого списку «Modes» (рис. 3.35) або командою виду %html;

2) за допомогою функції html().

Обравши перший спосіб, в комірці після команди %html можна вводити текстову частину з використанням тегів мови HTML або ж застосовуючи кнопки на панелі інструментів (рис. 3.36). Приклад змістової частини робочого аркуша подано на рис. 3.37.

Під час роботи в режимі мови HTML до форматованого тексту не має можливості додати значення змінної чи виразу.

Другий спосіб полягає в тому, що теги мови HTML (рис. 3.38) застосовуються всередині функції html(), синтаксис якої: html("Символьний рядок"). Якщо ж додамо форматування за рахунок тегів HTML: html("<Тег>Символьний рядок</Тег>").

При цьому комірку з програмним кодом буде приховано. Для її відображення достатньо двічі натиснути лівою кнопкою миші на результатах обчислень комірки.

Для того, щоб додати до символьного рядка значення змінної (рис. 3.39) треба вказати відповідний специфікатор (табл. 3.5) та змінну.

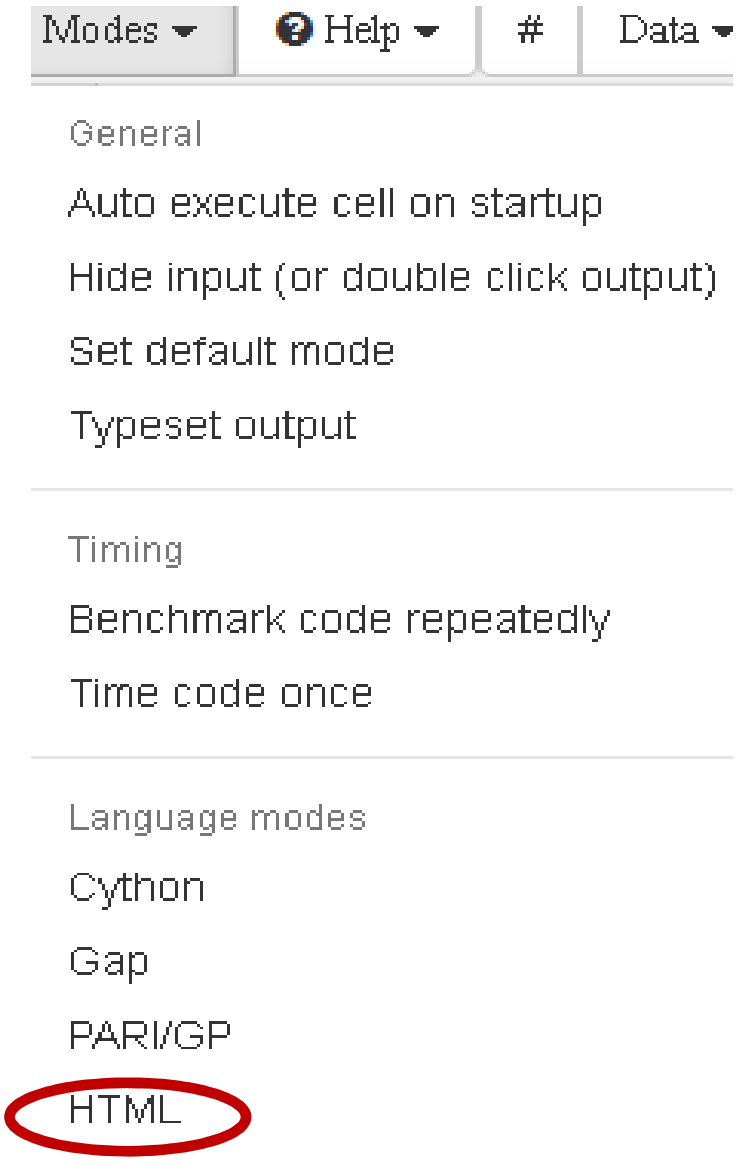

Рис. 3.35. Режим мови HTML

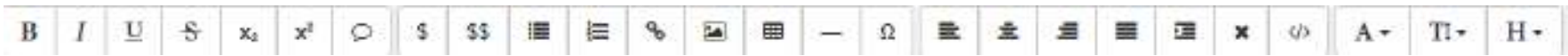

Рис. 3.36. Панель інструментів – Форматування: мова HTML

```
%html
<div align='justify'><font face='Times New Roman'
size=5><em>Завдання 12.</em> Ребро правильного тетраедера
<em>a</em>.  Побудувати рівносторнній конус, коло основи
якого вписано в грань тетраедера.</font></div>
```

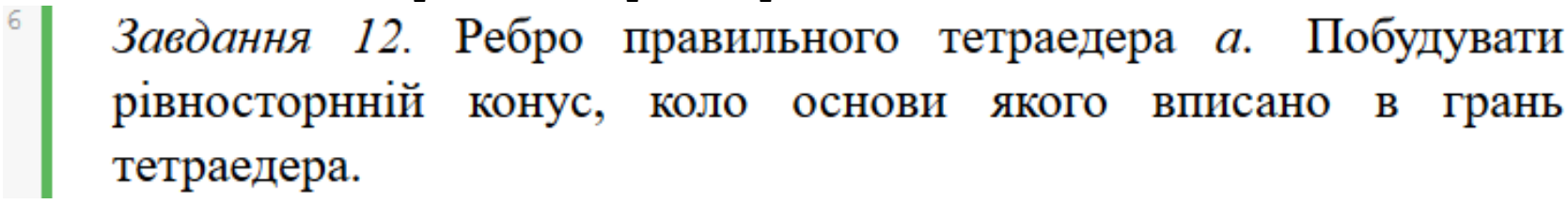

Рис. 3.37. Текстовий блок робочого аркуша

```
html("<div align='justify'><font face='Times New Roman'
size=5><em>Завдання 12.</em> Ребро правильного тетраедера
<em>a</em>. Побудувати рівносторнній конус, коло основи якого
вписано в грань тетраедера.</font></div>")
```

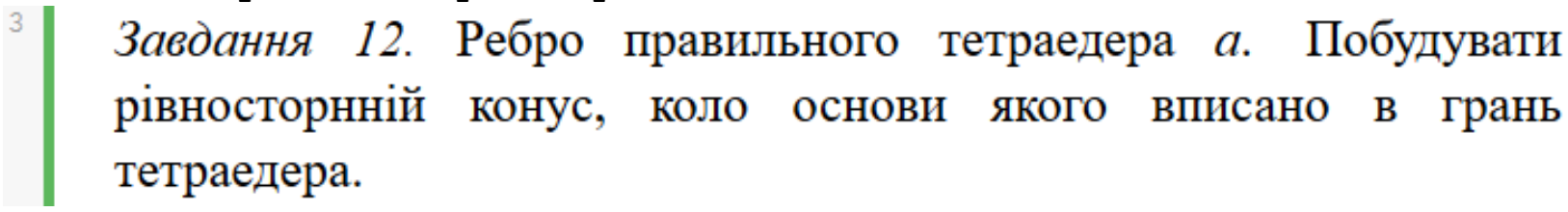

Рис. 3.38. Використання функції html()

```
%var x1
x1=7
html("<div align='justify'><font face='Times New Roman'
size=5><em>Завдання 4.</em> Обчислити значення виразу в точці
<em>x = %s</em>.</font></div>"%x1)
```

*Завдання 4.* Обчислити значення виразу в точці $x = 7$.

Рис. 3.39. Включення до символьного рядка значення змінної

*Таблиця 3.5*

**Основні специфікатори мови Python**

| Специфікатор | Формат подання значення змінної |
|---|---|
| `%d, %i, %u` | десяткове число |
| `%o` | число у вісімковій системі числення |
| `%x` | число у шістнадцятковій системі числення (літери у нижньому регістрі) |
| `%X` | число у шістнадцятковій системі числення (літери у верхньому регістрі) |
| `%e` | число з плаваючою точкою в експоненціальному форматі (експонента у нижньому регістрі) |
| `%E` | число з плаваючою точкою в експоненціальному форматі (експонента у верхньому регістрі) |
| `%f, %F` | число з плаваючою точкою |
| `%g` | число з плаваючою крапкою в експоненціальному форматі (експонента у нижньому регістрі), якщо порядок менше за -4 або більше за 4 |
| `%G` | число з плаваючою крапкою в експоненціальному форматі (експонента у верхньому регістрі), якщо порядок менше за -4 або більше за 4 |
| `%c` | символ |
| `%r` | рядок (літерал Python) |
| `%s` | рядок |

Якщо формат змінної важко визначити, за замовчуванням використовують `%s`. Даний специфікатор виводить будь-який тип змінної та символьний вираз (без математичного форматування).

Для того, щоб додати до символьного рядка декілька змінних, потрібно вказати таку кількість специфікаторів, скільки треба виводити змінних та вказати їх через кому в круглих дужках після рядка виведення (рис. 3.40).

Додати математичний текст до ресурсу типу sagews може знадобитись в двох випадках.

1) до змістової частини робочого аркуша;
2) в якості підпису до графічної побудови.

```
%var x1,x2
x1=7
x2=-3
html("<div align='justify'><font face='Times New Roman'
size=5><em>Завдання 4.</em> Обчислити значення виразу в точці
<em>x = %s</em> та <em>x = %s</em>.</font></div>"%(x1,x2))
```

6 *Завдання 4.* Обчислити значення виразу в точці *x* = *7* та *x* = *-3*.

Рис. 3.40. Включення до символьного рядка значень декількох змінних

Математичний текст формується за рахунок використання мови LaTeX. Команда мови LaTeX розміщується між символами `$...$` (формула в середині рядку) або `$$...$$` (формула окремо від тексту, з нового рядка, вирівнювання – по центру).

Для того, щоб вивести вираз з використанням математичного форматування, треба застосувати мову LaTeX, вказати специфікатор %s (оскільки вираз – це рядок) та скористатись функцією `latex()`, аргуметом якої виступає ім’я символьного виразу (рис. 3.41):

```
f=(x+5)/((x^2+2*x+3)*(x-1))
html("<font face='Times New Roman' size=5><i>Завдання 5.</i>
Розкласти дріб на суму простих дробів
<strong>$%s$</font></strong>."%latex(f))
```

7 *Завдання 5.* Розкласти дріб на суму простих дробів $\frac{x+5}{(x^2+2\,x+3)(x-1)}$

Рис. 3.41. Додавання математичного виразу до змістової частини

Додати математичний текст до ресурсу типу sagews можливо також в якості підпису до графічної побудови. В середині текстового рядка додаємо мову LaTeX між символами $...$ або $$...$$ (рис. 3.42):

```
vector([2,2/3]).plot(aspect_ratio=1)+
text(r"$z_1=2+\frac{2}{3}i$",(1,0.6),fontsize=20)
```

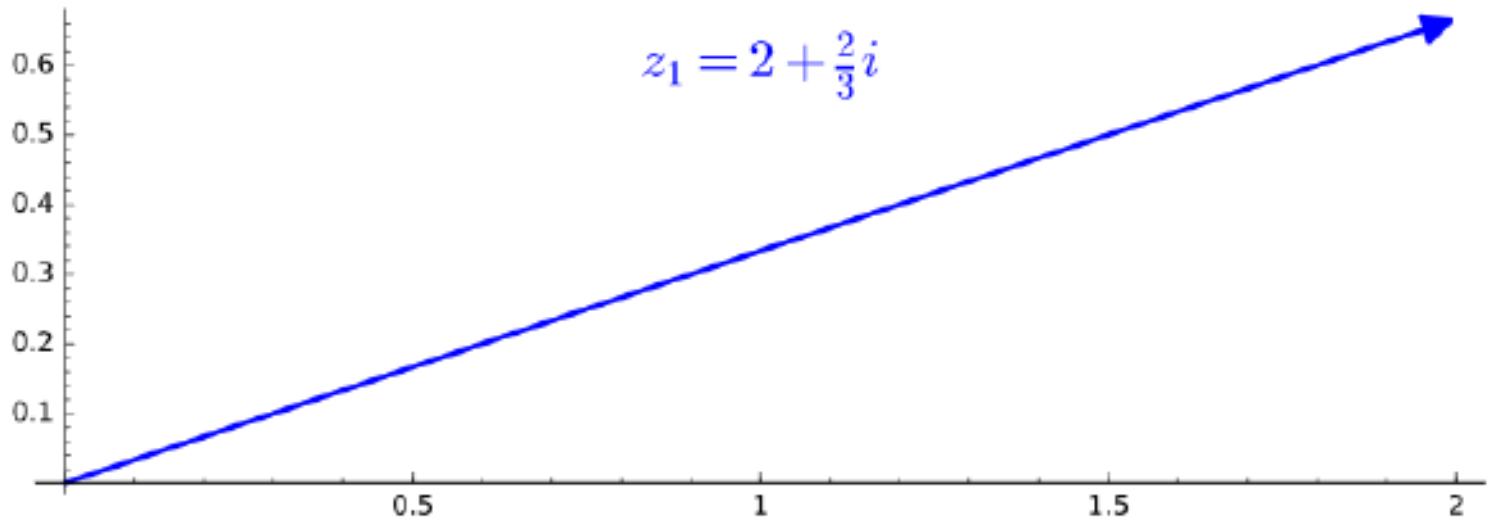

Рис. 3.42. Додавання математичного тексту до графічної побудови

Або без використання літери `r` перед рядком:

```
vector([2,2/3]).plot(aspect_ratio=1)+
text("$z_1=2+\\frac{2}{3}i$",(1,0.6),fontsize=20).
```

Для того, щоб використати в підписі символьний вираз потрібно так само, як було описано вище додати його скориставшись функцією `latex()` (рис. 3.43):

```
z=2+2/3*i
vector([2,2/3]).plot(aspect_ratio=1)+
text("$z_1=%s$"%latex(z),(1,0.6),fontsize=20)
```

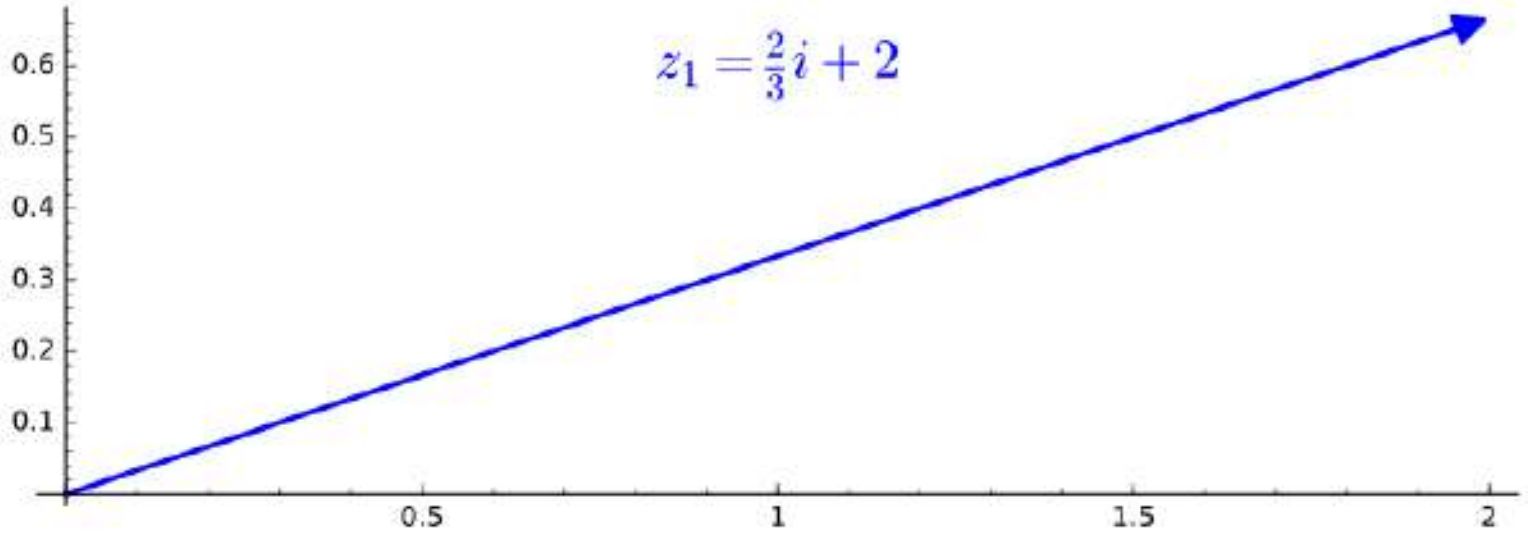

Рис. 3.43. Додавання символьного виразу до графічної побудови

Можна додати мову LaTeX до легенди графічної побудови.

Приклад додавання виразів до легенди графіків функцій (рис. 3.44):

```
y1=2*log_b(x+3,b=3);y2=x/2+1
show(plot(y1,0,6.5,linestyle=':',thickness=3,legend_label=
"$y_1=%s$"%latex(y1))+plot(y2,0,6.5,legend_label="$y_2=%s$"
%latex(y2),thickness=2),ymin=1,ymax=5,aspect_ratio=1)
```

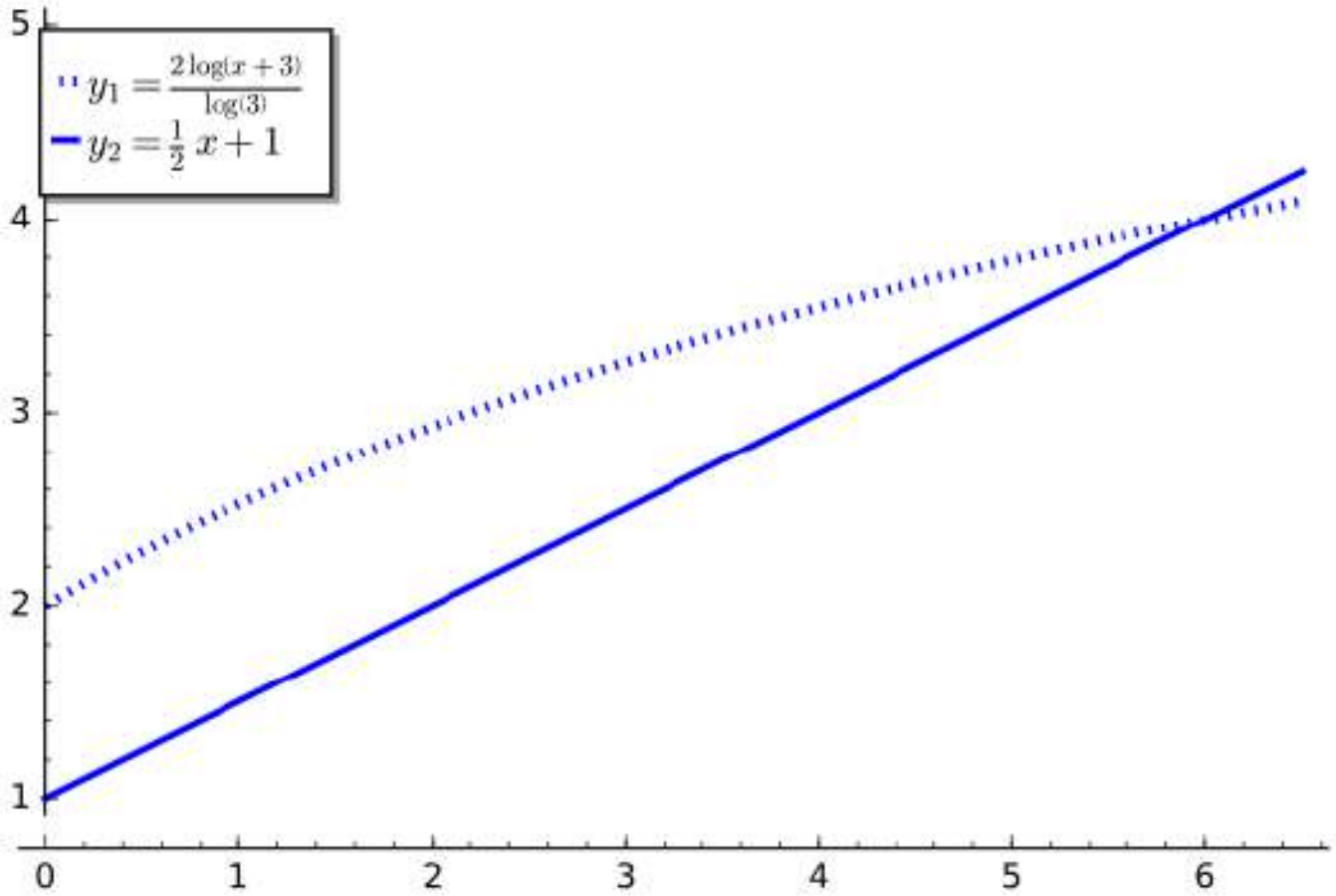

Рис. 3.44. Додавання виразів до легенди графіків функцій

Для того, щоб скористатись стандартними елементами управління треба перед функцією користувача помістити наступний рядок та вказати бажані значення параметрів (табл. 3.6):

```
@interact(self, f=None, layout=None, width=None, style=None,
update_args=None, auto_update=True, flicker=False,
output=True)
```

*Таблиця 3.6*

**Параметри @interact для CoCalc**

| Параметр | Значення |
|---|---|
| `f` | функція |
| `width` | ширина форми в одиницях HTML: '80%', '300px', '20em' |
| `css_style` | рядок CSS, за допомогою якого можна змінити такі параметри, як форму блоку стандартних елементів управління, межі форми, колір фону форми та ін. |
| `update_args` | рядок з перерахуванням тих змінних, для елементів форм яких треба автоматично виконувати обчислення після внесення змін в дані елементи. Для інших елементів форми не буде виконуватись автоматичне обчислення, після внесених змін. Значення за замовчуванням: None |
| `auto_update` | (за замовчуванням: True); якщо False, то з'явиться кнопка з написом «Оновити» яку ви можете натиснути для повторного виконання обчислення. |
| `layout` | (за замовчуванням: один елемент управління для кожного рядка) список кортежів із можливим зазначенням ширини та мітки:<br>`row0 = [(var_name, ширина, етикетка), ...],`<br>де `var_name` – рядки, ширина не повинна перевищувати 12, а мітка є необов'язковою |

Після налаштування зовнішнього вигляду блоку стандартних елементів управління, слідує опис функції користувача в якості параметрів якої виступають змінні-елементи управління (Додаток Ж):

- `button()` – кнопка;
- `checkbox()` – прапорець;
- `color_selector()` – поле вибору кольору;
- `input_box()` – поле для введення;
- `input_grid()` – комірки для введення;
- `range_slider()` – повзунок 2;
- `selector()` – меню вибору;
- `slider()` – повзунок 1;

– `text_control()` – коментар.

Деякі значення за замовчуванням створюють елементи управління автоматично без явного зазначення типу елементу управління.

Наприклад, можна оголосити змінну `x` типу «повзунок 2», просто вказавши `x=(u,v)` в списку аргументів.

Інші випадки:

– `u` – порожнє поле для введення;

– `u=elt` – за замовчуванням поле для введення, якщо в подальшому не вказано тип елементу;

– `u=(umin,umax)` – елемент управління типу «повзунок 1» з розмірністю 100 кроків;

– `u=(umin,umax,du)` – елемент управління типу «повзунок 1» з кроком du;

– `u=list` – меню вибору: якщо довжина списку не більше 5 елементів – у вигляді кнопок, більше – випадаючий список;

– `u=generator` – елемент управління типу «повзунок 1» з розмірністю 10000 кроків;

– `u=bool` – елемент управління типу «прапорець»;

– `u=Color('blue')` – елемент управління типу «поле вибору кольору»

– `u=matrix` – елемент управління типу «комірки для введення», за замовчуванням змінна типу матриця;

– `u=(default, v)` – `v` один з вище наведених варіантів зі значенням за замовчуванням;

– `u=(label, v)` – v один з вище наведених варіантів із підписом елемента управління.

Приклад розміщення стандартних елементів управління (рис. 3.45):

```
@interact(layout={'top': [['a', 'b']], 'left': [['c']],
'bottom': [['d']], 'right':[['e']]})
def _(a=x^2, b=(0..20), c=100, d=x+1, e=sin(2)):
    print a+b+c+d+e
```

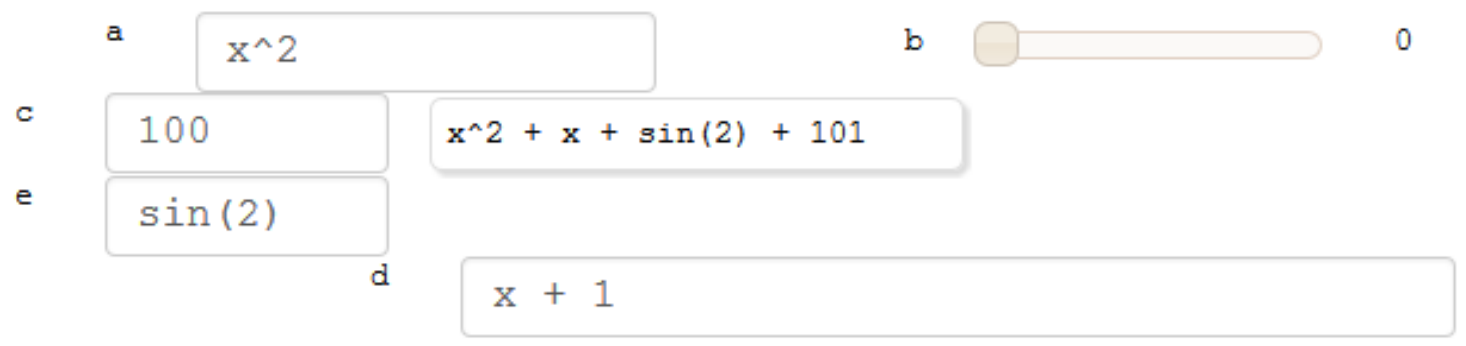

Рис. 3.45. Спосіб розміщення стандартних елементів управління

Приклад зміни зовнішнього вигляду форми (рис. 3.46):

```
@interact(width=25, style="background-color:lightgreen;
border:5px dashed red;")
def f(x=button('Merry ...',width=20)):
    pass
```

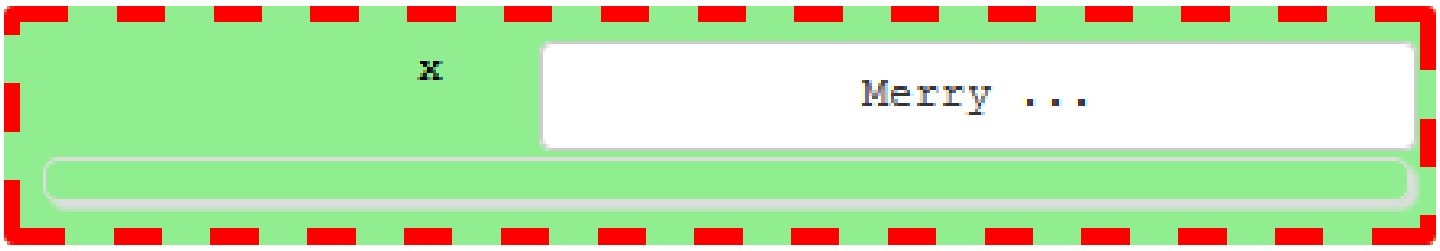

Рис. 3.46. Зміна зовнішнього вигляду форми

CoCalc може бути доповненням основних посібників навчальної дисципліни. Робочі аркуші можна структурувати таким чином, щоб проект містив у собі повний або частковий (окремі модулі, теми) навчальний план за певною дисципліною. Зазвичай можна використовувати комбіноване навчання з використанням наступних форм організації навчального процесу: частина – традиційні лекції, частина – лабораторні заняття з використанням хмарних сервісів, де значна частина часу витрачається на обговорення громіздких обчислень, методів, але усі рутинні обчислення організовані в хмарному сервісі [349].

### 3.5 Методи використання хмарного сервісу CoCalc у навчанні математичних дисциплін

Етапи, в процесі виконання яких було використано інструментарій CoCalc [336] у навчанні математичних дисциплін:

1. Підготовка документів з теми та виправлення зауважень.

На даному етапі, шляхом координації дій викладачів були розроблені найбільш доречні завдання на обчислення та тестові завдання, шкала оцінювання знань студентів. Разом з цим, усі завдання були реалізовані в програмному коді на відповідних робочих аркушах CoCalc. Аналіз правильних та неправильних відповідей дає підстави скорегувати програмний код в автоматичному режимі. На цьому етапі робота з файлами хмарного сервісу не завершилась. У процесі навчання робочі аркуші постійно оновлюються, доопрацьовуються та створюються нові. Таким чином перелік завдань постійно адаптують до нових досліджень, які виникають в процесі викладення матеріалу.

2. Координація викладачів та впровадження хмарного сервісу.

Виконання тих чи інших типів завдань обговорювалась заздалегідь, під час очних зустрічей. У процесі створення програмного коду системи завдань використовувались стандартні елементи управління CoCalc. Використання інтерфейсу в розроблених моделях дозволило багаторазово демонструвати приклади розв’язування, розкрити особливості групи завдань. Крім того, викладач мав змогу прослідкувати розуміння студентами теми, маючи напівавтоматичний режим роботи з моделлю математичної задачі. Використання одного і того ж програмного коду значно спрощується.

Методи навчання (за В. Л. Ортинським [175]):

– за джерелом знань: словесні, наочні та практичні;

– за етапом навчання: підготовка до вивчення нового матеріалу, вивчення нового матеріалу, закріплення вправ, контроль і оцінка;

– за способом керівництва навчальною діяльністю: безпосередні та опосередковані;

– за логікою навчального процесу: індуктивний та дедуктивний, аналітичний та синтетичний методи;

– за дидактичними цілями: організація навчальної діяльності, стимулювання і релаксація, контроль і оцінка;

– за характером пізнавальної діяльності: пояснювально-ілюстративні, репродуктивні, проблемного викладу, частково-пошукові, дослідницькі.

Проілюструємо застосування певних методів під час вивчення елементів комплексного аналізу майбутніх вчителів математики з використанням CoCalc.

Виконуючи перше завдання, студенти слідкують за діями викладача в спільному навчальному ресурсі (рис. 3.47), при цьому повторюючи дії у власних файлах (пояснювально-ілюстративний метод).

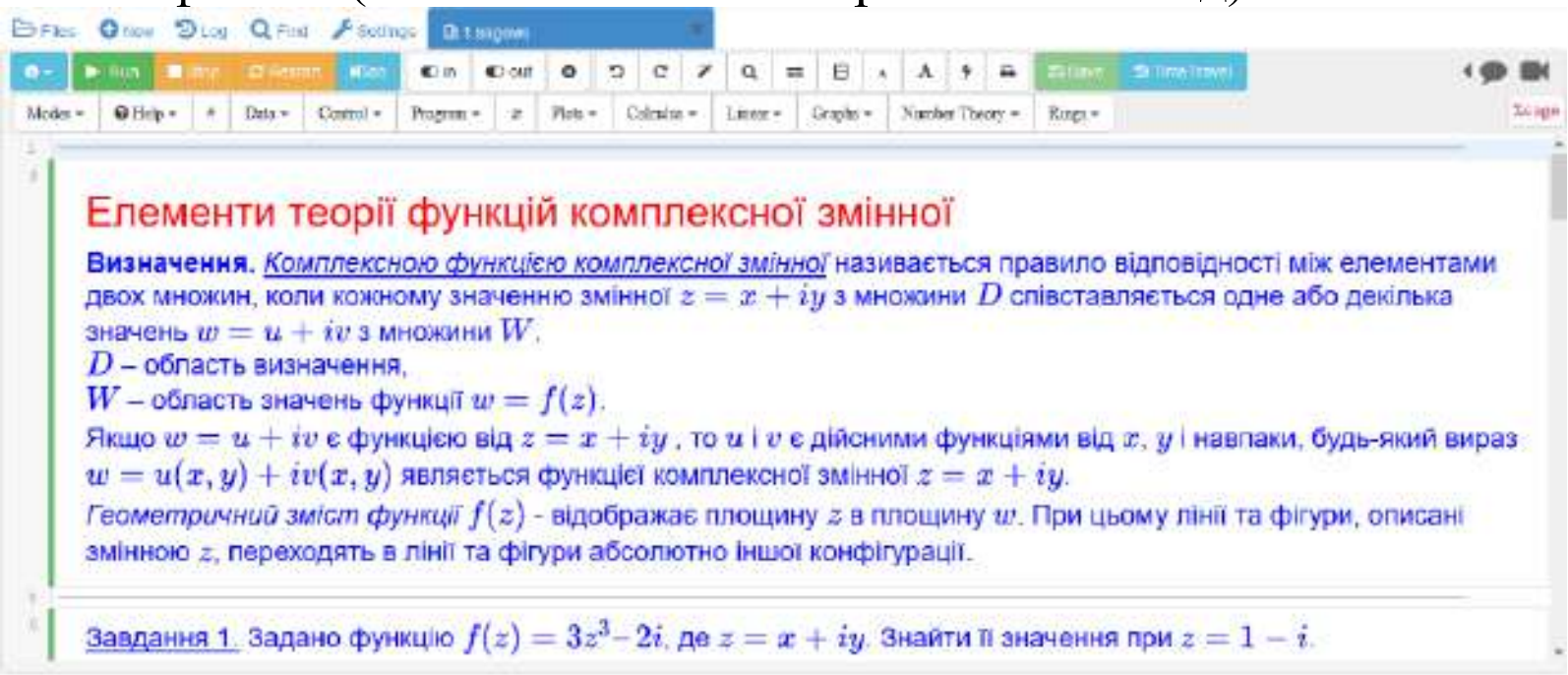

Рис. 3.47. Приклад текстового блоку в робочому аркуші

Вивчення основ комплексного аналізу доречно розпочати з повторення матеріалу про комплексні числа, що був розглянутий студентами на І курсі в рамках вивчення дисципліни алгебра та теорія чисел (метод підготовки до вивчення нового матеріалу). Акцентувати увагу студентів на понятті комплексного числа, алгебраїчній та тригонометричній формі запису, геометричній інтерпретації. На першому занятті доречно навести основний огляд системи CoCalc, розкрити основні характеристики. Безпосередньо роботу із системою краще розпочати лише на практичному занятті.

Під час оголошення змінних бажано використовувати як приклади, так і контр-приклади (табл. 3.7).

*Таблиця 3.7*

**Оголошення змінних в CoCalc**

| **Програмний код** | **Результат обчислення** |
|---|---|
| Приклади | |
| `var('x,y,p1')` | `var('x,y,p1')`<br>`(x, y, p1)` |
| `x,y,p1=var('x,y,p1')` | `x,y,p1=var('x,y,p1')` |
| `%var x, theta` | `%var x, theta` |
| Контр-приклади | |
| `var('x,y,p1')` | `var('x,y,p1')`<br>`(x, y, p1)` |
| `x,p1,y=var('x,y,p1')`<br>`f(x)=sqrt(x)`<br>`show(f(y))` | $\sqrt{p_1}$ |

Починати треба з найпростішого: ознайомити студентів із стандартними функціями CoCalc, які можна одразу застосовувати без елементів програмування (Додаток К, табл. К.1), представити способи задання комплексного числа в системі. Функції краще подавати за допомогою дедуктивного методу, без використання додаткових параметрів.

Наприклад, стандартну функцію `complex_plot()` можна представити як:

```
z,x,y=var('z,x,y')
f(z)=(3*z^3)-(2*I)
show(complex_plot(f(z),(x,-3,3),(y,-3,3)))
```

При цьому пояснити, що вказані параметри є основними, а значить обов’язковими, але їх достатньо для застосування вказаної функції. Потрібно підкреслити, що дана функція є стандартною для використання її у вивченні елементів комплексного аналізу. Запустивши вказаний програмний код одержимо графічну побудову (рис. 3.48).

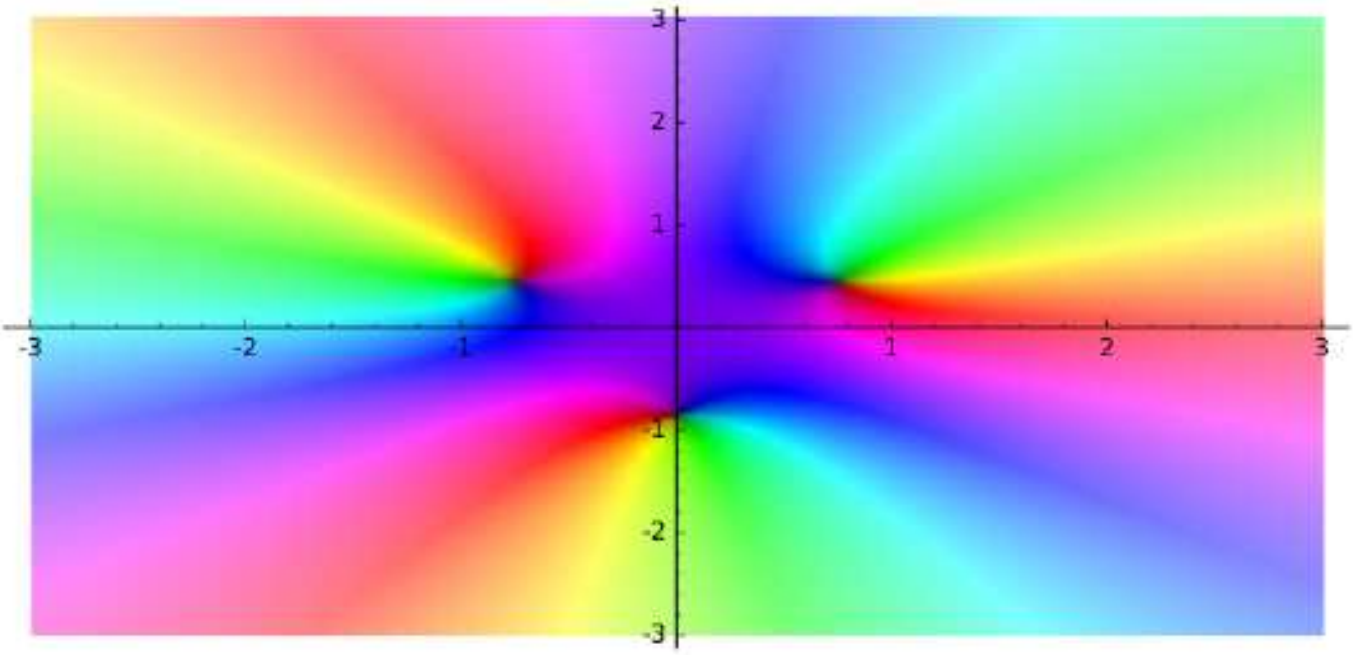

Рис. 3.48. Побудова з використанням функції `complex_plot()`

Функція `complex_plot()` інтерпретує модуль комплексної функції від комплексної змінної. При цьому: значенню 0 відповідає чорний колір, …, ∞ – білий.

В рамках проведення лабораторних занять бажано застосовувати один із методів проблемного навчання – моделювання. Використовуючи можливості запропоновані системою, моделювання інтегруватиме теоретико-методологічні знання та практичні вміння й навички студентів у єдиному процесі діяльності навчально-дослідницького характеру [91].

CoCalc надає змогу повністю організувати самостійну роботу студентів, як групову так і індивідуальну. Але самостійна робота буде більш ефективною, якщо в ній беруть участь декілька осіб. Студенти мають можливість самостійно оцінювати один одного, слідкувати за процесом виконання завдань своїх колег (рис. 3.49, рис. 3.51), задавати питання та самим давати відповіді, консультувати один одного з приводу певних моментів тощо. Викладач може приймати як пасивну (спостерігати за результатами та кроками виконання студентами робіт), так і активну участь (допомагати, консультувати online, в позааудиторний час) – рис. 3.50.

Допомагаючи студентам, викладач повинен розуміти, що основну частину роботи має виконати студент. Тому основна задача полягає саме в тому, щоб підштовхнути до правильного рішення. Не треба за студента самостійно виправляти помилки чи виконувати частину завдання. Студент має задавати якомога більше запитань, намагаючись самостійно розв'язати проблему. Викладач має навчити студентів самостійно діагностувати програмний код, з'ясовувати, де виникла помилка та як її можна виправити. Обговорюючи можливі варіанти розв'язання завдань, студенти вчаться оптимізувати програмний код, знаходять декілька шляхів вирішення питань.

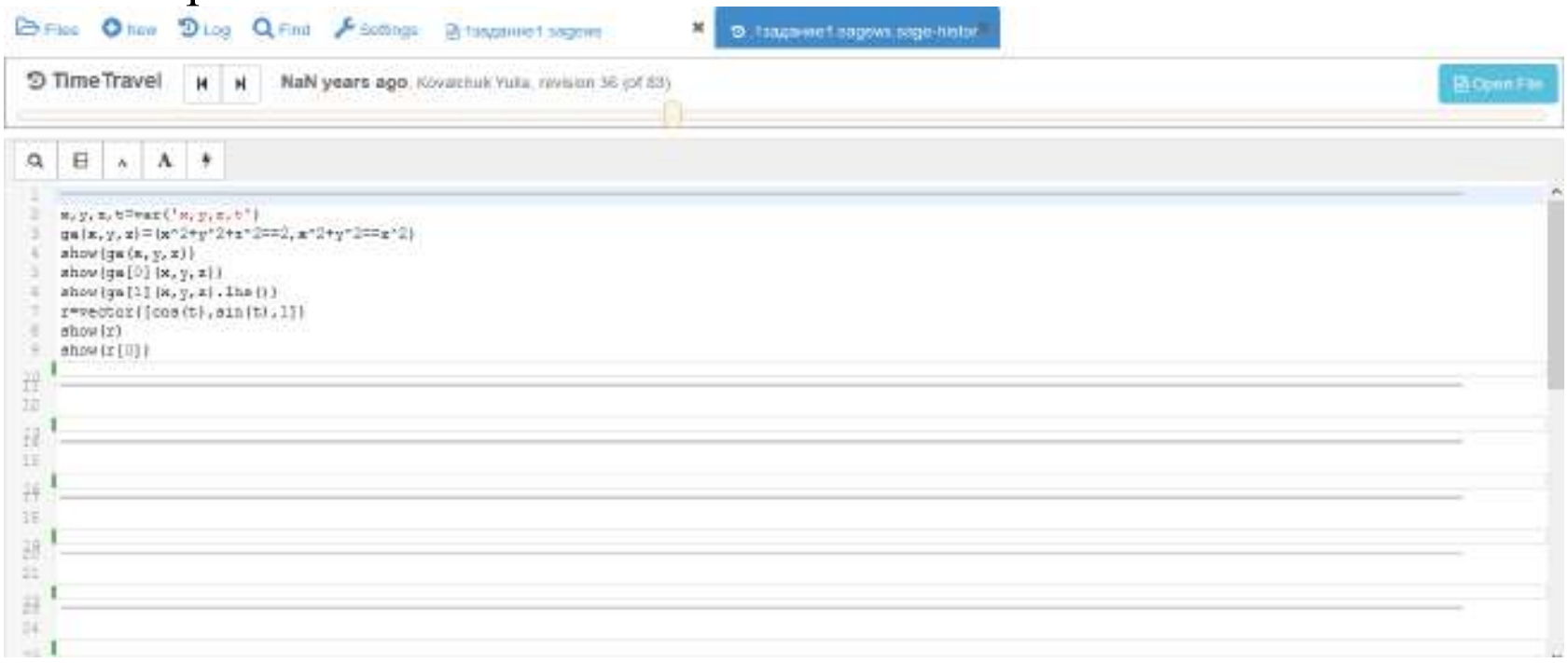

Рис. 3.49. Перегляд роботи з файлом: покрокове виконання

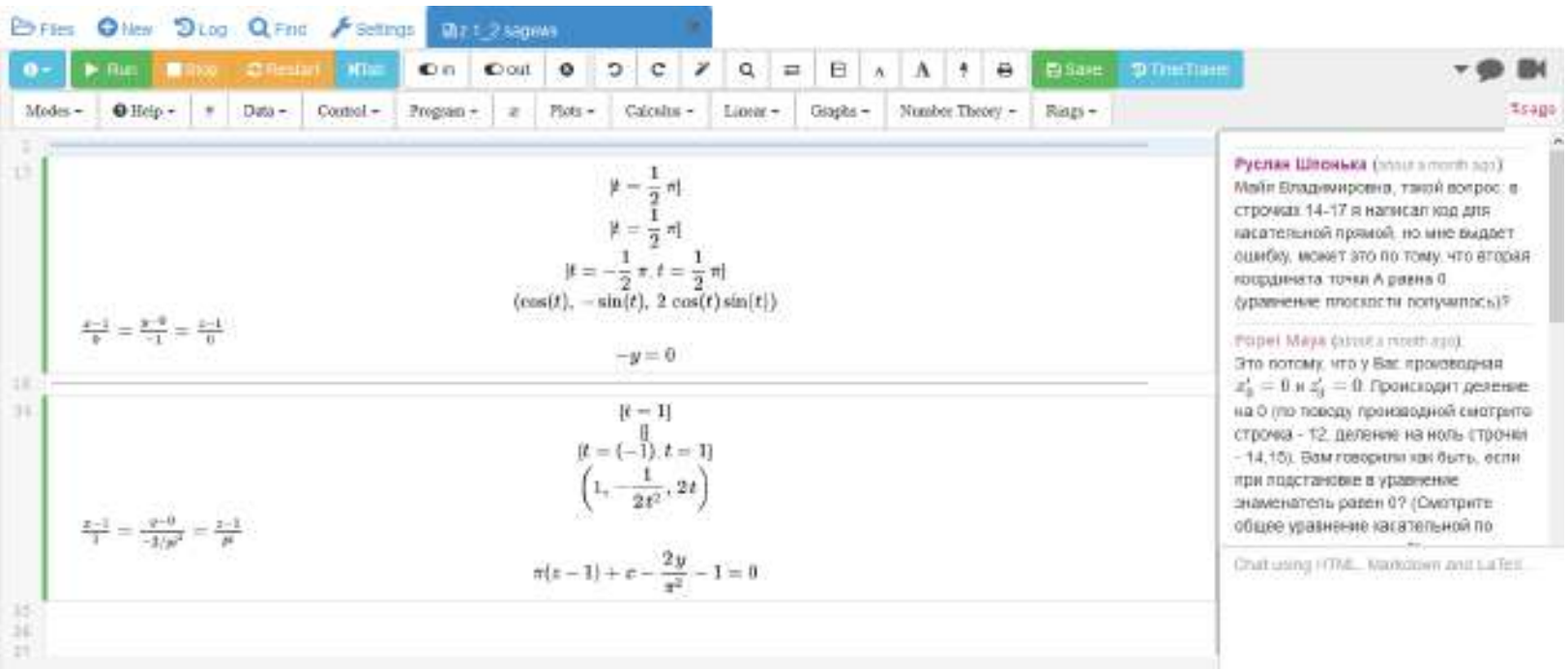

Рис. 3.50. Консультація з використанням чату

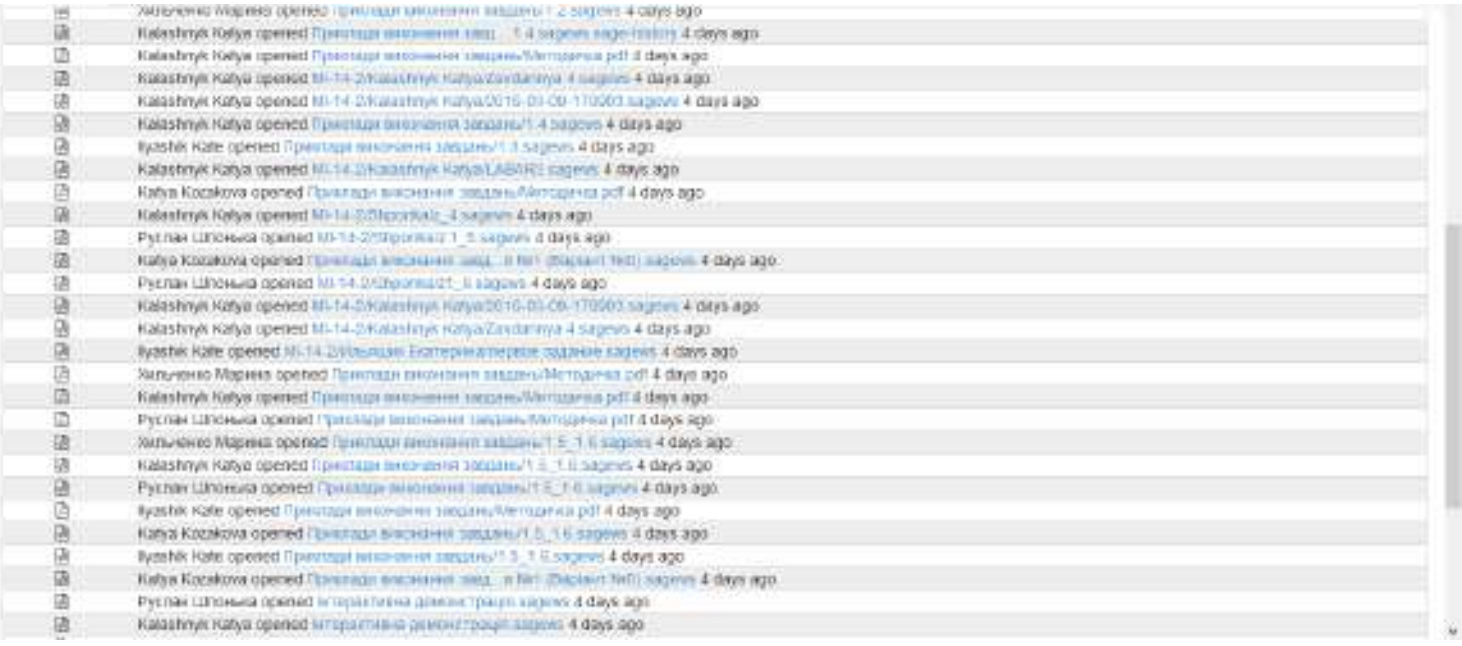

Рис. 3.51. Сторінка історії дій в проекті

Навіть якщо студентам здається, що вони одразу все зрозуміли, викладач має їх заохотити до обговорення шляху розв’язку завдання, допомогти своїм колегам зрозуміти найскладніші моменти, навчити співпрацювати.

Наведемо приклад вивчення тем з розділу «Вирази та їх перетворення» з курсу елементарної математики.

На початку проведення лабораторних занять, студенти відкривають робочий аркуш, який заздалегідь підготовлено викладачем. Студенти на лекції попередньо знайомі з теоретичними викладками, тому алгоритм та методи обчислення їм знайомі. На початку навчального курсу найрозповсюдженіші елементарні обчислення їм подаються на розгляд, а текст коду для виконання індивідуальних робіт є або аналогічним, або складається з частин вже відомих методів, попередньо розглянутих. У процесі викладення матеріалу курсу, студентам певні методи чи алгоритми обчислення доводиться складати самостійно, майже з нуля (спираючись на знання, які були одержані під час виконання практичних робіт з використанням хмарного сервісу). Крім того, змінюється специфіка інструкцій до кожної лабораторної роботи, оскільки більшість обчислень студенти можуть виконати самостійно за допомогою інструментарію CoCalc.

Якщо символьний вираз складається більше ніж з однієї змінної, або ж єдина змінна не являється `x`, тоді усі змінні треба оголосити одним з можливих способів.

1) `var('x,y,z');`

2) `x,y,z= var('x,y,z')` (суттєве значення має порядок слідування змінних як в лівій так і в правій частині);

3) `%var x,y,z.`

Над символьними виразами в системі CoCalc можна виконати операції спрощення, розкриття дужок, розкладання на множники та інші. Розглянемо основні з них.

Операцію спрощення (без розкриття дужок) виконує функція `simplify()`.

Спростити вираз: $x^2 - 2xy + y^2 + 2xy$ (рис. 3.52).

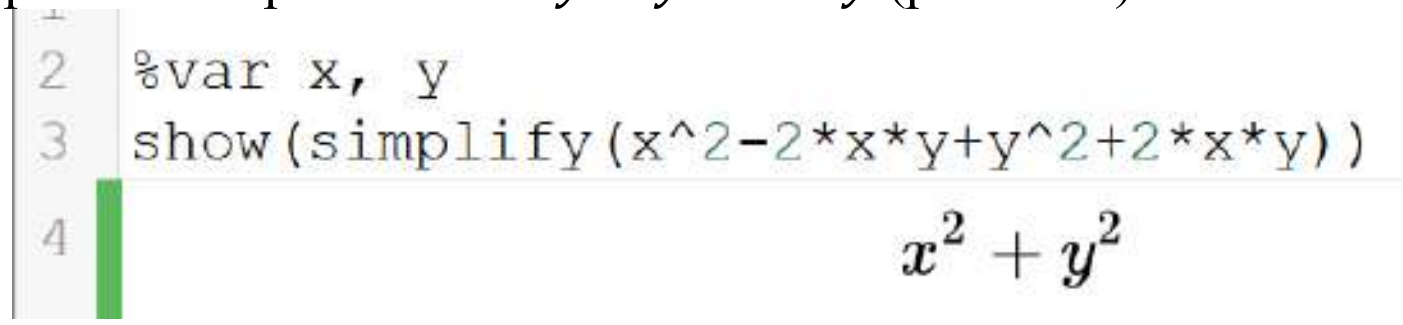

Рис. 3.52. Приклад спрощення символьного виразу із застосуванням функції `simplify()`

Зазначимо, що у CoCalc під час виведення результатів обчислень символьний вираз спрощується автоматично. Для порівняння (рис. 3.53).

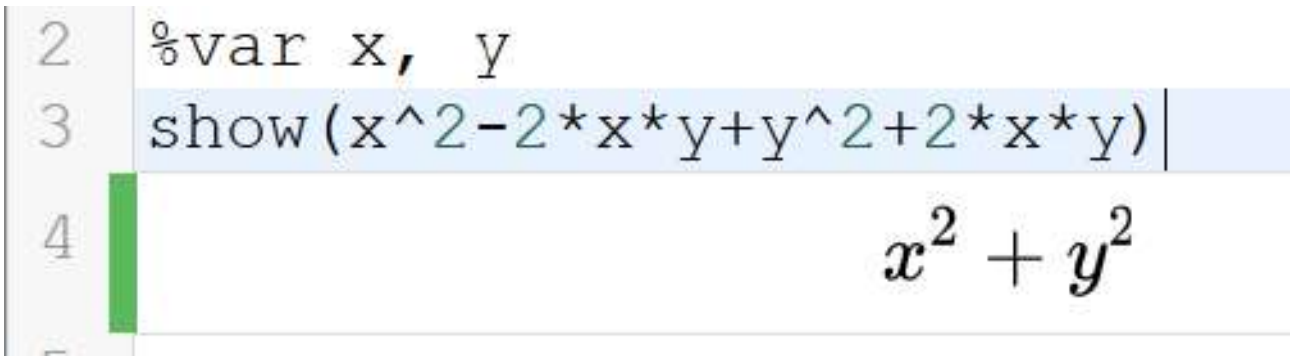

Рис. 3.53. Приклад автоматичного спрощення символьного виразу

Інколи, необхідно ініціалізувати функцію, залежну від декількох змінних (не треба плутати з функцією користувача). До такої функції, значенням якої виступає символьний вираз, можна застосовувати інші функції (вказуючи ім'я виразу як параметр), або як до об'єкту застосувати метод (рис. 3.54).

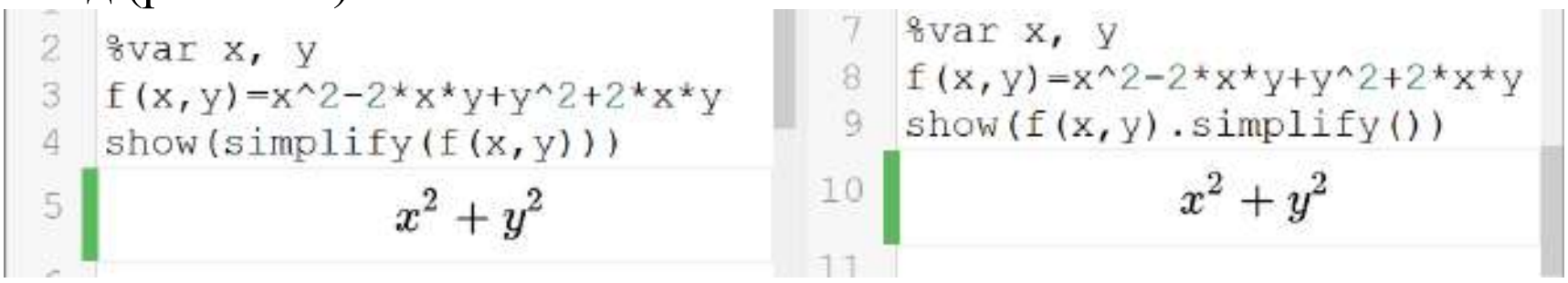

Рис. 3.54. Приклади ініціалізації функції від декількох змінних

Для розкриття дужок використовується функція `expand()`.

Розкрити дужки у виразі: *(x+y)(x–y)* (x + y)(x-y)(рис. 3.55: мітка а).

Якщо виконується лише одне обчислення/спрощення функцію для виведення результатів `show()` можна не вказувати. Але тоді відображення результатів виконання програмного коду не буде відформатовано в звичному для користувача вигляді(рис. 3.55: мітка б).

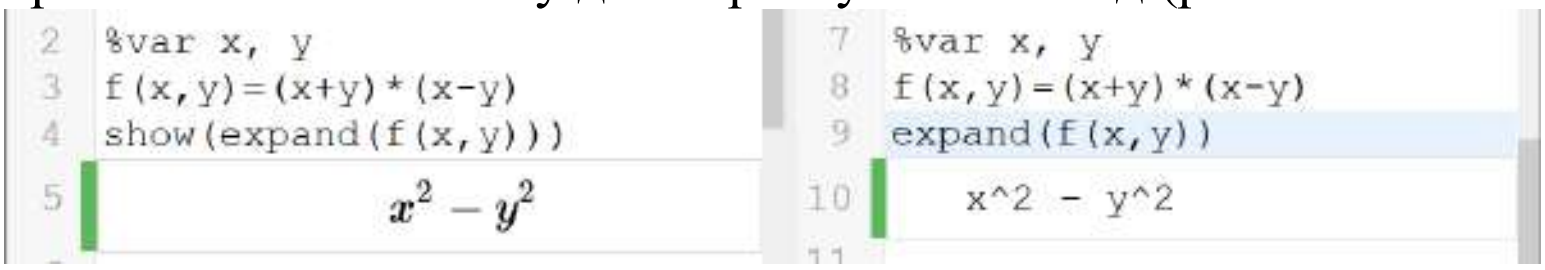

а) з використанням функції `show()`
б) без використання функції `show()`

Рис. 3.55. Приклад використання функції `expand()`

Розкласти вираз на множники можна, скориставшись функцією `factor()` (рис. 3.56).

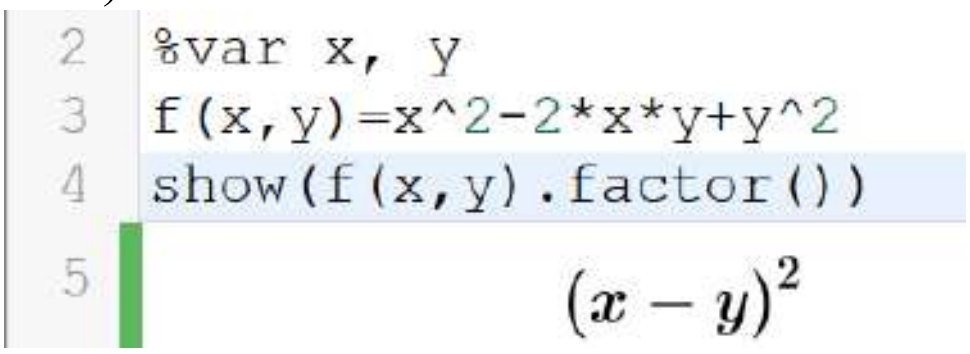

Рис. 3.56. Приклад розкладу виразу на множники

Для того, щоб обчислити значення виразу (рис. 3.57), у круглих дужках треба вказати значення змінної (або саму змінну, якій надано відповідне значення).

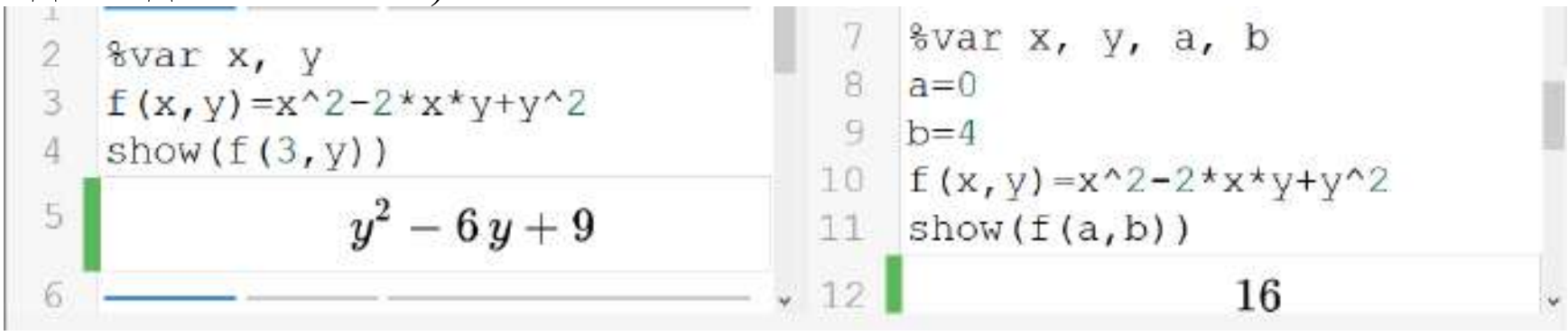

а) вказано значення однієї змінної б) вказано значення двох змінних

Рис. 3.57. Приклади обчислення значення виразу

Приклад вивчення теми «Функції і їх графіки». Ознайомлення з виконанням графічних примітивів (Додаток И, табл. И.1) рекомендується виконувати в рамках даної теми з курсу елементарної математики. Також наприкінці вивчення теми можна розглянути декілька прикладів створення анімації.

Графічні примітиви на площині:

`arrow2d()` – стрілка від мінімальної до максимальної точки. Червона стрілка: `arrow2d((-1, -1), (2, 3), color='red')`

Стрілка з потовщеною лінією та збільшеною стрілкою наприкінці (рис. 3.58).

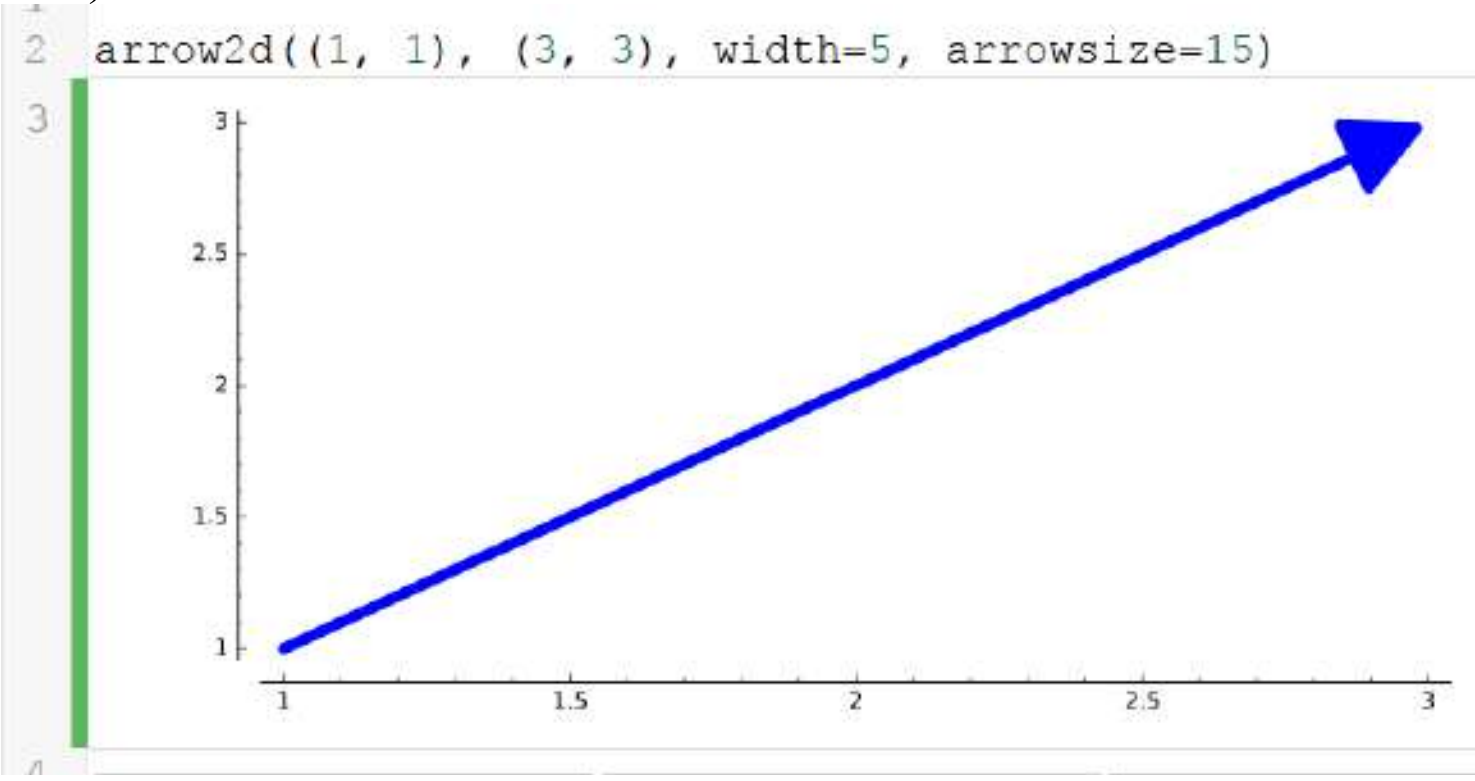

Рис. 3.58. Приклад побудови стрілки з потовщеною лінією

Круг жовтого кольору, радіусу 1, з центром в точці (4,5), з підписом легенди (параметр – `legend_label`), до якого застосовано метод `show()`, в якому зазначено, що вісі потрібно показати з початку координат (рис. 3.59).

Функція `show()` відображає результат виконання програмного коду,

а не лише показує обчислення. Це стосується і графічних побудов.

```
circle((4,5), 1, rgbcolor='yellow', fill=True,_
        legend_label='Kolo').show(xmin=0, ymin=0)
```

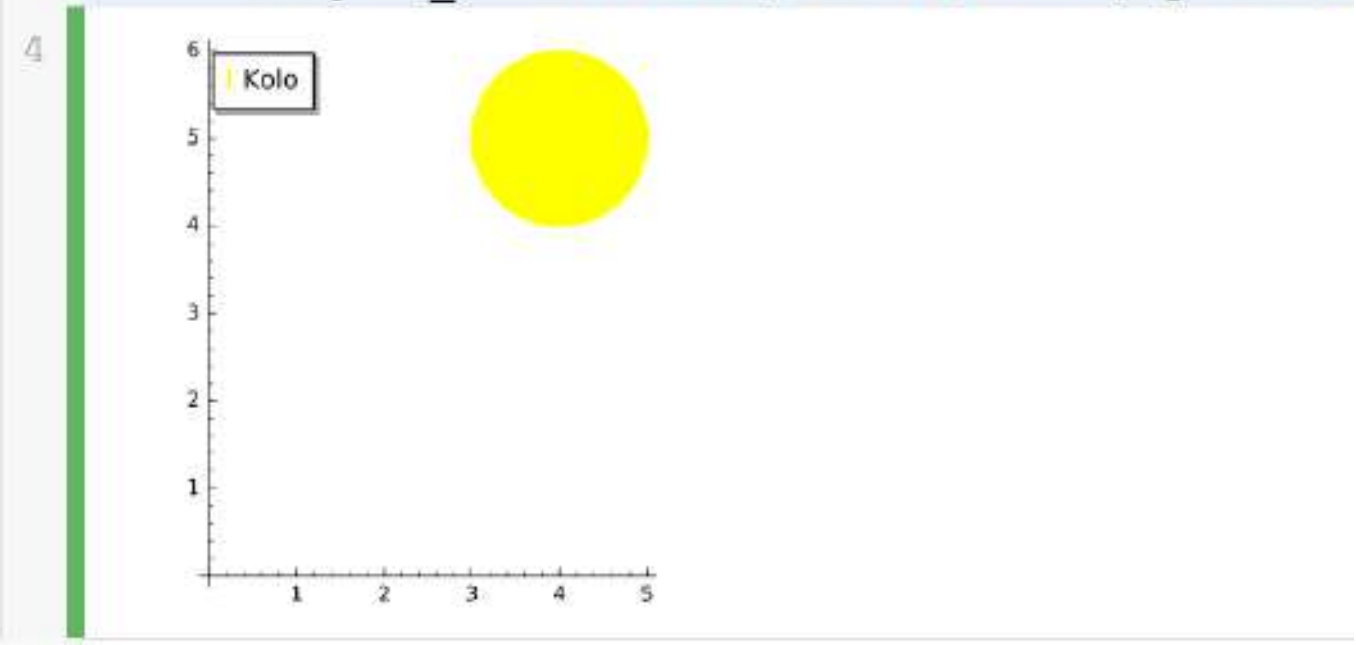

Рис. 3.59. Приклад побудови круга з легендою

Побудувати еліпс з центром в початку координат (рис. 3.60), радіусами 1 та 2, до якого зазначено легенду та колір легенди (параметр – `legend_color`).

```
ellipse((0,0),2,1,legend_label="My ellipse",_
        legend_color='green')
```

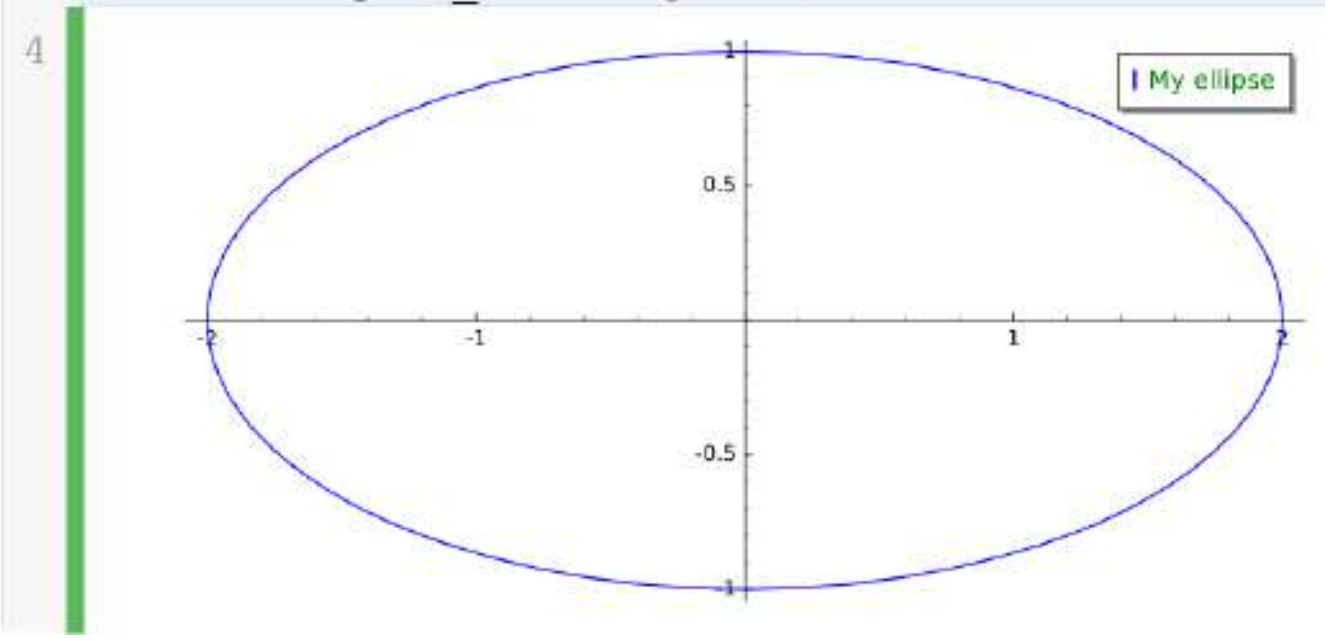

Рис. 3.60. Приклад побудови еліпса з текстовим підписом

Побудувати дугу, що являє собою частину еліпса, з центром в початку координат, з двома радіусами – 1 та 2, червоного кольору(рис. 3.61), тип лінії – штрих-пунктир.

Побудувати сектор круга $\left[\pi; \frac{3\pi}{2}\right]$, центр якого розташовано в точці (0, 0), радіуса 1, жовтого кольору (рис. 3.62).

`Line2d()` – лінія визначається послідовністю точок (не обов’язково пряма). Побудувати ламану (рис. 3.63), задану переліком точок: (1, 2), (2, 4), (3, 4), (4, 8), (4.5, 32), використовуючи логарифмічну шкалу з основою 2.

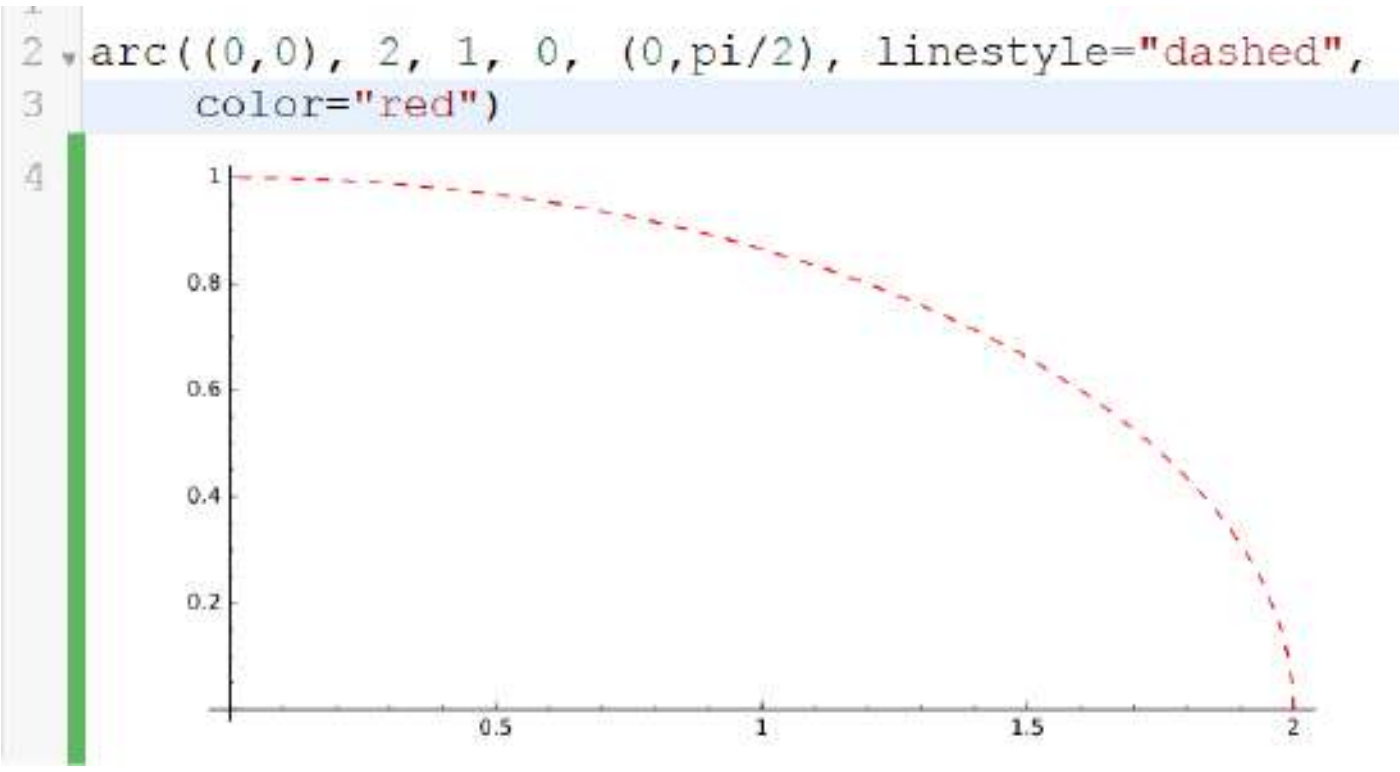

Рис. 3.61. Приклад побудови дуги червоного кольору

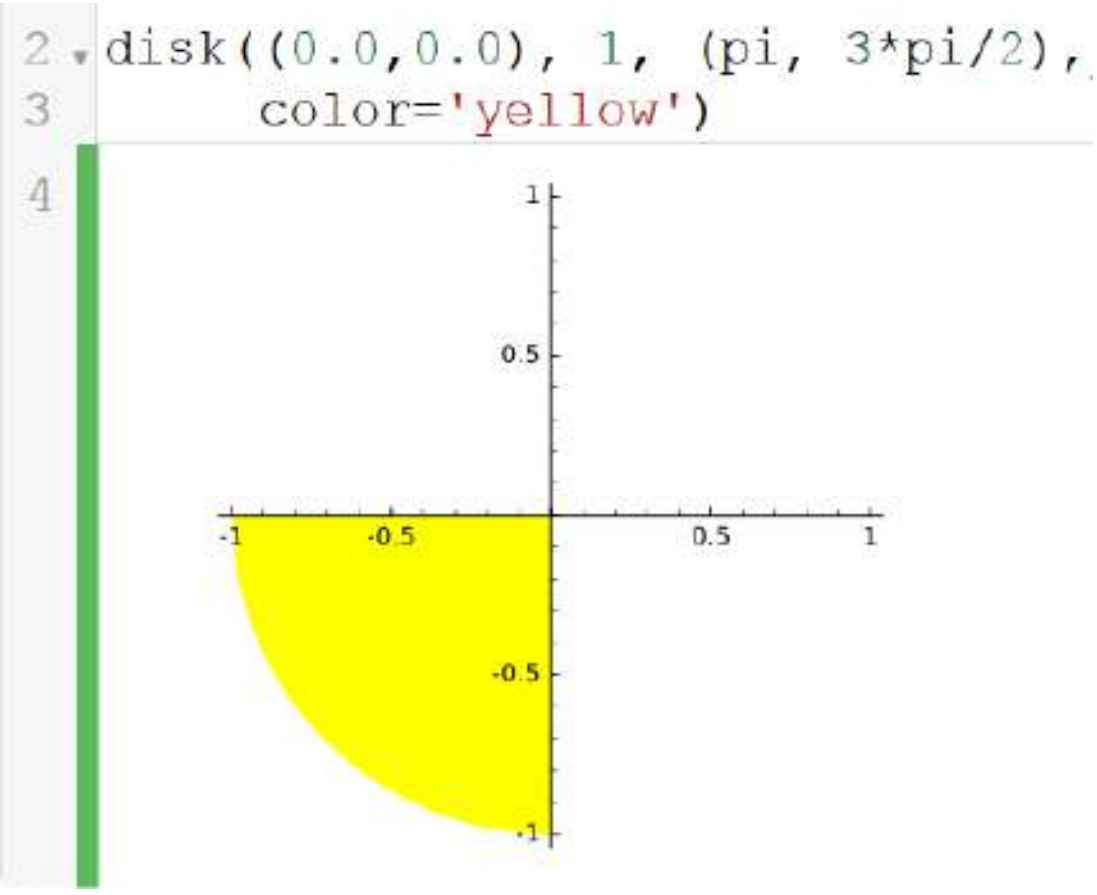

Рис. 3.62. Приклад побудови жовтого сектору круга

Точка, що розташована на початку системи координат, чорного кольору, трохи збільшена, з підписом:

```
point((0,0), rgbcolor='black', pointsize=40,
legend_label='origin')
```

`text()` – додати текст до побудови. Додати тест на червоному тлі, прив'язати його до точки з координатами (-2, 2).

```
text("So good", (-2,2), background_color='red')
```

Для того, щоб виконати дві побудови одночасно (рис. 3.64), потрібно їх додати (скористаємось двома останніми побудовами).

Можна окремо надати змінним значення графічного примітиву та виконати побудову використавши функцію `show()`:

```
A=point((0,0), rgbcolor='black', pointsize=40,
legend_label='origin')
B=text("So good", (-2,2), background_color='red')
show(A+B)
```

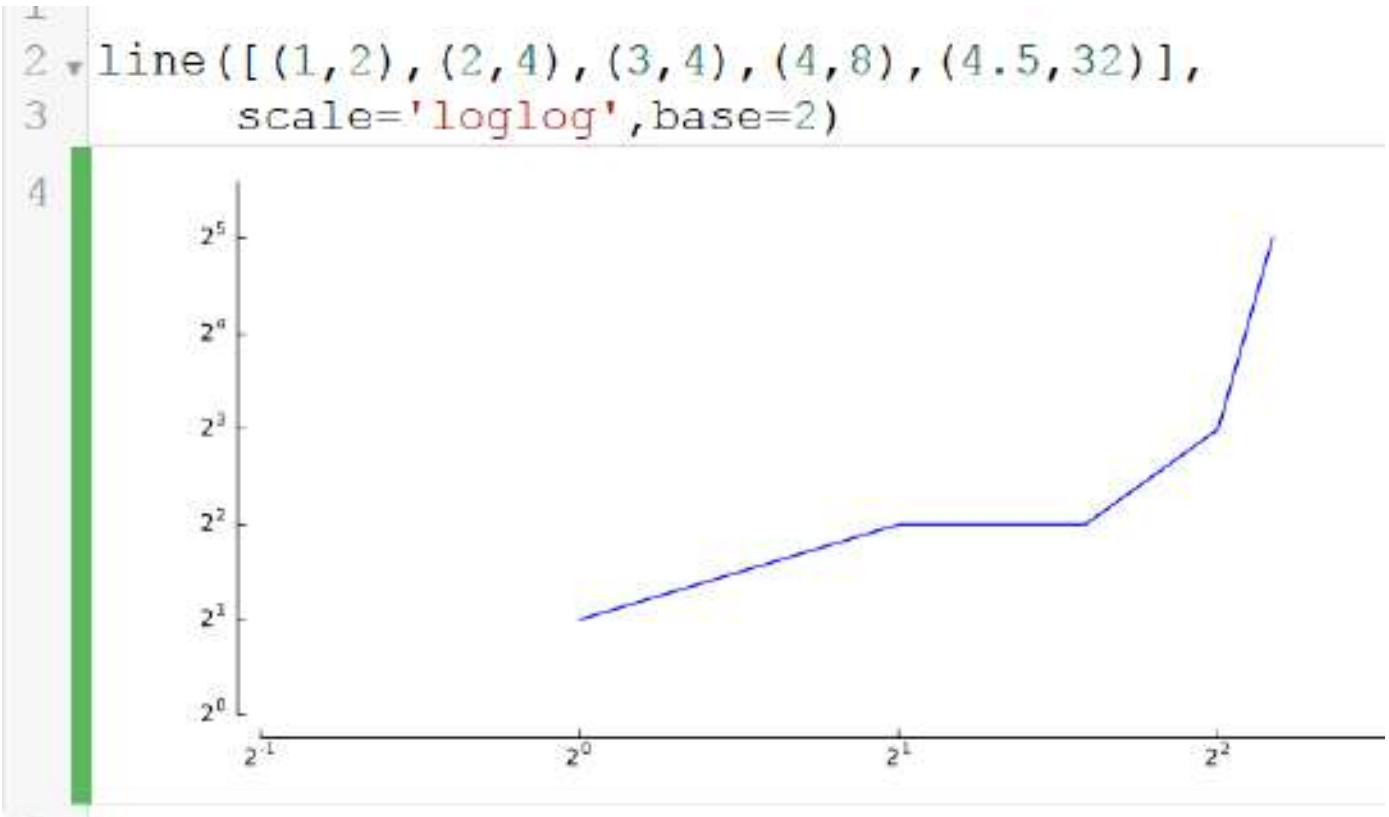

Рис. 3.63. Приклад побудови ламаної, заданої переліком точок

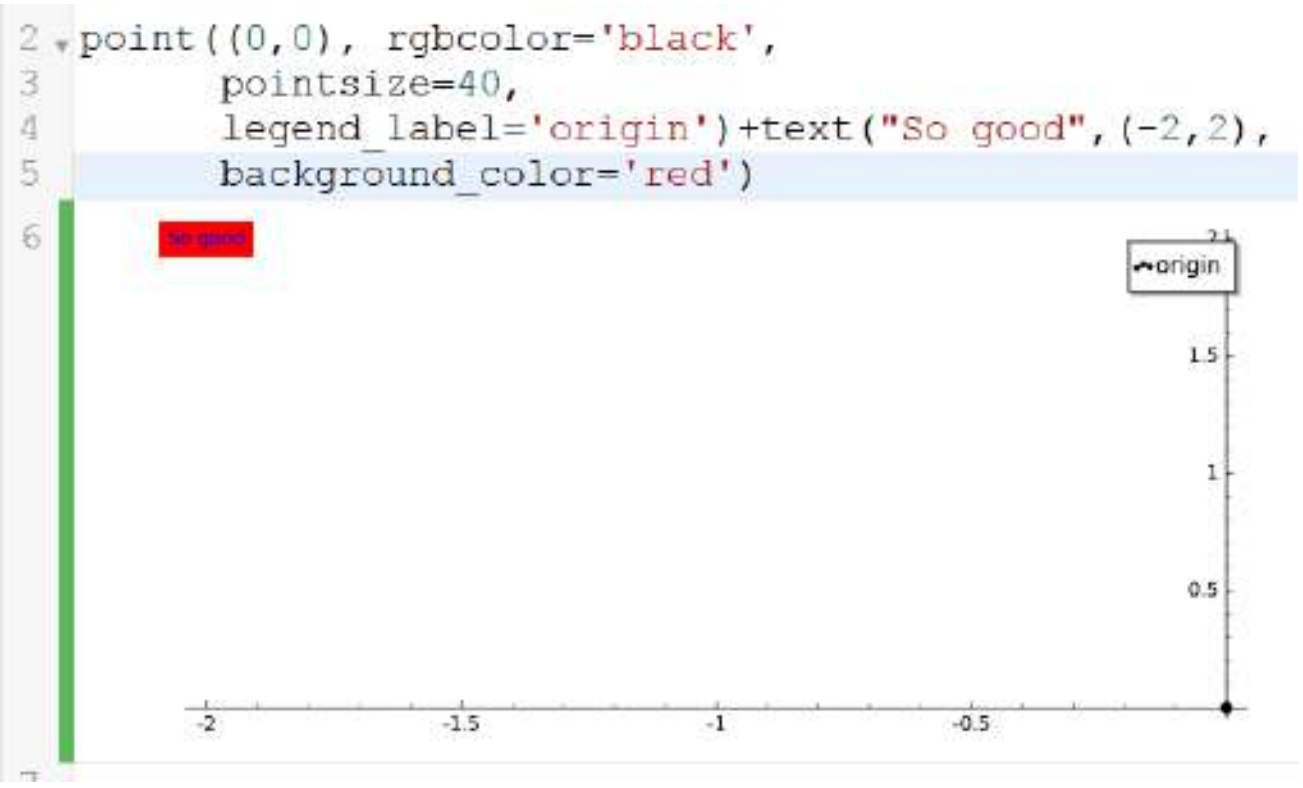

Рис. 3.64. Приклад побудови точки з текстовим підписом

Функція `show()` також має певні параметри. Корисно буде представити приклад вирівнювання значень на осях (рис. 3.65).

Для побудови многокутника треба перерахувати його вершини. Наприклад, трикутник не зафарбований, потовщений контур побудови помаранчевого кольору (рис. 3.66).

Побудови графіків функцій на площині:

Побудувати косинусоїду на певному проміжку $[0; 10]$ по точках, не сполучаючи їх лінією можна наступним чином (рис. 3.67).

За допомогою функції `parametric_plot()` можна виконати побудову параметрично заданої функції (рис. 3.68).

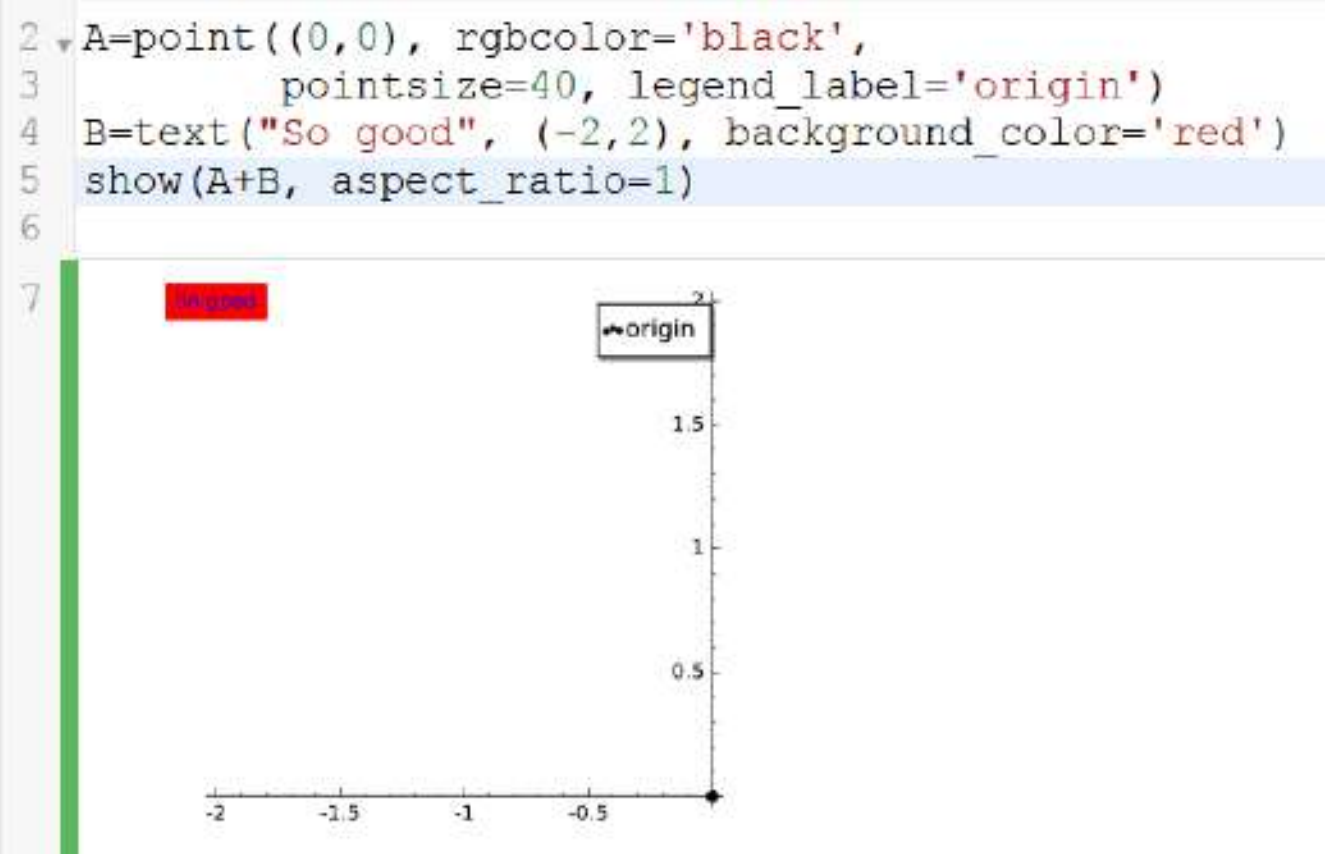

Рис. 3.65. Приклад вирівнювання значень на осях

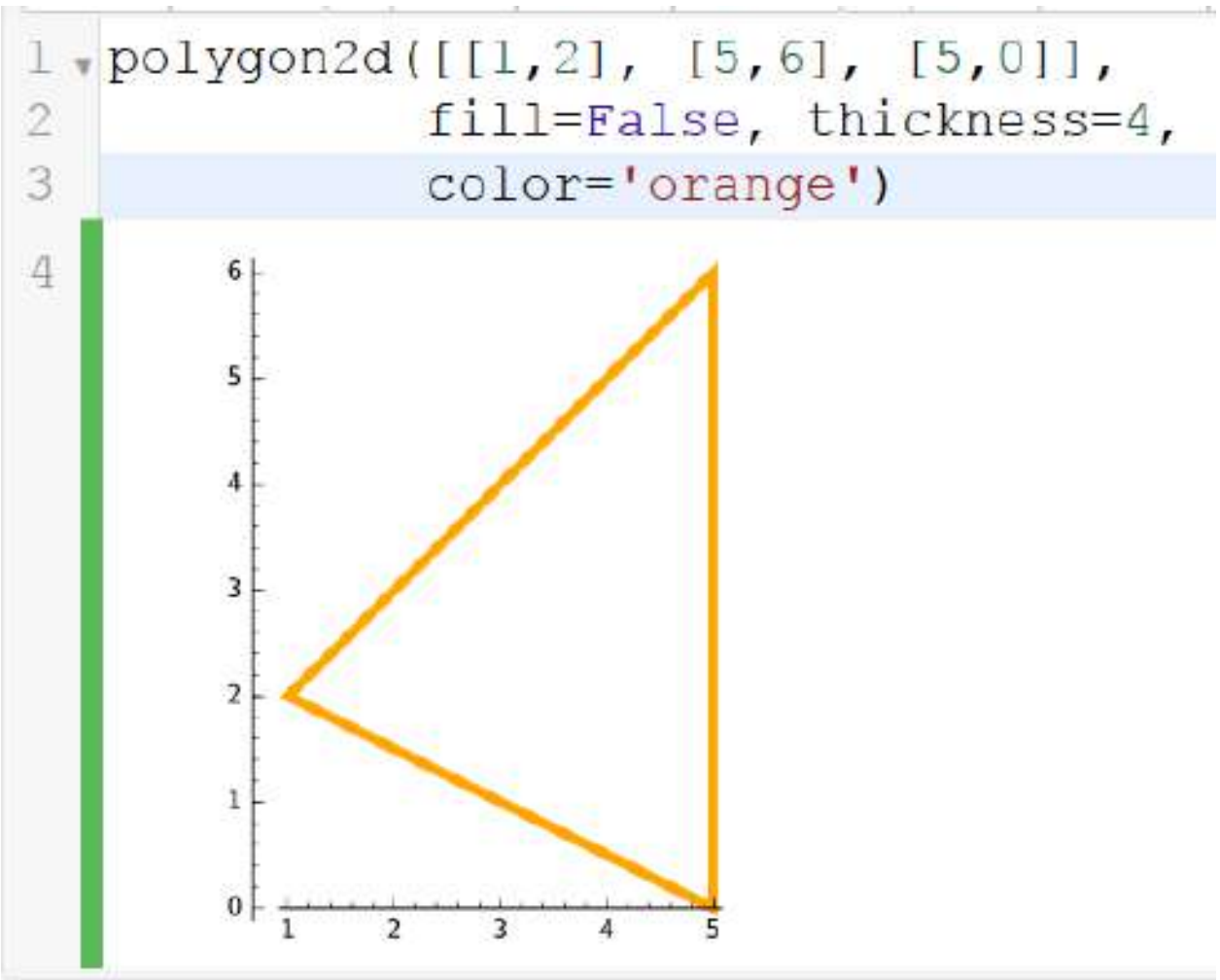

Рис. 3.66. Приклад побудови контуру трикутника

`implicit_plot()` – використовується для побудови неявно заданої функції. Побудуємо еліпс використавши його канонічне рівняння (рис. 3.69).

Можна створити анімацію (рис. 3.70) використавши функцію `animate()`, без використання функції `plot()`.

Використавши `plot()`, попередньо за допомогою циклу створивши список окремих графіків функції.

```
sines = [plot(c*sin(x), (-2*pi,2*pi), color=Color(c,0,0),
ymin=-1, ymax=1) for c in sxrange(0,1,.2)]
a = animate(sines)
a.show()
```

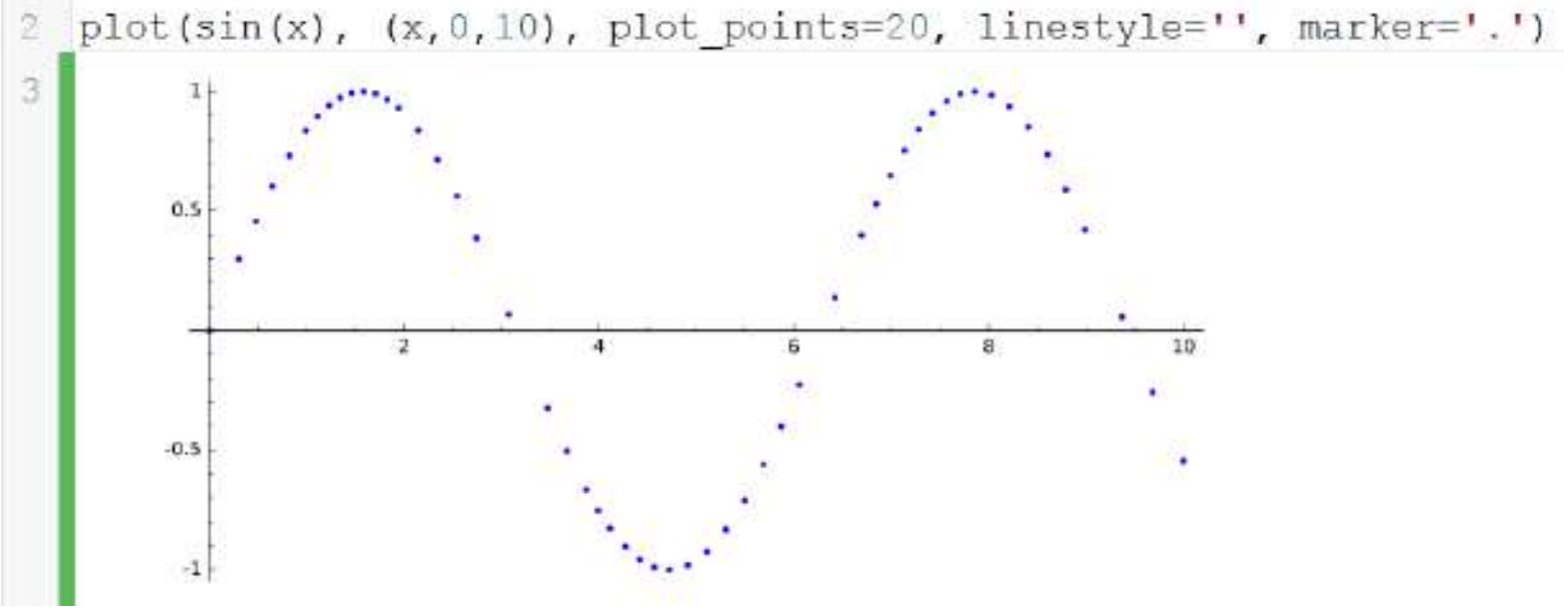

Рис. 3.67. Приклад побудови косинусоїди синього кольору

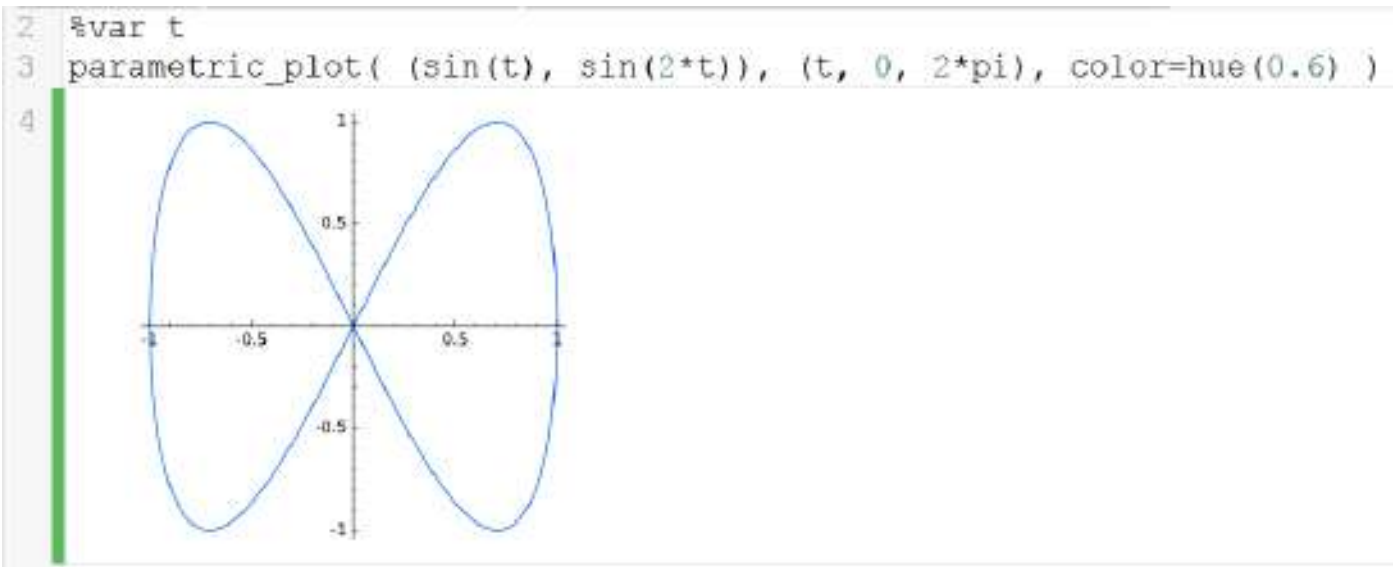

Рис. 3.68. Приклад побудови графіка параметрично заданої функції

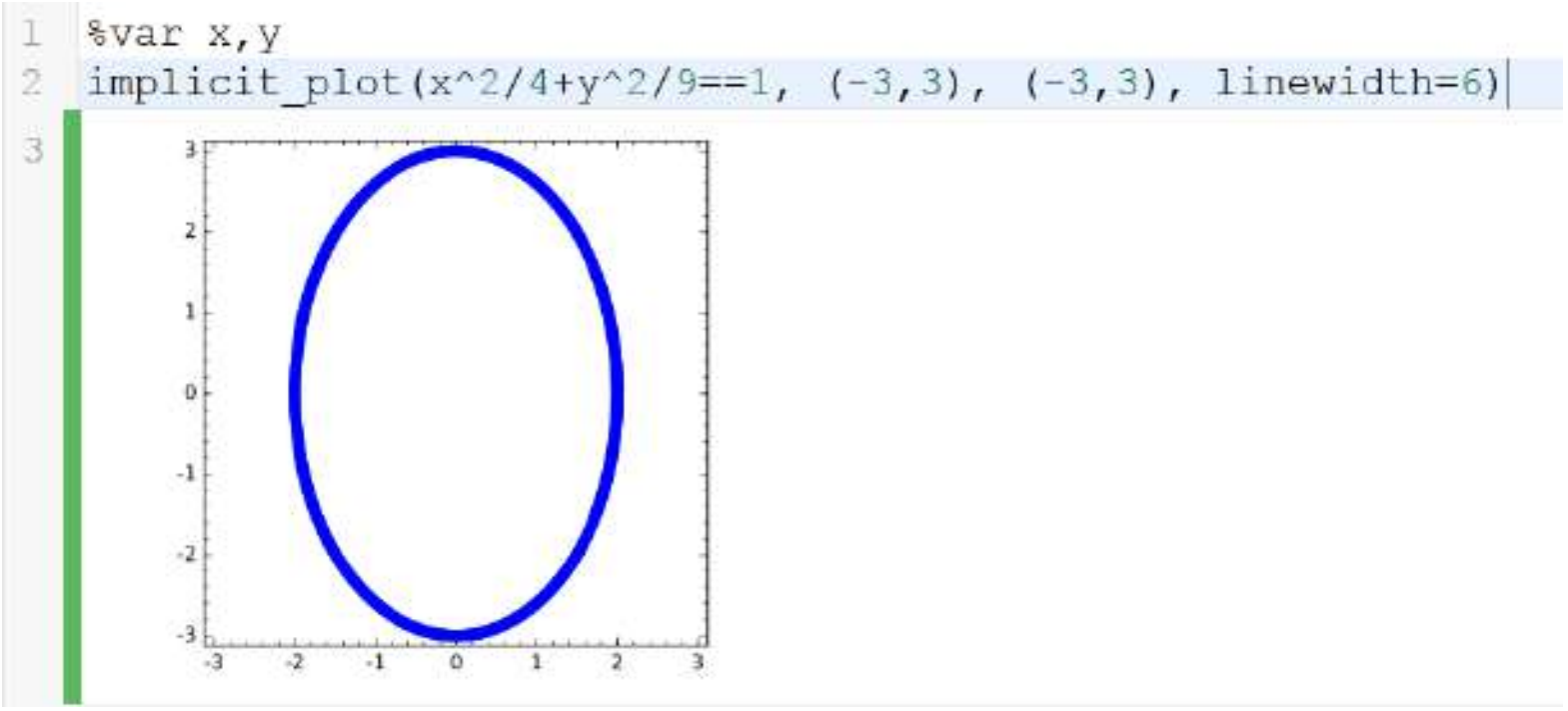

Рис. 3.69. Приклад побудови еліпса заданого канонічним рівнянням

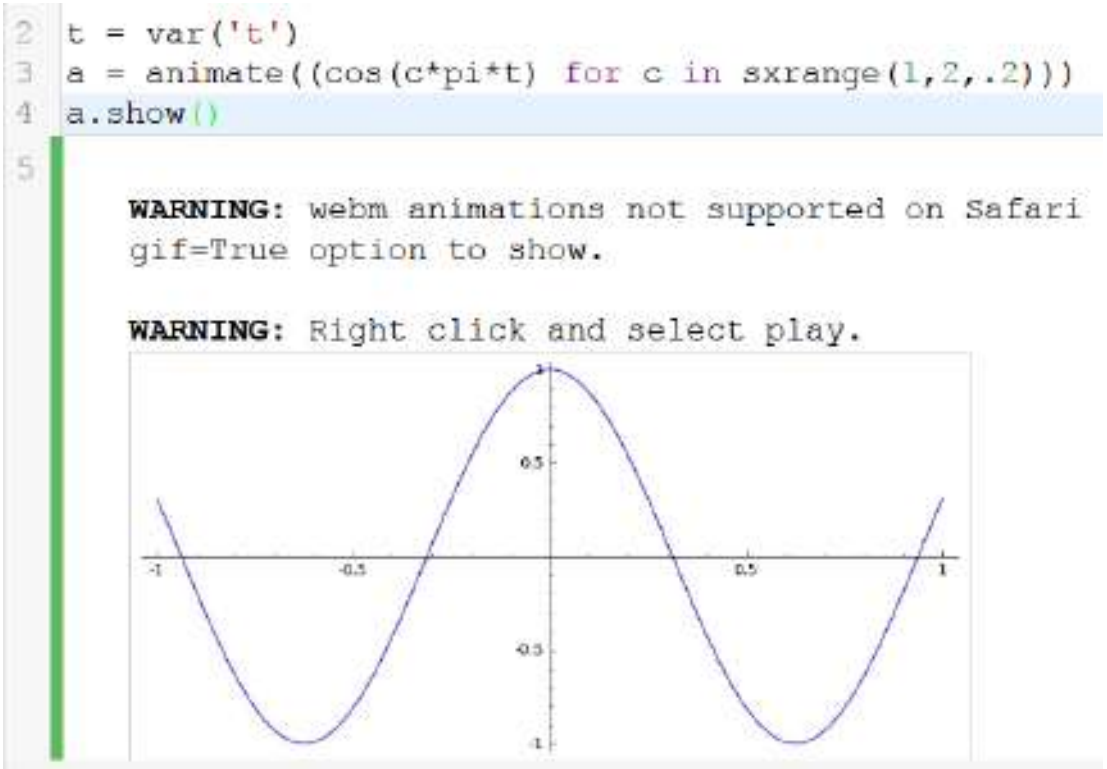

Рис. 3.70. Приклад представлення анімації як відео

Під час вивчення диференціальної геометрії і топології студенти будуть виконувати окрім обчислень побудови гладких ліній та поверхонь у просторі. Тому заздалегідь доречно розглянути виконання графічних примітивів у просторі, а вже потім – виконати покрокову побудову гладких ліній, поверхонь.

Дуже важливо, щоб студенти на лабораторні заняття приходили вже заздалегідь підготовленими, з опрацьованим конспектом лекцій. Програмний код, що розміщено в робочому аркуші їм не відомий, але вони в змозі зрозуміти основні прийоми обчислення самостійно. В процесі проведення практичних занять присутні й елементи дослідження, коли студенти на практиці, застосовуючи вже відомі методи, формули, алгоритми, відкривають для себе щось нове, узагальнюючи пройдений лекційний матеріал. Буде доречним, організовувати на початку занять засобами CoCalc короткі опитування, задля повторення основних теоретичних викладок. Це надасть додаткової мотивації для подальшого вивчення теми. Тестування, опитування можна проводити і для того, щоб переконатись що студенти знайомі з основними методами обчислення, що вони розуміють принцип не лише простих, але й громіздких обчислень, на які може знадобитись набагато більше часу. Подібними опитуваннями, можна завершити вивчення теми чи курсу, щоб систематизувати вивчений матеріал та перевірити якість засвоєння його студентами.

Побудови гладких ліній та поверхонь у просторі:

Основною функцією побудови у просторі є `plot3d()`. З її використанням можна виконати побудову поверхні заданої в явному вигляді (рис. 3.71).

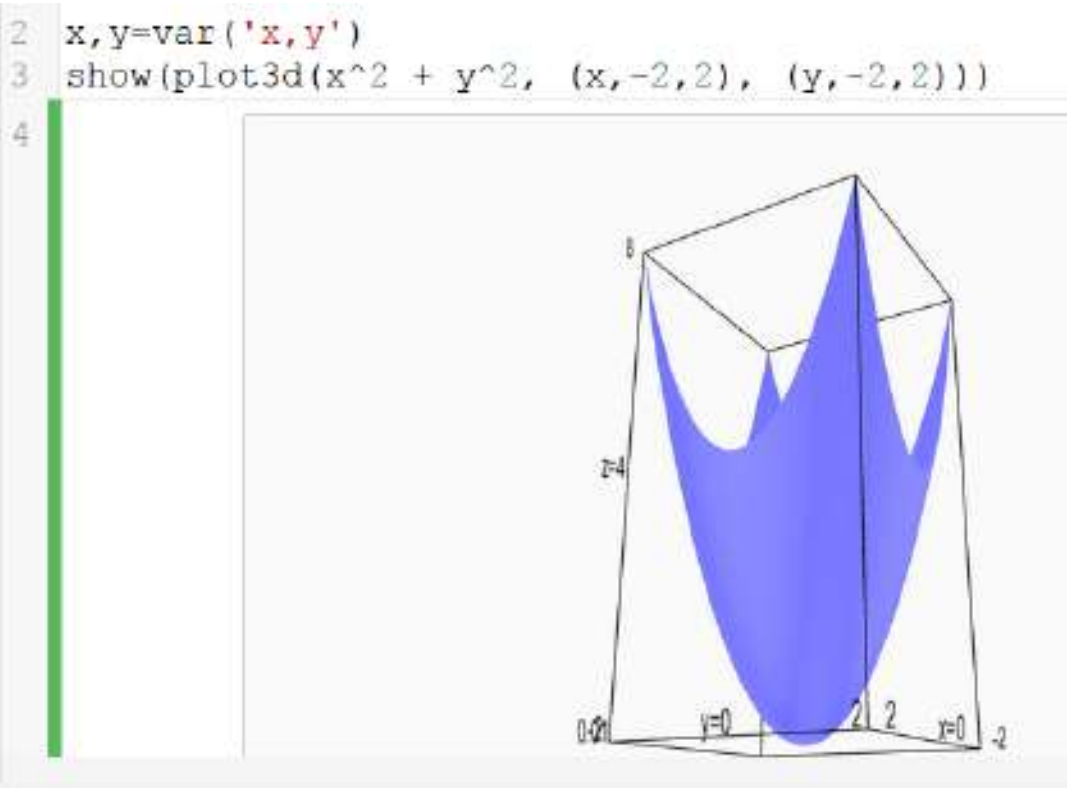

Рис. 3.71. Приклад побудови поверхні

Ще один варіант представлення:

```
x,y=var('x,y')
f(x,y)=x^2 + y^2
show(plot3d(f(x,y), (x,-2,2), (y,-2,2)))
```

Змінити колір можна за допомогою параметра `color='вказати бажаний колір англійською мовою'`

```
x,y=var('x,y')
f(x,y)=x^2 + y^2
show(plot3d(f(x,y), (x,-2,2), (y,-2,2),color='red'))
```

Зробити побудову прозорою можна за допомогою параметра `opacity='вказати десяткове число'`. 1 – не прозора фігура (значення 2, 3, і так ділі сприймаються як 1 – не прозора фігура). 0.9 – трохи прозора, 0.8 – більш прозора і так далі. Чим менший десятковий дріб – тим більш прозора фігура.

```
x,y=var('x,y')
show(plot3d(x^2 + y^2, (x,-2,2), (y,-2,2),opacity=0.5))
```

Показати сітку – параметр `mesh=True`.

```
x,y=var('x,y')
show(plot3d(x^2 + y^2, (x,-2,2), (y,-2,2), mesh=True))
```

Побудова двох поверхонь (рис. 3.72) виконується за рахунок додавання двох графічних зображень поверхонь.

Функцію побудови лінії представленої рівнянням у параметричній формі (рис. 3.73) можна побудувати скориставшись `parametric_plot3d()`.

Поверхню, утворену кривою, що обертається (рис. 3.74) можна побудувати використавши функцію`revolution_plot3d()`. Параметр `show_curve=True` – показати задану криву.

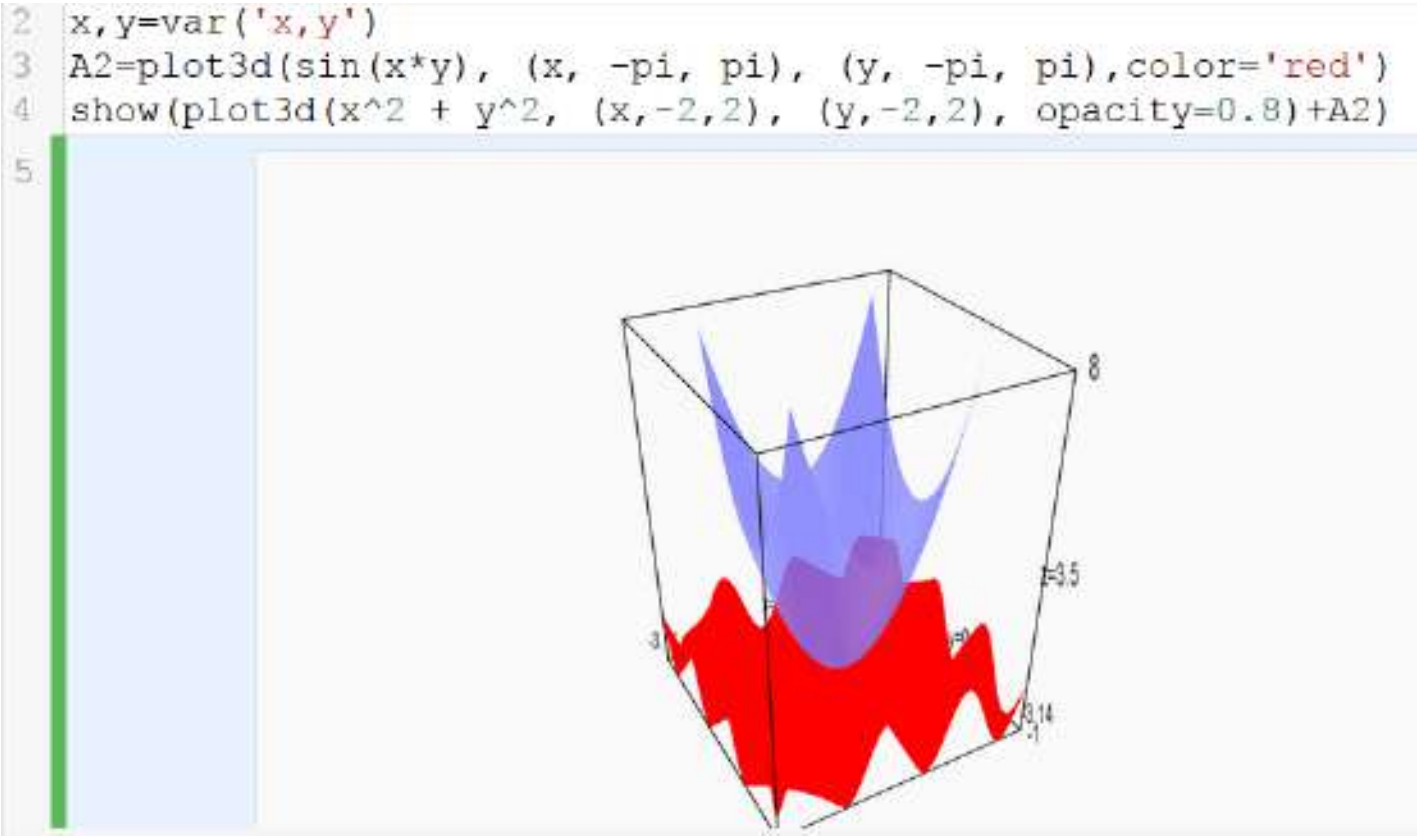

Рис. 3.72. Приклад побудови декількох поверхонь

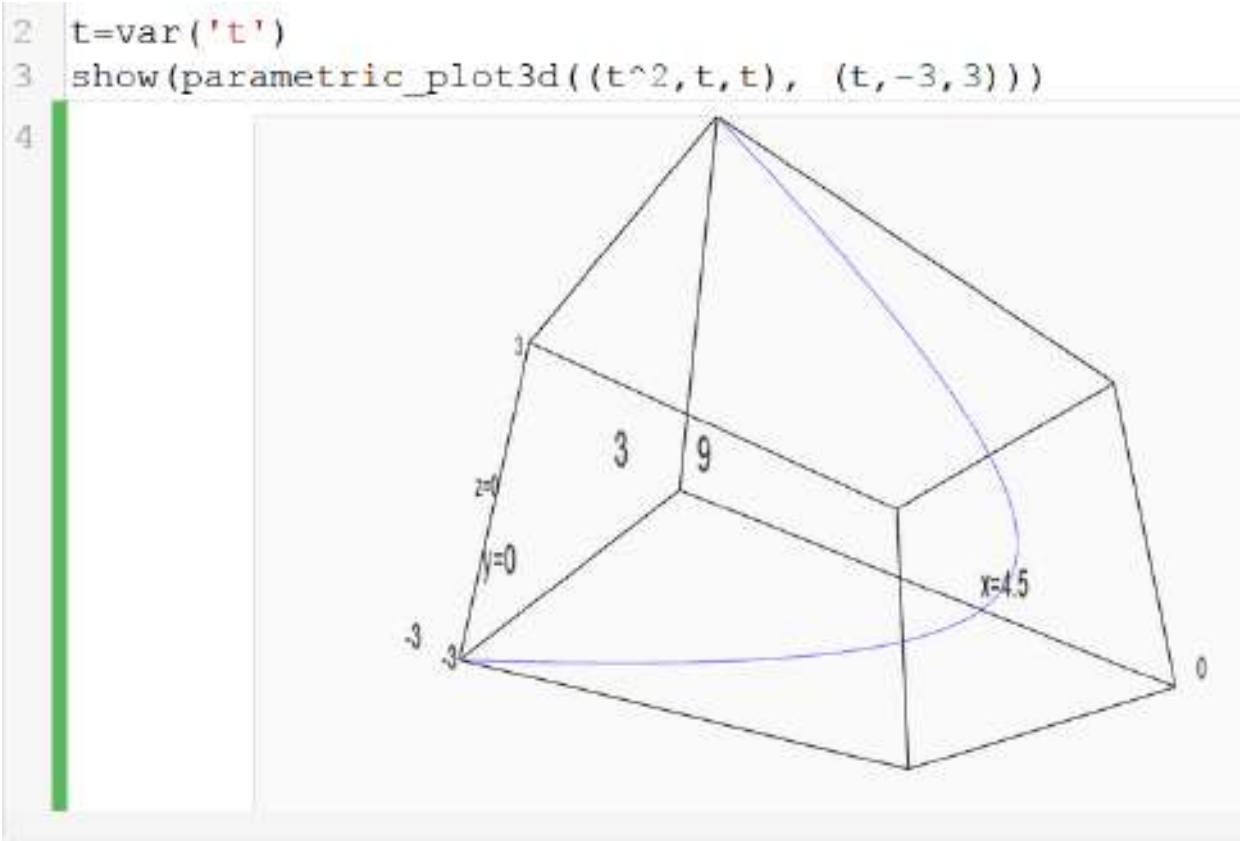

Рис. 3.73. Приклад побудови гладкої лінії

Для того, щоб вирівняти масштаб, потрібно в функції `show()` вказати параметр `aspect_ratio=(1,1,1)`:

```
x,y=var('x,y')
show(plot3d(x^2 + y^2, (x,-2,2), (y,-2,2)),
aspect_ratio=(1,1,1))
```

Побудова неявно заданої функції відбувається за рахунок застосування функції `implicit_plot3d()`. Наприклад, побудову сфери можна використати наступним чином (рис. 3.75).

Виконаємо комбінацію декількох побудов (рис. 3.76).

```
u = var('u')
f=u^2
show(revolution_plot3d(f,(u,0,1),show_curve=True,opacity=0.7))
```

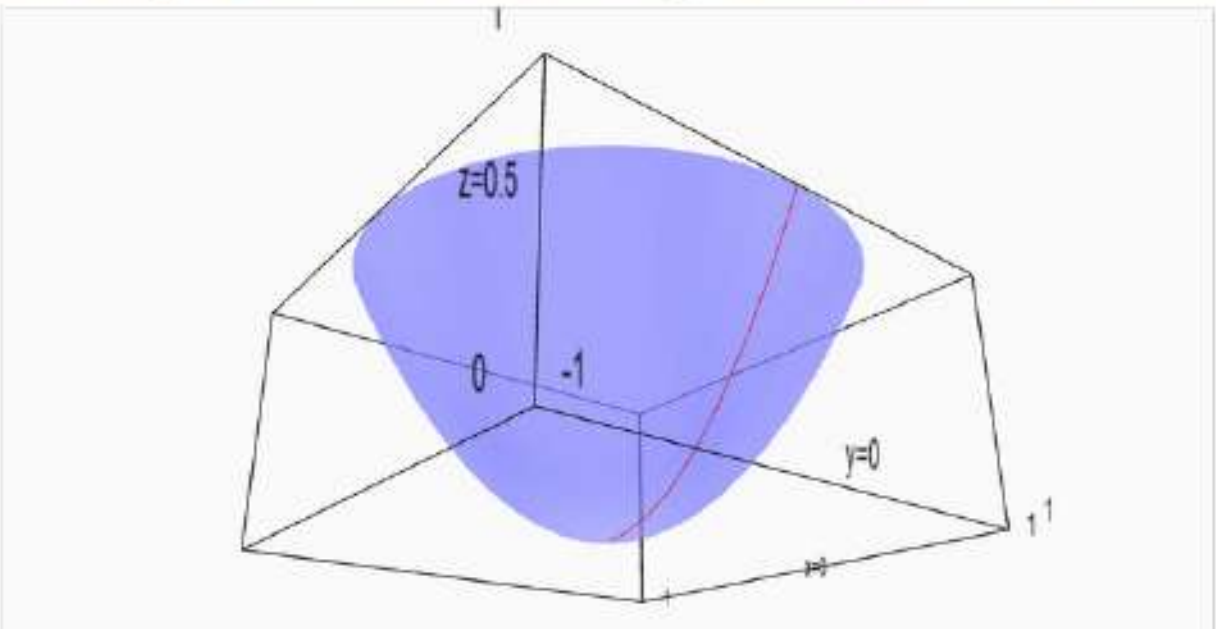

Рис. 3.74. Приклад побудови поверхні утвореної обертанням кривої навколо вісі

```
x,y,z=var('x,y,z')
implicit_plot3d(x^2+y^2+z^2==4, (x, -3, 3), (y, -3,3), (z, -3,3))
```

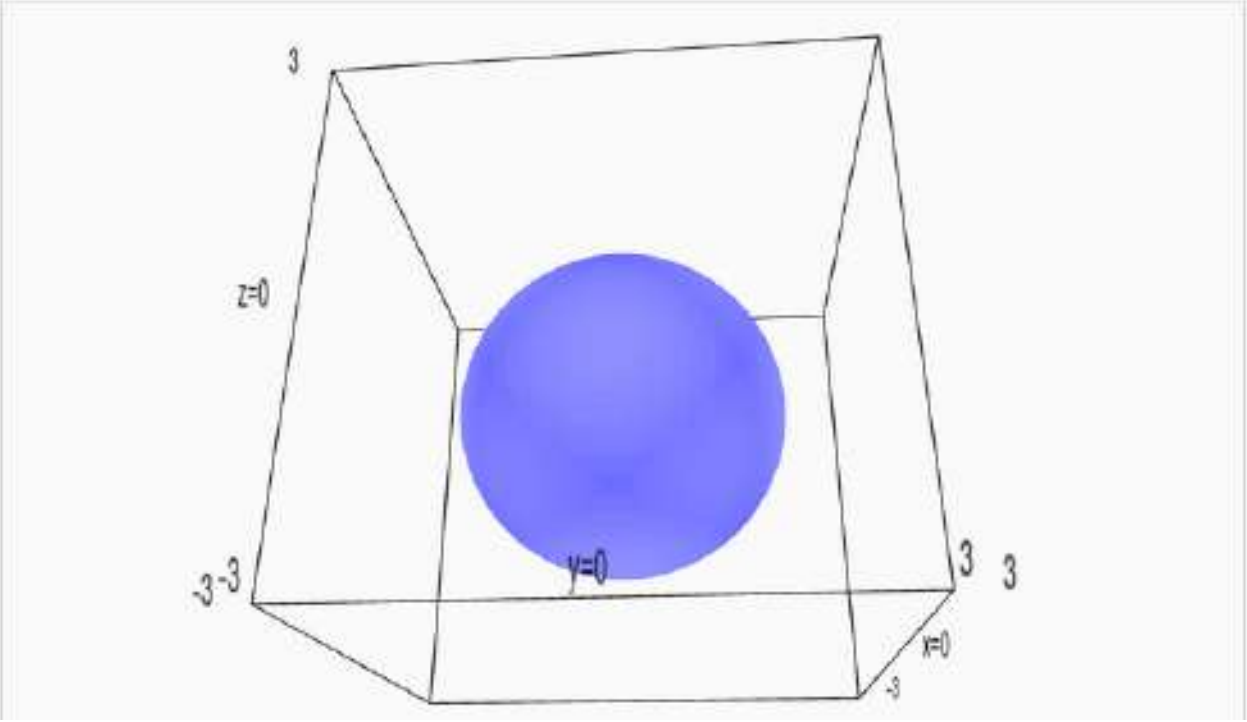

Рис. 3.75. Приклад побудови сфери

### 3.6 Виконання навчально-дослідницьких проектів з використанням CoCalc

Виконання навчально-дослідницьких проектів, курсових робіт та випускної роботи бакалавра з використанням CoCalc передбачає два шляхи:

1. Використанням окремих засобів, що представлені в CoCalc.
2. Виконанням, написанням та оформленням результатів навчально-дослідницької роботи в CoCalc без залучення допоміжних програмних засобів.

*Інтерпретатор IPython* у процесі навчання майбутніх учителів математики може бути використаний для розробки динамічних моделей з напівавтоматичним / автоматичним режимами демонстрації [196].

```
x,y,z,t=var('x,y,z,t')
A1=implicit_plot3d(y^2==x, (x, -3, 3), (y, -3,3), (z, -3,3),opacity=0.7)
A2=implicit_plot3d(z==y, (x, -3, 3), (y, -3,3), (z, -3,3),color='green')
A3=parametric_plot3d((t^2,t,t), (t,-2,2),color='red')
show(A1+A2+A3,aspect_ratio=(1,1,1))
```

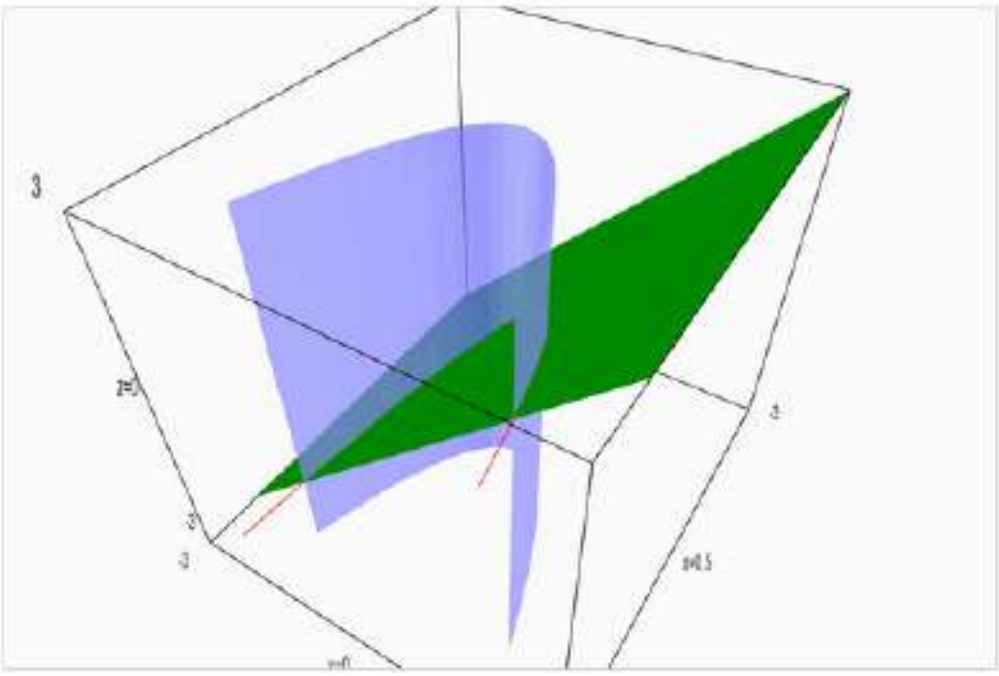

Рис. 3.76. Перетин поверхні площиною

Перший шлях передбачає створення моделі (моделей) досліджуваного явища на робочому аркуші з використанням стандартних елементів управління, тегів мови HTML, команд LaTeX та використанням мови CSS (рис. 3.77).

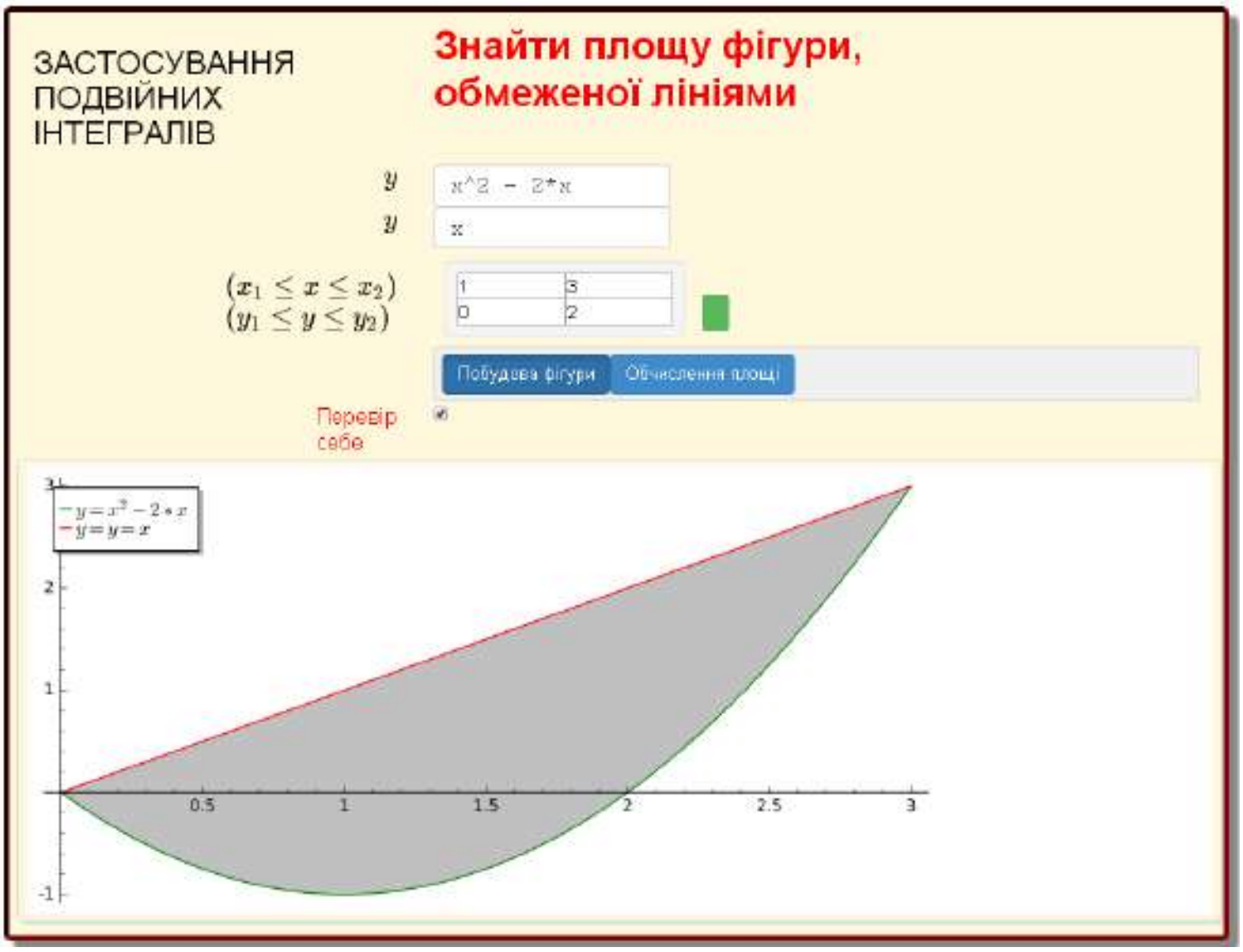

Рис. 3.77. Приклад моделі, розробленої стеденткою

Недоліками такого використання є те, що в процесі оформлення одержаних результатів доводиться залучати інші програмні засоби: текстовий редактор, програмне забезпечення для створення презентацій, відео редактор (за необхідністю). В результаті, лише певний пункт навчально-дослідної роботи виконано з використанням інструментарію CoCalc. Крім того, в процесі представлення наукових наробок студенту доведеться окрім презентації демонструвати своїм колегам розроблену модель залучаючи браузер (або відео редактор). Цього можна уникнути, якщо залучити інструментарій CoCalc не лише для виконання навчально-дослідницької частини певної роботи.

Другий шлях використання CoCalc у навчально-дослідницьких проектах, курсових роботах та випускній роботі бакалавра передбачає створення (рис. 3.78) та використання ресурсу типу tex.

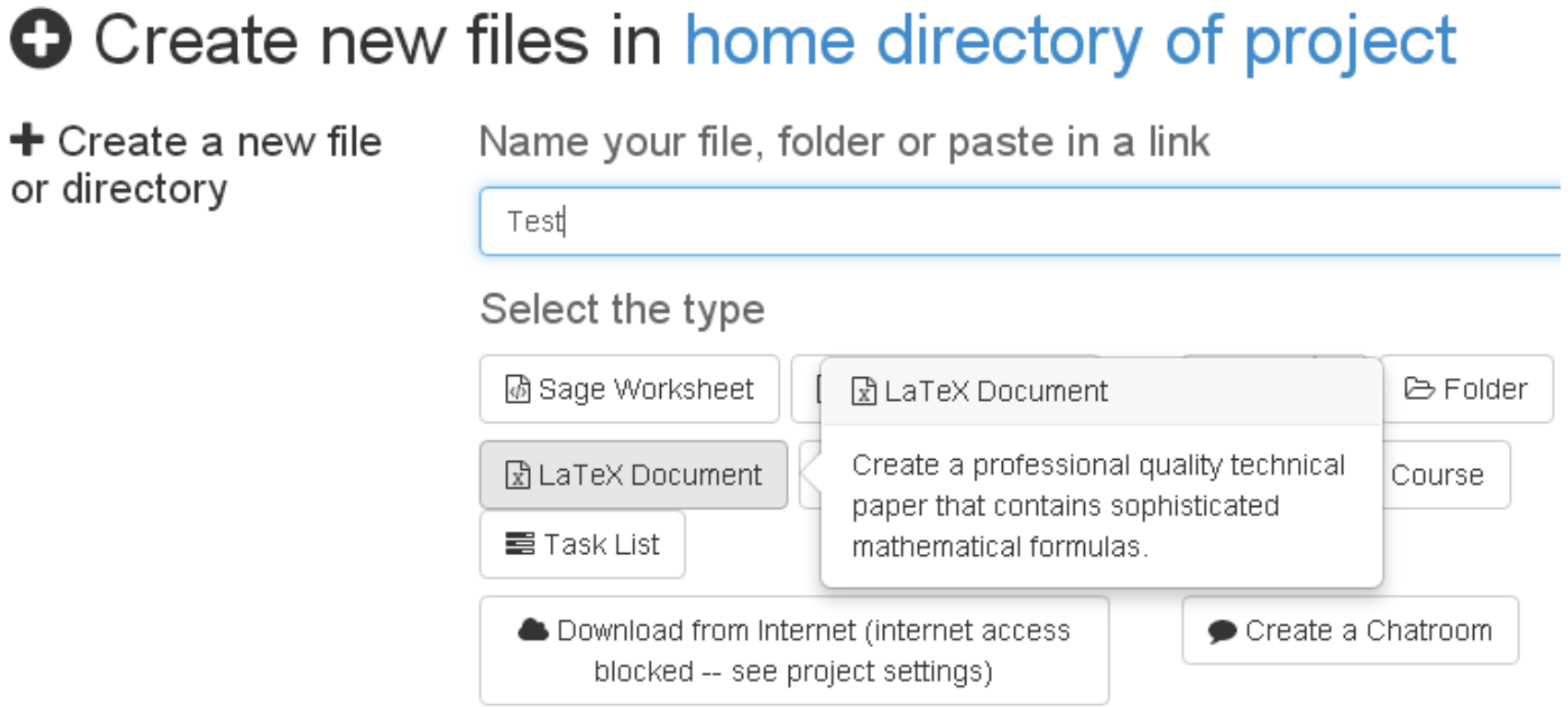

Рис. 3.78. Створення ресурсу типу tex

Використання *редактора LaTeX* (рис. 3.79) дозволить студенту підготувати якісні навчально-дослідницькі матеріали, матеріали математичного змісту (презентації, якісно оформити навчально-дослідницький проект, курсову роботу або випускну роботу бакалавра) [197].

Наступне представлення є більш вдалим та зручним, оскільки студент може вбудовувати обчислення та виконувати побудови (рис. 3.80) одразу ж в текст роботи:

```
\documentclass{article}
\usepackage[a5paper]{geometry}
\usepackage[utf8]{inputenc}
\usepackage[ukrainian]{babel}
\usepackage{sagetex}
\title{Спільне використання Sage та LaTeX}
```

```
\author{М. В. Попель}
\date {13 січня 2015 року}
\begin{document}
\maketitle
Найпростіший спосіб убудування результатів виконання команд
Sage у методичні матеріали, створювані у LaTeX, - використання
тегів sage та sageplot:
а) знаходження похідної:
$(x^3)'=$$\sage{diff(x^3,x)}$
б) побудова графіка:
\sageplot{plot(sin(x),-pi,pi)}
\end{document}
```

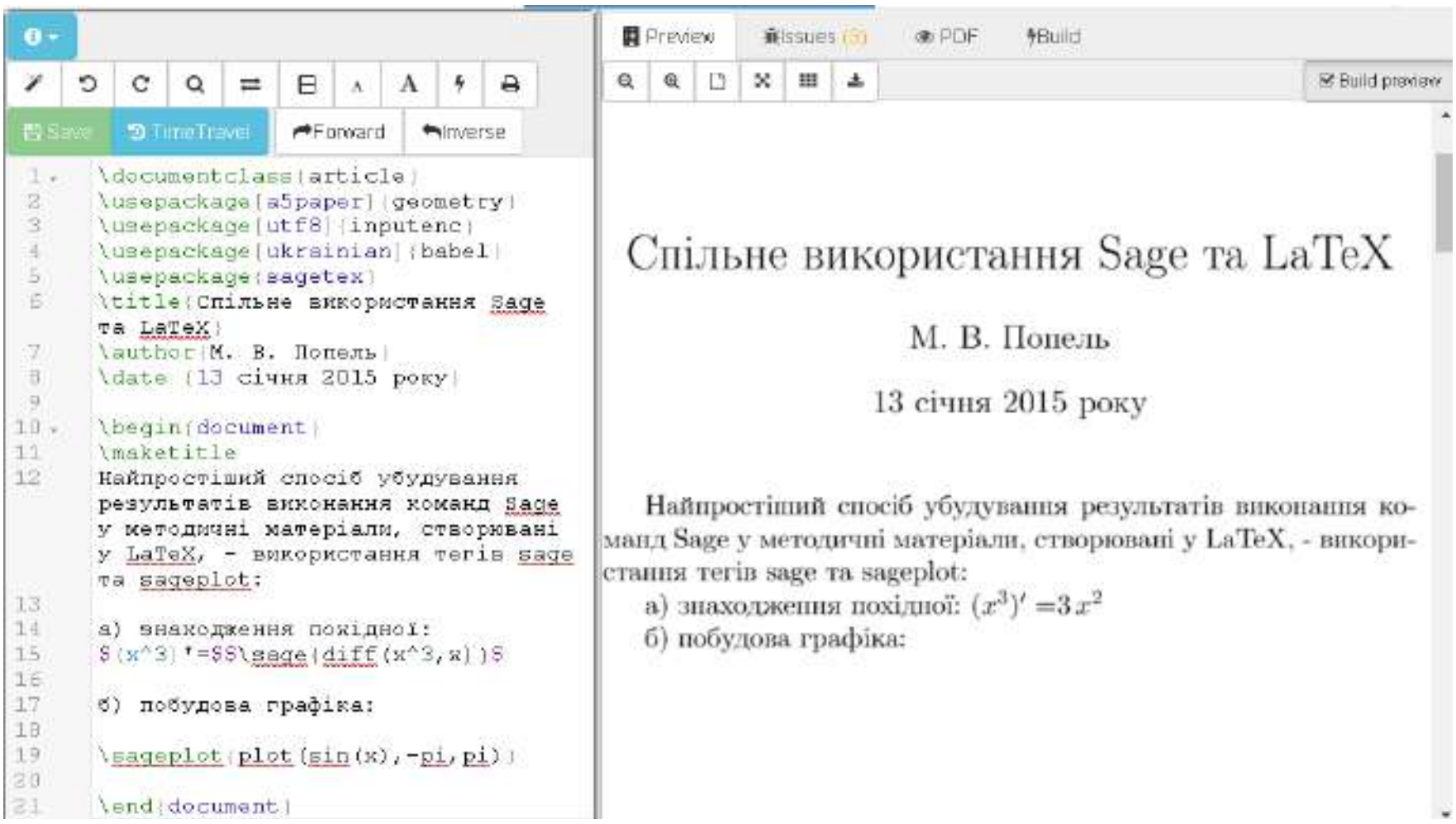

Рис. 3.79. Приклад використання редактора LaTeX

Тобто одночасно відбувається процес оформлення одержаних результатів, виконання обчислень, представлення та подання основних положень проведеного дослідження (з використанням презентації, що розроблена в редакторі LaTeX) та демонстрація створеної моделі. Студенту не треба долучати додаткові програмні засоби для виконання, оформлення чи представлення одержаних результатів, адже повністю вся робота уніфікована в межах одного хмарного сервісу – CoCalc.

Виконуючи курсову роботу чи випускну роботу бакалавра в редакторі LaTeX студент має можливість її роздрукувати, попередньо сформувавши на основі ресурсу типу tex PDF-документ (рис. 3.81).

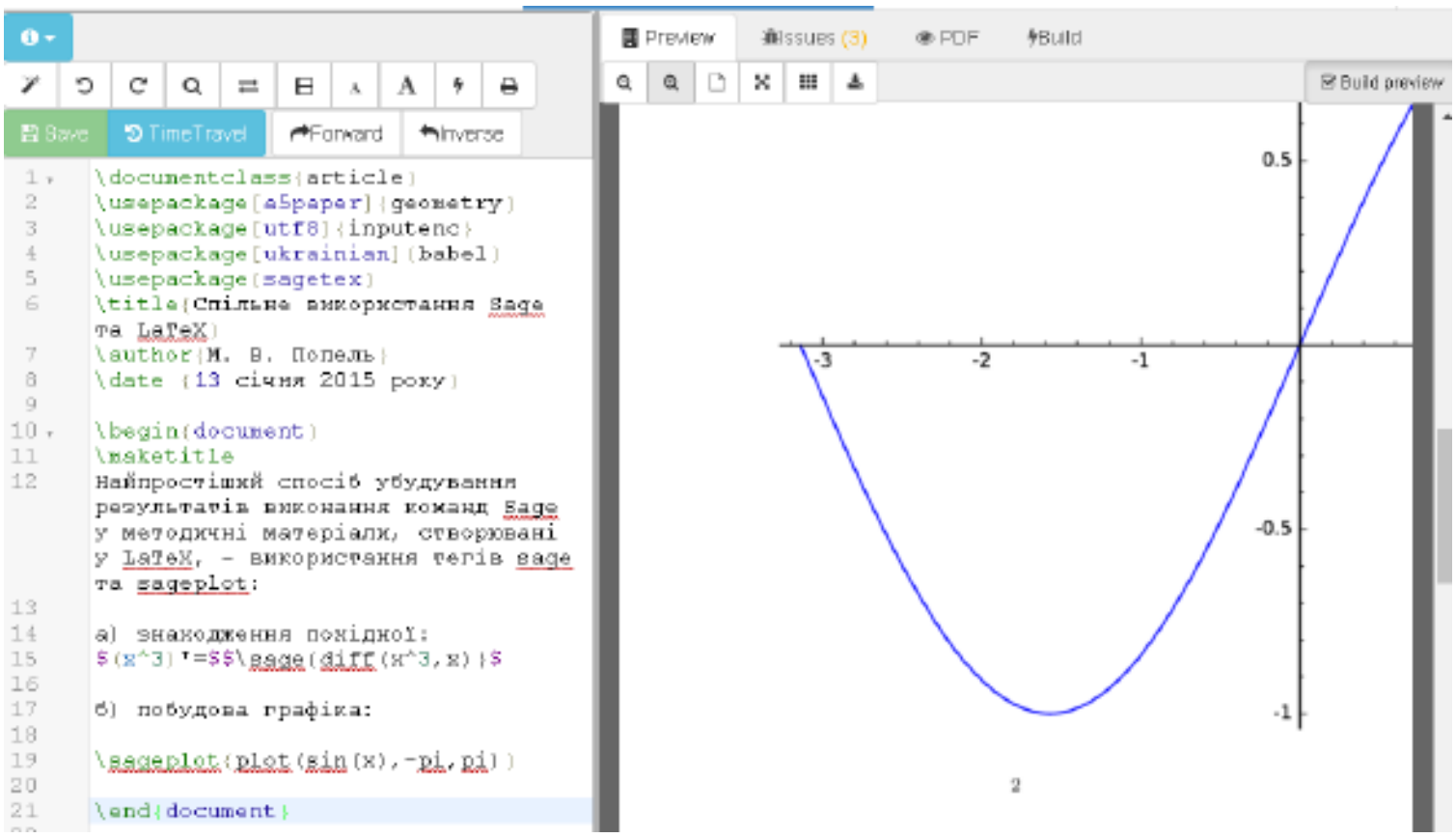

Рис. 3.80. Приклад використання редактора LaTeX для побудови графіка функції

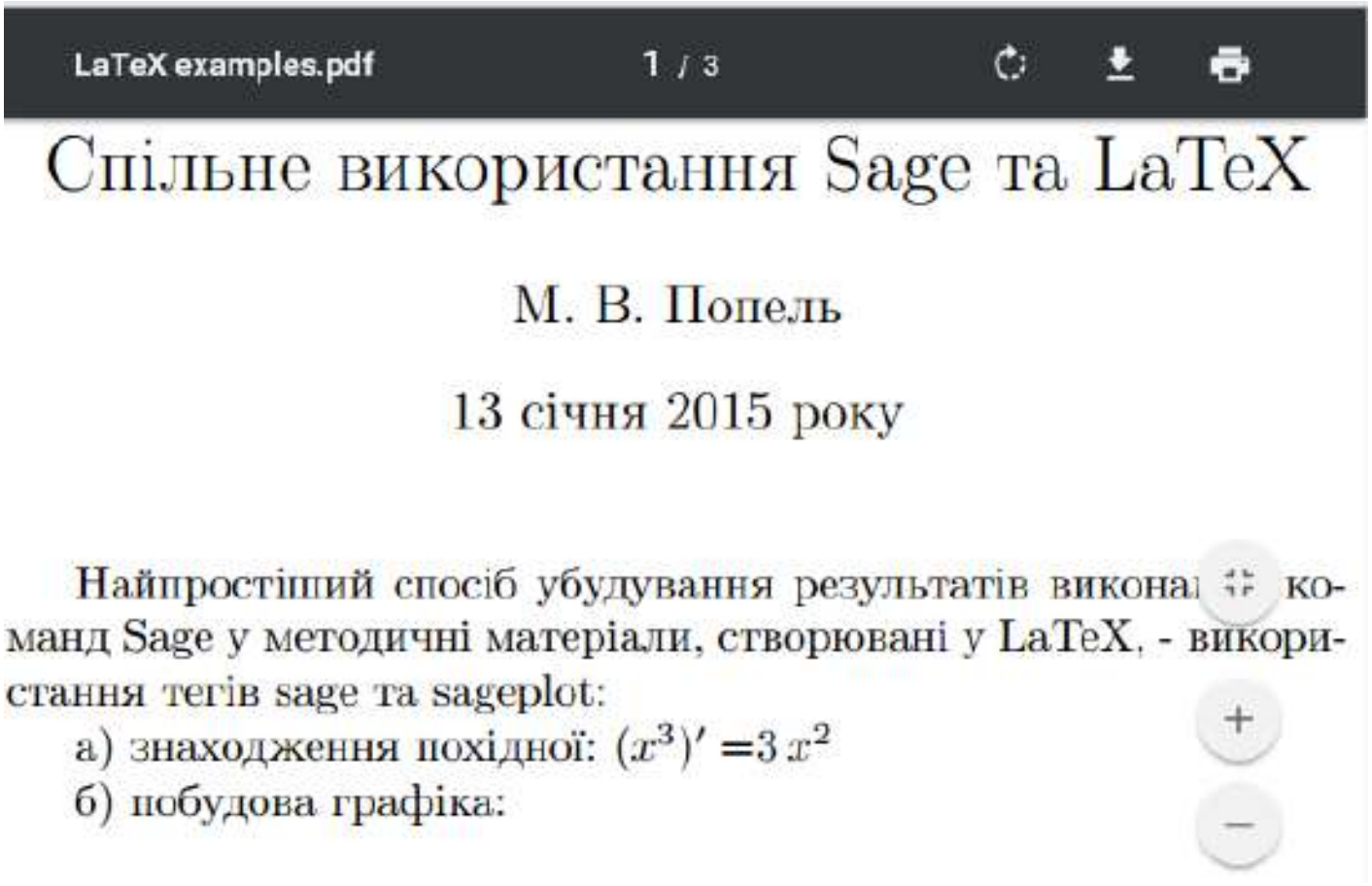

Рис. 3.81. Приклад PDF-документу, створеного на основі ресурсу типу tex

## Висновки до розділу 3

1. Запропонована методика використання CoCalc як засобу формування професійних компетентностей учителя математики передбачає два етапи запровадження CoCalc у навчальний процес:

I. Навчання за програмою факультативу «Використання CoCalc у процесі вивчення математичних дисциплін», як елементу змісту підготовки, перепідготовки, підвищення кваліфікації наукових і науково-педагогічних кадрів.

II. Запровадження системи тренінгів, семінарів, вебінарів, індивідуальних консультацій, що можуть здійснюватися у ході пілотного експериментального дослідження (проекту) з розгортання хмаро орієнтованого інформаційно-освітнього середовища у навчальному закладі.

На I етапі формування професійних компетентностей згідно запропонованої методики до варіативної частини освітньо-професійної програми підготовки бакалавра математики доцільно включити факультативний курс «Використання CoCalc у процесі вивчення математичних дисциплін». Факультатив спрямований на врахування міждисциплінарних зв'язків (математичних та інформатичних дисциплін професійно науково-предметної підготовки). Для роботи у CoCalc студенти повинні володіти наступними вміннями та навичками, які набуваються на пропедевтичному етапі: реєструватись та авторизуватись у системі; створювати навчальні ресурси; працювати з sagews-аркушами (включаючи найпоширеніші режими, знати основи мов: LaTeX, Python, HTML); вбудовувати відео, аудіо, анкети, графічні файли в ресурс «sagews»; спілкуватись у чатах навчальних ресурсів та в ресурсі типу «sage-chat»; працювати з навчальним ресурсом типу «tex»; завантажувати нові ресурси з електронних носіїв. На II етапі формування професійних компетентностей відбувається у межах вивчення нормативних математичних навчальних дисциплін. III етап охоплює виконання навчально-дослідницьких проектів з використанням CoCalc, курсових робіт, випускної роботи бакалавра.

2. Змістовий компонент методики використання хмарного сервісу CoCalc включає предметне навчання цього сервісу, педагогічно обґрунтовані, логічно впорядковані та текстуально зафіксовані в навчальних програмах наукові відомості про матеріал, що доцільно вивчати із застосуванням CoCalc.

3. Форми організації навчання із використанням хмарного сервісу CoCalc: діалогічні форми, індивідуальні та групові консультації, самостійна робота, практична робота, індивідуальна робота, парна робота, фронтально-колективна робота, диференціально-групова робота, колективні та індивідуальні проекти.

4. Провідні методи навчання математичних дисциплін з використанням хмарного сервісу CoCalc: методи організації й здійснення навчальної діяльності (словесні, наочні, практичні репродуктивні й проблемні, самостійної роботи); методи стимулювання й мотивації навчання (методи формування обов'язковості й відповідальності в навчанні: пред'явлення педагогічних вимог); методи контролю й самоконтролю (письмовий контроль, лабораторні й практичні роботи, фронтальний і диференційований контроль, поточний і підсумковий контролі).

5. Засоби формування професійних компетентностей учителя математики, що передбачені із використанням хмарного сервісу:

– робочі аркуші, на яких студенти виконують дії з побудови та дослідження математичних моделей;

– чат-кімнати, що використовуються для обговорення процесу та результатів моделювання;

– засоби підтримки навчальної діяльності (ресурси типу course, tasks);

– засоби для створення математичних текстів (tex) та гіпертекстів (html);

– мобільний доступ до інших засобів підтримки математичної діяльності.

Додатковими засобами є:

– навчальний посібник «Організація навчання математичних дисциплін у CoCalc»;

– web-сайт з методичними рекомендаціями для майбутніх учителів математики з використання хмарного сервісу CoCalc у навчанні різних математичних дисциплін та проекти з використанням CoCalc для підтримки навчання.

# РОЗДІЛ 4
# ОРГАНІЗАЦІЯ ПРОВЕДЕННЯ ТА РЕЗУЛЬТАТИ ЕКСПЕРИМЕНТАЛЬНОЇ РОБОТИ

## 4.1 Основні етапи дослідно-експериментальної роботи

Задля перевірки ефективності методики використання CoCalc як засобу формування професійних компетентностей учителя математики було проведено педагогічний експеримент.

*Педагогічний експеримент* – сукупність методів збору педагогічних фактів в природніх або спеціально створених умовах, які дають можливість перевірити вірогідність сформульованої гіпотези [118, с. 6].

При цьому педагогічний експеримент прийнято поділяти на лабораторний та природний. Останній вид експерименту включає в себе перехресний та лонгітюдний.

Експериментальна робота щодо моделювання та впровадження авторської методики використання CoCalc як засобу формування професійних компетентностей учителя математики проходила як природний, перехресний педагогічний експеримент у чотири етапи:

1) підготовчий (2013-2014 рр.),
2) констатувальний (друга половина 2014 р.),
3) формувальний (2015 р.),
4) контрольний (2016 р.).

Основні завдання педагогічного експерименту:

– виявлення вимог до професійної підготовки учителів математики у ЗВО України;

– дослідження процесу формування професійних компетентностей учителя математики;

– виявлення місця хмарного сервісу CoCalc у системі засобів навчання математичних дисциплін;

– проектування системи професійних компетентностей учителя математики та створення методики використання CoCalc як засобу формування професійних компетентностей учителя математики;

– на основі аналізу та узагальнення результатів формувального експерименту підтвердження гіпотези або її спростування.

Етапи педагогічного експерименту мали наступну послідовність дій:

– підготовчий етап експерименту – вибір теми, дослідження її актуальності та наявних педагогічних досліджень;

– створення програми дослідження – визначення мети, об’єкта та предмета дослідження, постановка задач, формулювання гіпотези, окреслення методів дослідження, збору та аналізу даних та складання індивідуального плану;

– укладання договору з експериментальними майданчиками про проведення експериментального дослідження;

– збір педагогічних фактів, їх кількісне та якісне опрацювання;

– оформлення результатів, висновків і методичних рекомендацій науково-педагогічного дослідження;

– впровадження результатів експерименту у навчальний процес майбутніх вчителів математики у педагогічних ЗВО України.

У процесі проведення експериментальної роботи було використано наступні методи:

– аналіз праць науковців з досліджуваної проблеми;

– дослідження досвіду роботи викладачів ЗВО, вивчення окремих педагогічних досліджень;

– спостереження, бесіда, опитування та анкетування студентів і викладачів;

– аналіз дидактичних можливостей використання хмарного сервісу CoCalc процесі навчання математичних дисциплін майбутніх вчителів математики;

– метод статистичної обробки результатів науково-педагогічного експерименту;

– узагальнення результатів діяльності викладачів та студентів.

Експериментальною базою дослідження на констатувальному етапі проведення експерименту були Криворізький державний педагогічний університет (м. Кривий Ріг) та Національний педагогічний університет імені М. П. Драгоманова (м. Київ), а на формувальному етапі – Криворізький державний педагогічний університет (м. Кривий Ріг). Загальна кількість учасників експерименту склала 16 викладачів та 224 студентів.

В якості мети *першого етапу* експерименту (2013-2014 рр.) було обрати галузь педагогіки, в якій планується експеримент, сформулювати тему дослідження, визначення нерозв'язаних проблем, окреслення з переліку проблем, основні, які можна вирішити в рамках проведення дослідження, окреслити ланку освіти та вікова категорія піддослідних, з'ясувати ступінь наявних наукових досліджень в літературі. Задля досягнення мети була проаналізована психолого-педагогічна, навчально-методична та наукова література; розглянуті проекти стандартів вищої освіти, кваліфікаційні характеристики, та освітньо-професійні програми попередніх років підготовки вчителів математики; вивчався досвід викладачів педагогічних ЗВО щодо використання хмарних сервісів у навчанні математичних дисциплін; вивчалися й аналізувалися шляхи формування професійних компетентностей майбутніх вчителів математики у процесі навчання математичних дисциплін, визначалися напрями та завдання наступних етапів науково-педагогічного експерименту.

В якості мети *другого етапу* експерименту (друга половина 2014 р.) було визначити наявний стан матеріально-технічного забезпечення експериментальних майданчиків, використання хмарних сервісів викладачами в навчальному процесі та визначення ставлення студентів та викладачів до хмарних сервісів, їх готовності використовувати CoCalc в процесі вивчення математичних дисциплін та рівень сформованості професійних компетентностей майбутніх вчителів математики.

Дослідження здійснювалося шляхом аналізу умов для проведення експерименту: визначення стану матеріально-технічного забезпечення, готовності та зацікавленості викладачів кафедри математики та методики її навчання Криворізького державного педагогічного університету та кафедри математики і теорії та методики навчання математики Національного педагогічного університету імені М. П. Драгоманова і студентів фізико-математичних факультетів цих університетів використовувати хмарний сервіс CoCalc. Проведене анкетування мало включати дослідження умов роботи як викладачів так і студентів.

Задля визначення рівня сформованості професійних компетентностей майбутніх вчителів математики були обрані складники предметних, технологічних та професійно-практичних компетентностей системи професійних компетентностей учителя математики: предметно-педагогічні компетентності, інформаційно-технологічні та математичні компетентності. Кожна складова розглядалась окремо та обчислювалась за рівнями: високий, достатній, середній та низький.

Узагальнюючи одержані результати констатувального етапу науково-педагогічного експерименту, можна стверджувати, що:

– переважна більшість студентів і викладачів мають можливість роботи з хмарним сервісом CoCalc як у ЗВО, так і вдома;

– викладачі в більшості випадків не використовують у навчальному процесі хмарні сервіси, за винятком використання хмарних сервісів в якості хмарного сховища;

– викладачі зацікавлені у впровадженні в навчальний процес хмарного сервісу CoCalc, в той час, як студенти не готові до цього;

– студенти на початку експерименту показали низький рівень сформованості інформаційно-технологічних та предметно-педагогічних компетентностей, достатній – математичних компетентностей (рівні сформованості відповідних компетентностей статистично підтверджені);

– студенти та викладачі користуються лише вільно поширюваними програмними засобами (переважно локальними СКМ).

Отже, виявлено протиріччя між:

– достатнім матеріально-технічним забезпеченням ЗВО та відсутністю досвіду викладачів використання хмарного сервісу CoCalc в навчальному процесі майбутніх вчителів математики;

– зацікавленістю та готовністю викладачів використовувати хмарний сервіс CoCalc в навчальній діяльності та відсутністю методик використання хмарних сервісів для майбутніх вчителів математики;

– необхідністю формування професійних компетентностей, та не достатньою увагою, що приділяється формуванню предметних та технологічних компетентностей.

Аналізуючи результати констатувального етапу експерименту стає зрозумілим, що усунення зазначених протиріч можливе за рахунок використання хмарних сервісів, як засобів навчання математичних дисциплін, зокрема хмарного сервісу CoCalc.

Тобто, є проблема розробки методики використання CoCalc як засобу формування професійних компетентностей учителя математики. В процесі проведення констатувального етапу науково-педагогічного експерименту було реалізовано І етап методики використання CoCalc як засобу формування професійних компетентностей учителя математики, який полягав у проведенні факультативного курсу «Використання CoCalc у навчанні математичних дисциплін».

Мета *третього етапу* дослідження (2015 р.) полягала в перевірці ефективності застосування методики використання CoCalc як засобу формування професійних компетентностей учителя математики та порівнянні рівнів сформованості окремих складників системи професійних компетентностей учителя математики експериментальних і контрольних груп. Досягненню даної мети сприяла реалізація ІІ етапу методики використання CoCalc як засобу формування професійних компетентностей учителя математики, а саме: використання CoCalc у навчанні математичних дисциплін. У процесі проведення формувального етапу експерименту було виконано два проміжні зрізи, які дозволили зробити висновок стосовно сформованості наукових компетентностей та компетентності навчання математики у майбутніх вчителів математики. Окрім отого задля порівняння з рівнем сформованості професійних компетентностей на констатутвальному етапі проведення експерименту було виконано, по завершенню формувального етапу, повторне вимірювання рівнів предметно-педагогічних компетентностей, інформаційно-технологічних та математичних компетентностей.

В якості мети *четвертого етапу* (2016 р.) було виявлення результатів формувального впливу. Задля цього було перевірено статистично однорідність контрольної та експериментальної груп, статистично опрацьовані результати та проаналізовано рівні сформованості окремих складників професійних компетентностей майбутніх вчителів математики. В рамках даного етапу було реалізовано останній етап авторської методики та проаналізовано результати державного екзамену за спеціальністю груп МІ-11 та МІ-12.

### 4.2 Статистичне опрацювання та аналіз результатів констатувального етапу педагогічного експерименту

Метою констатувального етапу експерименту було визначити наявний стан матеріально-технічного забезпечення експериментальних майданчиків, використання хмарних сервісів викладачами в навчальному процесі та визначення ставлення студентів та викладачів до хмарних сервісів, їх готовності використовувати CoCalc в процесі вивчення математичних дисциплін та рівень сформованості професійних компетентностей майбутніх вчителів математики.

У ряді випадків констатувальний етап експерименту ефективно проводити методом анкетування. Для досліджуваного процесу складають ретельно продуману методику. Основні дані збирають методом опитування за попередньо складеною анкетою. Цей метод дозволяє зібрати дуже велику кількість даних спостережень, або вимірювань по досліджуваному питанню. Однак до результатів анкетних даних потрібно ставитися з особливою ретельністю, оскільки вони не завжди містять достатньо достовірні відомості [132, с. 27].

На етапі вибору експериментального майданчика нами було розроблено анкету «Експериментальний майданчик. Матеріальна база» в двох варіантах: для студентів та викладачів. Анкета складається з 9 закритих дихотомічних питань та одного відкритого, короткого. На меті було: визначити рівень матеріально-технічного забезпечення експериментальних баз дослідження. Розглянемо спочатку анкету, складену для викладачів. Респонденту спочатку потрібно вказати назву навчального закладу та кафедру, на якій він працює. Зрозуміло, що згідно обраної теми дослідження, в першу чергу у центр уваги потрапляють педагогічні ЗВО, у яких готують учителів математики. Крім того, викладацький склад кафедри має читати математичні дисципліни. Питання в більшій мірі спрямовані на визначення рівня комплектації матеріальної бази майбутнього експериментального майданчика.

Для робити з CoCalc необхідно, щоб робоче місце викладача було обладнане комп’ютером (ноутбуком, нетбуком, планшетом), або він мав би власний пристрій. Для роботи на практичному занятті достатньо буде використання смартфону, але для підготовки до лекційного заняття, попередньої роботи з моделями, їх застосуванням під час проведення заняття потрібно забезпечити викладача комп’ютером (ноутбуком, нетбуком, планшетом) з доступом до мережі Інтернет. Якщо це буде підключення за допомогою Wi-Fi, то параметри швидкості Інтернету та технічних характеристик роутера також відіграють важливу роль, бо одночасне підключення цілої групи студентів до мережі Інтернет сповільнить його роботу. Експериментальний майданчик має бути забезпечений достатньою кількістю комп'ютерних аудиторій, щоб мати можливість частину практичних занять проводити в них.

Останнє питання анкети: «Чи змогли б Ви проводити практичні заняття, лабораторні роботи в комп'ютерних аудиторіях?» спрямоване на визначення готовності викладача використовувати на практичних заняттях комп'ютери, працювати в комп'ютерних аудиторіях. Хоча цим питанням, звичайно, не можна охарактеризувати особисте ставлення до використання CoCalc у навчальному процесі. В даній анкеті не ставилася мета визначити проблеми, які можуть виникнути підчас використання CoCalc. Також не перевірявся рівень обізнаності викладачів стосовно доступу до мережі Інтернет, кількості комп'ютерної техніки. Основне завдання анкети дослідити матеріальну базу експериментального майданчика, наявні умови роботи, можливі труднощі, взагалі можливість проведення експерименту в даному ЗВО.

Подібна анкета була створена і для студентів. Частина питань дублюються з анкети, складеної для викладачів. Анкета складається також з 10 питань. На меті було визначити умови навчання студентів вдома та у ЗВО з використанням комп'ютерів (ноутбуків, нетбуків, планшетів та ін.). Також за допомогою питань визначається можливість використання мережі Інтернет як у ЗВО, так і під час підготовки до наступного заняття. Головне, щоб експериментальні групи студентів у своїй більшості мали вільний доступ до мережі Інтернет, щоб у них були всі необхідні умови для роботи з CoCalc. Зрозуміло, що академічна група студентів, в якій нараховується 50 % і більше, не забезпечених комп'ютерною технікою (чи хоча б смартфоном) або ж які не мають постійного вільного доступу до мережі Інтернет, не може брати участь в проведенні експерименту. Питання на зразок: «У Вашому навчальному закладі є мережа Wi-Fi?», які повторюються в обох анкетах, спрямовані на визначення обізнаності студентів стосовно підключення до наявної мережі Інтернет. Зрозуміло, що студенти, які не знали паролю до того ж Wi-Fi, чи можливості підключення до мережі Інтернет їх освітньої установи, будуть цікавитися з цього приводу. Останнє, питання анкети: «Проведення практичного заняття в комп'ютерній аудиторії не відволікатиме Вас від теми заняття?», спрямоване на визначення готовності студентів використовувати на занятті комп'ютерні технології. На нашу думку, дане питання в жодній мірі не може претендувати на визначення готовності студентів взагалі використовувати CoCalc.

Загальна кількість вибірки склала 112 респондентів.

Задля зручності усі питання можна віднести до двох категорій: власне матеріальне забезпечення та матеріальне забезпечення ЗВО. Лише останнє питання показує готовність використовувати на заняттях комп'ютери.

У першу чергу нас цікавило достатня кількість комп'ютерних аудиторій у ЗВО, оснащеність кожної аудиторії комп'ютерами та вільний доступ до мережі Інтернет. По-перше, це одна з основних вимог проведення експерименту, а по-друге, треба забезпечити усіх студентів можливістю виконання індивідуальних та самостійних завдань відведених на самопідготовку. Адже вдома не в усіх студентів буде така можливість. Хоча, як ми бачимо з діаграм (рис. 4.1 та рис. 4.2) власна матеріальна база студентів навіть краща, ніж та, що надає ЗВО. Лише у деяких студентів відсутні смартфони чи не має доступу до мобільного Інтернет. Треба звернути увагу, що у всіх студентів є хоча б один з пристроїв, за допомогою якого можна працювати з CoCalc.

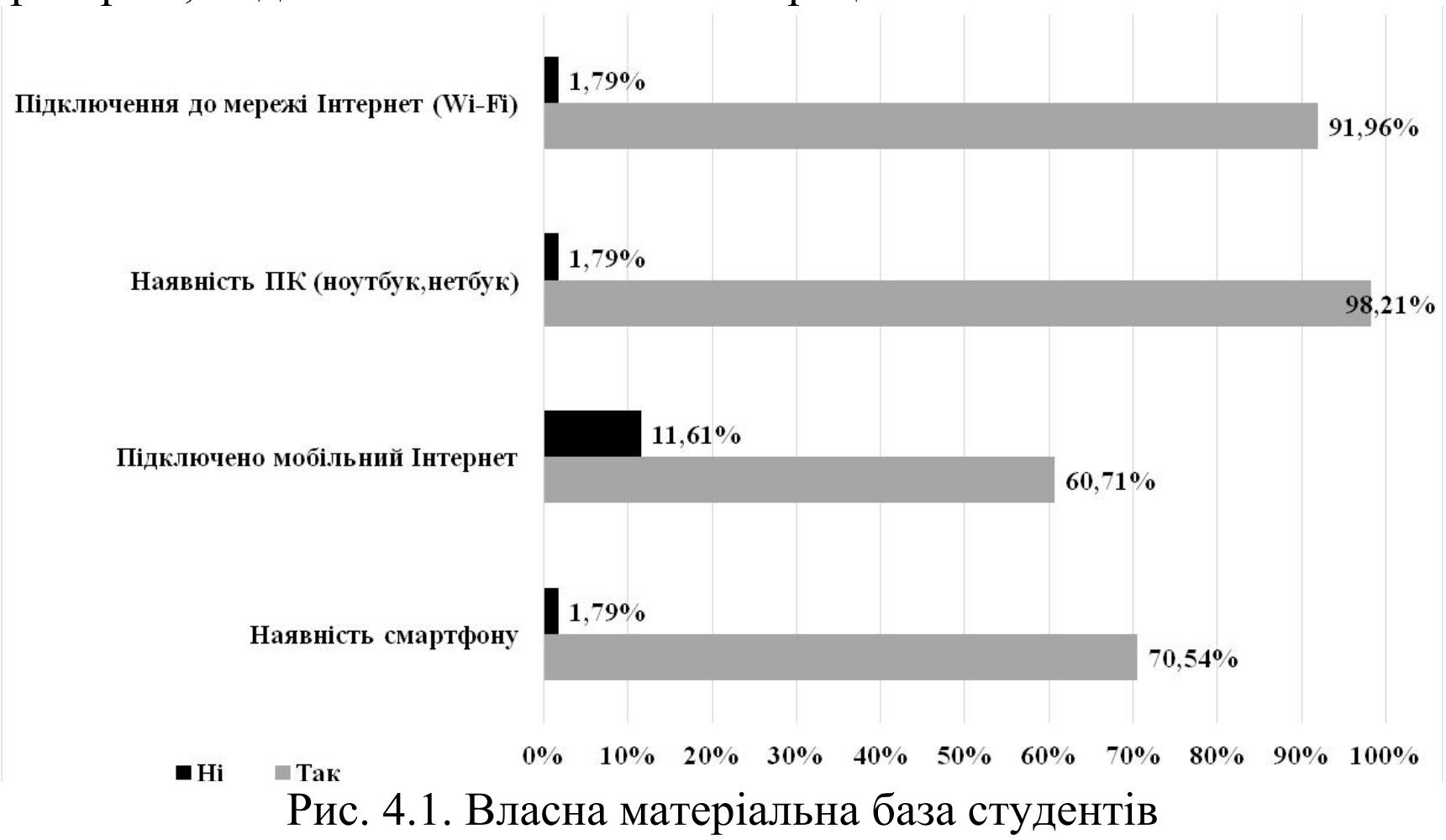

Рис. 4.1. Власна матеріальна база студентів

Під час проведення анкетування зіткнулись з певними проблемами. Більшість студентів не знають, скільки в їх корпусі знаходиться комп'ютерних аудиторій. І може скластись враження, що ЗВО в недостатній мірі обладнаний задля проведення експерименту.

Що ж стосується готовності використовувати на заняттях комп'ютер, то більше половини студентів (52 %) досліджуваних груп вважають, що використання комп'ютеру не відволікатиме їх від теми заняття.

При опитуванні викладачів (16 респондентів) анкета відрізнялась питаннями, що стосувались робочого місця викладача (за рахунок чого забезпечувалась можливість підготовки до заняття) та підключенням пристрою до мережі Інтернет (рис. 4.3, рис. 4.4). Відповіді, що віднесені до категорії «Власні», можна розглядати як ті, що представляють матеріальне забезпечення ЗВО (в дужках вказана позначка «роб.»), але закріплені за робочим місцем окремого викладача, та власні пристрої (в дужках вказана позначка «вл.»).

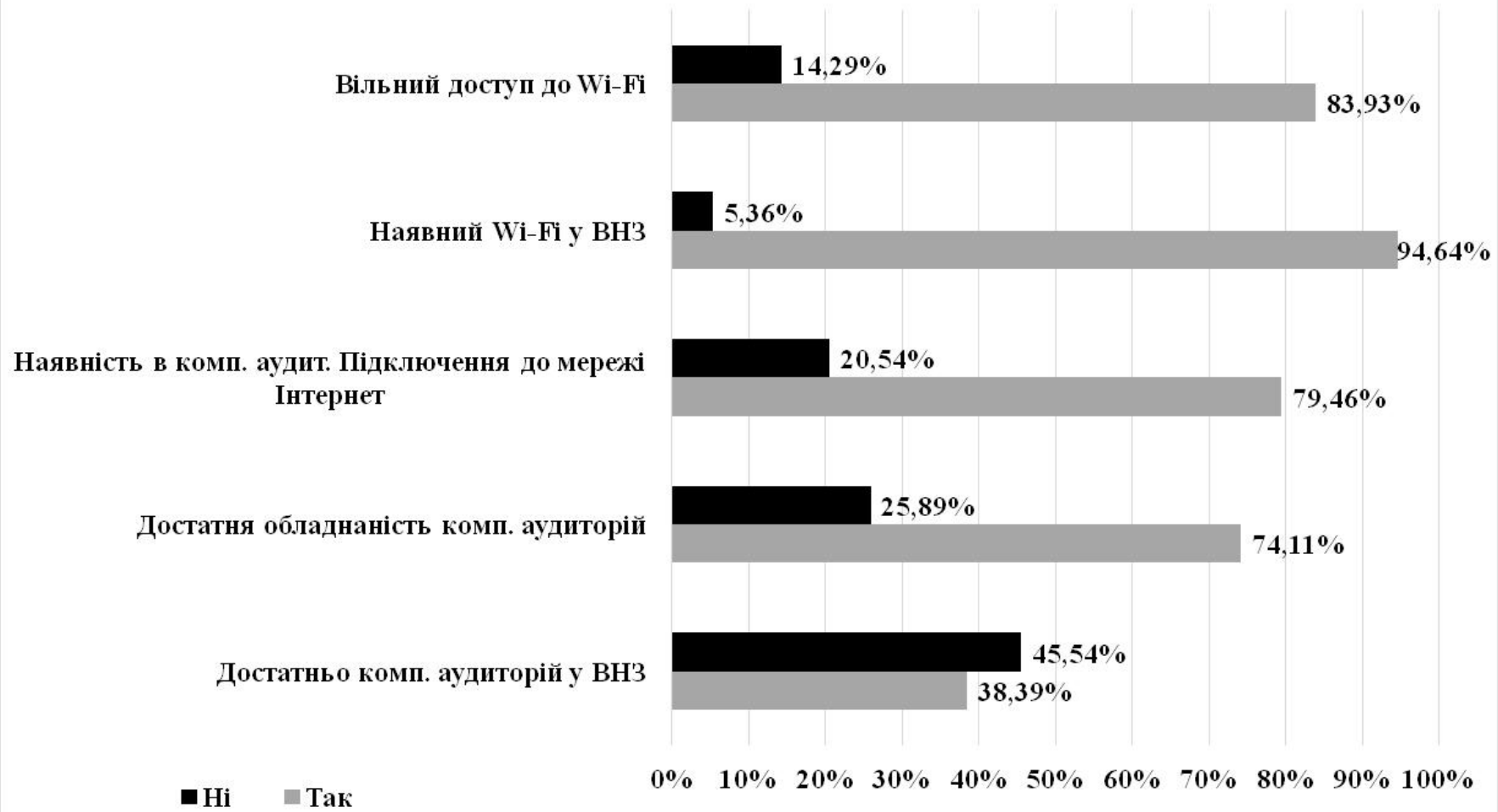

Рис. 4.2. Експериментальний майданчик: матеріальне забезпечення ЗВО

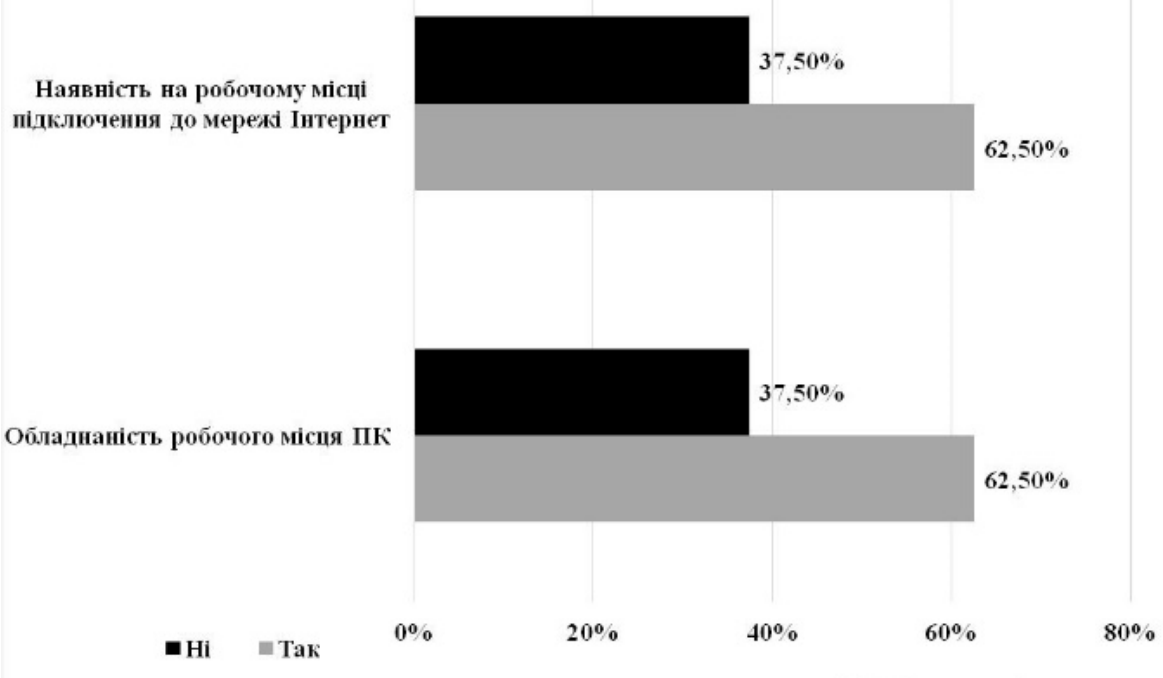

Рис. 4.3. Матеріальна база робочого місця викладача ЗВО

Як показують результати анкетування, не у всіх викладачів робоче місце обладнане персональним комп'ютером (рис. 4.3). Проте, майже у всіх є можливість працювати з хмарним сервісом вдома (наявний пристрій та підключення мережі Інтернет, рис. 4.5).

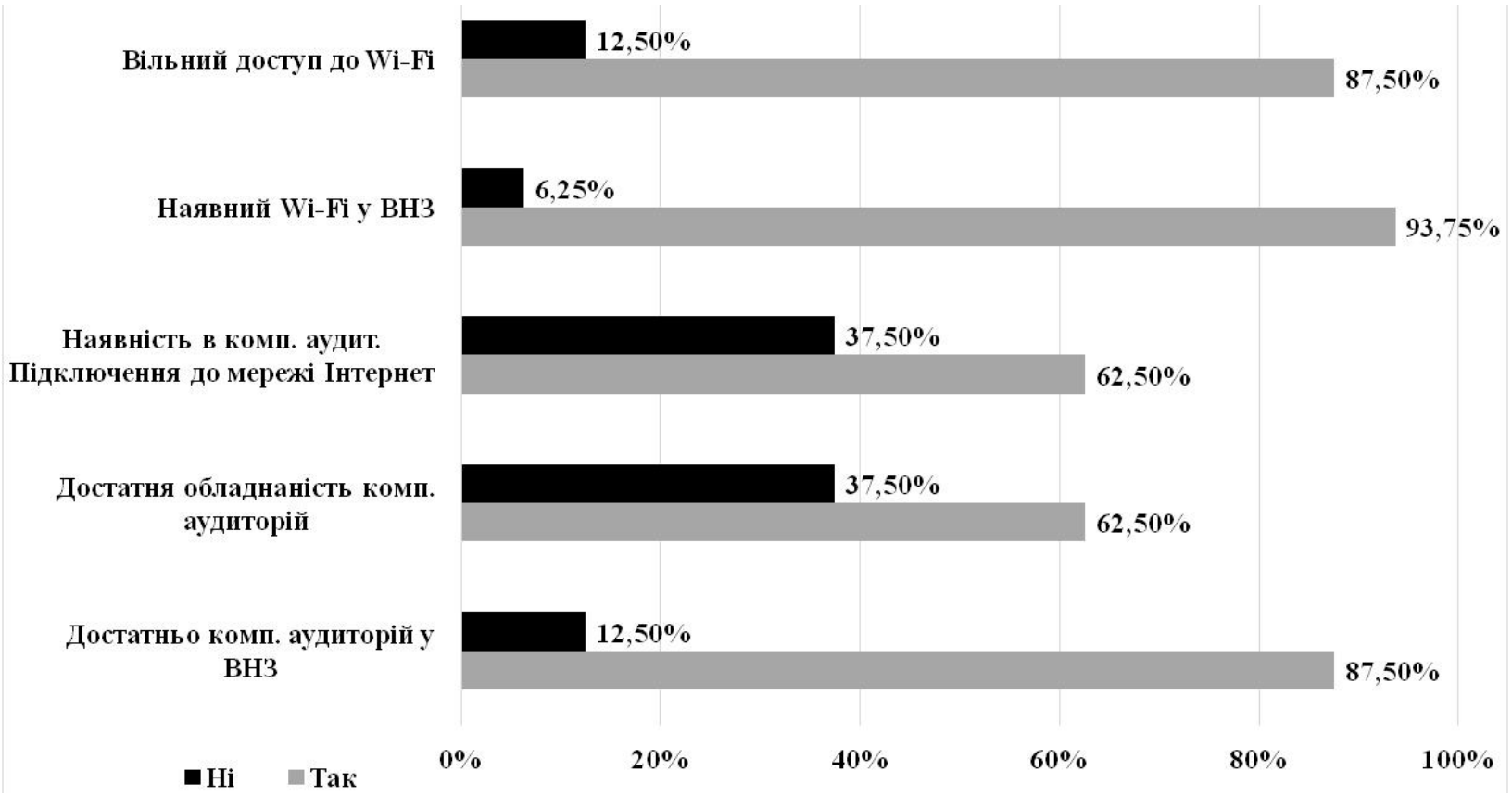

Рис. 4.4. Експериментальний майданчик: матеріальне забезпечення ЗВО (відповіді викладачів)

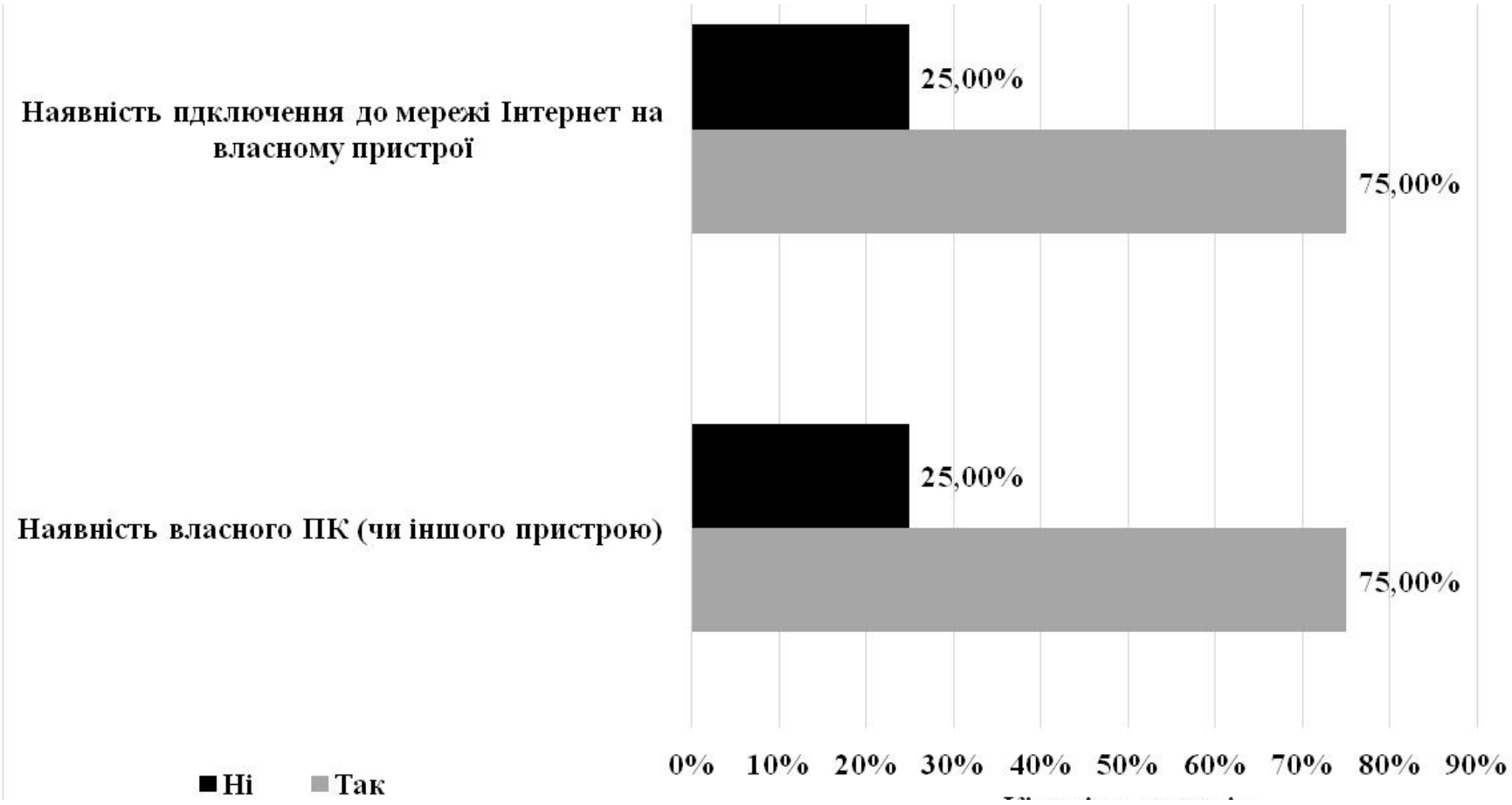

Рис. 4.5. Власна матеріальна база викладачів

Однією із загальних вимог щодо проведення педагогічного експерименту є наявність бажання та готовності (готовність викладачів складає 62,5 %) до впровадження запропонованої методики, власне до використання хмарного сервісу (зокрема CoCalc) педагогічним колективом кафедри математики та методики її навчання, кафедри математики і теорії та методики навчання математики. Задля визначення стану обізнаності викладачів математичних дисциплін на обраних експериментальних майданчиках було проведено бесіди з викладачами та анкетування. Анкета складалась з питань, які містили закриті відповіді, переважно дихотомічного типу. Два питання – відкритого типу задля визначення з яким програмними забезпеченням працювали викладачі в підтримку математичних дисциплін. Виявилось, що більша частина респондентів знайомі з СКМ, хмарними сервісами та висловлюють намір використовувати в навчальному процесі хмарні сервіси (рис. 4.6).

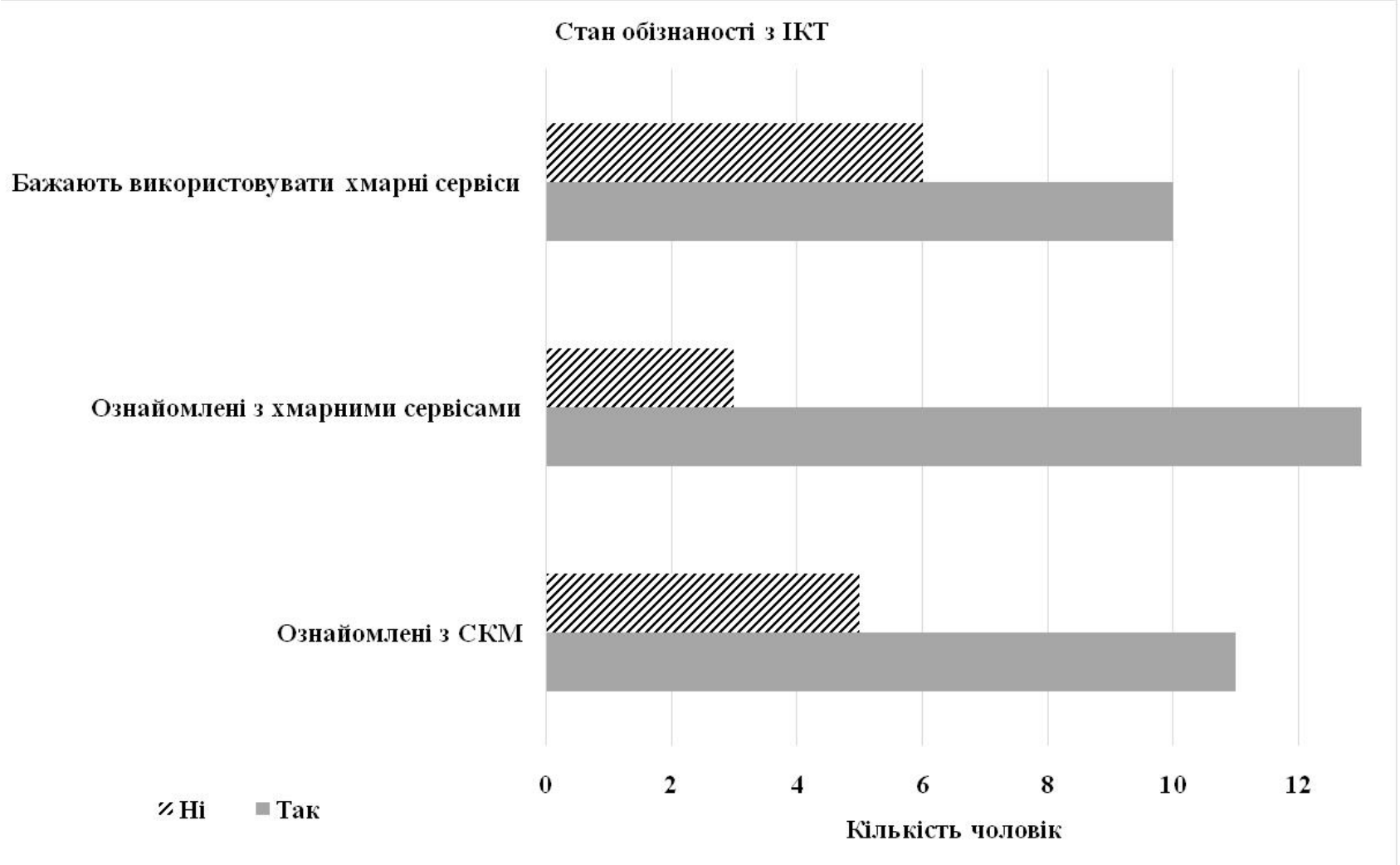

Рис. 4.6. Стан обізнаності викладачів з ІКТ

Проте, далеко не всі респонденти мають досвід використання СКМ та хмарних сервісів (лише 7 осіб відповідно). Тому, лише деякі з них радять студентам в процесі виконання громіздких обчислень користуватись СКМ чи хмарними сервісами (рис. 4.7). Крім того, в процесі бесіди було виявлено, що викладачі, які використовують той чи інший хмарний сервіс в навчальному процесі в повній мірі залучають весь можливий інструментарій за браком методичних розробок використання хмарного сервісу як засобу вивчення математичних дисциплін.

Більшість викладачів використовують хмарний сервіс лише як хмарне сховище електронних навчальних ресурсів. Використання хмарного сервісу CoCalc як засобу навчання математичних дисциплін викладачами кафедр взагалі не розглядалось. Після проведення серії навчальних семінарів та тренінгів для викладачів було виконано додаткове опитування, метою якого було виявлення перспектив використання хмарних сервісів в процесі вивчення математичних дисциплін та визначення форм організації навчального процесу математичних дисциплін, які в першу чергу потребують активної підтримки CoCalc. Викладачі вбачають перспективи використання хмарних сервісів під час вивчення математичних дисциплін в наступному (рис. 4.8): індивідуалізація навчання; економія часу викладача; різноманітність навчання.

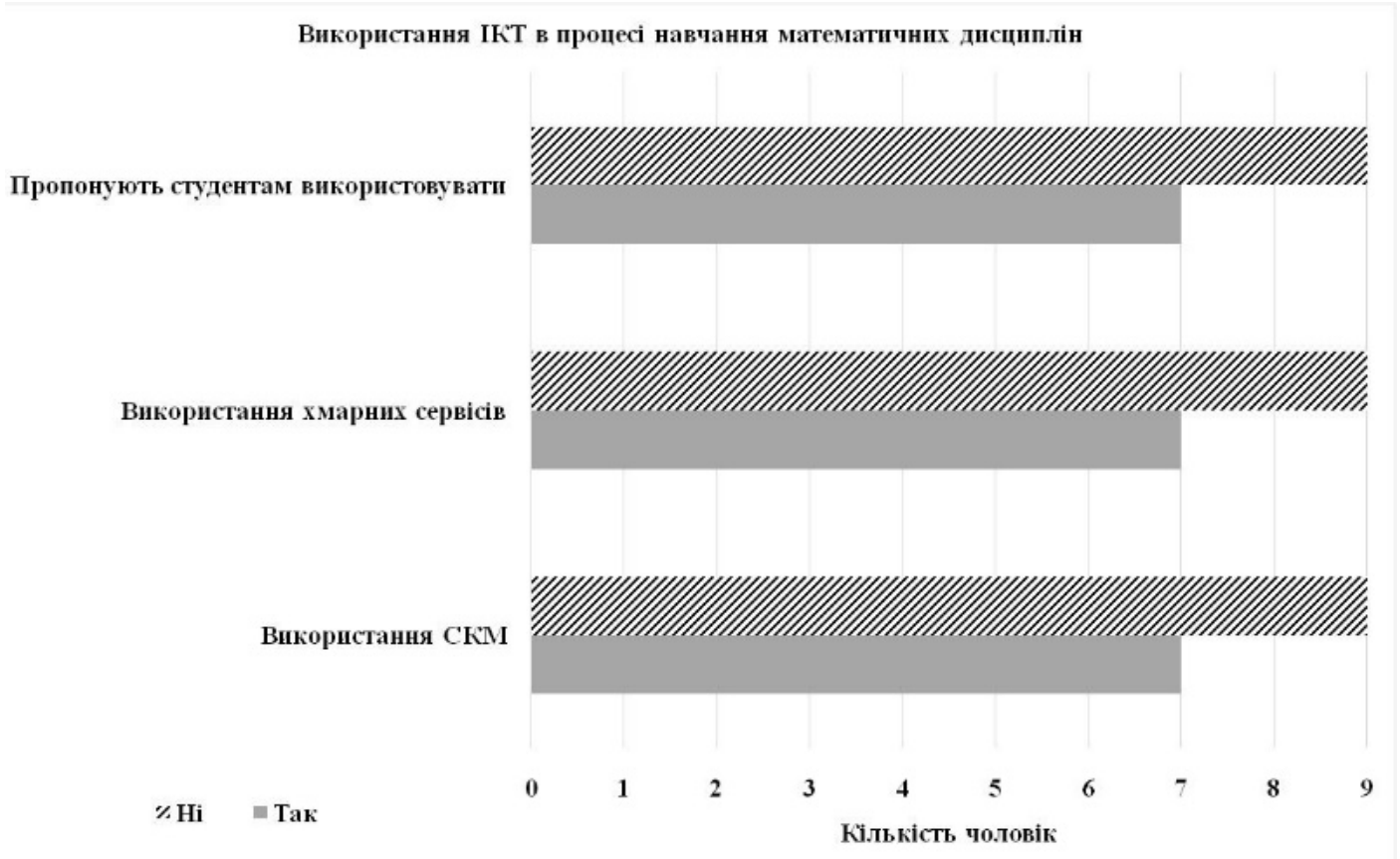

Рис. 4.7. Використання ІКТ викладачами в процесі навчання математичних дисциплін

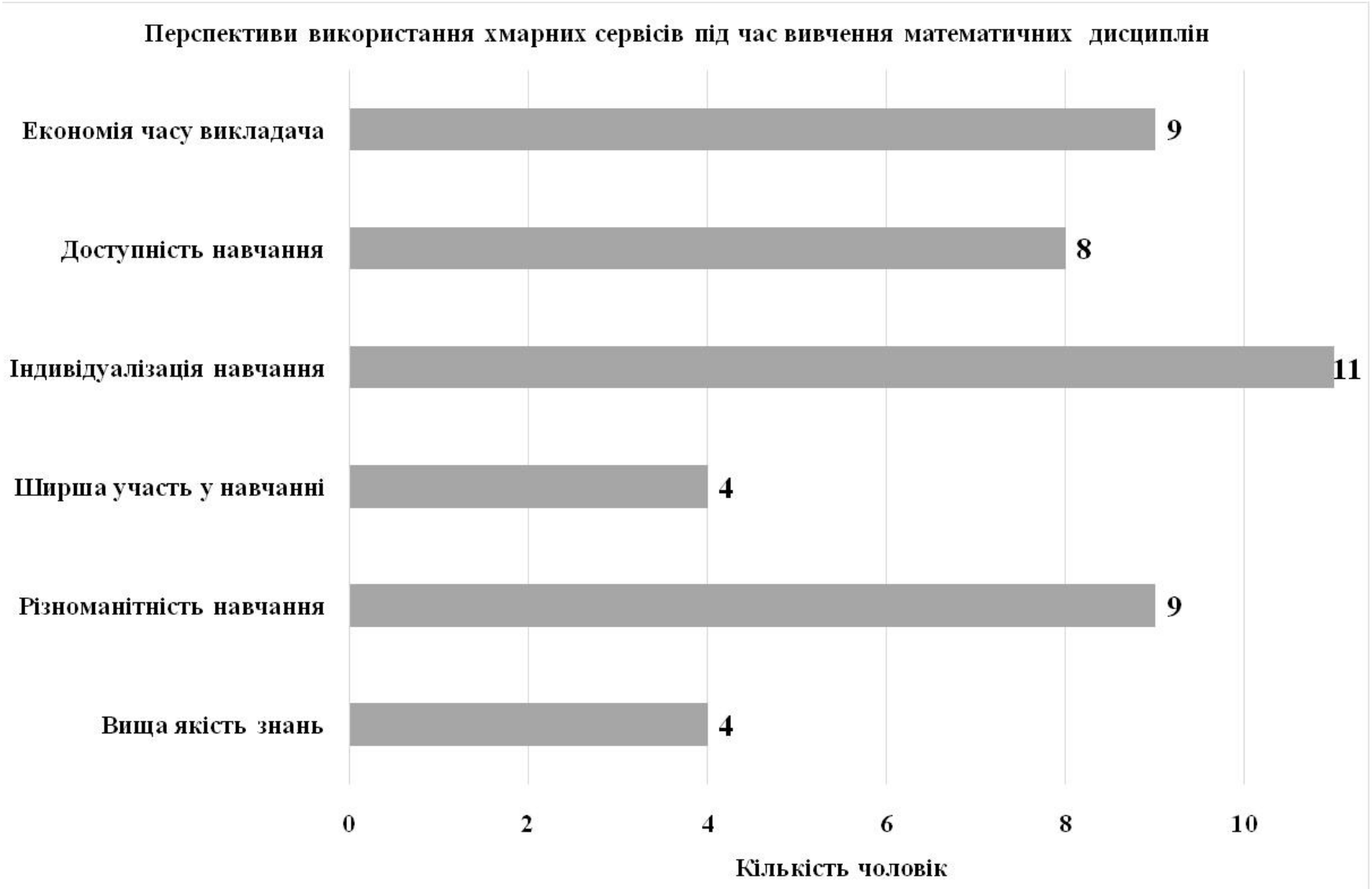

Рис. 4.8. Перспективи використання хмарних сервісів під час вивчення математичних дисциплін

Щодо форм організації навчального процесу, то на думку більшості респондентів, в першу чергу потребують активної підтримки (рис. 4.9): лекції та практичні заняття. В процесі проведення консультації викладачі просто не уявляють яким чином можна організувати роботу із залученням інструментарію хмарного сервіса CoCalc, що виступає додатковим підтвердженням запровадження методики використання вказаного хмарного сервісу.

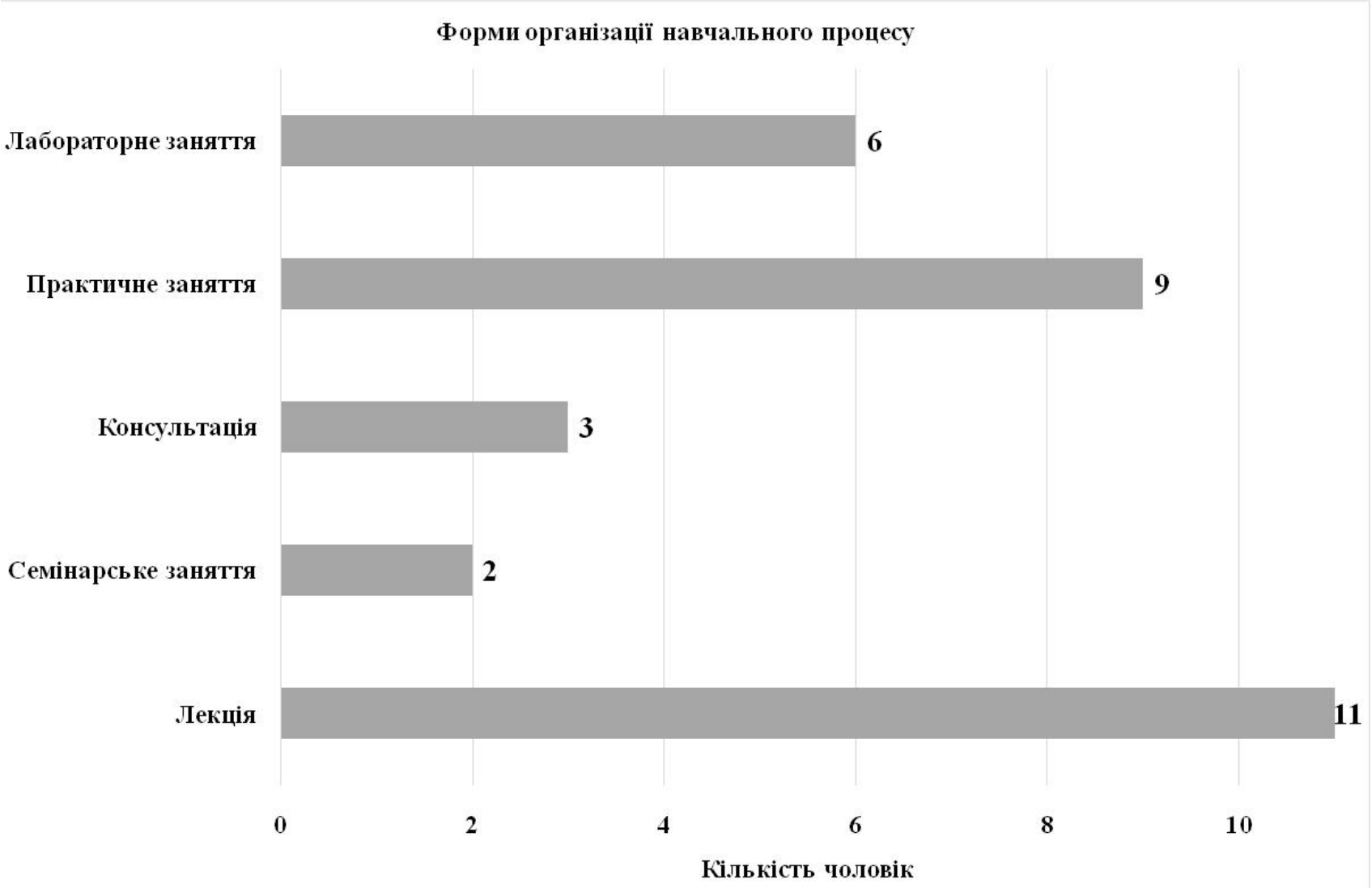

Рис. 4.9. Форми організації навчального процесу, що потребують використання хмарних сервісів

Контрольні та експериментальні групи формувалися наступним чином:

– до складу *контрольних груп* (КГ) були включені студенти груп МІ-11, МІ-13-1, МІ-14-1 (друга половина 2014 р.) та МІ-15 (друга половина 2015 р.) фізико-математичного факультету Криворізького державного педагогічного університету (всього 61 студент). Студенти контрольних груп навчалися за традиційною методикою формування професійних компетентностей учителя математики (без використання хмарного сервісу CoCalc);

– до складу *експериментальних груп* (ЕГ) були включені студенти груп МІ-12, МІ-13-2, МІ-14-2 (друга половина 2014 р.) та МІ-15 (друга половина 2015 р.) фізико-математичного факультету Криворізького державного педагогічного університету (всього 61 студент). Студенти експериментальних груп навчалися за авторською методикою використання CoCalc як засобу формування професійних компетентностей учителя математики (з використанням хмарного сервісу CoCalc).

Склад контрольних та експериментальних груп наведено в таблиці 4.1.

*Таблиця 4.1*

**Склад контрольних та експериментальних груп**

| **Групи** | **Назва групи та кількість студентів за навчальними роками** | | **Разом** |
|---|---|---|---|
| | **друга половина 2014–2016** | **друга половина 2015–2016** | |
| Контрольні | МІ-11 (26)<br>МІ-13-1 (11)<br>МІ-14-1 (16) | МІ-15 / перша частина (8) | 61 |
| Експериментальні | МІ-12 (26)<br>МІ-13-2 (11)<br>МІ-14-2 (16) | МІ-15 / друга частина (8) | 61 |
| Разом | 106 | 16 | 122 |

Була спроба урівняти фактори, що впливають на процес навчання: кількісний склад студентів у експериментальних та контрольних групах істотно не відрізнявся; заняття проводилися одними і тими ж викладачами; у контрольних групах застосовувалися СКМ у підтримку вивчення математичних дисциплін.

З метою з’ясування *стану сформованості професійних компетентностей* учителів інформатики протягом другої половини 2014 р. виконувалися констатувальні зрізи наступних складників предметних, технологічних та професійно-практичних компетентностей системи професійних компетентностей учителя математики: предметно-педагогічні компетентності (рис. 4.12), інформаційно-технологічні (рис. 4.10) та математичні компетентності (рис. 4.11). Кожна складова розглядалась окремо та обчислювалась за рівнями: високий, достатній, середній та низький.

Аналізуючи констатувальні зрізи, можна зробити висновок, про те, що студенти мають низький рівень сформованості інформаційно-технологічних та предметно-педагогічних компетентностей, достатній – математичних компетентностей.

На основі даних, наведених на рис. 4.10, спочатку перевіримо достовірність гіпотези про відсутність, з статистичної точки зору, відмінностей між рівнями сформованості інформаційно-технологічних компетентностей експериментальних і контрольних груп за результатами констатувального зрізу. Для цього скористаємося критерієм Фішера

[165].

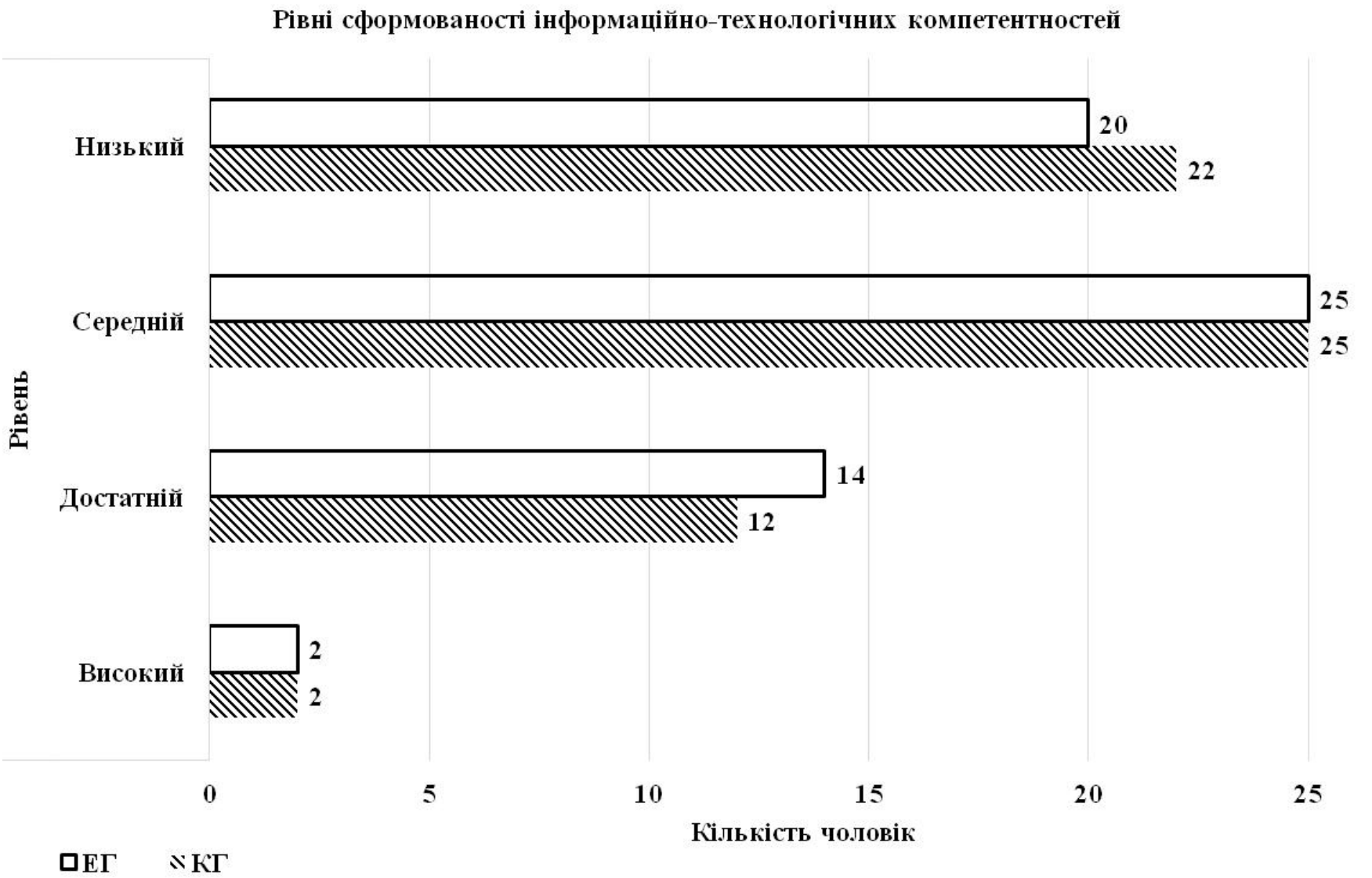

Рис. 4.10. Порівняння розподілів експериментальних та контрольних груп студентів за рівнями сформованості інформаційно-технологічних компетентностей на констатувальному етапі експерименту

Сформулюємо гіпотези:

$H_0$: Частка студентів, які за результатами дослідження рівнів сформованості інформаційно-технологічних компетентностей показали високий та достатній рівень не більше, ніж у контрольних групах;

$H_1$: Частка студентів, які за результатами дослідження рівнів сформованості інформаційно-технологічних компетентностей показали високий та достатній рівень більше, ніж у контрольних групах.

Побудуємо таблицю, яка фактично є таблицю емпіричних частот за двома значеннями ознаки: якщо рівні сформованості інформаційно-технологічних компетентностей зазначені високий та достатній, то «ефект має місце», у протилежному випадку – «ефект відсутній» (табл. 4.2). При цьому в обрахунках використовуються лише частки, що відповідають спостереженням, для яких ефект має місце.

Експериментальні дані повністю задовольняють обмеження, що накладаються кутовим перетворенням Фішера:

а) жодна з часток, що порівнюються, не дорівнює нулю;

б) кількість спостережень у обох вибірках більше 5, що дозволяє будь-які співставлення.

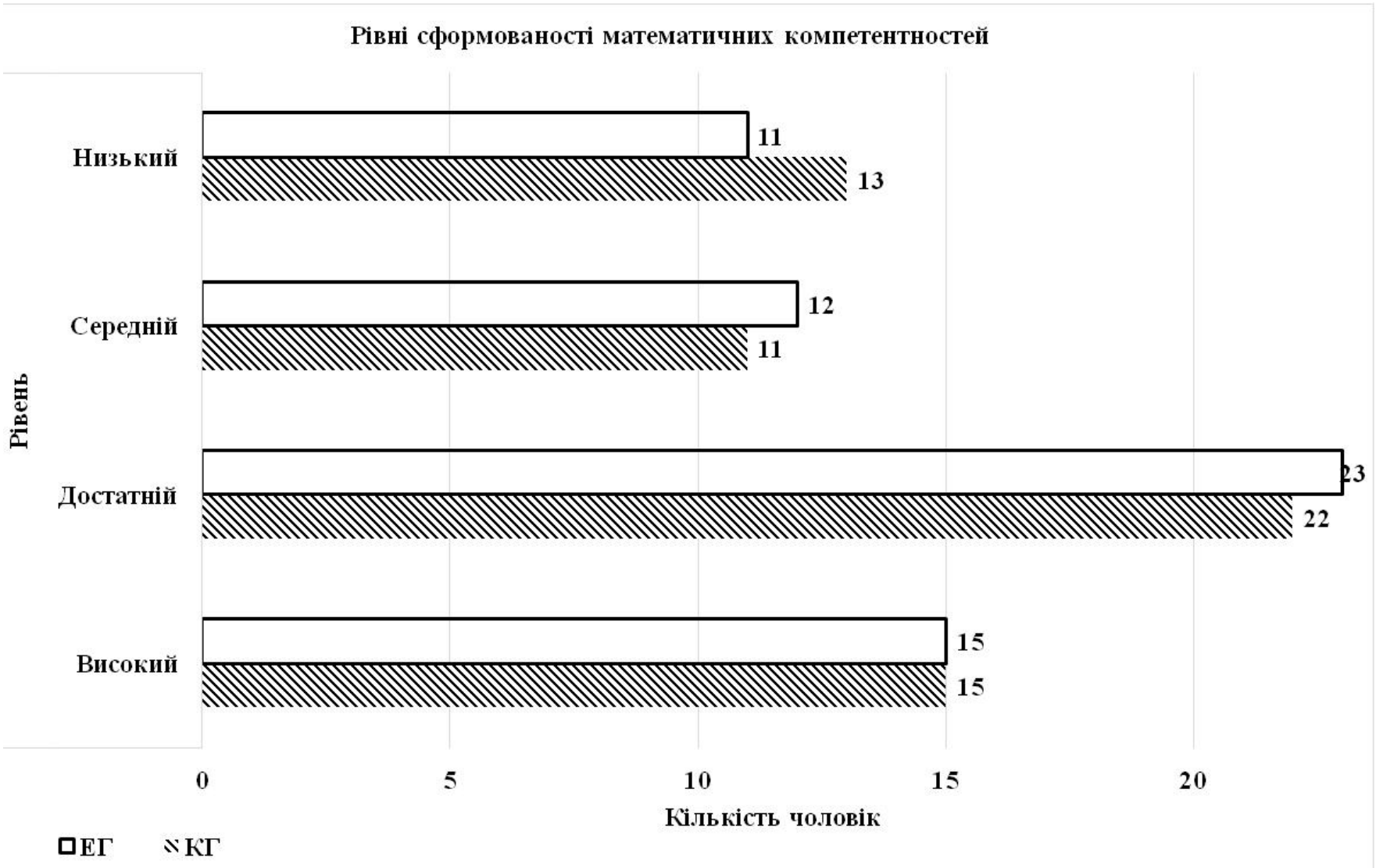

Рис. 4.11. Порівняння розподілів експериментальних та контрольних груп студентів за рівнями сформованості математичних компетентностей на констатувальному етапі експерименту

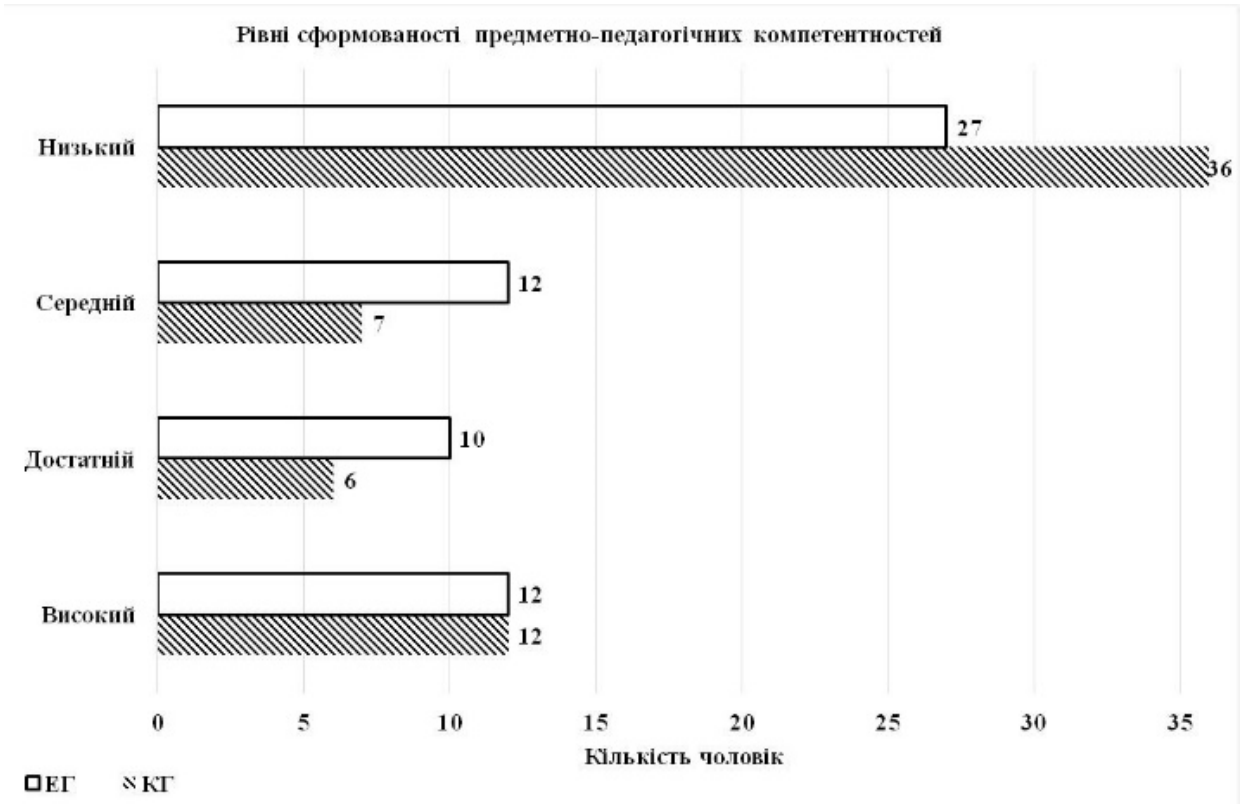

Рис. 4.12. Порівняння розподілів експериментальних та контрольних груп студентів за рівнями сформованості предметно-педагогічних компетентностей на констатувальному етапі експерименту

За критичне значення критерію Фішера для кожного із зазначених складників системи оберемо – 1,64.

Емпіричне значення критерію Фішера – 0,49. Характеристики порівнюваних вибірок збігаються на рівні значущості 0,05.

*Таблиця 4.2*

**Таблиця для розрахунків за критерієм Фішера при порівнянні двох груп за часткою студентів, які мають високий та достатній рівні сформованості інформаційно-технологічних компетентностей на констатувальному етапі експерименту**

| Групи | Ефект має місце | | Ефект відсутній | | Всього |
|---|---|---|---|---|---|
| | Кількість студентів | % | Кількість студентів | % | |
| Контрольні | 14 | 23% | 47 | 77% | 61 |
| Експериментальні | 16 | 26% | 45 | 74% | 61 |
| Всього | 25 | | 12 | | 122 |

Тобто емпіричне значення 0,49 знаходиться у *зоні незначущості* і гіпотеза $H_0$ приймається. Це означає, що достовірно, з рівнем значущості 0,05, що частка студентів, які за результатами дослідження рівнів сформованості інформаційно-технологічних компетентностей показали високий та достатній рівень не більше, ніж у контрольних групах.

На основі даних, наведених на рис. 4.11, перевіримо достовірність гіпотези про відсутність, з статистичної точки зору, відмінностей між рівнями сформованості математичних компетентностей експериментальних і контрольних груп. Для цього скористаємося критерієм Фішера.

Сформулюємо гіпотези:

$H_0$: Частка студентів, які за результатами дослідження рівнів сформованості математичних компетентностей показали високий та достатній рівень не більше, ніж у контрольних групах;

$H_1$: Частка студентів, які за результатами дослідження рівнів сформованості математичних компетентностей показали високий та достатній рівень більша, ніж у контрольних групах.

Побудуємо таблицю, яка фактично є таблицею емпіричних частот за двома значеннями ознаки: якщо рівні сформованості інформаційно-технологічних компетентностей зазначені високий та достатній, то «ефект має місце», у протилежному випадку – «ефект відсутній» (табл. 4.3). При цьому в обрахунках використовуються лише частки, що відповідають спостереженням, для яких ефект має місце.

Експериментальні дані повністю задовольняють обмеження, що накладаються кутовим перетворенням Фішера.

Емпіричне значення критерію Фішера – 0,19, критичне – 1,64. Характеристики порівнюваних вибірок збігаються на рівні значущості 0,05. Тобто емпіричне значення 0,19 знаходиться у *зоні незначущості* і гіпотеза $H_0$ приймається. Це означає, що достовірно, з рівнем значущості 0,05, що частка студентів, які за результатами дослідження рівнів сформованості математичних компетентностей показали високий та достатній рівень не більше, ніж у контрольних групах.

*Таблиця 4.3*

**Таблиця для розрахунків за критерієм Фішера при порівнянні двох груп за часткою студентів, які мають високий та достатній рівні сформованості математичних компетентностей на констатувальному етапі експерименту**

| Групи | Ефект має місце | | Ефект відсутній | | Всього |
|---|---|---|---|---|---|
| | Кількість студентів | % | Кількість студентів | % | |
| Контрольні | 37 | 61% | 24 | 39% | 61 |
| Експериментальні | 38 | 62% | 23 | 38% | 61 |
| Всього | 25 | | 12 | | 122 |

На основі даних, наведених на рис. 4.12, перевіримо достовірність гіпотези про відсутність, з статистичної точки зору, відмінностей між рівнями сформованості предметно-педагогічних компетентностей експериментальних і контрольних груп. Для цього скористаємося критерієм Фішера. Сформулюємо гіпотези:

$H_0$: Частка студентів, які за результатами дослідження рівнів сформованості предметно-педагогічних компетентностей показали високий та достатній рівень не більше, ніж у контрольних групах;

$H_1$: Частка студентів, які за результатами дослідження рівнів сформованості предметно-педагогічних компетентностей показали високий та достатній рівень більша, ніж у контрольних групах.

Побудуємо таблицю, яка фактично є таблицю емпіричних частот за двома значеннями ознаки: якщо рівні сформованості інформаційно-технологічних компетентностей зазначені високий та достатній, то «ефект має місце», у протилежному випадку – «ефект відсутній» (табл. 4.4). При цьому в обрахунках використовуються лише частки, що відповідають спостереженням, для яких ефект має місце.

Експериментальні дані повністю задовольняють обмеження, що накладаються кутовим перетворенням Фішера.

Емпіричне значення критерію Фішера – 0,77, критичне – 1,64. Характеристики порівнюваних вибірок збігаються на рівні значущості 0,05.

Тобто емпіричне значення 0,77 знаходиться у *зоні незначущості* і гіпотеза $H_0$ приймається. Це означає, що достовірно, з рівнем значущості 0,05, що частка студентів, які за результатами дослідження рівнів сформованості предметно-педагогічних компетентностей показали високий та достатній рівень не більше, ніж у контрольних групах.

*Таблиця 4.4*

**Таблиця для розрахунків за критерієм Фішера при порівнянні двох груп за часткою студентів, які мають високий та достатній рівні сформованості предметно-педагогічних компетентностей на констатувальному етапі експерименту**

| Групи | Ефект має місце | | Ефект відсутній | | Всього |
|---|---|---|---|---|---|
| | Кількість студентів | % | Кількість студентів | % | |
| Контрольні | 18 | 30% | 43 | 70% | 61 |
| Експериментальні | 22 | 36% | 39 | 64% | 61 |
| Всього | 25 | | 12 | | 122 |

Нами було проведено додатково анкетування та тестування академічних груп студентів МІ-11 (26 студентів), МІ-12-1 (17 студентів), МІ-12-2 (18 студентів). Загальна кількість вибірки становить – 61 студент. Анкета «Ставлення до використання інформаційно-комунікаційних технологій (ІКТ)» та тест, який мав за мету визначити рівень самостійності опанування студентами матеріалу, що винесено на самостійне опрацювання з математичних дисциплін. Анкетування проводилося повне. Анкета була напівзакритою, загалом містила в собі прямі запитання. В анкетах присутні питання дихотомічні та зі стандартизованою низкою відповідей [221, с. 70-71].

Анкета та тест було створено інструментами Google Диску, а саме Google Форми для кожної окремої групи [190]. Для групи студентів МІ-11 анкета та тест було розміщено на сайті https://sites.google.com/a/kramarenko.com.ua/grupa-mi-11 де в рамках експерименту створено сторінку «Майя Попель, Sagemath Cloud» для ознайомлення з основними відомостями, що стосується CoCalc: https://sites.google.com/a/kramarenko.com.ua/grupa-mi-11/maja-popel-eksperimenti. На вказаній сторінці розміщені посилання на офіційний сайт Sage та підручники, представлена загальна інформація стосовно хмарного сервісу та презентація «Знайомство з CoCalc». Сторінка носить суто інформативний характер. Нижче – розташована анкета та тест.

Студенти групи МІ-11 заповнили анкету та пройшли тест у вересні-жовтні 2014 року. Студенти академічних груп МІ-12-1 та МІ-12-2 заповнили анкету у вересні, а тест пройшли в жовтні 2014 року [190].

Студенти груп МІ-12-1 та МІ-12-2 отримали посилання на анкету та тест електронною поштою разом з поясненнями стосовно CoCalc в яких були подані основні відомості.

Анкета «Ставлення до використання інформаційно-комунікаційних технологій (ІКТ)» націлена на визначення попереднього досвіду роботи студентів з ІКТ, хмарними сервісами та окресленням переліку математичних дисциплін, які потребують, на думку студентів, активного використання ІКТ .

Результати даної анкети показали, що група студентів МІ-11 активно використовують в своїй практиці ІКТ як на заняттях з математичних дисциплін, так і під час тем винесених на самостійне опрацювання.

Студенти ж груп МІ-12-1 та МІ-12-2 в більшій мірі взагалі дуже рідко (6 %) використовують в своїй практиці засоби ІКТ. Про хмарні сервіси знають лише 17 % студентів. Активно використовують хмарні сервіси – 6 %. 33 % студентів вбачають у використанні ІКТ позитивні зміни під час вивчення математичних дисциплін.

За допомогою тесту вдалось встановити, що студенти в більшій мірі оцінюють свій рівень знань як «середній» (92 %). Усі студенти групи МІ-11 використовують під час підготовки до тієї чи іншої математичної дисципліни засоби ІКТ [190]. Найбільше проблем у студентів викликають завдання на доведення, виконанні побудов у просторі та під час роботи з графіками. На думку студентів, найінтенсивнішого використання ІКТ та хмарних сервісів потребують наступні дисципліни: «Математичний аналіз», «Теорія ймовірностей та математична статистика», «Елементарна математика». 92 % студентів групи МІ-11 сподіваються, що практичні заняття будуть в подальшому супроводжуватись більшою кількістю прикладів, більш різноманітним ілюстративним матеріалом.

Студенти груп МІ-12-1 та МІ-12-2 вважають, що не в змозі опанувати матеріал самостійно на високому рівні (82 %). До речі, в цих групах студенти не так активно використовують засоби ІКТ під час підготовки до математичних дисциплін (лише 47 %). Майже всі студенти хотіли б покращити свій рівень знань з математичного аналізу, причому або з усіх розділів, або ж тем, які відносяться до інтегрального числення.

Ми не обмежувались результатами, одержаними, лише на основі заповнення анкети та проходження тесту. Ми використали в своєму дослідженні метод бесіди з викладачами. Дані, одержані нами в процесі анкетування, підтвердились.

Аналізуючи усі одержані результати, ми можемо зробити висновки, що більшість проблем студенти вбачають під час вивчення наступних математичних дисциплін: математичного аналізу, теорії ймовірностей та математичної статистики, елементарної математики. Студенти хотіли б підвищити свій рівень знань за рахунок більшого використання ІКТ на практичних заняттях з обраних математичних дисциплін. Крім того, визначено, які саме завдання викликають найбільш труднощів: завдання на доведення, побудови у просторі та робота з графіками функцій. Якщо ж розглянути на конкретній дисципліні, наприклад обрати математичний аналіз, то більшої уваги потребують тема «Інтегральне числення». Таким чином, застосування хмарного сервісу CoCalc є актуальним і бажаним, і його доцільно зорієнтувати на опрацювання визначених завдань.

### 4.3 Статистичне опрацювання та аналіз результатів формувального етапу педагогічного експерименту

В якості мети формувального етапу дослідження було обрано перевірку ефективності застосування методики використання CoCalc як засобу формування професійних компетентностей учителя математики та порівнянні рівнів сформованості окремих складників системи професійних компетентностей учителя математики експериментальних і контрольних груп.

Розподіл рівнів сформованості окремих компонентів системи професійних компетентностей учителя математики в контрольних і експериментальних групах за результатами контрольного зрізу та кінцевого (по завершенню формувального етапу науково-експериментальної роботи) подано у табл. 4.5, табл. 4.6 та табл. 4.7 (рівні сформованості інформаційно-технологічних, математичних та предметно-педагогічних компетентностей відповідно).

Гістограми порівняльного розподілу рівнів сформованості інформаційно-технологічних, математичних та предметно-педагогічних компетентностей за результатами констатувального та кінцевого зрізу представлено на рис. 4.13, рис. 4.14 та рис. 4.15 відповідно.

*Таблиця 4.5*

**Розподіл рівнів сформованості інформаційно-технологічних компетентностей у контрольних і експериментальних групах за результатами констатувального та кінцевого зрізу**

| Рівень | Констатувальний зріз (кількість студентів) | | Кінцевий зріз (кількість студентів) | |
|---|---|---|---|---|
| | Контрольна група (КГ) | Експериментальна група (ЕГ) | Контрольна група (КГ) | Експериментальна група (ЕГ) |
| Високий | 2 | 2 | 2 | 5 |
| Достатній | 12 | 14 | 12 | 28 |
| Середній | 25 | 25 | 28 | 21 |
| Низький | 22 | 20 | 19 | 7 |
| Всього | 61 | 61 | 61 | 61 |

*Таблиця 4.6*

**Розподіл рівнів сформованості математичних компетентностей у контрольних і експериментальних групах за результатами констатувального та кінцевого зрізу**

| **Рівень** | **Констатувальний зріз (кількість студентів)** | | **Кінцевий зріз (кількість студентів)** | |
|---|---|---|---|---|
| | Контрольна група (КГ) | Експеримен-тальна група (ЕГ) | Контрольна група (КГ) | Експериме-нтальна група (ЕГ) |
| Високий | 15 | 15 | 8 | 15 |
| Достатній | 22 | 23 | 18 | 25 |
| Середній | 11 | 12 | 20 | 16 |
| Низький | 13 | 11 | 15 | 5 |
| Всього | 61 | 61 | 61 | 61 |

*Таблиця 4.7*

**Розподіл рівнів сформованості предметно-педагогічних компетентностей у контрольних і експериментальних групах за результатами констатувального та кінцевого зрізу**

| **Рівень** | **Констатувальний зріз (кількість студентів)** | | **Кінцевий зріз (кількість студентів)** | |
|---|---|---|---|---|
| | Контрольна група (КГ) | Експеримен-тальна група (ЕГ) | Контрольна група (КГ) | Експеримен-тальна група (ЕГ) |
| Високий | 12 | 12 | 11 | 16 |
| Достатній | 6 | 10 | 6 | 13 |
| Середній | 7 | 12 | 9 | 14 |
| Низький | 36 | 27 | 35 | 18 |
| Всього | 61 | 61 | 61 | 61 |

На основі даних, наведених на рис. 4.13, спочатку перевіримо достовірність гіпотези про наявність, з статистичної точки зору, відмінностей між рівнями сформованості інформаційно-технологічних компетентностей експериментальних і контрольних груп за результатами кінцевого зрізу. Для цього скористаємося критерієм Фішера. Сформулюємо гіпотези:

$H_0$: Частка студентів, які за результатами дослідження рівнів сформованості інформаційно-технологічних компетентностей показали високий та достатній рівень більша, ніж у контрольних групах;

$H_1$: Частка студентів, які за результатами дослідження рівнів сформованості інформаційно-технологічних компетентностей показали

високий та достатній рівень не більша, ніж у контрольних групах.

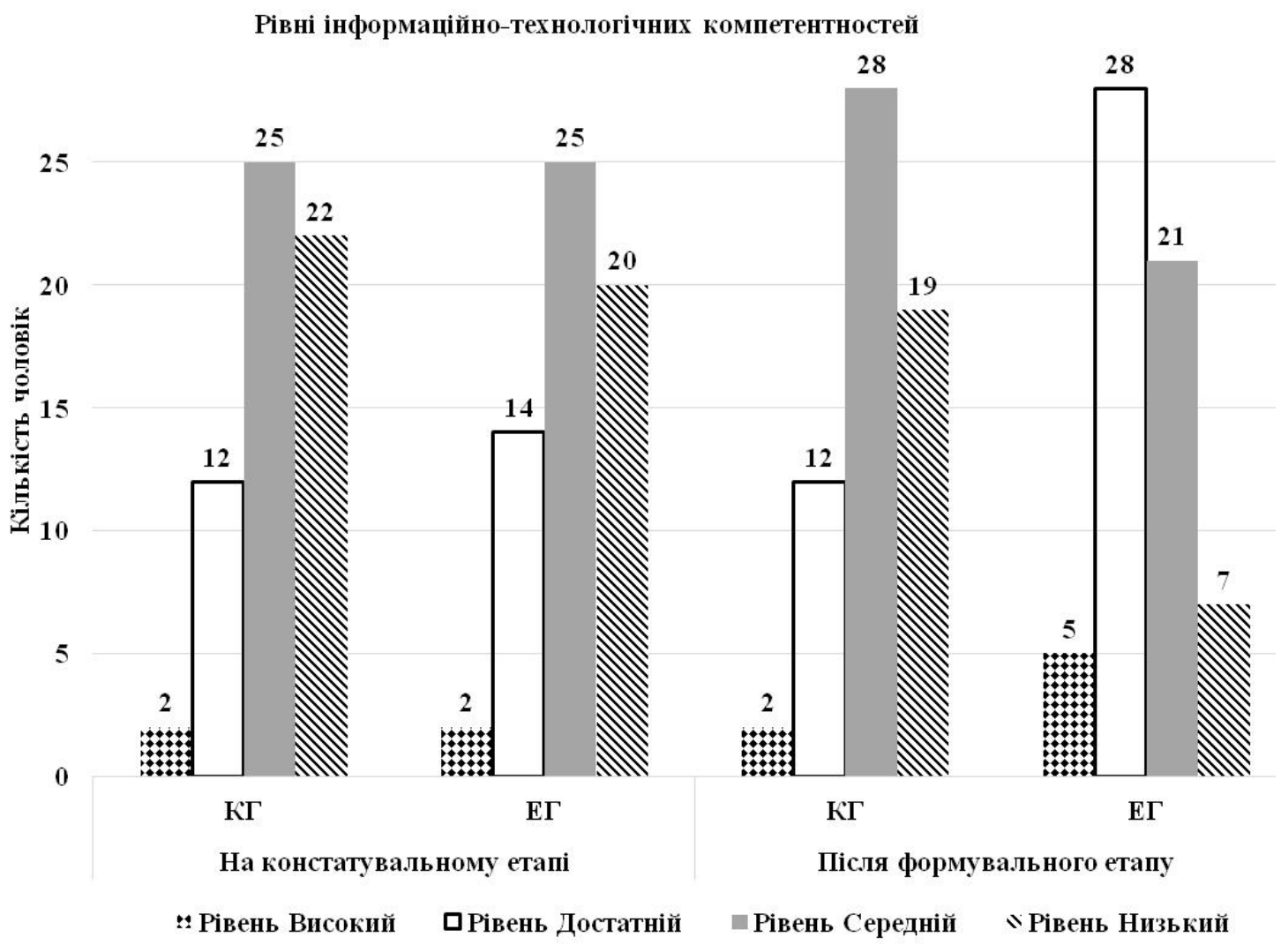

Рис. 4.13. Гістограми порівняльного розподілу рівнів сформованості інформаційно-технологічних компетентностей за результатами констатувального та кінцевого зрізу

Побудуємо таблицю, яка фактично є таблицею емпіричних частот за двома значеннями ознаки: якщо рівні сформованості інформаційно-технологічних компетентностей зазначені високий та достатній, то «ефект має місце», у протилежному випадку – «ефект відсутній» (табл. 4.8).

При цьому в обрахунках використовуються лише частки, що відповідають спостереженням, для яких ефект має місце. Експериментальні дані повністю задовольняють обмеження, що накладаються кутовим перетворенням Фішера:

а) жодна з часток, що порівнюються, не дорівнює нулю;

б) кількість спостережень у обох вибірках більше 5, що дозволяє будь-які співставлення.

За критичне значення критерія Фішера для кожного із зазначених складників системи оберемо – 1,64.

Емпіричне значення критерію Фішера – 3,61, критичне – 1,64. Достовірність відмінностей характеристик експериментальної і контрольної груп за статистичним критерієм Фішера дорівнює 95%.

Отже, якщо характеристики експериментальної і контрольної груп до початку експерименту збігаються з рівнем значущості 0,05, і, одночасно з цим, достовірність відмінностей характеристик експериментальної і контрольної груп після експерименту дорівнює 95%, то можна зробити висновок, що застосування методики використання CoCalc як засобу формування професійних компетентностей учителя математики призводить до статистично значущих відмінностей результатів.

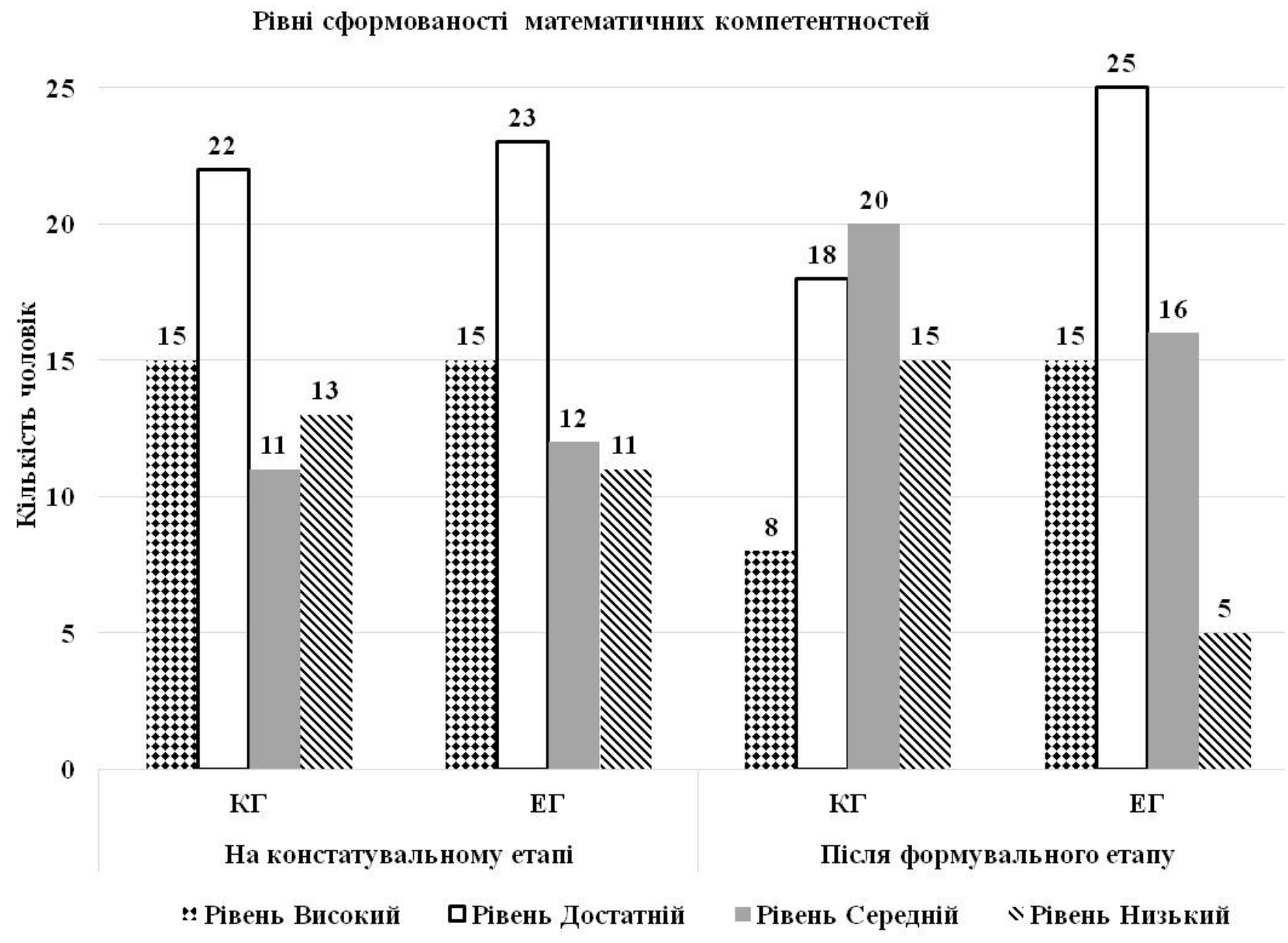

Рис. 4.14. Гістограми порівняльного розподілу рівнів сформованості математичних компетентностей за результатами констатувального та кінцевого зрізу

На основі даних, наведених на рис. 4.14, спочатку перевіримо достовірність гіпотези про наявність, з статистичної точки зору, відмінностей між рівнями сформованості математичних компетентностей експериментальних і контрольних груп за результатами кінцевого зрізу. Для цього скористаємося критерієм Фішера. Сформулюємо гіпотези:

$H_0$: Частка студентів, які за результатами дослідження рівнів сформованості математичних компетентностей показали високий та достатній рівень більша, ніж у контрольних групах;

$H_1$: Частка студентів, які за результатами дослідження рівнів сформованості математичних компетентностей показали високий та достатній рівень не більша, ніж у контрольних групах.

Побудуємо таблицю, яка фактично є таблицею емпіричних частот за двома значеннями ознаки: якщо рівні сформованості математичних компетентностей зазначені високий та достатній, то «ефект має місце», у протилежному випадку – «ефект відсутній» (табл. 4.9). При цьому в обрахунках використовуються лише частки, що відповідають спостереженням, для яких ефект має місце. Експериментальні дані повністю задовольняють обмеження, що накладаються кутовим перетворенням Фішера.

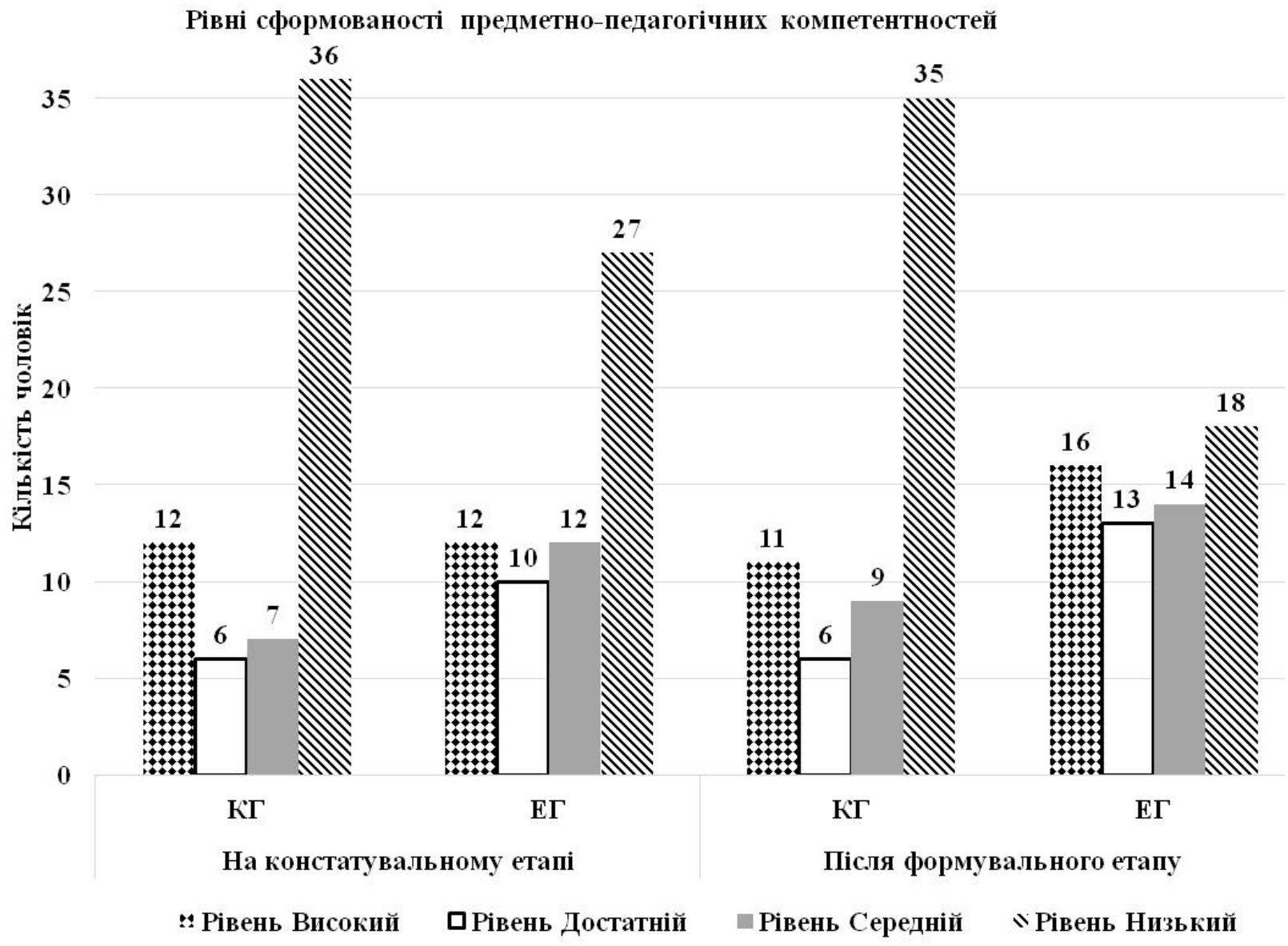

Рис. 4.15. Гістограми порівняльного розподілу рівнів сформованості предметно-педагогічних компетентностей за результатами констатувального та кінцевого зрізу

*Таблиця 4.8*

**Таблиця для розрахунків за критерієм Фішера при порівнянні двох груп за часткою студентів, які мають високий та достатній рівні сформованості інформаційно-технологічних компетентностей після формувального етапу експерименту**

| Групи | Ефект має місце | | Ефект відсутній | | Всього |
|---|---|---|---|---|---|
| | Кількість студентів | % | Кількість студентів | % | |
| Контрольні | 14 | 23% | 47 | 77% | 61 |
| Експериментальні | 33 | 54% | 28 | 46% | 61 |
| Всього | 25 | | 12 | | 122 |

*Таблиця 4.9*

**Таблиця для розрахунків за критерієм Фішера при порівнянні двох груп за часткою студентів, які мають високий та достатній рівні сформованості математичних компетентностей після формувального етапу експерименту**

| Групи | Ефект має місце | | Ефект відсутній | | Всього |
|---|---|---|---|---|---|
| | Кількість студентів | % | Кількість студентів | % | |
| Контрольні | 26 | 43% | 35 | 57% | 61 |
| Експериментальні | 40 | 66% | 21 | 34% | 61 |
| Всього | 25 | | 12 | | 122 |

Емпіричне значення критерію Фішера – 2,57, критичне – 1,64. Достовірність відмінностей характеристик експериментальної і контрольної груп за статистичним критерієм Фішера дорівнює 95%.

Отже, якщо характеристики експериментальної і контрольної груп до початку експерименту збігаються з рівнем значущості 0,05, і, одночасно з цим, достовірність відмінностей характеристик експериментальної і контрольної груп після експерименту дорівнює 95 %, то можна зробити висновок, що застосування методики використання CoCalc як засобу формування професійних компетентностей учителя математики призводить до статистично значущих відмінностей результатів.

На основі даних, наведених на рис. 4.15, спочатку перевіримо достовірність гіпотези про наявність, з статистичної точки зору, відмінностей між рівнями сформованості предметно-педагогічних компетентностей експериментальних і контрольних груп за результатами кінцевого зрізу. Для цього скористаємося критерієм Фішера. Сформулюємо гіпотези:

$H_0$: Частка студентів, які за результатами дослідження рівнів сформованості предметно-педагогічних компетентностей показали високий та достатній рівень більша, ніж у контрольних групах;

$H_1$: Частка студентів, які за результатами дослідження рівнів сформованості предметно-педагогічних компетентностей показали високий та достатній рівень не більша, ніж у контрольних групах.

Побудуємо таблицю, яка фактично є таблицю емпіричних частот за двома значеннями ознаки: якщо рівні сформованості предметно-педагогічних компетентностей зазначені високий та достатній, то «ефект має місце», у протилежному випадку – «ефект відсутній» (табл. 4.10). При цьому в обрахунках використовуються лише частки, що відповідають спостереженням, для яких ефект має місце.

Експериментальні дані повністю задовольняють обмеження, що накладаються кутовим перетворенням Фішера.

*Таблиця 4.10*

**Таблиця для розрахунків за критерієм Фішера при порівнянні двох груп за часткою студентів, які мають високий та достатній рівні сформованості предметно-педагогічних компетентностей після формувального етапу експерименту**

| Групи | Ефект має місце | | Ефект відсутній | | Всього |
|---|---|---|---|---|---|
| | Кількість студентів | % | Кількість студентів | % | |
| Контрольні | 26 | 43% | 35 | 57% | 61 |
| Експериментальні | 40 | 66% | 21 | 34% | 61 |
| Всього | 25 | | 12 | | 122 |

Емпіричне значення критерію Фішера – 2,26, критичне – 1,64. Достовірність відмінностей характеристик експериментальної і контрольної груп за статистичним критерієм Фішера дорівнює 95 %.

Отже, якщо характеристики експериментальної і контрольної груп до початку експерименту збігаються з рівнем значущості 0,05, і, одночасно з цим, достовірність відмінностей характеристик експериментальної і контрольної груп після експерименту дорівнює 95 %, то можна зробити висновок, що застосування методики використання CoCalc як засобу формування професійних компетентностей учителя математики призводить до статистично значущих відмінностей результатів.

Задля визначення стану професійних компетентностей по завершенню формувального етапу проведення експерименту були розглянуті також результати державного екзамену за спеціальністю, оскільки саме державний екзамен відображає комплексний стан сформованості професійних компетентностей з урахуванням усіх складників. Так, за результатами державного екзамену в експериментальній групі відсоток студентів, які одержали оцінку «відмінно» більший, ніж у контрольній (на 7 %). Відсоток студентів, які склали державний екзамен на оцінку «добре» в експериментальній групі дорівнює 58 %, а в контрольній – 31 %.

Державний екзамен зазвичай складається з перевірки професійних знань, що формують уміння та навичок випускників певного ЗВО, виконання комплексного кваліфікаційного завдання, що формують професійні уміння та передбачає співбесіду з членами державної екзаменаційної комісії (ДЕК). Державний екзамен виступає в якості загальнодержавного методу комплексної діагностики сформованості професійних знань, умінь та навичок.

На основі даних, наведених на рис. 4.16, перевіримо достовірність гіпотези про статистично значущі відмінностей між результатами державного екзамену, а отже і сформованістю професійних компетентностей студентів експериментальної (МІ-12) і контрольної груп (МІ-11). Експериментальні дані повністю задовольняють обмеження, що накладаються на застосування критерію Вілкоксона-Манна-Уїтні (кожна вибірка має містити не менше трьох елементів).

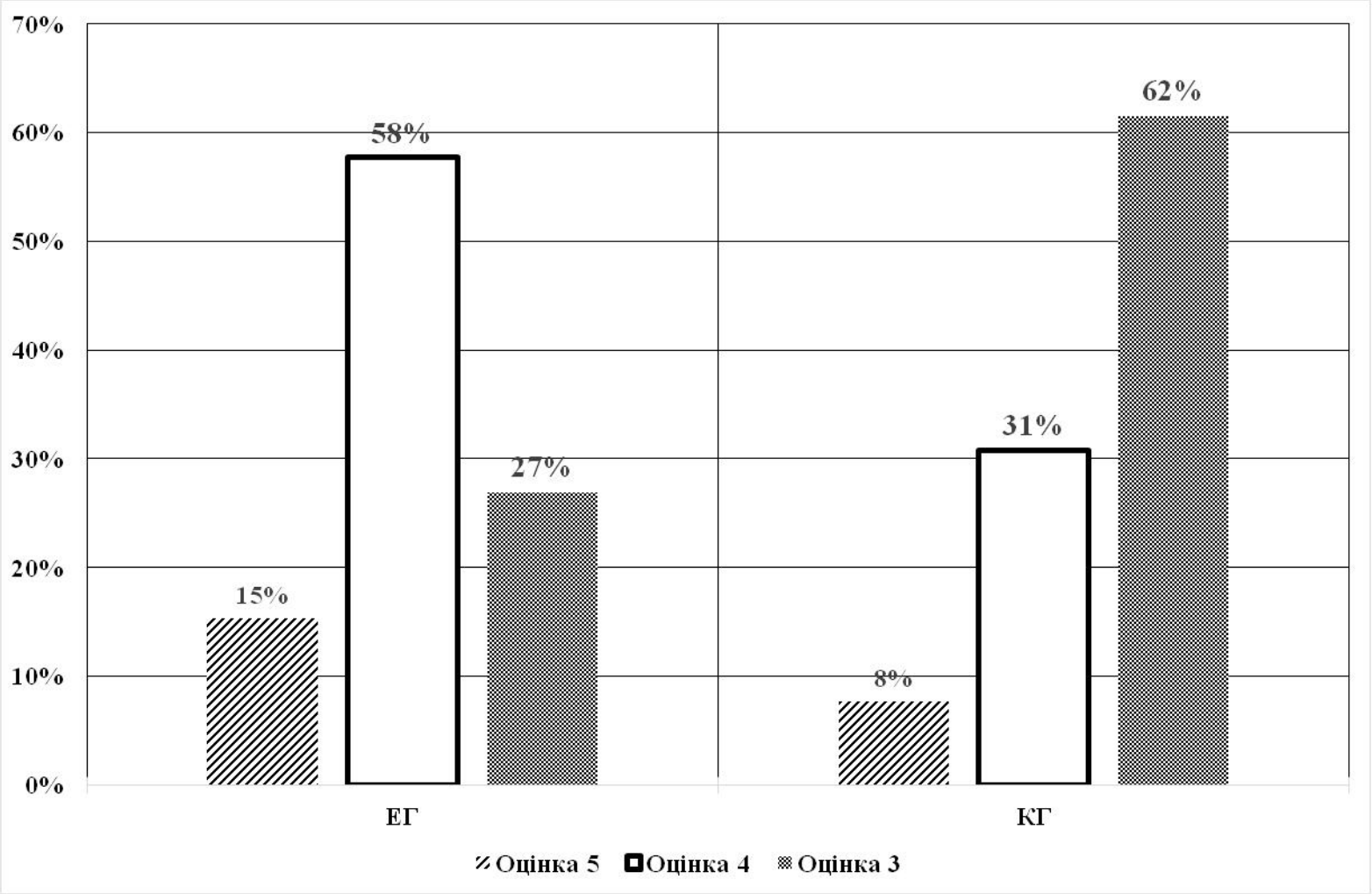

Рис. 4.16. Процентне співвідношення результатів державного екзамену у контрольній та експериментальній групах

Емпіричне значення критерію Вілкоксона-Манна-Уїтні – 2,1595, критичне – 1,96. Достовірність відмінностей характеристик порівнюваних вибірок складає 95 %. Тобто емпіричне значення 2,1595 знаходиться у *зоні значущості*. Це означає, що достовірно, з рівнем значущості 0,05, що результати державного екзамену за спеціальністю в експериментальних групах відрізняються від результатів в контрольних групах на 95%, а отже рівні сформованості професійних компетентностей студентів у експериментальних групах за результатами констатувальних зрізів відрізняються від рівнів сформованості професійних компетентностей студентів у контрольних групах.

Порівнюючи рівні сформованості професійних компетентностей у контрольних та експериментальних групах на початку формувального етапу та наприкінці експерименту, можна спостерігати збільшення частки студентів, які мають високий та середній рівні сформованості професійних компетентностей.

Аналіз результатів формувального етапу педагогічного експерименту показав, що розподіли рівнів сформованості професійних компетентностей в експериментальній та контрольній групах майбутніх учителів математики мають статистично значущі відмінності, зумовлені упровадженням розробленої методики, що підтверджує гіпотезу дослідження.

**Висновки до розділу 4**

1. У процесі експериментальної роботи контрольні та експериментальні групи формувалися наступним чином:

– склад контрольних груп охоплював студентів груп, що навчались за традиційною методикою формування професійних компетентностей учителя математики;

– склад експериментальних груп охоплював студентів груп, які навчалися за авторською методикою використання CoCalc як засобу формування професійних компетентностей учителя математики.

2. Узагальнюючи одержані результати констатувального етапу педагогічного експерименту, можна стверджувати:

– переважна більшість студентів і викладачів мають можливість роботи з хмарним сервісом CoCalc як у ЗВО, так і вдома;

– викладачі в більшості випадків не використовують у навчальному процесі хмарні сервіси, за винятком застосування їх в якості хмарного сховища;

– викладачі зацікавлені у впровадженні в навчальний процес хмарного сервісу CoCalc, але студенти не готові до цього;

– студенти на початку експерименту показали низький рівень сформованості інформаційно-технологічних та предметно-педагогічних компетентностей, достатній – математичних компетентностей;

– студенти та викладачі користуються лише вільно поширюваними програмними засобами (переважно локальними системами комп’ютерної математики).

3. З метою з’ясування стану сформованості професійних компетентностей та оцінювання ефективності методики використання CoCalc виконувалися констатувальні зрізи наступних складників предметних, технологічних та професійно-практичних компетентностей системи професійних компетентностей учителя математики: предметно-педагогічні, інформаційно-технологічні та математичні компетентності. Кожний складник розглядався окремо та обчислювались значення за рівнями: високий, достатній, середній та низький.

Задля аналізу даних було встановлено збіги (на констатувальному етапі експерименту) та відмінності (після формувального етапу експерименту) характеристик експериментальної і контрольної груп за критерієм Фішера. Для цього були сформульовані статистичні гіпотези: про відсутність відмінностей між рівнями сформованості окремих складників системи професійних компетентностей та про значимість відмінностей між рівнями сформованості обраних складників.

Аналізуючи одержані результати на констатувальному етапі експерименту, можна зробити висновок, що рівні сформованості професійних компетентностей майбутніх вчителів математики контрольної та експериментальної груп співпадають з рівнем значущості $\alpha=0,05$.

Аналізуючи одержані результати після формувального етапу експерименту, можна зробити висновок, що достовірність відмінностей рівнів сформованості професійних компетентностей майбутніх вчителів математики контрольної та експериментальної груп складає 95 %.

4. Задля підтвердження відмінностей між рівнями сформованості професійних компетентностей по завершенню формувального етапу експерименту були розглянуті результати державного екзамену за спеціальністю, оскільки саме державний екзамен відображає комплексний стан сформованості професійних компетентностей з урахуванням усіх складників.

Так, за результатами державного екзамену в експериментальній групі відсоток студентів, які одержали оцінку «відмінно» більший, ніж у контрольній (на 7 %). Відсоток студентів, які склали державний екзамен на оцінку «добре» в експериментальній групі дорівнює 58 %, а в контрольній – 31 %.

Задля перевірки достовірності відмінностей у контрольній та експериментальній групах було застосовано критерій Вілкоксона-Манна-Уїтні. Порівнюючи рівні сформованості професійних компетентностей у контрольній та експериментальній групах на початку формувального етапу та наприкінці експерименту, можна спостерігати збільшення частки студентів, які мають високий та середній рівні сформованості професійних компетентностей.

Аналіз результатів формувального етапу педагогічного експерименту показав, що розподіли рівнів сформованості професійних компетентностей в експериментальній та контрольній групах майбутніх учителів математики мають статистично значущі відмінності, зумовлені впровадженням розробленої методики використання хмарного сервісу CoCalc, що підтверджує гіпотезу дослідження.

# ВИСНОВКИ

1. У результаті аналізу стану проблеми запровадження у навчальний процес хмарних сервісів на основі вітчизняного і зарубіжного досвіду виявлено, що нині вже існують хмарні версії різних систем комп'ютерної математики, що породжує тенденції розвитку програмного забезпечення математичного призначення, що полягають у переході до використання хмаро орієнтованих платформ його постачання, віртуалізації сервісів, а також використання їх як послуги. Виокремлені напрями використання сервісу CoCalc у процесі навчання математичних дисциплін, зокрема, підтримування індивідуальних та групових форм організації навчальної діяльності; забезпечення наочності; підвищення часової та просторової мобільності та ін.; обґрунтовано переваги використання хмарних сервісів: економія ресурсів, мобільність доступу, еластичність.

2. Формування професійних компетентностей учителя математики передбачає набуття ним компетентностей у галузі математики та суміжних з нею дисциплін, методики навчання та дидактики, психологічних і педагогічних основ здійснення навчально-виховного процесу, дослідницької діяльності та педагогічного спілкування, що визначає якість його професійної діяльності.

У складі професійних компетентностей майбутнього вчителя математики, що набуваються з використанням CoCalc, виокремлено наступні компетентності: предметно-педагогічні, інформаційно-технологічні та математичні компетентності, які входять до складу предметних, технологічних та професійно-практичних компетентностей. Для кожного складника були визначені показники сформованості професійних компетентностей згідно до чотирьох рівнів (високий, достатній, середній, низький).

3. На основі спроектованої системи професійних компетентностей майбутнього вчителя математики розроблено модель використання хмарного сервісу CoCalc як засобу формування професійних компетентностей учителя математики, в якій враховано зв'язки між компонентами професійних компетентностей та усіма циклами дисциплін програми підготовки учителя математики у педагогічному ЗВО. Модель охоплює три етапи формування професійних компетентностей із використанням хмарного сервісу CoCalc. Було виявлено, що використання цього хмарного сервісу у процесі навчання майбутніх учителів математики впливає, в першу чергу, на формування спеціально професійних компетентностей.

4. У складі методики використання хмарного сервісу CoCalc як засобу формування професійних компетентностей учителя математики

виокремлено взаємозв'язані мету, зміст, форми організації, методи і засоби навчання та результати. Її запровадження відбувається у три етапи: I етап – пропедевтичний, II етап – формувальний, III етап – розвивальний. Експериментально підтверджено, що рівень сформованості професійних компетентностей майбутніх учителів математики буде вищим, якщо у процес навчання педагогічно обґрунтовано запроваджувати розроблену методику використання хмарного сервісу CoCalc як засобу формування професійних компетентностей учителя математики.

Результати педагогічного експерименту, перевірені із застосуванням критеріїв Фішера та Вілкоксона-Манна-Уітні, дають підстави вважати, що гіпотеза дослідження дістала підтвердження.

Виконане дослідження не вичерпує всіх аспектів поставленої проблеми. Продовження наукового пошуку за даною проблематикою доцільно у таких напрямах: розроблення теоретико-методичних засад проектування хмаро орієнтованого середовища навчання математичних дисциплін майбутніх учителів математики у педагогічному ЗВО; розробка методики використання хмарного сервісу CoCalc у процесі підвищення кваліфікації викладачів математики.

# СПИСОК ВИКОРИСТАНИХ ДЖЕРЕЛ

# ДОДАТОК А. ТИПИ ДІЯЛЬНОСТІ, ТИПОВІ ЗАВДАННЯ ДІЯЛЬНОСТІ ТА УМІННЯ

*Таблиця А.1*

**Типи діяльності, типові завдання діяльності та уміння майбутніх вчителів математики**

| Тип діяльності | Назва типового завдання діяльності | Зміст уміння |
|---|---|---|
| Дослідження математичних відображень ідеалізованих об'єктів | Аналіз сучасних математичних теорій | Володіти уявленням про математику як науку та як про навчальний предмет, її місце в сучасному світі та системі наук |
| | | Володіти поняттями даної математичної теорії |
| | | Вміти підготувати огляд літератури з тематики математичного дослідження |
| | | Вміти з'ясовувати склад і структуру теорії: поняття, наукові факти, закони, принципи та зв'язки між ними |
| | | Вміти аналізувати теорії на предмет зв'язку з досліджуваним об'єктом та проблемою |
| | | Вміти аналізувати методи теорій на предмет їх придатності на розв'язування існуючої проблеми |
| | | Вміти обирати предмет та об'єкт дослідження, визначати мету дослідження та його основні завдання |
| | Постановка математичної задачі | Вміти раціонально і повно використовувати закони логіки |
| | | Вміти аналізувати математичні факти, закономірності і теорії на предмет логічної строгості та повноти |
| | | Вміти бачити логічні прогалини в обґрунтуванні математичних фактів, побудові математичних теорій |
| | | Вміти використовувати методи пізнання (моделювання, аналіз, синтез, узагальнення, конкретизація, порівняння, аналогія тощо) для постановки математичної задачі |
| | | Вміти будувати приклади та контрприклади, зокрема з використанням інформаційних технологій |
| | | Вміти формулювати нові коректно поставлені задачі |
| | | Вміти усвідомлювати застосовність існуючих методів до розв'язування поставлених проблем |
| | | Вміти оцінювати перспективність розв'язування математичної задачі |

*Продовження табл. А.1*

| **Тип діяльності** | **Назва типового завдання діяльності** | **Зміст уміння** |
|---|---|---|
| **Дослідження математичних відображень ідеалізованих об'єктів** | **Аналіз математичної проблеми (задачі)** | Вміти досліджувати коректність постановки математичної задачі |
| | | Вміти аналізувати до якої галузі математичних знань належить досліджуваний об'єкт і проблема з ним пов'язана |
| | | Вміти аналізувати чи має теорія, якій належить проблема, ізоморфні теорії |
| | | Вміти аналізувати чи нерозв'язана дана проблема в ізоморфній теорії |
| | | Вміти аналізувати взаємозв'язки досліджуваного математичного об'єкта з відомими об'єктами, та математичної проблеми з науковими фактами |
| | | Вміти встановлювати ізоморфність математичних об'єктів |
| | **Формулювання гіпотетичного твердження** | Вміти виділяти математичний об'єкт і виділяти його суттєві властивості |
| | | Вміти обирати понятійний апарат, адекватний математичному об'єкту |
| | | Вміти формулювати твердження в імплікативній формі |
| | | Вміти формулювати твердження в еквівалентній формі |
| | | Вміти формулювати твердження в формі необхідних, достатніх, необхідних і достатніх умов |
| | | Вміти встановлювати протиріччя між твердженнями |
| | | Вміти проводити комп'ютерні експерименти з метою встановлення нових закономірностей |
| | | Вміти наводити приклади математичних об'єктів, що задовольняють умови гіпотетичного твердження |

| Тип діяльності | Назва типового завдання діяльності | Зміст уміння |
|---|---|---|
| **Дослідження математичних відображень ідеалізованих об'єктів** | **Доведення гіпотетичного твердження, спростування гіпотетичного твердження** | Вміти формулювати твердження, що є окремим випадком гіпотетичного твердження, і твердження більш загальне, ніж розглядуване гіпотетичне |
| | | Вміти відбирати знання, необхідні для доведення або спростування гіпотетичного твердження |
| | | Вміти аналізувати гіпотетичне твердження і у разі можливості розкладати його на простіші |
| | | Вміти побудувати логічну схему доведення |
| | | Вміти використовувати метод від супротивного при доведенні гіпотетичного твердження |
| | | Вміти використовувати аналітичний метод доведення гіпотетичного твердження |
| | | Вміти використовувати синтетичний метод доведення гіпотетичного твердження |
| | | Вміти використовувати аналітико-синтетичний метод доведення гіпотетичного твердження |
| | | Вміти обирати раціональні методи (способи, прийоми) доведення або спростування гіпотетичного твердження |
| | | Вміти реалізовувати побудовану логічну схему доведення |
| | | Вміти будувати контрприклади для спростування гіпотетичного твердження |
| | | Вміти проводити комп'ютерне моделювання та чисельні експерименти для перевірки гіпотетичного твердження та його окремих випадків |
| | | Вміти використовувати засоби комп'ютерної алгебри при доведенні теорем |

*Продовження табл. А.1*

| Тип діяльності | Назва типового завдання діяльності | Зміст уміння |
|---|---|---|
| Математичне моделювання природничих, технічних, економічних та соціальних явищ і процесів | Аналіз створеної моделі реального об'єкта | Вміти добирати критерії оцінювання математичної моделі на предмет її досконалості (у відповідності до цілей моделювання) |
| | | Вміти встановлювати адекватність побудованої математичної моделі досліджуваному об'єкту |
| | | Вміти наводити приклади реалій, моделями яких є математичні об'єкти, наводити приклади задач з реальним змістом |
| | | Вміти конструювати моделі проблемної (задачної) ситуації (предметні, схематичні, графічні імітаційні та ін.) |
| | | Вміти інтерпретувати математичні залежності в термінах конкретних математичних теорій |
| | Дослідження математичної моделі з використанням засобів комп'ютерної техніки | Вміти добирати ефективний метод дослідження математичної моделі для розв'язування поставленої задачі |
| | | Володіти поняттями «конкретність», «стійкість», «обумовленість» задач |
| | | Вміти добирати ефективні методи чисельного аналізу математичних моделей різних задач |
| | | Вміти добирати та використовувати готові програмні засоби (математичні пакети прикладних програм) для символьно-формульного, графічного чисельного аналізу математичних моделей реальних об'єктів |
| | | Вміти при необхідності розробити алгоритм і програму для розв'язування математичної задачі, яка є математичною моделлю |
| | | Вміти виконувати чисельний експеримент, в тому числі з використанням комп'ютера |
| | | Вміти аналізувати похибки при чисельному розв'язуванні задач |
| | | Вміти інтерпретувати, аналізувати та узагальнювати результати розрахунків чисельного експерименту |

*Продовження табл. А.1*

| Тип діяльності | Назва типового завдання діяльності | Зміст уміння |
|---|---|---|
| **Математичне моделювання природничих, технічних, економічних та соціальних явищ і процесів** | **Створення аксіоматичної теорії та її аналіз** | Володіти сучасними поглядами на аксіоматичну теорію на аксіоматичний метод побудови математичної теорії |
| | | Вміти створювати моделі аксіоматичних теорій: інтерпретувати основні (неозначувані) поняття, положення та відношення систем аксіом в термінах конкретних математичних теорій (знаходити конкретні множини і відношення на них, які мають задані властивості) |
| | | Вміти обґрунтовувати еквівалентність тверджень, зокрема аксіом |
| | | Вміти перевіряти несуперечливість, повноту, категоричність систем аксіом, незалежність аксіом |
| | | Вміти конструювати математичні об'єкти із заданими властивостями |

# ДОДАТОК Б
# ХРОНОЛОГІЧНА СТАТИСТИКА ВИКОРИСТАННЯ COCALC. ВИКОРИСТАННЯ І ЗАВАНТАЖЕНІСТЬ ОБЧИСЛЮВАЛЬНИХ СЕРВЕРІВ (СТАНОМ НА 10.03.2016 Р.)

На графіку (рис. А.1) показана кількість одночасно підключених користувачів до системи (активне з’єднання між службою SMC і веб-браузер користувача).

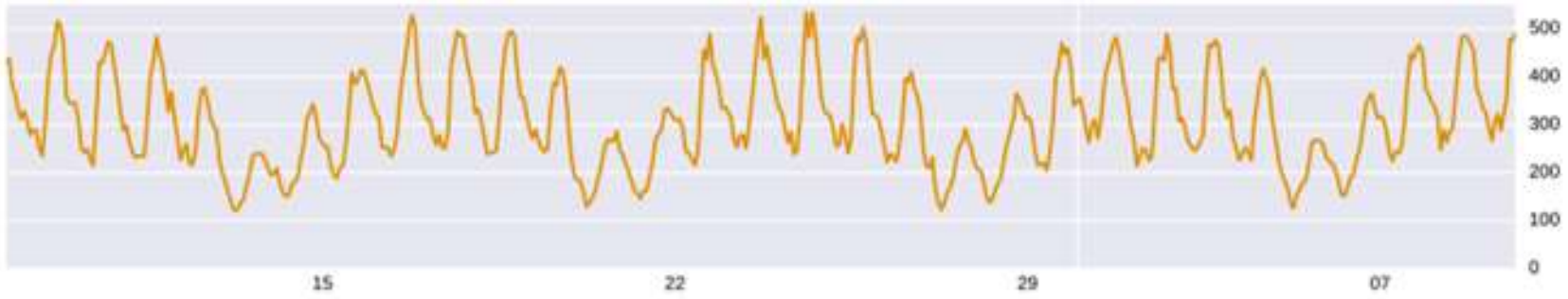

Рис. А.1. Загальна кількість активних користувачів

На графіку нижче (рис. А.2) показано кількість проектів (зі станом – «running») за останні 30 днів (станом на 10.03.2016 р.).

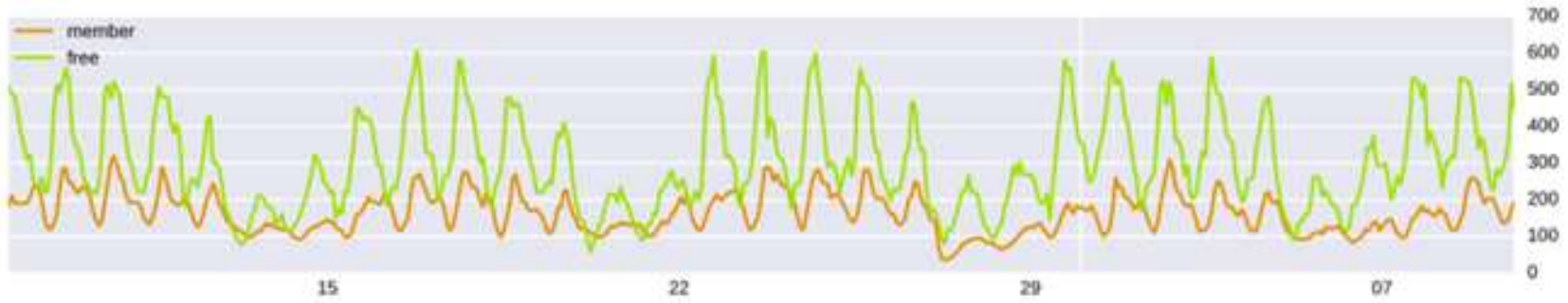

Рис. А.2. Кількість запущених проектів

Порівнюючи кількість змін, які були виконані в проектах користувачів, починаючи з березня 2014 року (рис. А.3), можна спостерігати, як з кожним місяцем їх стає все більше. На сьогодні їх кількість становить понад 4000.

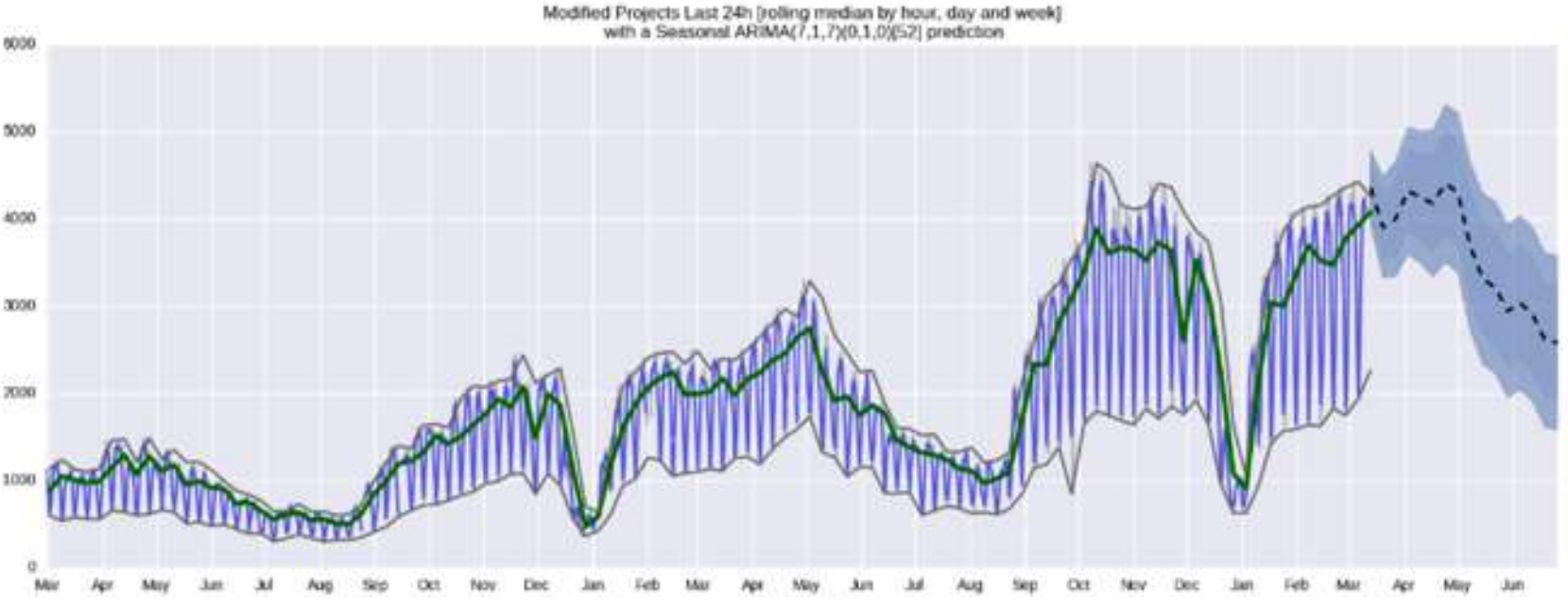

Рис. А.3. Кількість змін в проектах

У CoCalc передбачено одночасне виконання декількох проектів. У

період з січня 2014 р. по лютий 2015 р. найбільша кількість проектів, що одночасно використовувались, зафіксовано в лютому 2015 року (одночасно виконувались обчислення майже в 340 проектах). У середньому ж їх значення коливається в межах 100 активних проектів (статус – «виконується»).

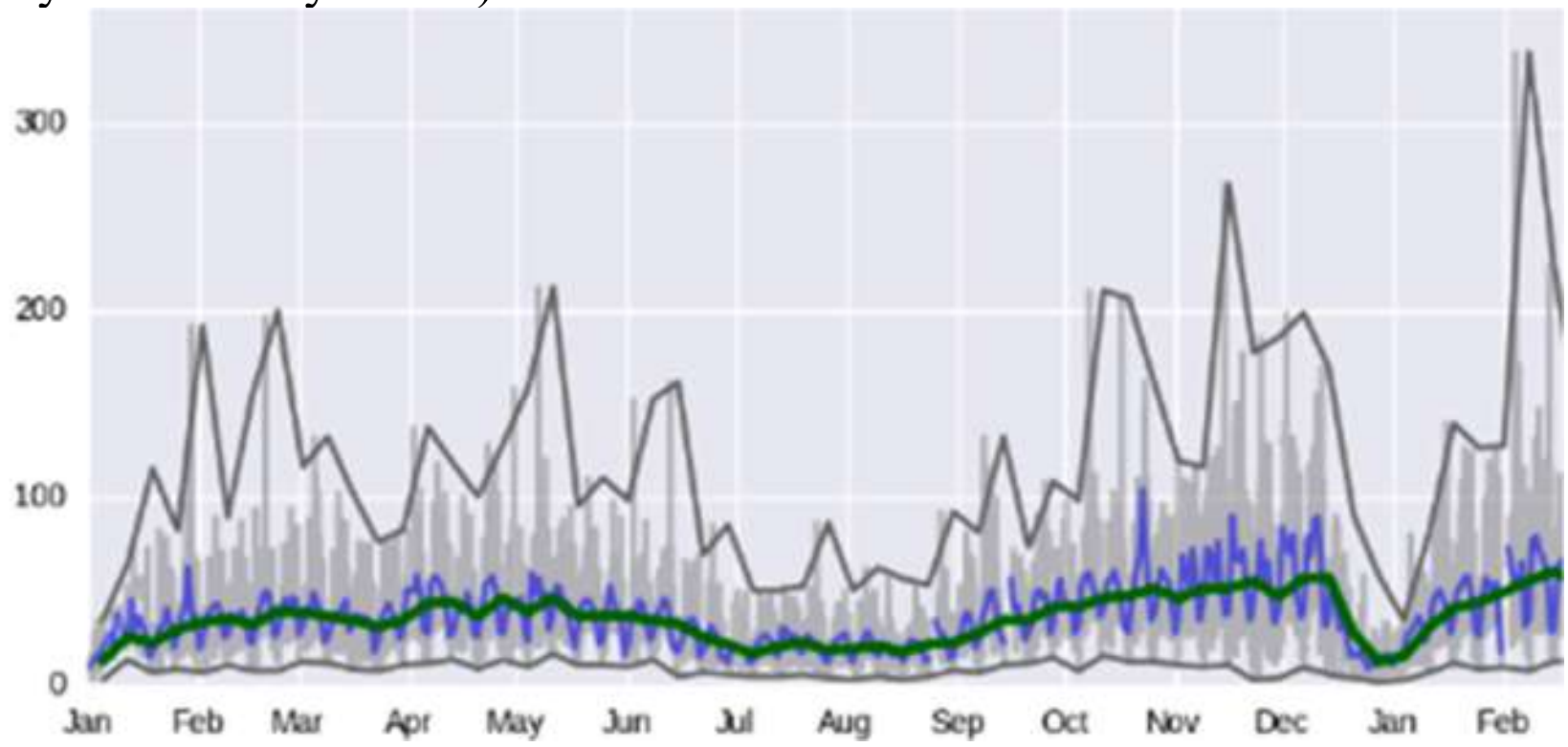

Рис. А.4. Одночасний запуск проектів

На рис. А.5 показано кількість одночасного використання системи користувачами. Значення постійно зростає. Порівнюючи середнє значення в січні 2014 року (приблизно 100 користувачів) та лютому 2015 року (400 користувачів), можна зробити висновок, що технічні показники системи з часом вдосконалюються, забезпечують більшу кількість користувачів одночасним доступом до системи.

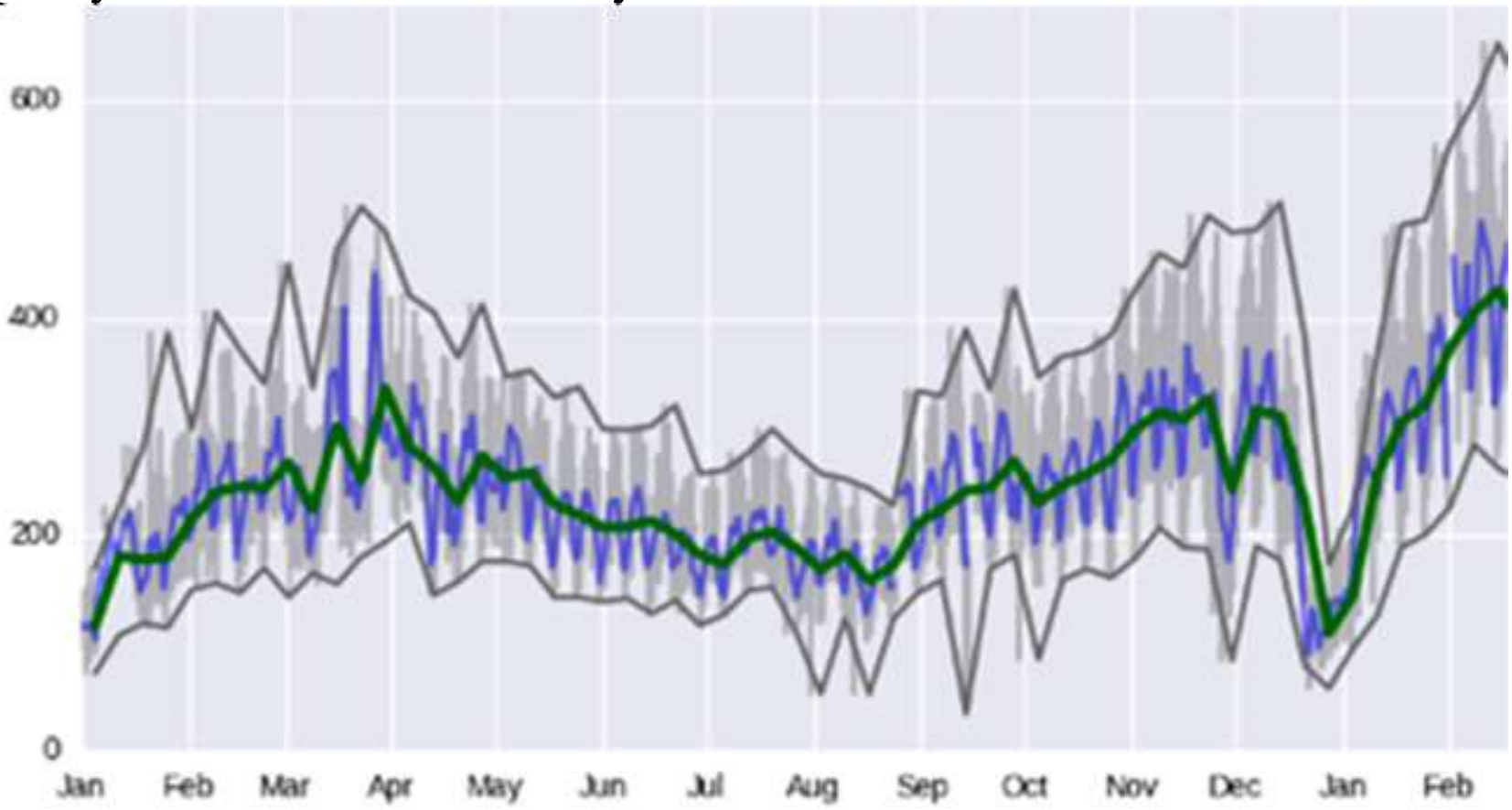

Рис. А.5. Одночасне з'єднання користувачів з системою

Але зростання кількості облікових записів користувачів, ще не свідчить про активне використання інструментарію CoCalc. Співвідношення між кількістю облікових записів користувачів та створених проектів можна прослідкувати на рис. А.6. Зі збільшенням кількості нових облікових записів збільшується і кількість створених користувачами проектів (при чому мова йде про активні проекти, які використовуються).

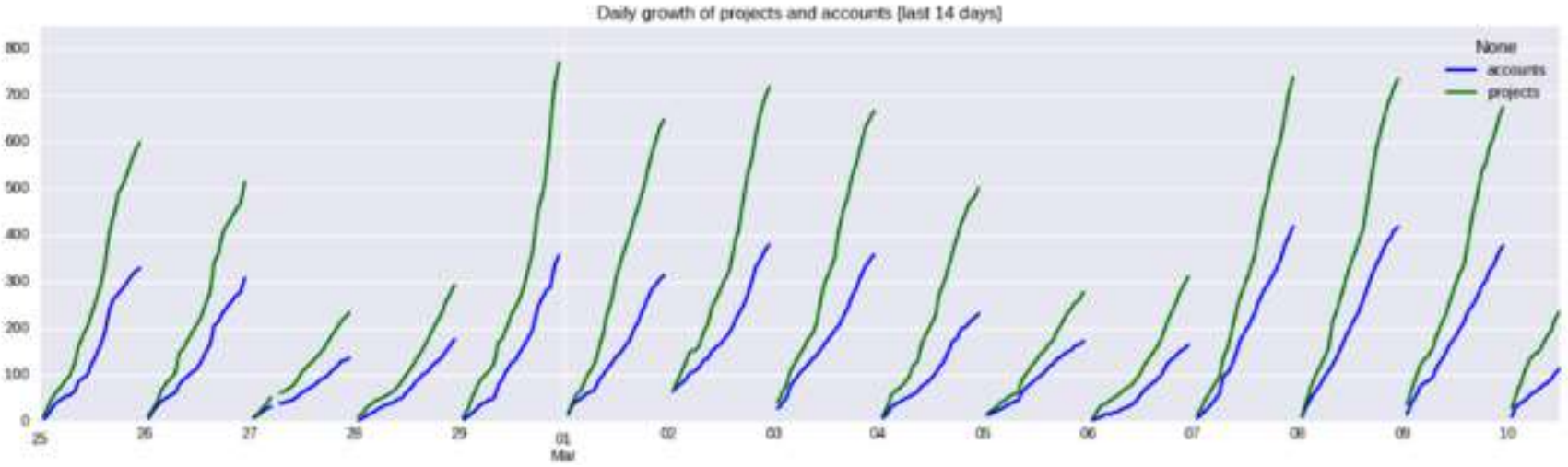

Рис. А.6. Співвідношення кількості проектів та облікових записів

Якщо ж розглянути кількість нових облікових записів, то можна побачити тенденцію, що з початком навчального року, вона збільшується (рис. А.7). Наприклад, у липні 2015 року – приблизно 800 нових облікових записів, а вже у вересні спостерігаємо – 2500. Це говорить про те, що дана система активно використовується викладачами, студентами, науковими співробітниками.

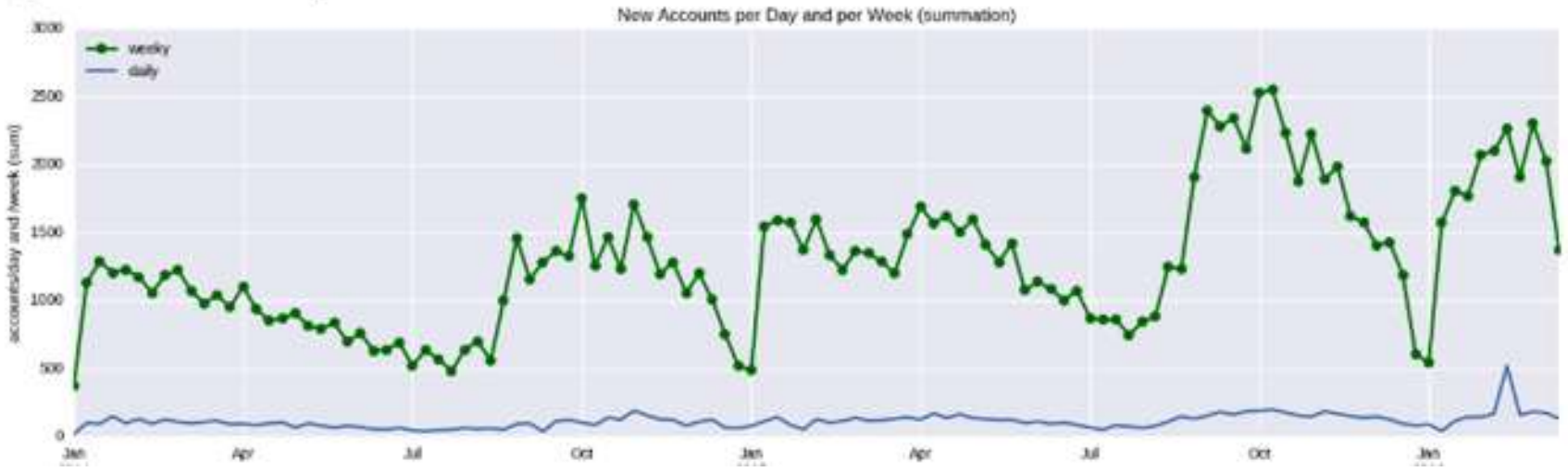

Рис. А.7. Реєстрація нових облікових записів

Відомості про зростання загальної кількості облікових записів та проектів в хронологічному порядку показано на рис. А.8.

При цьому з часом збільшується і загальна кількість проектів створених одним користувачем. Середнє значення на сьогодні це – майже 2 проекти у кожного користувача системи. В той час, як в 2014 році – 1 проект в середньому.

Інтерес до хмарних сервісів з часом зростає, про що свідчить збільшення інтенсивності їх використання (рис. А.10).

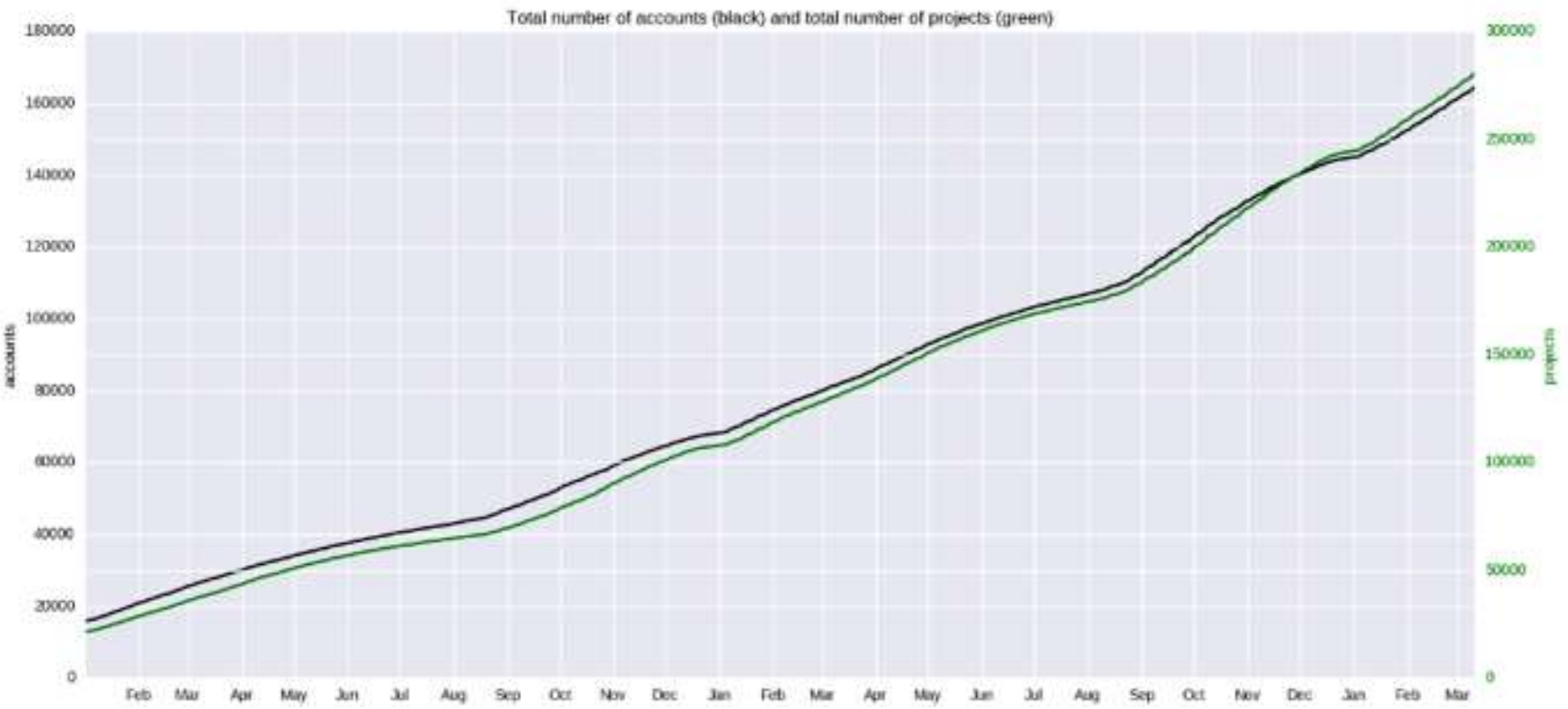

Рис. А.8. Загальна кількість облікових записів та проектів

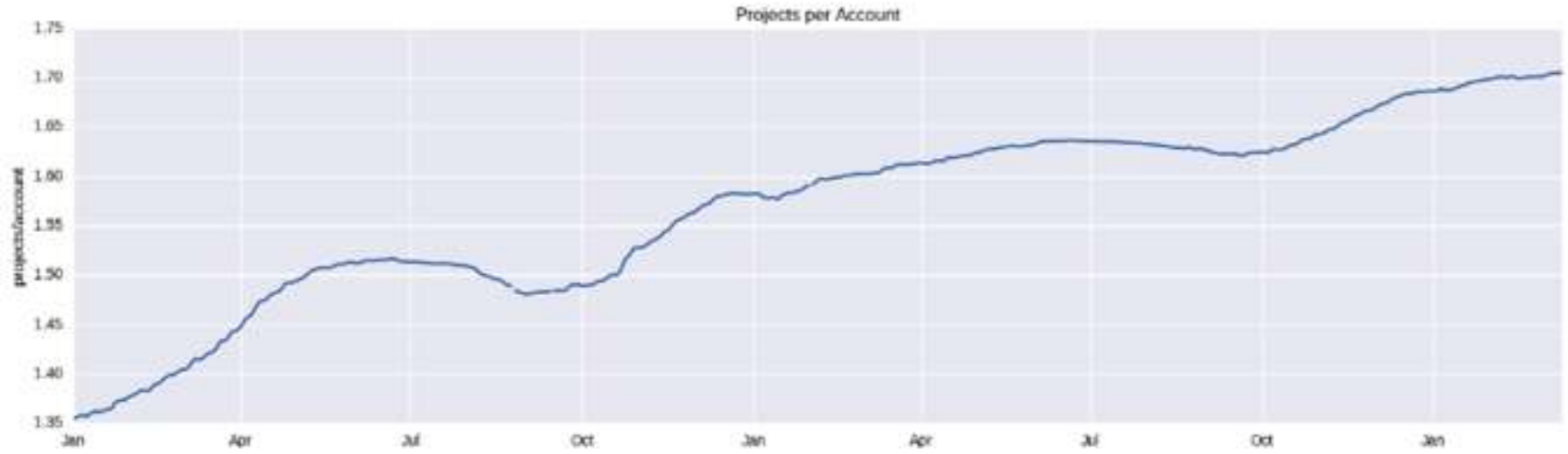

Рис. А.9. Середня кількість проектів в одному обліковому записі

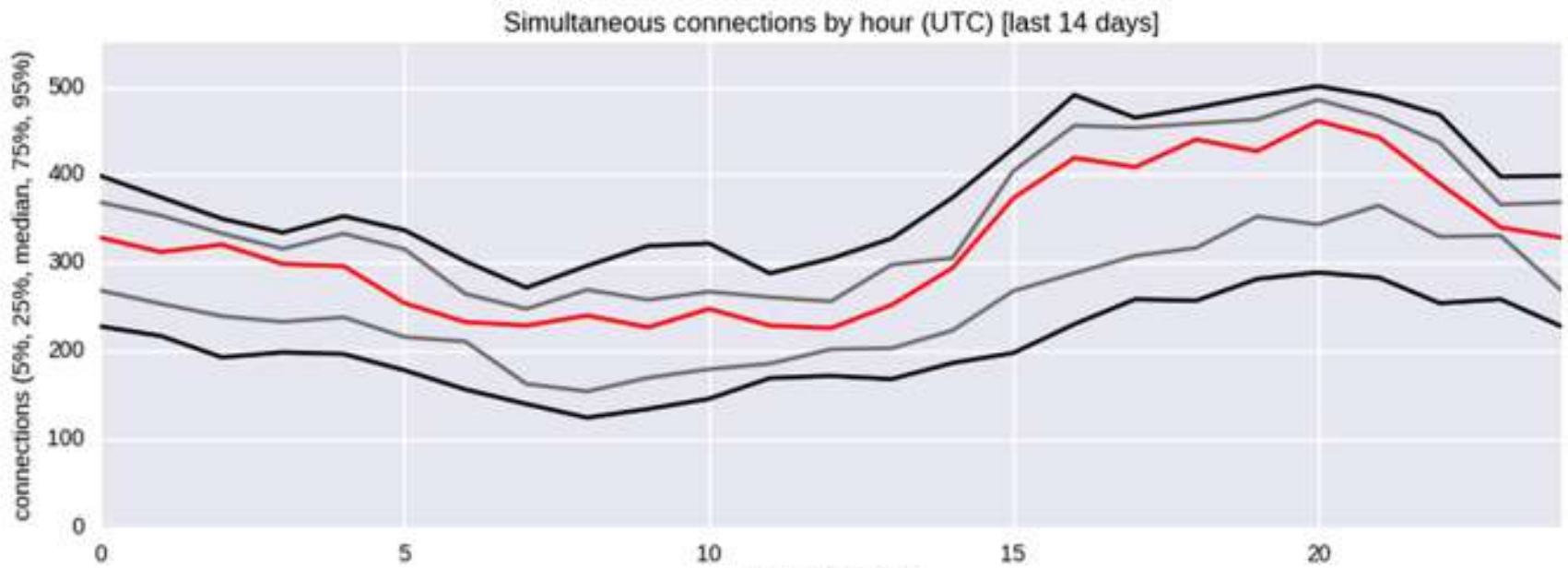

Рис. А.10. Одночасне підключення протягом доби

## ДОДАТОК В
## ЗАГАЛЬНІ НАЛАШТУВАННЯ ОБЛІКОВГО ЗАПИСУ. БЛОК «EDITOR»:ПЕРЕМИКАЧІ ТИПУ «ПРАПОРЕЦЬ»

– «Line wrapping: wrap long lines» – переносити довгі рядки;

– «Line numbers: show line numbers» – показувати номери рядків;

– «Code folding: fold code using control+Q» – показувати кнопки згортання коду (згорнути код можна, скориставшись гарячими клавішами Ctrl + Q);

– Smart indent: context sensitive indentation» – ввімкнути автоматичний відступ програмного коду(в залежності від контексту);

– «Electric chars: sometimes reindent current line» – змінювати вирівнювання рядків у залежності від попередньо вживаних символів (групування та ін.);

– «Match brackets: highlight matching brackets near cursor» – виділяти кольором дужки поруч з курсором (зеленим – правильна кількість закриваючих/відкриваючих дужок, червоним – кількість відкритих та закритих дужок не співпадає);

– «Auto close brackets: automatically close brackets» – автоматично закривати дужки (закриваюча дужка має з’явиться автоматично);

– «Match XML tags: automatically match XML tags» – перевіряти відповідність XML-тегів;

– «Auto close XML tags: automatically close XML tags» – автоматично закривати XML-теги;

– «Strip trailing whitespace: remove whenever file is saved» – видаляти пропуски на кінцях рядків(щоразу, коли файл зберігається);

– «Show trailing whitespace: show spaces at ends of lines» – показати пропуски в кінці рядка;

– «Spaces instead of tabs: send 4 spaces when the tab key is pressed» – замінювати табуляцію чотирма пропусками;

– «Track revisions: record history of changes when editing files» – зберігати історію змін при редагуванні файлів;

– «Extra button bar: more editing functions (mainly in Sage worksheets)» – показати додаткову панель інструментів (переважно на робочих аркушах).

# ДОДАТОК Г
# ГАРЯЧІ КЛАВІШІ РЕДАКТОРУ ФАЙЛІВ

| Сполучення клавіш | Інтерпретація | Кнопка стандартної панелі інструментів (або ілюстрація дії) |
|---|---|---|
| *Ctrl* + < | зменшити шрифт |   |
| *Ctrl* + > | збільшити шрифт |   |
| *Ctrl* + *L* | перейти до рядка за вказаним номером | 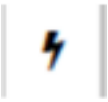 |
| *Ctrl* + *F* | виконати пошук за ключовим словом | 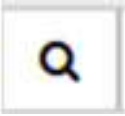 |
| *Ctrl* + *Q* | згорнути/розгорнути логічну конструкцію (цикл, функцію користувача тощо) | 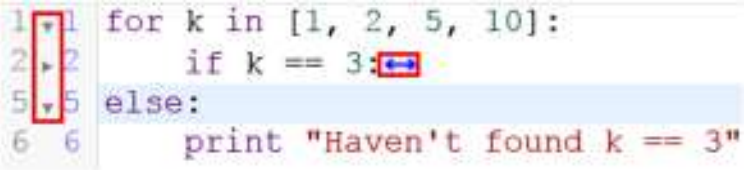  |
| *Tab* | збільшити відступ праворуч | – |
| *Shift* + *Tab* | збільшити відступ ліворуч | – |
| *Ctrl* + *I* | дублювати вікно редактора (вертикально, горизонтально) /відмінити режим дублювання вікна | 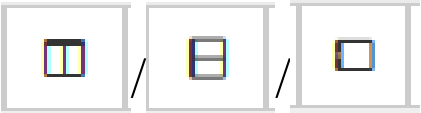  |
| *Ctrl* + *ліва кнопка миші* | одночасно редагувати декілька виділених фрагментів тексту | 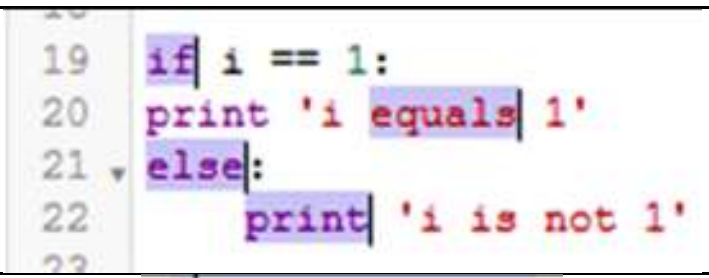  |
| *Ctrl* + *пропуск* | автоматичне доповнення назви функції | 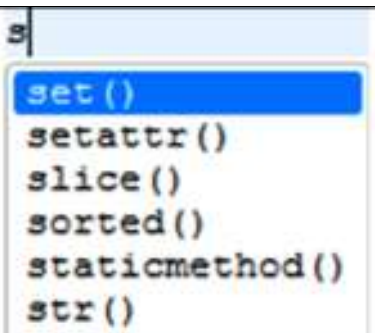  |
| *Tab* | автоматичне доповнення команд Sage | 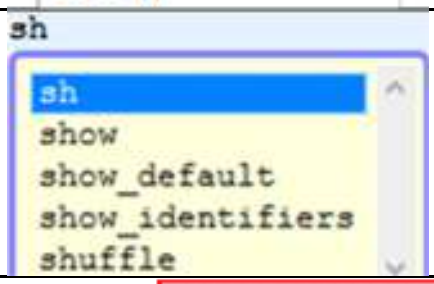  |
| *Ctrl+;* | додати нову комірку | 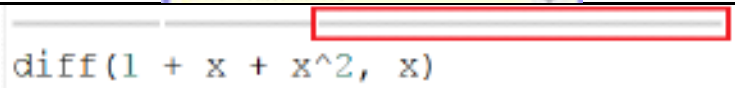  |

## ДОДАТОК Д
## РОЗДІЛИ АНГЛОМОВНОЇ ЗАГАЛЬНОЇ ДОВІДКИ SAGE

**L-функції**
http://doc.sagemath.org/pdf/en/reference/lfunctions/lfunctions.pdf
**P-адичні числа**
http://doc.sagemath.org/pdf/en/reference/padics/padics.pdf
**Алгебраїчні функціональні поля**
http://doc.sagemath.org/pdf/en/reference/function_fields/function_fields.pdf
**Алгебри**
http://doc.sagemath.org/pdf/en/reference/algebras/algebras.pdf
**Алгебри Гекке**
http://doc.sagemath.org/pdf/en/reference/hecke/hecke.pdf
**Арифметичні групи**
http://doc.sagemath.org/pdf/en/reference/arithgroup/arithgroup.pdf
**Асимптотичні розширення**
http://doc.sagemath.org/pdf/en/reference/asymptotic/asymptotic.pdf
**Бази даних**
http://doc.sagemath.org/pdf/en/reference/databases/databases.pdf
**Блокнотний сервер**
http://doc.sagemath.org/pdf/en/reference/notebook/notebook.pdf
**Геометрія**
http://doc.sagemath.org/pdf/en/reference/geometry/geometry.pdf
**Гомологія**
http://doc.sagemath.org/pdf/en/reference/homology/homology.pdf
**Графи**
http://doc.sagemath.org/pdf/en/reference/graphs/graphs.pdf
**Групи**
http://doc.sagemath.org/pdf/en/reference/groups/groups.pdf
**Дискретна динаміка**
http://doc.sagemath.org/pdf/en/reference/dynamics/dynamics.pdf
**Діофантові наближення**
http://doc.sagemath.org/pdf/en/reference/diophantine_approximation/diophantine_approximation.pdf
**Довідка Sage**
http://doc.sagemath.org/pdf/en/reference/reference.pdf
**Документація Sage**
http://doc.sagemath.org/pdf/en/website/sage_documentation.pdf
**Загальні відомості: посібник**
http://doc.sagemath.org/pdf/en/prep/prep_tutorials.pdf
**Здійсненність бульових формул**

http://doc.sagemath.org/pdf/en/reference/sat/sat.pdf
**Ігри**
http://doc.sagemath.org/pdf/en/reference/games/games.pdf
**Інтерфейс із C/C++**
http://doc.sagemath.org/pdf/en/reference/libs/libs.pdf
**Інтерфейси**
http://doc.sagemath.org/pdf/en/reference/interfaces/interfaces.pdf
**Історія та ліцензія**
http://doc.sagemath.org/pdf/en/reference/history_and_license/history_and_license.pdf
**Категорії класів**
http://doc.sagemath.org/pdf/en/reference/categories/categories.pdf
**Квадратичні форми**
http://doc.sagemath.org/pdf/en/reference/quadratic_forms/quadratic_forms.pdf
**Кватерніони**
http://doc.sagemath.org/pdf/en/reference/quat_algebras/quat_algebras.pdf
**Кільця**
http://doc.sagemath.org/pdf/en/reference/rings/rings.pdf
**Командний рядок Sage**
http://doc.sagemath.org/pdf/en/reference/repl/repl.pdf
**Комбінаторика**
http://doc.sagemath.org/pdf/en/reference/combinat/combinat.pdf
**Константи**
http://doc.sagemath.org/pdf/en/reference/constants/constants.pdf
**Криптографія**
http://doc.sagemath.org/pdf/en/reference/cryptography/cryptography.pdf
**Логіка**
http://doc.sagemath.org/pdf/en/reference/logic/logic.pdf
**Математична статистика**
http://doc.sagemath.org/pdf/en/reference/stats/stats.pdf
**Матриці**
http://doc.sagemath.org/pdf/en/reference/matrices/matrices.pdf
**Матроїди**
http://doc.sagemath.org/pdf/en/reference/matroids/matroids.pdf
**Многовиди**
http://doc.sagemath.org/pdf/en/reference/manifolds/manifolds.pdf
**Модулі**
http://doc.sagemath.org/pdf/en/reference/modules/modules.pdf
**Модульні абелеві групи**
http://doc.sagemath.org/pdf/en/reference/modabvar/modabvar.pdf
**Модульні символи**

http://doc.sagemath.org/pdf/en/reference/modsym/modsym.pdf
**Модульні форми**
http://doc.sagemath.org/pdf/en/reference/modfrm/modfrm.pdf
**Моноїди**
http://doc.sagemath.org/pdf/en/reference/monoids/monoids.pdf
**Мультиграфи**
http://doc.sagemath.org/pdf/en/reference/quivers/quivers.pdf
**Найрозповсюдженіші питання**
http://doc.sagemath.org/pdf/en/faq/faq.pdf
**Напівкільця**
http://doc.sagemath.org/pdf/en/reference/semirings/semirings.pdf
**Огляд можливостей Sage**
http://doc.sagemath.org/pdf/en/a_tour_of_sage/a_tour_of_sage.pdf
**Паралельні обчислення**
http://doc.sagemath.org/pdf/en/reference/parallel/parallel.pdf
**Перетворення типів**
http://doc.sagemath.org/pdf/en/reference/coercion/coercion.pdf
**Побудова графіків функцій**
http://doc.sagemath.org/pdf/en/reference/plotting/plotting.pdf
**Плоскі криві**
http://doc.sagemath.org/pdf/en/reference/plane_curves/plane_curves.pdf
**Побудови у просторі**
http://doc.sagemath.org/pdf/en/reference/plot3d/plot3d.pdf
**Поліноміальні кільця**
http://doc.sagemath.org/pdf/en/reference/polynomial_rings/polynomial_rings.pdf
**Посібник початківця**
http://doc.sagemath.org/pdf/en/tutorial/SageTutorial.pdf
**Різні модульні форми**
http://doc.sagemath.org/pdf/en/reference/modmisc/modmisc.pdf
**Ріманова геометрія**
http://doc.sagemath.org/pdf/en/reference/riemannian_geometry/riemannian_geometry.pdf
**Символьні обчислення**
http://doc.sagemath.org/pdf/en/reference/calculus/calculus.pdf
**Скінчені кільця**
http://doc.sagemath.org/pdf/en/reference/finite_rings/finite_rings.pdf
**Стандартні кільця**
http://doc.sagemath.org/pdf/en/reference/rings_standard/rings_standard.pdf
**Створення та використання об’єктів Sage**
http://doc.sagemath.org/pdf/en/constructions/constructions.pdf

**Степеневі ряди**
http://doc.sagemath.org/pdf/en/reference/power_series/power_series.pdf
**Структури**
http://doc.sagemath.org/pdf/en/reference/structure/structure.pdf
**Структури даних**
http://doc.sagemath.org/pdf/en/reference/data_structures/data_structures.pdf
**Схеми**
http://doc.sagemath.org/pdf/en/reference/schemes/schemes.pdf
**Тематичні посібники**
http://doc.sagemath.org/pdf/en/thematic_tutorials/thematic_tutorials.pdf
**Тензори**
http://doc.sagemath.org/pdf/en/reference/tensor_free_modules/tensor_free_modules.pdf
http://doc.sagemath.org/pdf/en/reference/tensor/tensor.pdf
**Теорія вузлів**
http://doc.sagemath.org/pdf/en/reference/knots/knots.pdf
**Теорія ігор**
http://doc.sagemath.org/pdf/en/reference/game_theory/game_theory.pdf
**Теорія ймовірностей**
http://doc.sagemath.org/pdf/en/reference/probability/probability.pdf
**Теорія кодування**
http://doc.sagemath.org/pdf/en/reference/coding/coding.pdf
**Установка**
http://doc.sagemath.org/pdf/en/installation/installation.pdf
**Утиліти**
http://doc.sagemath.org/pdf/en/reference/misc/misc.pdf
**Фінансові дані**
http://doc.sagemath.org/pdf/en/reference/finance/finance.pdf
**Форми Гекке для трикутних груп**
http://doc.sagemath.org/pdf/en/reference/modfrm_hecketriangle/modfrm_hecketriangle.pdf
**Функції**
http://doc.sagemath.org/pdf/en/reference/functions/functions.pdf
**Чисельна оптимізація**
http://doc.sagemath.org/pdf/en/reference/numerical/numerical.pdf
**Числові кільця**
http://doc.sagemath.org/pdf/en/reference/rings_numerical/rings_numerical.pdf
**Числові поля**
http://doc.sagemath.org/pdf/en/reference/number_fields/number_fields.pdf

# ДОДАТОК Е
# ОСНОВНІ КОМАНДИ SAGE

| Команда | Призначення |
|---|---|
| *Базові команди* | |
| <Shift> + <Enter> | розпочати виконання командного коду у комірці |
| com<tab> | вивести список команд, що починаються з послідовності символів com, із можливістю подальшого автодоповнення введення команди |
| command? | надати контекстну довідку з команди command |
| command?<tab> | |
| command?? | вихідний текст команди command |
| _ | звернутися до результату виконання попередньої команди |
| %package_name | увімкнути режим роботи package_name (наприклад: %sage, %html, %maxima, %latex тощо) |
| show() | відобразити результат виконання команди |
| *Основні математичні сталі та функції* | |
| π = pi, e = e, i = I (або I), ∞ = oo | |
| sin(), cos(), tan(), cot(), asin(), acos(), atan(), acot(), sec(), csc(), sinh(), cosh(), tanh(), sech(), csch(), coth(), log(), ln(), exp(), abs(), sqrt(), factorial(), floor(), ceil() | |
| var('x1 x2 … xn') | оголосити x1,…,xn як змінні величини |
| var('x1,…,xn') | |
| x1,…,xn=<br>var('x1,…,xn') | |
| %var x1,…,xn | |
| f=x^2-25 | визначити вираз |
| f(x)=x^2-25 | визначити явно задану функцію |
| f(x,y)=x^2+sqrt(y) | |
| f(x)=x^2==25 | визначити неявно задану функцію |
| f=lambda x: x^2-25 | визначити лямбда-функцію |
| def f(x):<br>return x^2-25 | визначити функцію користувача |
| def nothing(a,b):<br>  skip | |
| *Команди для роботи з виразами* | |

| Команда | Призначення |
|---|---|
| factor(exp) | розкласти на множники (примітка: від цілочисельного аргументу – факторизація цілого числа) |
| expand(exp) | розкрити дужки |
| simplify(exp) | звести подібні без розкриття дужок (для алгебраїчних виразів) |
| RR(exp) | подати значення у полі дійсних чисел |
| n(digits=кількість_значущих_цифр) | округлити до вказаної кількості значимих цифр |
| simplify_full() | спростити вираз |
| simplify_trig() | спростити тригонометричний символьний вираз |
| *Команди для роботи з графікою* | |
| point(((x1,y1),…,(xn, yn)),options) point2d([(x1,y1),…,(xn,yn)],options) | побудувати множину точок на площині |
| arrow2d((x1,y1),(x2, y2),options) | побудувати стрілку |
| line2d([(x1,y1),…,(xn,yn)],options) | побудувати лінію на площині по точках, заданих у вигляді списку |
| polygon2d([(x1,y1),…, (xn,yn)],options) | побудувати зафарбований многокутник |
| circle((x,y),r, options) | побудувати коло |
| disk((x,y),r,(angle1, angle2),options) | побудувати круговий сектор |
| plot(f(x),xmin,xmax, options) plot(f(x),(x,xmin, xmax),options) parametric_plot((f(t),g(t)),(t,tmin,tmax), options) polar_plot(f(t),(t, tmin,tmax),options) | побудувати графік функціональної залежності, заданої аналітично, параметрично та у полярних координатах |
| bar_chart([дані]) | побудувати стовпчикову діаграму |
| contour_plot(f,(x, xmin,xmax),(y,ymin, ymax), options) | побудувати контурні лінії для функції двох змінних |
| plot_vector_field(( | побудувати векторне поле для двох функцій |

| Команда | Призначення |
|---|---|
| `f, g),(x,xmin,xmax),(y, ymin,ymax))` | двох змінних |
| `circle((1,1),1)+ line2d([(0,0),(2,2)])` | задати комбінацію графіків |
| `animate([об'єкт1, …, об'єктn], options).show(delay=20)` | із заданим інтервалом циклічно показати перераховані об'єкти |
| `point3d(((x1,y1,z1),…,(xn,yn,zn)),options)` | побудувати множину точок у просторі |
| `line3d([(x1,y1,z1), …, (xn,yn,zn)], options)` | побудувати ламану у просторі |
| `sphere((x,y,z),r, options)` | побудувати сферу |
| `tetrahedron((x,y,z), size,options)` | побудувати тетраедр |
| `cube((x,y,z),size, options)` | побудувати куб |
| `octahedron((x,y,z), size,options)` | побудувати октаедр |
| `dodecahedron((x,y,z), size,options)` | побудувати додекаедр |
| `icosahedron((x,y,z), size,options)` | побудувати ікосаедр |
| `plot3d(f(x,y), (x,xb, xe), (y,yb,ye), options)` `parametric_plot3d((f(t),g(t),h(t)),(t, tb, te),options)` `list_plot3d([(x1,y1),…,(xn,yn)],options)` | побудувати поверхню, задану функцією двох змінних, параметрично та переліком точок |
| *Команди для роботи з рівняннями та їх системами* | |
| `f(x)==g(x)` | задати рівняння |
| `solve(f(x)==g(x),x)` | розв'язати рівняння аналітично |
| `solve([f(x,y)==0, g(x,y)==0],x,y)` | розв'язати систему рівнянь аналітично |

| Команда | Призначення |
|---|---|
| `find_root(f(x),a,b)` або `find_root(f(x)==0,a,b)` | знайти наближені значення коренів рівняння |
| *Команди для роботи з векторами та матрицями* | |
| `v=vector([1,2])` | задати вектор |
| `a.cross_product(b)` | обчислити векторний добуток векторів *a* і *b* |
| `A=matrix([[1,2,3],[4,5,6]])` | задати матрицю |
| `identity_matrix(n)` | задати одиничну матрицю *n-го* порядку |
| `zero_matrix(n,m)` | задати нульову матрицю розмірності *n*x*m* |
| `diagonal_matrix([a11, a22,…,ann])` | задати діагональну матрицю *n-го* порядку з діагональними елементами $a_{11}$, $a_{22}$,…, $a_{nn}$ |
| `random_matrix(Ring, n, m)` | задати матрицю випадкових чисел з кільця *Ring* (ZZ, QQ, RR, CC) |
| `det(A), A.det()` | обчислити визначник (детермінант) матриці А |
| `A^-1, A.inverse()` | знайти матрицю, обернену до матриці А |
| `A.transpose()` | транспонувати матрицю |
| *Початки аналізу* | |
| `limit(f(x),x=a)`<br>`limit(f(x),x=a, dir='minus')`<br>`limit(f(x),x=a, dir='plus')` | обчислити границю<br>обчислити ліву границю<br>обчислити праву границю |
| `diff(f(x),x)`<br>`derivative(f(x,y),x)` | обчислити похідну |
| `integral(f(x),x)`<br>`integrate(f(x),x,a,b)` | обчислити невизначений інтеграл |
| `sum(f(x), x, 1, oo)` | обчислити суму елементів ряду |
| `prod(f(x), x, 1, oo)` | обчислити добуток елементів ряду |
| `taylor(f(x),x,x0,n)` | виконати розкладання у ряд Тейлора |
| *Диференційні рівняння* | |
| `desolve(de,vars)` | знайти загальний розв’язок звичайного диференційного рівняння *de* відносно змінних *vars* |
| `desolve_laplace(de, vars,ics)` | знайти розв’язок задачі Коші для звичайного диференційного рівняння *de* за початкових |

| Команда | Призначення |
|---|---|
| | умов *ics*, використовуючи перетворення Лапласа |
| `desolve_system(de, vars,ics)` | знайти розв'язок задачі Коші для системи звичайних диференційних рівнянь |
| *Елементи комбінаторики* | |
| `tuples(set,k)` | визначити множину кортежів |
| `number_of_unordered_tuples (S, k)` | обчислити кількість невпорядкованих кортежів |
| `number_of_tuples(S, k)` | обчислити кількість кортежів |
| `fibonacci(n)` | визначити *n*-ий елемент послідовності чисел Фібоначі |
| `euler_number(n)` | визначити *n*-ий елемент послідовності чисел Ейлера |

# ДОДАТОК Ж
# СТАНДАРТНІ ЕЛЕМЕНТИ УПРАВЛІННЯ У COCALC

## Кнопка

**Функція:** `button(default=None, label=None, classes=None, width=None, icon=None),`

`default` – значення, що повертається функцією за замовчуванням;

`label` – підпис, що з'являється ліворуч від елемента;

`classes` – вказує тип кнопки (`btn-primary`, `btn-info`, `btn-success`, `btn-warning`, `btn-danger`, `btn-link`, `btn-large`, `btn-small`, `btn-mini`);

`width` – довжина елементу управління;

`icon` – піктограма на кнопці.

Приклади:

Можна не вказувати параметри `default` та `label`:

```
@interact
def f(n=button('Повторити', icon='fa-repeat'),
m=button('Видно?', icon="fa-eye", classes="btn-
large")):
  print interact.changed()
```

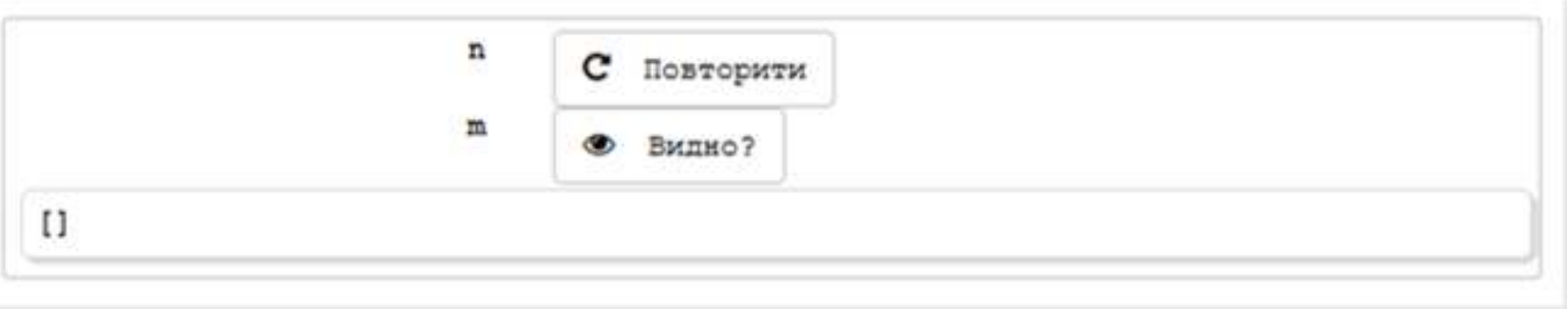

Змінимо зовнішній вигляд однієї з кнопок:

```
@interact
def f(hi=button('Привіт', label='', classes="btn-
primary btn-large"), by=button("Усього
найкращого!")):
  if 'hi' in interact.changed():
    print "Вітаю!"
  if 'by' in interact.changed():
    print "До побачення!"
```

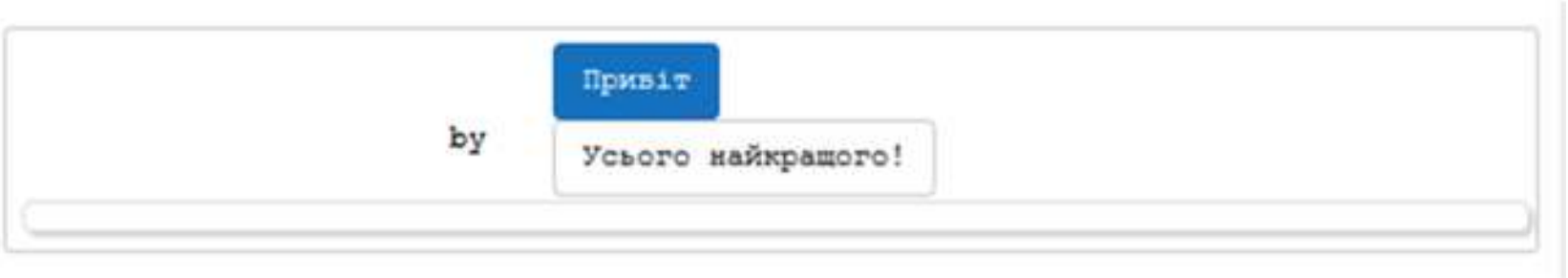

## Поле для введення

**Функція:** `input_box(default=None, label=None, type=None, nrows=1, width=None, readonly=False,`

submit_button=None),
default – значення, що повертається функцією за замовчуванням;
label – підпис, що з'являється ліворуч від елемента;
type – тип даних;
nrows – кількість рядків;
width – довжина елемента управління;
readonly – логічне значення (за замовчуванням – False), що вказує, чи можна буде змінити значення за замовчуванням в елементі управління;
submit_button – показувати кнопку, що вносить вказане значення в поле.
Приклади:

```
@interact
def primer(a=input_box(default="2+2",
           label='Введіть значення', width=10)):
show(a)
```

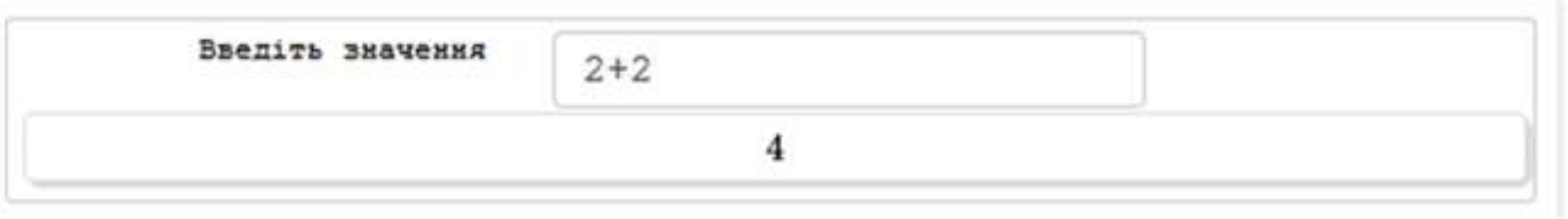

Можна не вказувати параметри default та label:

```
@interact
def primer(a=input_box("2+2", 'Вираз')):
  show(a)
```

Якщо поле для введення має містити строку, тоді матимемо наступний програмний код:

```
@interact
def primer(a=input_box('sage', label="Введіть ім'я:")):
  show(a)
```

Шкала з одним повзунком

Функція: slider(start, stop=None, step=None, default=None, label=None, display_value=True, max_steps=500, step_size=None, range=False, width=None, animate=True),
start – мінімальне значення повзунка;
stop – максимальне значення повзунка;
step – крок значення повзунка;
default – значення, що повертається функцією за замовчуванням;
label – підпис, що з'являється ліворуч від елемента;
display_value – чи слід відображати в праворуч від елемента управління поточне значення;

`max_steps` – максимальний крок значення повзунка;
`step_size` – крок зміни числових значень (може бути як цілим так і дійсним числом);
`range` – якщо значення True, використовується подвійний повзунок;
`width` – довжина елемента управління;
`animate` – швидкість зміни значень повзунка (`"fast", "slow"` або значення в мілісекундах).
Приклади:

```
@interact
def primer(a=slider(vmin=0, vmax=6, step_size=1, default=3, label="Число: ")):
  show(a)
```

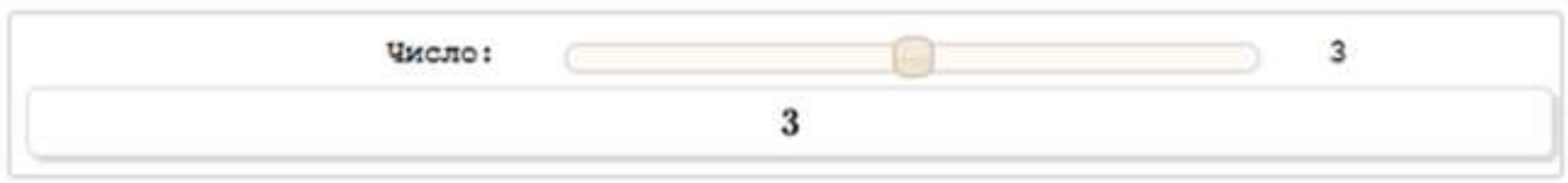

Скорочено даний елемент управління в програмному коді матиме наступний вигляд:

```
@interact
def primer(a=slider(2, 5, 3/17, 3, 'Число:')):
  show(a)
```

Значеннями шкали з одним повзунком можуть бути елементи списку:

```
@interact
def primer(a=slider([1..10], None, None, 3, 'Число:')):
  show(a)
```

Елементами цього списку можуть бути різноманітні об’єкти:

```
@interact
def primer(a=slider([1, 'x', 'abc', 2/3], None, None, 'x', 'alpha')):
  show(a)
```

### Шкала з двома повзунками

Функція: `range_slider(start, stop=None, step=None, default=None, label=None, display_value=True, max_steps=500, step_size=None, width=None, animate=True),`
`start` – мінімальне значення повзунка;
`stop` – максимальне значення повзунка;
`step` – крок значення повзунка;
`default` – значення, що повертається функцією за замовчуванням;
`label` – підпис, що з’являється ліворуч від елемента;
`display_value` – чи слід відображати в праворуч від елемента управління теперішнє значення;

max_steps – максимальний крок значення повзунка;
step_size – крок зміни числових значень (може бути як цілим так і дійсним числом);
width – довжина елемента управління;
animate – швидкість зміни значень повзунка ("fast", "slow", або значення в мілісекундах).
Приклади:

```
@interact
def      primer(a=range_slider(start=0,       stop=6,
step_size=1,
           default=(4,5), label="Число: ")):
  show(a)
```

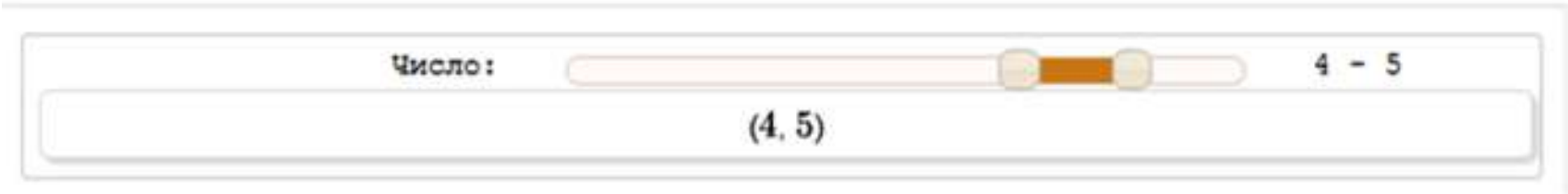

Скорочений запис можна представити в наступному вигляді:

```
@interact
def primer(a=range_slider(0, 6, 1, (4,5), "Число: ")):
  show(a)
```

Крім того,шкалу з двома повзунками можна задати за допомогою списку:

```
@interact
def primer(a=range_slider([0..6], None, None, (4,5),
"Число: ")):
  show(a)
```

Прапорець

Функція: checkbox(default=True, label=None, readonly=False),
default – значення, що повертається функцією за замовчуванням (даний елемент управління може приймати в даному параметрі лише значення False чи True);
label – підпис, що з’являється ліворуч від елемента;
readonly – логічне значення (за замовчуванням – False), що вказує, чи можна буде змінити значення за замовчуванням;
Приклади:

```
@interact
def primer(a=checkbox(default=False, label="Точки")):
  show(a)
```

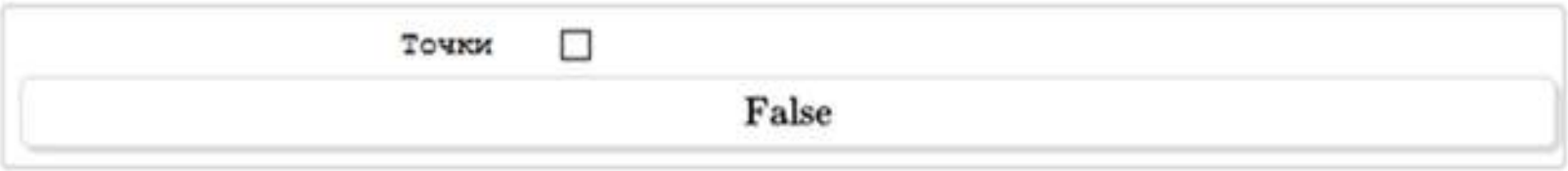

Скорочений варіант застосування даного елементу управління виглядає наступним чином:

```
@interact
def primer(a=checkbox(False, "Точки")):
  show(a)
```

Коли параметр `label` не вказано, за замовчуванням підпис «прапорця» буде збігатись зі змінною, яка визначається функцією `checkbox()`:

```
@interact
def primer(a=checkbox(True)):
  show(a)
```

Якщо ж функція не має параметру `default`, то за замовчуванню змінній надаватись значення `True`:

```
@interact
def primer(a=checkbox()):
  show(a)
```

### Меню вибору

Функція: `selector(values, label=None, default=None, nrows=None, ncols=None, width=None, buttons=False, button_classes=None)`,

`values` – значення, яких може набувати змінна;

`label` – підпис, що з’являється ліворуч від елемента;

`default` – значення, що повертається функцією за замовчуванням;

`nrows` – число рядків, у яких розташовуються кнопки;

`ncols` – число стовпців, у які розміщуються кнопки;

`width` – ширина усіх кнопок;

`buttons` – визначає вигляд меню: кнопки (`True`) чи випадаючий список (`False`);

`button_classes` – вказує тип кнопки (`btn-primary, btn-info, btn-success, btn-warning, btn-danger, btn-link, btn-large, btn-small, btn-mini`).

Приклади:

```
@interact
def primer(a=selector([1,5,8,14], label="Меню", nrows=2, ncols=2, buttons=True)):
  show(a)
```

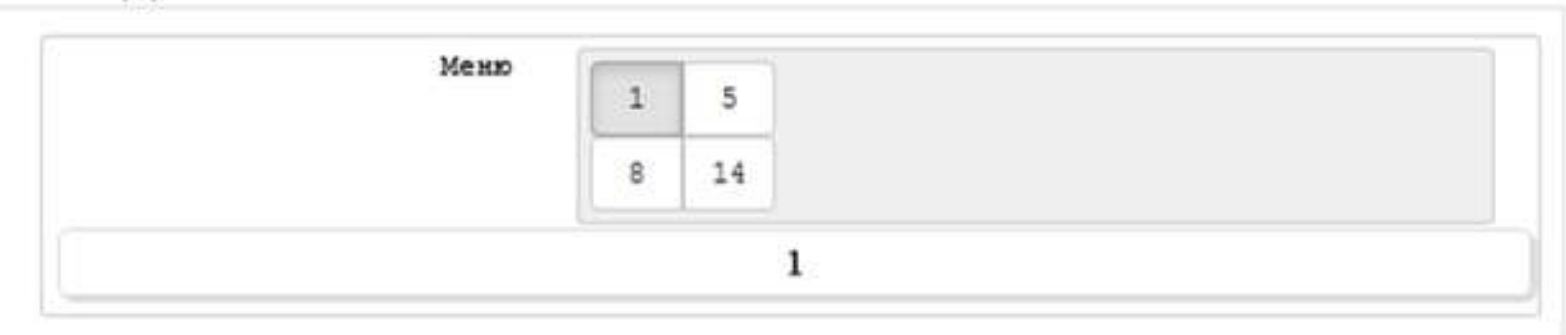

Скорочене визначення елементу управління у вигляді випадаючого

списку:

```
@interact
def primer(a=selector([1..5])):
  show(a)
```

Використання параметру `default`:

```
@interact
def primer(a=selector([1,2,7], default=2)):
  show(a)
```

Розташування трьох кнопок у два ряди:

```
@interact
def primer(a=selector([1,2,7], nrows=2)):
  show(a)
```

Зміна стандартного розміру кнопок на бажаний:

```
@interact
def primer(a=selector([1,2,7], width=10,
buttons=True)):
  show(a)
```

## Комірки для введення

**Функція:** `input_grid(nrows, ncols, default=None, label=None, to_value=lambda x:x, width=4),`

`nrows` – параметр для задання кількості рядків;

`ncols` – параметр для задання кількості стовпців;

`default` – задання початкових значень у комірках;

`label` – підпис, що з’являється ліворуч від елемента;

`to_value` – формування та виведення заданих даних у вигляді комірок;

`width` – загальна ширина однієї комірки.

Приклади:

```
@interact
def obernena(a = input_grid(nrows=3, ncols=3, default=
[1,2,3,4,5,6,7,8,9], label='Матриця A',
to_value=lambda x:x, width=2)):
  show(a)
```

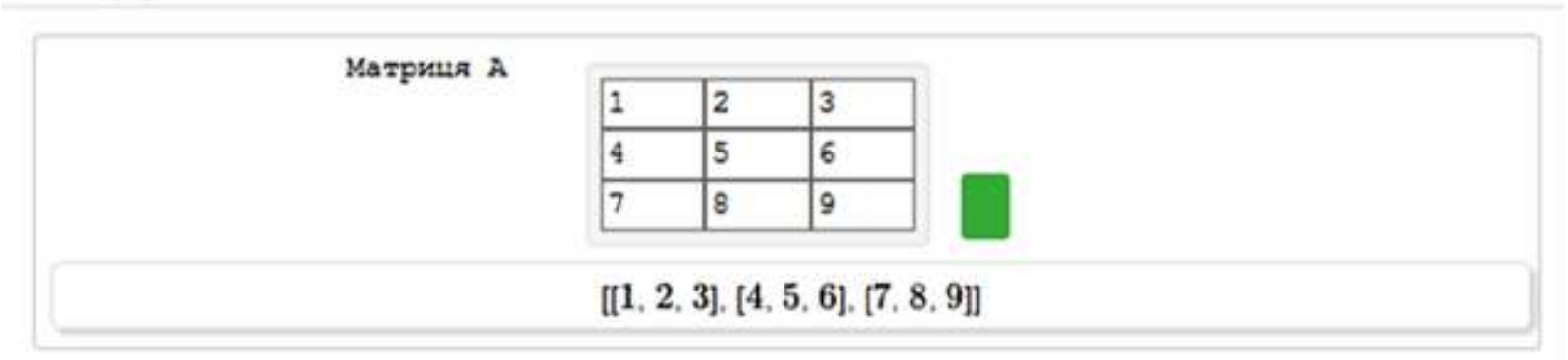

Скорочений запис передбачає наступне:

```
@interact
def primer(a = input_grid(2,2, default = 0,
label='M')):
```

```
  show(a)
```

Комірки для введення значення вектора можна оголосити наступним чином:

```
@interact
def primer(a = input_grid(1, 3, default=[[1,2,3]],
           to_value=lambda x: vector(flatten(x)))):
  show(a)
```

## Поле вибору кольору

Функція: `color_selector(default=(0,0,1), label=None, readonly=False, widget='farbtastic', hide_box=False),`

`default` – значення кольору за замовчуванняму форматі RGB, наприклад: `(0,0,1)`, `‘#abcdef’`;

`label` – підпис, що з’являється ліворуч від елемента;

`readonly` – логічне значення (за замовчуванням – `False`), що вказує, чи можна буде змінити значення кольору за замовчуванням;

`widget` – основний параметр для задання вигляду діалогового вікна, за замовчуванням присвоюється значення `jpicker`, але може набувати значень `colorpicker` та `farbtastic`;

`hide_box` – параметр для відображення вікна вводу кольору.

Приклади:

```
@interact
def primer(a = color_selector(default=(1,0,0), label="Колір",
           widget='farbtastic', hide_box=False)):
  show(a)
```

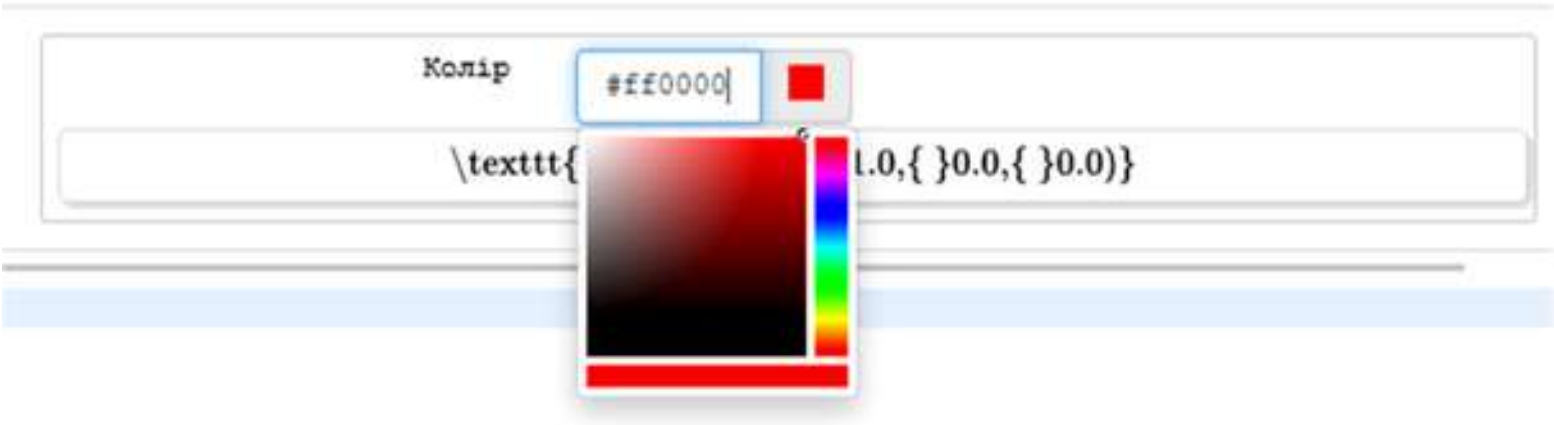

```
@interact
def primer(a = color_selector(default=(1,0,0), label="Колір",
           widget='jpicker', hide_box=True)):
  show(a)
```

## Статичне текстове поле

*Функція:* `text_control (default='', label=None, classes=None),`

`default` – текст у форматі HTML;

`label` – підпис, що з’являється ліворуч від елемента;

`classes` – розділений пробілами рядок класів CSS.

Приклади:

```
@interact
def f(n=text_control("Текст <b>спеціальний</b>",
      classes='btn')):
  pass
```

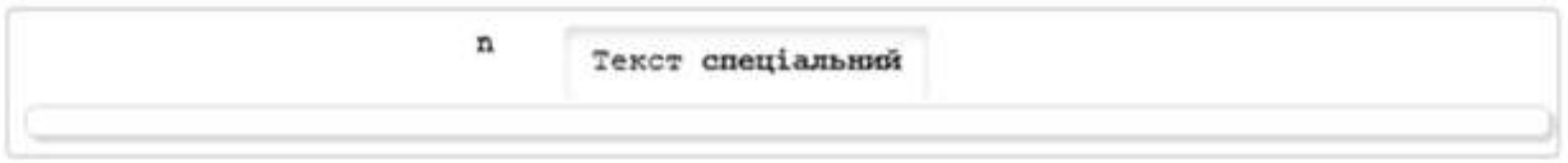

Як що ж використовувати скорочений варіант, то одержимо наступні результати:

```
@interact
def primer(a = text_control(
           "<h1><i>Скорочений варіант</i></h1>")):
  show(a)
```

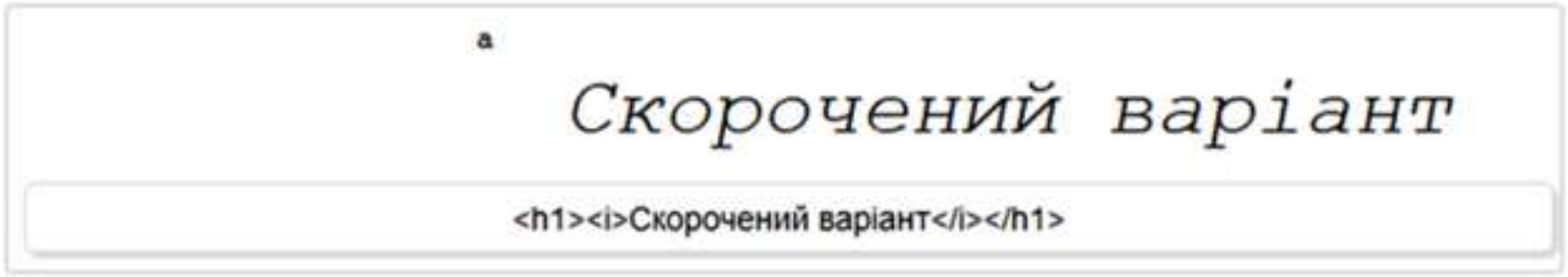

**М. В. Попель**

**Хмарний сервіс CoCalc як засіб формування професійних компетентностей учителя математики**

E-mail: semerikov@ccjournals.eu